\documentstyle[10pt,epsf,epsfig,hangcaption,xspace,amssymb,amsfonts,amsmath,amsthm,cite,
dp_delphititle,lineno]{dp_delphi}
\makeindex
\def\DpPaperGroup{EP}
\def\DpPaperRef{2003-007}
\def\DpDate{21 July 2003}
\def\DpAuthors{DELPHI Collaboration}
\def\DpTitle{{Searches for supersymmetric particles
in e$^+$e$^-$ collisions up to 208~GeV and
interpretation of the results within the MSSM}}
\def\DpSubmit{Accepted by Eur. Phys. J. C}
\def\DpComment{ }

\def\leqsim{\mathbin{\;\raise1pt\hbox{$<$}\kern-8pt\lower3pt\hbox{$\sim$}\;}}
\def\geqsim{\mathbin{\;\raise1pt\hbox{$>$}\kern-8pt\lower3pt\hbox{$\sim$}\;}}

\def\MXN#1{\mbox{$ M_{\tilde{\chi}^0_#1}                                $}}
\def\MXNN#1#2{\mbox{$ M_{\tilde{\chi}^0_{#1,#2}}                        $}}
\def\MXC#1{\mbox{$ M_{\tilde{\chi}^{\pm}_#1}                            $}}
\def\XP#1{\mbox{$ \tilde{\chi}^+_#1                                     $}}
\def\XM#1{\mbox{$ \tilde{\chi}^-_#1                                     $}}
\def\XPM#1{\mbox{$ \tilde{\chi}^{\pm}_#1                                $}}
\def\XN#1{\mbox{$ \tilde{\chi}^0_#1                                     $}}
\def\XNN#1#2{\mbox{$ \tilde{\chi}^0_{#1,#2}                             $}}
\def\MXn{\mbox{$ M_{\tilde{\chi}^0}                                     $}}
\def\MXc{\mbox{$ M_{\tilde{\chi}^{\pm}}                                 $}}

\def\Xpm{\mbox{$ \tilde{\chi}^{\pm}                                     $}}
\def\Xn{\mbox{$ \tilde{\chi}^0                                          $}}

\def\p#1{\mbox{$ \mbox{\bf p}_1                                         $}}
\newcommand{\Ptmis}   {\mbox{$/\mkern-11mu P_t \,                          $}}

\newcommand{\tanb}    {\mbox{$ \tan \beta                                  $}}
\newcommand{\smu}     {\mbox{$ \tilde{\mu}                                 $}}
\newcommand{\smur}    {\mbox{$ \tilde{\mu}_{\mathrm R}                     $}}
\newcommand{\msmu}    {\mbox{$ M_{\tilde{\mu}}                             $}}
\newcommand{\msmur}   {\mbox{$ M_{\tilde{\mu}_{\mathrm R}}                 $}}

\newcommand{\sell}    {\mbox{$ \tilde{\mathrm e}_{\mathrm L}               $}}
\newcommand{\selr}    {\mbox{$ \tilde{\mathrm e}_{\mathrm R}               $}}
\newcommand{\sellb}   {\mbox{$ \overline{\tilde{\mathrm e}_{\mathrm L}}    $}}
\newcommand{\selrb}   {\mbox{$ \overline{\tilde{\mathrm e}_{\mathrm R}}    $}}
\newcommand{\msel}    {\mbox{$ M_{\tilde{\mathrm e}}                       $}}
\newcommand{\snu}     {\mbox{$ \tilde\nu                                   $}}
\newcommand{\msnu}    {\mbox{$ m_{\tilde\nu}                               $}}

\newcommand{\mselr}   {\mbox{$ M_{\tilde{\mathrm e}_{\mathrm R}}           $}}
\newcommand{\sfe}     {\mbox{$ \tilde{\mathrm f}                           $}}
\newcommand{\sfel}    {\mbox{$ \tilde{\mathrm f}_{\mathrm L}               $}}
\newcommand{\sfer}    {\mbox{$ \tilde{\mathrm f}_{\mathrm R}               $}}
\newcommand{\sfelb}   {\mbox{$ \overline{\tilde{\mathrm f}_{\mathrm L}}    $}}
\newcommand{\sferb}   {\mbox{$ \overline{\tilde{\mathrm f}_{\mathrm R}}    $}}

\newcommand{\stau}    {\mbox{$ \tilde{\tau}                                $}}
\newcommand{\stone}   {\mbox{$ \tilde{\tau}_1                              $}}

\newcommand{\staur}   {\mbox{$ \tilde{\tau}_{\mathrm R}                    $}}
\newcommand{\mstau}   {\mbox{$ M_{\tilde{\tau}}                            $}}
\newcommand{\mstaur}  {\mbox{$ M_{\tilde{\tau}_{\mathrm R}}                $}}
\newcommand{\mstone}  {\mbox{$ M_{\tilde{\tau}_1}                          $}}

\newcommand{\stq}     {\mbox{$ \tilde {\mathrm t}                          $}}
\newcommand{\stqone}  {\mbox{$ \tilde {\mathrm t}_1                        $}}

\newcommand{\mstq}    {\mbox{$ M_{\tilde {\mathrm t}}                      $}}
\newcommand{\sbq}     {\mbox{$ \tilde {\mathrm b}                          $}}
\newcommand{\sbqone}  {\mbox{$ \tilde {\mathrm b}_1                        $}}

\newcommand{\msbqone}    {\mbox{$ M_{\tilde {\mathrm b}_1}                 $}}

\newcommand{\mstqone}    {\mbox{$ M_{\tilde {\mathrm t}_1}                 $}}

\newcommand{\hn}      {\mbox{$ {\, \mathrm h}^0                               $}}
\newcommand{\Zn}      {\mbox{$ {\, \mathrm Z}                                 $}}
\newcommand{\Zstar}   {\mbox{$ {\, \mathrm Z}^*                               $}}

\newcommand{\WW}      {\mbox{$ {\, \mathrm W}^+{\mathrm W}^-                  $}}
\newcommand{\ZZ}      {\mbox{$ {\, \mathrm{Z Z}}$}}

\newcommand{\Zg}      {\mbox{$ \Zn \gamma                                  $}}
\newcommand{\sqs}     {\mbox{$ \sqrt{s}                                    $}}

\newcommand{\ee}      {\mbox{$ {\, \mathrm e}^+ {\mathrm e}^-                 $}}
\newcommand{\mumu}    {\mbox{$ \, \mu^+ \mu^-                                 $}}
\newcommand{\tautau}  {\mbox{$ \, \tau^+ \tau^-                               $}}
\newcommand{\eeto}    {\mbox{$ {\, \mathrm e}^+ {\mathrm e}^- \to             $}}
\newcommand{\ellell}  {\mbox{$ \, \ell^+ \ell^-                               $}}

\newcommand{\MeVcc}   {\mbox{$ {\mathrm{MeV}}/c^2                          $}}
\newcommand{\GeV}     {\mbox{$ {\mathrm{GeV}}                              $}}
\newcommand{\GeVc}    {\mbox{$ {\mathrm{GeV}}/c                            $}}
\newcommand{\GeVcc}   {\mbox{$ {\mathrm{GeV}}/c^2                          $}}

\newcommand{\TeVc}    {\mbox{$ {\mathrm{TeV}}/c                            $}}
\newcommand{\TeVcc}   {\mbox{$ {\mathrm{TeV}}/c^2                          $}}
\newcommand{\pbi}     {\mbox{$ {\mathrm{pb}}^{-1}                          $}}
\newcommand{\MZ}      {\mbox{$ m_{{\mathrm Z}}                             $}}
\newcommand{\MW}      {\mbox{$ m_{\mathrm W}                               $}}
\newcommand{\MA}      {\mbox{$ m_{\mathrm A}                               $}}

\newcommand{\dgree}   {\mbox{$ ^\circ                                      $}}

\newcommand{\Wev}     {\mbox{$ {\mathrm W} {\mathrm e} \, \nu_e            $}}

\newcommand{\Zee}     {\mbox{$ \Zn \ee                                     $}}

\def    \ll           {\mbox{$\ell \ell                                    $}}
\def    \jjl          {\mbox{$j j \ell                           $}}

\def   \jjjj          {\mbox{${\it jets}                                   $}}

\newcommand{\pT}      {\mbox{$ p_{\mathrm{T}}                              $}}
\newcommand{\pt}      {\mbox{$ p_{\mathrm{t}}                              $}}

\newcommand{\gamgam}  {\mbox{$ \gamma \gamma                               $}}

\newcommand{\ffbar}   {\mbox{$ {\, \mathrm f} \, \bar{\mathrm f}           $}}
\newcommand{\ffbarp}  {\mbox{$ {\, \mathrm f} \, \bar{\mathrm f}'          $}}
\newcommand{\qqbar}   {\mbox{$\mathrm q \, \bar{\mathrm q}                 $}}

\newcommand{\charm}   {\mbox{$\mathrm c                                    $}}
\newcommand{\bottom}  {\mbox{$\mathrm b                                    $}}
\newcommand{\topq}    {\mbox{$\mathrm t                                    $}}

\newcommand{\Etvis}    {\mbox{$ E^{T}_{\mathrm{\small vis}}                      $}}
\newcommand{\Evis}    {\mbox{$ E_{\mathrm{\small vis}}                      $}}

\newcommand{\Mvis}    {\mbox{$ M_{\mathrm{\small vis}}                      $}}

\newcommand{\etal}  {\mbox{\it et al.}}

\def\PLB#1#2#3{{\rm Phys.~Lett.} {\bf{B#1}} (#2) #3}
\def\PRD#1#2#3{{\rm Phys.~Rev.} {\bf{D#1}} (#2) #3}

\def\ZPC#1#2#3{{\rm Z.~Phys.} {\bf C#1} (#2) #3}

\def\PR#1#2#3{{\rm Phys.~Rep.} {\bf#1} (#2) #3}

\def\NIMA#1#2#3{{\rm Nucl.~Instr.~and~Meth.} {\bf#1} (#2) #3} 
\def\CPC#1#2#3{{\rm Comp.~Phys.~Comm.} {\bf#1} (#2) #3}
\def\EPJC#1#2#3{{\rm E.~Phys.~J.} {\bf{C#1}} (#2) #3}
\def    \DM          {\mbox{$\Delta$M}}
\def    \missEt      {\ifmmode{/\mkern-11mu E_t}\else{${/\mkern-11mu E_t}$}\fi}
\def    \missE       {\ifmmode{/\mkern-11mu E}\else{${/\mkern-11mu E}$}\fi}
\def    \missp       {\ifmmode{/\mkern-11mu p}\else{${/\mkern-11mu p}$}\fi}
\def    \misspt      {\ifmmode{/\mkern-11mu p_t}\else{${/\mkern-11mu p_t}$}\fi}

\def    \rs          {\mbox{$\sqrt{s}$}}
\def    \msneu       {\mbox{$m_{\tilde{\nu}}$}}
\def    \rad         {\mbox{$\it{rad}$}}

\begin{document}
\makeatletter
\newcount\@tempcntc
\def\@citex[#1]#2{\if@filesw\immediate\write\@auxout{\string\citation{#2}}\fi
  \@tempcnta\z@\@tempcntb\m@ne\def\@citea{}\@cite{\@for\@citeb:=#2\do
    {\@ifundefined
       {b@\@citeb}{\@citeo\@tempcntb\m@ne\@citea\def\@citea{,}{\bf ?}\@warning
       {Citation `\@citeb' on page \thepage \space undefined}}%
    {\setbox\z@\hbox{\global\@tempcntc0\csname b@\@citeb\endcsname\relax}%
     \ifnum\@tempcntc=\z@ \@citeo\@tempcntb\m@ne
       \@citea\def\@citea{,}\hbox{\csname b@\@citeb\endcsname}%
     \else
      \advance\@tempcntb\@ne
      \ifnum\@tempcntb=\@tempcntc
      \else\advance\@tempcntb\m@ne\@citeo
      \@tempcnta\@tempcntc\@tempcntb\@tempcntc\fi\fi}}\@citeo}{#1}}
\def\@citeo{\ifnum\@tempcnta>\@tempcntb\else\@citea\def\@citea{,}%
  \ifnum\@tempcnta=\@tempcntb\the\@tempcnta\else
   {\advance\@tempcnta\@ne\ifnum\@tempcnta=\@tempcntb \else \def\@citea{--}\fi
    \advance\@tempcnta\m@ne\the\@tempcnta\@citea\the\@tempcntb}\fi\fi}
 
\makeatother
\begin{titlepage}
\pagenumbering{roman}
\CERNpreprint{\DpPaperGroup}{\DpPaperRef} 
\date{{\small\DpDate}} 
\title{\DpTitle} 
\address{\DpAuthors} 
\begin{shortabs} 
\noindent
DELPHI data collected at centre-of-mass energies up to 208 GeV have been
analysed to search for
charginos, neutralinos and sfermions 
in the framework of the Minimal Supersymmetric Standard Model (MSSM) 
with R-parity conservation.
No evidence for a signal was found in any of the channels.
The results of each search were used to derive  limits on production 
cross-sections and   particle masses. 
In addition, the combined result of all searches excludes
regions in the parameter
space of the constrained MSSM, leading
to limits on the mass of the 
Lightest Supersymmetric Particle and other supersymmetric particles.


\end{shortabs}
\vfill
\begin{center}
\DpSubmit \ \\ 
\DpComment \ \\
\end{center}
\vfill
\clearpage
\headsep 10.0pt
\addtolength{\textheight}{10mm}
\addtolength{\footskip}{-5mm}
\begingroup
%
\newcommand{\DpName}[2]{\hbox{#1$^{\ref{#2}}$},\hfill}
\newcommand{\DpNameTwo}[3]{\hbox{#1$^{\ref{#2},\ref{#3}}$},\hfill}
\newcommand{\DpNameThree}[4]{\hbox{#1$^{\ref{#2},\ref{#3},\ref{#4}}$},\hfill}
\newskip\Bigfill \Bigfill = 0pt plus 1000fill
\newcommand{\DpNameLast}[2]{\hbox{#1$^{\ref{#2}}$}\hspace{\Bigfill}}
%
\footnotesize
\noindent
\DpName{J.Abdallah}{LPNHE}
\DpName{P.Abreu}{LIP}
\DpName{W.Adam}{VIENNA}
\DpName{P.Adzic}{DEMOKRITOS}
\DpName{T.Albrecht}{KARLSRUHE}
\DpName{T.Alderweireld}{AIM}
\DpName{R.Alemany-Fernandez}{CERN}
\DpName{T.Allmendinger}{KARLSRUHE}
\DpName{P.P.Allport}{LIVERPOOL}
\DpName{U.Amaldi}{MILANO2}
% ...
\normalsize
\endgroup
\titlefoot{Department of Physics and Astronomy, Iowa State
     University, Ames IA 50011-3160, USA
    \label{AMES}}
\titlefoot{Physics Department, Universiteit Antwerpen,
     Universiteitsplein 1, B-2610 Antwerpen, Belgium \\
     \indent~~and IIHE, ULB-VUB,
     Pleinlaan 2, B-1050 Brussels, Belgium \\
     \indent~~and Facult\'e des Sciences,
     Univ. de l'Etat Mons, Av. Maistriau 19, B-7000 Mons, Belgium
    \label{AIM}}
\titlefoot{Physics Laboratory, University of Athens, Solonos Str.
     104, GR-10680 Athens, Greece
    \label{ATHENS}}
\titlefoot{Department of Physics, University of Bergen,
     All\'egaten 55, NO-5007 Bergen, Norway
    \label{BERGEN}}
\titlefoot{Dipartimento di Fisica, Universit\`a di Bologna and INFN,
     Via Irnerio 46, IT-40126 Bologna, Italy
    \label{BOLOGNA}}
\titlefoot{Centro Brasileiro de Pesquisas F\'{\i}sicas, rua Xavier Sigaud 150,
     BR-22290 Rio de Janeiro, Brazil \\
     \indent~~and Depto. de F\'{\i}sica, Pont. Univ. Cat\'olica,
     C.P. 38071 BR-22453 Rio de Janeiro, Brazil \\
     \indent~~and Inst. de F\'{\i}sica, Univ. Estadual do Rio de Janeiro,
     rua S\~{a}o Francisco Xavier 524, Rio de Janeiro, Brazil
    \label{BRASIL}}
\titlefoot{Coll\`ege de France, Lab. de Physique Corpusculaire, IN2P3-CNRS,
     FR-75231 Paris Cedex 05, France
    \label{CDF}}
\titlefoot{CERN, CH-1211 Geneva 23, Switzerland
    \label{CERN}}
\titlefoot{Institut de Recherches Subatomiques, IN2P3 - CNRS/ULP - BP20,
     FR-67037 Strasbourg Cedex, France
    \label{CRN}}
\titlefoot{Now at DESY-Zeuthen, Platanenallee 6, D-15735 Zeuthen, Germany
    \label{DESY}}
\titlefoot{Institute of Nuclear Physics, N.C.S.R. Demokritos,
     P.O. Box 60228, GR-15310 Athens, Greece
    \label{DEMOKRITOS}}
\titlefoot{FZU, Inst. of Phys. of the C.A.S. High Energy Physics Division,
     Na Slovance 2, CZ-180 40, Praha 8, Czech Republic
    \label{FZU}}
\titlefoot{Dipartimento di Fisica, Universit\`a di Genova and INFN,
     Via Dodecaneso 33, IT-16146 Genova, Italy
    \label{GENOVA}}
\titlefoot{Institut des Sciences Nucl\'eaires, IN2P3-CNRS, Universit\'e
     de Grenoble 1, FR-38026 Grenoble Cedex, France
    \label{GRENOBLE}}
\titlefoot{Helsinki Institute of Physics, P.O. Box 64,
     FIN-00014 University of Helsinki, Finland
    \label{HELSINKI}}
\titlefoot{Joint Institute for Nuclear Research, Dubna, Head Post
     Office, P.O. Box 79, RU-101 000 Moscow, Russian Federation
    \label{JINR}}
\titlefoot{Institut f\"ur Experimentelle Kernphysik,
     Universit\"at Karlsruhe, Postfach 6980, DE-76128 Karlsruhe,
     Germany
    \label{KARLSRUHE}}
\titlefoot{Institute of Nuclear Physics,Ul. Kawiory 26a,
     PL-30055 Krakow, Poland
    \label{KRAKOW1}}
\titlefoot{Faculty of Physics and Nuclear Techniques, University of Mining
     and Metallurgy, PL-30055 Krakow, Poland
    \label{KRAKOW2}}
\titlefoot{Universit\'e de Paris-Sud, Lab. de l'Acc\'el\'erateur
     Lin\'eaire, IN2P3-CNRS, B\^{a}t. 200, FR-91405 Orsay Cedex, France
    \label{LAL}}
\titlefoot{School of Physics and Chemistry, University of Lancaster,
     Lancaster LA1 4YB, UK
    \label{LANCASTER}}
\titlefoot{LIP, IST, FCUL - Av. Elias Garcia, 14-$1^{o}$,
     PT-1000 Lisboa Codex, Portugal
    \label{LIP}}
\titlefoot{Department of Physics, University of Liverpool, P.O.
     Box 147, Liverpool L69 3BX, UK
    \label{LIVERPOOL}}
\titlefoot{Dept. of Physics and Astronomy, Kelvin Building,
     University of Glasgow, Glasgow G12 8QQ
    \label{GLASGOW}}
\titlefoot{LPNHE, IN2P3-CNRS, Univ.~Paris VI et VII, Tour 33 (RdC),
     4 place Jussieu, FR-75252 Paris Cedex 05, France
    \label{LPNHE}}
\titlefoot{Department of Physics, University of Lund,
     S\"olvegatan 14, SE-223 63 Lund, Sweden
    \label{LUND}}
\titlefoot{Universit\'e Claude Bernard de Lyon, IPNL, IN2P3-CNRS,
     FR-69622 Villeurbanne Cedex, France
    \label{LYON}}
\titlefoot{Dipartimento di Fisica, Universit\`a di Milano and INFN-MILANO,
     Via Celoria 16, IT-20133 Milan, Italy
    \label{MILANO}}
\titlefoot{Dipartimento di Fisica, Univ. di Milano-Bicocca and
     INFN-MILANO, Piazza della Scienza 2, IT-20126 Milan, Italy
    \label{MILANO2}}
\titlefoot{IPNP of MFF, Charles Univ., Areal MFF,
     V Holesovickach 2, CZ-180 00, Praha 8, Czech Republic
    \label{NC}}
\titlefoot{NIKHEF, Postbus 41882, NL-1009 DB
     Amsterdam, The Netherlands
    \label{NIKHEF}}
\titlefoot{National Technical University, Physics Department,
     Zografou Campus, GR-15773 Athens, Greece
    \label{NTU-ATHENS}}
\titlefoot{Physics Department, University of Oslo, Blindern,
     NO-0316 Oslo, Norway
    \label{OSLO}}
\titlefoot{Dpto. Fisica, Univ. Oviedo, Avda. Calvo Sotelo
     s/n, ES-33007 Oviedo, Spain
    \label{OVIEDO}}
\titlefoot{Department of Physics, University of Oxford,
     Keble Road, Oxford OX1 3RH, UK
    \label{OXFORD}}
\titlefoot{Dipartimento di Fisica, Universit\`a di Padova and
     INFN, Via Marzolo 8, IT-35131 Padua, Italy
    \label{PADOVA}}
\titlefoot{Rutherford Appleton Laboratory, Chilton, Didcot
     OX11 OQX, UK
    \label{RAL}}
\titlefoot{Dipartimento di Fisica, Universit\`a di Roma II and
     INFN, Tor Vergata, IT-00173 Rome, Italy
    \label{ROMA2}}
\titlefoot{Dipartimento di Fisica, Universit\`a di Roma III and
     INFN, Via della Vasca Navale 84, IT-00146 Rome, Italy
    \label{ROMA3}}
\titlefoot{DAPNIA/Service de Physique des Particules,
     CEA-Saclay, FR-91191 Gif-sur-Yvette Cedex, France
    \label{SACLAY}}
\titlefoot{Instituto de Fisica de Cantabria (CSIC-UC), Avda.
     los Castros s/n, ES-39006 Santander, Spain
    \label{SANTANDER}}
\titlefoot{Inst. for High Energy Physics, Serpukov
     P.O. Box 35, Protvino, (Moscow Region), Russian Federation
    \label{SERPUKHOV}}
\titlefoot{J. Stefan Institute, Jamova 39, SI-1000 Ljubljana, Slovenia
     and Laboratory for Astroparticle Physics,\\
     \indent~~Nova Gorica Polytechnic, Kostanjeviska 16a, SI-5000 Nova Gorica, Slovenia, \\
     \indent~~and Department of Physics, University of Ljubljana,
     SI-1000 Ljubljana, Slovenia
    \label{SLOVENIJA}}
\titlefoot{Fysikum, Stockholm University,
     Box 6730, SE-113 85 Stockholm, Sweden
    \label{STOCKHOLM}}
\titlefoot{Dipartimento di Fisica Sperimentale, Universit\`a di
     Torino and INFN, Via P. Giuria 1, IT-10125 Turin, Italy
    \label{TORINO}}
\titlefoot{INFN,Sezione di Torino, and Dipartimento di Fisica Teorica,
     Universit\`a di Torino, Via P. Giuria 1,\\
     \indent~~IT-10125 Turin, Italy
    \label{TORINOTH}}
\titlefoot{Dipartimento di Fisica, Universit\`a di Trieste and
     INFN, Via A. Valerio 2, IT-34127 Trieste, Italy \\
     \indent~~and Istituto di Fisica, Universit\`a di Udine,
     IT-33100 Udine, Italy
    \label{TU}}
\titlefoot{Univ. Federal do Rio de Janeiro, C.P. 68528
     Cidade Univ., Ilha do Fund\~ao
     BR-21945-970 Rio de Janeiro, Brazil
    \label{UFRJ}}
\titlefoot{Department of Radiation Sciences, University of
     Uppsala, P.O. Box 535, SE-751 21 Uppsala, Sweden
    \label{UPPSALA}}
\titlefoot{IFIC, Valencia-CSIC, and D.F.A.M.N., U. de Valencia,
     Avda. Dr. Moliner 50, ES-46100 Burjassot (Valencia), Spain
    \label{VALENCIA}}
\titlefoot{Institut f\"ur Hochenergiephysik, \"Osterr. Akad.
     d. Wissensch., Nikolsdorfergasse 18, AT-1050 Vienna, Austria
    \label{VIENNA}}
\titlefoot{Inst. Nuclear Studies and University of Warsaw, Ul.
     Hoza 69, PL-00681 Warsaw, Poland
    \label{WARSZAWA}}
\titlefoot{Fachbereich Physik, University of Wuppertal, Postfach
     100 127, DE-42097 Wuppertal, Germany \\
\noindent
{$^\dagger$~deceased}
    \label{WUPPERTAL}}
\addtolength{\textheight}{-10mm}
\addtolength{\footskip}{5mm}
\clearpage
\headsep 30.0pt
\end{titlepage}
%
\pagenumbering{arabic} 
\setcounter{footnote}{0} %
\large

\section{Introduction}
\label{sec:introduction}

Supersymmetry (SUSY)~\cite{susy}
is at present one of the most attractive possible extensions of the 
Standard Model (SM) and its signatures
could be observed at LEP through a large 
variety of different channels.
This paper presents searches for the pair-production of 
charginos, neutralinos, sleptons and squarks.
The searches were performed and interpreted in the most model-independent way 
possible in terms of production cross-sections and masses. 
The results were interpreted in the framework of constrained SUSY models, with 
the different search channels complementing each other in constraining 
the parameter space.

The data collected by the DELPHI experiment in \ee\  collisions 
at centre-of-mass energies (\sqs) up to 208 GeV were used. 
No signal was found in any of the channels, and limits were set at 95\% confidence level (CL).

The paper is organised as follows. 
In section~\ref{sec:framework} the basic supersymmetry framework is described:
the general phenomenology is discussed in section \ref{sec:phenomenology},
and implications of more constrained models used for interpreting the data are given in
section \ref{sec:constraining}. 
The DELPHI detector is  described in section~\ref{sec:detector},
and in section~\ref{sec:samples} the data sets and event generators are reviewed. 
In section~\ref{sec:analysis} the general analysis framework is described, the analysis
methods are briefly mentioned and the specific searches for 
sleptons, squarks, charginos and neutralinos are discussed. 
The results of each search are separately presented and interpreted in section~\ref{sec:results}.
In section~\ref{sec:combined} the results are combined and interpreted in the framework
of constrained SUSY scenarios with gravity-induced breaking of supersymmetry (SUGRA).
A brief summary is given in section~\ref{sec:summary}.

Previous results published by DELPHI 
can be found in
references~\cite{slep} to~\cite{osakalsp}.


\nopagebreak

\section{SUSY framework} 
\label{sec:framework}

The searches presented in this paper were performed in the framework of the Minimal
Supersymmetric extension of the Standard Model (MSSM)~\cite{susy}.
$R$-parity~\footnote{$R$-parity is a multiplicative quantum number defined as $R=(-1)^{3(B-L)+2S}$
where $B$, $L$ and $S$ are the baryon number, the lepton number and the spin
of the particle, respectively. SM particles have R=+1 while their SUSY partners have $R=-1$.} conservation is assumed,
implying that the Lightest Supersymmetric Particle (LSP) is stable and SUSY particles (``sparticles'') 
are pair-produced. In addition, they decay directly or 
indirectly into the LSP. In this paper the lightest neutralino,
 $\tilde{\chi}^0_1$, is assumed to be the weakly interacting LSP,
 which escapes detection giving signatures of missing energy and momentum.

The searches for sparticle production were developed with minimal assumptions,
and the selections employed depended 
primarily on the masses of the particles involved. 
In particular, the sensitivity of the searches depends on the visible energy
released in the decay process. In direct decays into the LSP, this visible energy is 
largely determined by the mass difference (\DM) 
between the decaying sparticle and the LSP.
In indirect (cascade) decays, other mass differences can also be important.

The MSSM has a large number of free parameters in addition to the SM ones. The most
model-independent interpretation of the results is in terms of the masses and
cross-sections 
explicitly involved,  for 
each production channel. 
A common interpretation of the results from the various searches can also be 
performed 
and used to exclude regions of the model parameter space.
This, however,
requires a manageable number of free parameters. 
For this reason, assumptions must be made and specific scenarios defined for 
such an interpretation.
The general phenomenology of the searches will be discussed
in section \ref{sec:phenomenology}, followed by
a description of the implications of more specific scenarios
in section \ref{sec:constraining}.

\subsection{General phenomenology}
\label{sec:phenomenology}

\vspace{0.15cm}
\noindent
\underline{Squarks and Sleptons}
\vspace{0.25cm}

The ``sfermions'', squarks and sleptons, are the scalar partners of the SM fermions. 
The left- and right-handed chiral states of each SM fermion, f$_{\rm L}$ and
f$_{\rm R}$, have as SUSY partners two scalars, usually labelled \sfel\ and \sfer. 

Sleptons and squarks could be pair-produced at LEP via \ee\ annihilation 
into $\Zn/\gamma$, leading to \sfer\sferb\ or \sfel\sfelb\
final states.
Selectrons could also be produced through $t$-channel neutralino exchange.
The selectron cross-section 
depends critically on the neutralino
mass, and destructive interference can make it very small.
The $t$-channel contribution also 
introduces the possibility of \sell\selrb\ and \selr\sellb\ production. 

If the unification of sfermion masses at a high mass scale typical of 
Grand Unified Theories (GUT) is assumed,
smaller masses and cross-sections for a given universal 
mass parameter are typically expected for 
the partners of right-handed fermions.
Under this assumption,
the kinematic accessibility of first and
second family sfermions at LEP depends only on their assumed common mass at the
unification scale. Squarks are in general 
expected to be heavier than sleptons.
However, for sfermions of the third family
the large Yukawa couplings lower the masses, as large mixing between 
left and right states may occur. In this case, the lighter mass states 
of third family sleptons and squarks, \stone\ (stau), \sbqone\ (sbottom) and \stqone (stop),
are candidates for the lightest charged supersymmetric 
particle.



In large regions of the SUSY parameter space the dominant decay of the sfermions is
to the corresponding fermion and the lightest neutralino, 
$\sfe \rightarrow {\mathrm f} \XN{1} $. 
In the case of the stop, the decay 
$\stq \rightarrow \topq \tilde{\chi}^0_1 $ 
is not kinematically allowed at LEP, and the dominant two-body decay channel is expected
to be $\stq \rightarrow \charm \XN{1} $  
($\stq \rightarrow \bottom \XPM{1} $ being disfavoured by existing
limits on the chargino mass). If $\msnu\! <\! \mstqone $, the three-body
decay $\stqone\! \to$ b$\ell\snu$ may compete with the c\XN{1} decay.

Thus final state topologies with a pair of acoplanar~\footnote{In this
context, acoplanar means that the direction of one of the leptons/jets 
is not in the plane defined by the direction of the other lepton/jet and
the beam line. Throughout this paper, the acoplanarity angle between 
two particles or jets is defined as the complementary (with respect
to 180\dgree) of the angle between their
directions in the plane transverse to the beam direction.} 
leptons or jets and missing energy are the 
relevant ones in the search for sleptons and squarks, respectively, and the 
total energy of the detectable final state particles 
(and thus the sensitivity of the search)
depends primarily on the mass difference between the sfermion and the LSP.

\vspace{0.35cm}
\noindent
\underline{Charginos and Neutralinos}
\vspace{0.25cm}

In the MSSM there are four neutralinos, $\XN{i},i=1,4$ (numbered in order of 
increasing mass) and two charginos $\XPM{i},i=1,2$ which are linear combinations
of the SUSY partners of neutral and charged gauge and Higgs bosons (gauginos
and higgsinos). The lightest states can be mainly gaugino or 
higgsino, or strongly mixed (for similar gaugino and higgsino mass parameters). 

Neutralinos (charginos) could be pair-produced at LEP via $s$-channel \Zn\ (\Zn/$\gamma$)
exchange or $t$-channel exchange of a selectron (sneutrino).
The $t$-channel contribution can be important if the slepton is light. 
The interference with the $s$-channel diagram is constructive in the case of 
neutralinos, but destructive in the case of charginos.

Expected decays 
are  
\XN{i} $\to$ \XN{1}\ffbar\ and
\XPM{j} $\to$ \XN{1}\ffbarp.
If the sfermions are heavy, these decays proceed via \Zn\ or W 
exchange.
However, sfermion emission may also contribute if the sfermions are light,
increasing the partial width for decays into the corresponding
fermions, and two-body decays into $\bar{\rm f}\, \tilde{\rm f}$
can dominate if kinematically allowed. 
The one-loop decay \XN{2} $\to$ \XN{1}$\gamma$ can be important
in specific  regions of the parameter space 
when other decays are suppressed.

Thus the sfermion mass spectrum may significantly 
affect both the production cross-section
and the decay modes of charginos and neutralinos, and many final state
topologies are possible.


In the case of chargino pair-production, the final state is expected to be
four jets if both charginos decay hadronically, two jets and one lepton if
one chargino decays into $\ell \nu \XN{1}$, and leptons only if both charginos
decay into leptons.
The branching ratio of $\XPM{1} \to \XN{2} \ffbarp$ can 
be sizable, in particular in the regions of the parameter space where 
 $\XN{2} \rightarrow \XN{1} \gamma$ is important. In this case, the above
topologies are accompanied by photons.

If the mass difference \DM\ between the chargino and the LSP is very small 
the visible energy released in the decay is very small, making the 
signal hard to detect. 
The simultaneous production of a photon by initial state radiation (ISR) can be used to
explore such regions, as this allows a very efficient background rejection (at 
the expense of a small signal cross-section).
Still smaller mass
differences imply a long lifetime of the chargino, which can then be
identified as a heavy stable charged particle or one with a displaced decay
vertex.



In the case of the detectable \XN{1}\XN{k} neutralino production channels 
($i.e.$ excluding \XN{1}\XN{1}), the most important signatures are 
expected to be acoplanar pairs of jets or leptons
with
high missing energy and momentum. Although \XN{1}\XN{2} and \XN{1}\XN{3}  
are expected to dominate in most of the parameter space, 
channels like \XN{2}\XN{3} and \XN{2}\XN{4} must also be considered 
for a complete coverage. 
These give rise to cascade decays with multiple jets or leptons in the 
final state, possibly accompanied by photons from $\XN{2} \to \XN{1} \gamma$.

\subsection{Constraining the parameter space}
\label{sec:constraining}

In this paper the results are combined and interpreted in the framework of constrained
SUSY scenarios 
with gravity-induced SUSY breaking. 
To make the model more predictive, the unification of some parameters at a 
high mass scale typical 
of Grand Unified Theories (GUT) can be assumed~\cite{MSUGRA}. 
The MSSM parameters and the assumptions that can be relevant
in the interpretation of the results 
are listed below:

\begin{itemize}
\item {\bf tan}~\mbox{\boldmath$\beta$}, 
the ratio of the vacuum expectation values of the two Higgs doublets; 

\item
\mbox{\boldmath$\mu$}, the Higgsino mixing mass parameter;

\item
\mbox{\boldmath $M_1,M_2,M_3$}, the  $U(1)\times SU(2)\times SU(3)$ gaugino masses 
at the electroweak (EW) scale:
when gaugino mass unification at the GUT scale is assumed, with a common gaugino mass
 \mbox{\boldmath$m_{1/2}$}, the relation between $M_1$ and $M_2$ is
$M_1 = \frac{5}{3}M_2\tan^2\theta_W \sim 0.5 M_2$, with the assumption
that $M_i/\alpha_i$~($i=1,2,3$) is renormalization group invariant, fulfilled
at the one-loop level;

\item
\mbox{\boldmath $M_{\tilde{\mathrm f}}$}, the sfermion masses:
under the assumption of sfermion mass unification, \mbox{\boldmath${m_0}$} 
is the common sfermion mass at the GUT scale;

\item
the pseudoscalar Higgs mass, \mbox{\boldmath$m_{\mathrm A}$}, on which the masses in the Higgs sector depend: 
if scalar mass  unification is assumed, \MA\ at the EW scale can be derived from $m_0$;

\item
the trilinear couplings, \mbox{\boldmath$A_{\mathrm f}$}, determining the 
mixing in the sfermion families:
the third family trilinear couplings are the most relevant ones,   
\mbox{\boldmath$A_\tau, A_{\mathrm b}, A_{\mathrm t}$}, and 
under the assumption of universal parameters at high mass scale there is
a common trilinear coupling \mbox{\boldmath$A$}, to which $A_\tau$, $A_{\mathrm b}$
and $A_{\mathrm t}$ at the EW scale can be related.  

\end{itemize}

Mass 
mixing 
terms at the EW scale 
given by
$m_{\tau}(A_\tau - \mu \tanb)$, $m_{b}(A_b - \mu \tanb)$ and
$m_{\mathrm t}(A_t - \mu / \tanb)$ are considered for 
\stau, \sbq\ and \stq, respectively.
The mass splitting 
grows with the mixing terms, and
for large $|\mu|$ this can give light \stone\ and \sbqone\ states
if \tanb\ is large, or a light \stqone\ for small \tanb.

In the model referred to as the ``Constrained MSSM'' (CMSSM) in the following, 
sfermion and gaugino mass unification are assumed. 
The parameter set is then reduced to $M_{2}, m_0, \tanb, \mu, m_A$
and three $A_f$ couplings.
This is the model considered in section~\ref{sec:combined}.

Further tightening the assumptions, the individual $A_f$ couplings can be
replaced by a universal coupling $A$ and $m_A$ can be related to the other
parameters by assuming scalar (including Higgs bosons)
mass unification. Requiring in addition the correct reproduction of the EW symmetry 
scale,
which fixes the 
absolute value of $\mu$, defines the minimal 
MSSM with gravity-induced SUSY breaking
(mSUGRA).

%




The direct results of the searches are first derived and presented in the most 
model-independent way possible. 
Under the assumptions described above, the results
are then used to constrain the SUSY parameters.
Presently, the strongest constraints on SUSY models come from the MSSM Higgs 
searches~\cite{higgs}.  

Chargino production is the most important direct SUSY detection channel 
for large regions of the parameter space. 
However, if sfermions are light (corresponding to a small 
$m_0$ scenario),
or if the parameters take particular values, the chargino production
cross-section can be greatly
suppressed or undetectable final states can dominate (in particular for small
mass splittings). 
The most relevant of these cases are the following:

\vspace{0.25cm}
\noindent
\underline{Large $m_0$}
\vspace{0.25cm}

For large $m_0$, the sfermions are heavy and have little influence on
the observable phenomenology.
The chargino pair-production cross-section  is large and the chargino is excluded nearly up to the kinematic limit.
The search for charginos in non-degenerate scenarios has been applied down to 
$\DM\! =\! \MXC{1}\! -\! \MXN{1}\! =\! 3~\GeVcc$. 
The region $\DM\! <\! 3~\GeVcc$ is covered by the search requiring an 
ISR photon and by the searches for long-lived charginos. Moreover, small values of \DM\ occurring in scenarios without gaugino 
mass unification~\cite{cdg} have also been investigated.  

At small \tanb\ ($\tanb\! <\! 1.2$), neutralino searches can extend the 
sensitivity of the searches beyond the kinematic limit for chargino production. This concerns 
the region of (small) negative $\mu$ and $M_2\! >\! 60~\GeVcc$. In particular,
searches for neutralino 
cascade decays are crucial
for investigating the $M_2\! <\! 120~\GeVcc$ region, where the \XN{1}\XN{2} 
cross-section is small.

\vspace{0.25cm}
\noindent
\underline{Small $m_0$}
\vspace{0.25cm}

If $m_0$ is small, light 
sfermions affect the chargino and neutralino production
cross-sections.
In particular, the decrease of $m_0$ causes the chargino 
production cross-section to drop in the region where the
gaugino components dominate (small $M_2$ and large $|\mu|$). 
Down to $m_0\! \simeq\! 200~\GeVcc$, the cross-section remains large enough to
allow chargino exclusion nearly up to the kinematic limit.
For smaller $m_0$, the neutralino production cross-section is very much
enhanced, and neutralino searches become sensitive instead. 

If $m_0$ is very small ($m_0\! \simeq\! 100~\GeVcc$), sleptons can be 
sufficiently light to affect drastically the decay patterns of charginos and 
neutralinos, and nearly invisible final states can become dominant 
in some cases. 
However, for such small $m_0$ and small $M_2$ (below 200~\GeVcc) sleptons 
can also be searched for in direct pair-production.

If $\MXC{1}\! >\! \msnu$ and 
the mass difference $ \MXC{1}\! -\! \msnu$ is small,
the chargino decay chain
$\XPM{1}\! \to\! \snu\ell\! \to\! \nu\XN{1}\ell$
is dominant, and leads to an experimentally undetectable final state
(the only visible final state lepton has very low momentum). However, 
in this case the search for selectrons can be used to constrain the sneutrino
mass (under the assumption of unification) and thus the chargino mass.

It can also happen, in scenarios with large mixing among sfermions 
and large \tanb\ and $M_2$,
that ``blind spots'' occur in the chargino detection sensitivity due to
$\XPM{1}\! \to\!  \stau\nu\! \to\! \tau\XN{1}\nu$ 
with a small mass difference $ \mstau\! -\! \MXN{1}$. 
In this case, 
\XN{1}\XN{2} or \XN{2}\XN{2}
production with $\XN{2}\! \to\! \stau\tau$
are the only detectable channels.
A specific search was designed for this case.


\section{Detector description}
\label{sec:detector}

The DELPHI detector is described in detail in~\cite{delphi}. The central
tracking system consisted of
a Time Projection Chamber (TPC), supplemented
by a system of silicon tracking detectors and drift chambers.
These included the Vertex Detector (VD), closest to the beam pipe,
the Inner Detector (ID) and the Outer Detector (OD).
These were situated inside a solenoidal magnetic field of 1.2 T, parallel to the
beam axis. 
The average momentum resolution for charged particles in hadronic final states 
was in the range $\Delta p/p^2 \simeq 0.001-0.01$~(GeV/$c$)$^{-1}$.

The electromagnetic (EM)
calorimeters were symmetric around the plane perpendicular to the
beam ($\theta=90^\circ$)~\footnote{
In DELPHI, a right-handed Cartesian coordinate system
is used with the $z$ direction defined by the direction of the electron beam, 
and the $x$-axis pointing towards the centre of the LEP ring. The origin is at
the centre of the detector. 
The polar and azimuthal angles $\theta$ and $\phi$ are defined with respect 
to the $z$ axis and $\phi\! =\! 0$ corresponds to the $x$-direction, and 
the coordinate $r$ is defined in the usual way as $r=\sqrt{x^2+y^2}$.
In this paper, polar angle ranges are always assumed to be symmetric with 
respect to the $\theta=90^\circ$ axis.},
with the High density Projection Chamber (HPC)
extending from 88.7$^\circ$ to 43.1$^\circ$ (barrel region), the
Forward Electromagnetic Calorimeter (FEMC) from 36$^\circ$ down to 9$^\circ$,
overlapping with the Small angle Tile Calorimeter (STIC), the DELPHI luminometer, 
which covered the range $1.7^\circ \le \theta \le 10.6^\circ$. 
The region of poor
electromagnetic calorimetry at a polar angle close to 40$^\circ$ was
instrumented by scintillators (hermeticity taggers)~\cite{tag} to tag photons.

The Hadron Calorimeter (HCAL) covered 98\% of the solid angle. Muons with momenta
above 2~\GeVc\ penetrated the HCAL and were recorded in sets of Muon drift chambers
located in the barrel (MUB), forward (MUF) and surround~\footnote{The region between the barrel and end-cap 
parts of DELPHI not covered by the MUF and the MUB  chambers.} (MUS) regions of the detector.


\section{Data samples and  event generators}
\label{sec:samples}

During the year 
2000 DELPHI collected data in the centre-of-mass energy range from
201.5 to 208.8 GeV. The average centre-of-mass energy  was 
$<\!\sqrt{s}\! > \simeq 206$~GeV 
and the total integrated luminosity amounted to about 224 pb$^{-1}$.
In 1999 (1998) about 227~pb$^{-1}$ (158~pb$^{-1}$) were collected
at centre-of-mass energies around 192, 196, 200 and 202 GeV (189 GeV).

The data collected in the years 1999-2000 were analysed in the 
searches presented in this paper.
In some cases the 1998 data were re-analysed for consistency with the 
improved methods now presented, as detailed in 
the description of the analyses.
Combination with earlier results was performed to obtain cross-section upper limits
and excluded regions in the model parameter space.
Details for each channel are given in 
section~\ref{sec:results}.

Quality requirements on the status of the subdetectors most 
relevant for each analysis were applied, 
generically based on the status of the main 
tracking devices. Muon chambers or calorimeters were considered in channels where 
muon or electromagnetic shower detection was crucial. 
The luminosity loss was at most of the order of a few percent and 
was taken into account in the 
analyses. 

On September 1$^{\rm st}$ 2000, sector 6 of the TPC (corresponding to 
1/12 of the TPC acceptance) failed beyond repair.
This required modifications of the pattern recognition, and
affected the quality of charged track reconstruction. Thus
special care had to be taken for each search when analysing
the data collected without TPC sector 6.
The accumulated
integrated luminosities with and without a working sector 6 are respectively
164~\pbi\ and 60~\pbi at average centre-of-mass energies around 206 \GeV .

In order to increase the sensitivity for a discovery, the data collected in 2000 
were divided into 4 regions of centre-of-mass 
energy as given in table \ref{tab:ENERGY2000}.

\begin{table}[ht]
 \begin{center}
   \begin{tabular}{|c|c|c|c|c|}
 \hline
\multicolumn{3}{|c|}{ \sqs\ region of analysis}   & ${\cal L}$ ($pb^{-1}$) & $< \sqs>$ (\GeV) \\
 \hline
 1   &  Sector 6 on & \sqs $\leq$ 205.75~\GeV   & 78.3  & 204.9  \\
 \hline
 2   &  Sector 6 on & 205.75 $<$\sqs $\leq$ 207.5~\GeV  & 78.8 & 206.7  \\
 \hline
 3   &  Sector 6 on & 207.5~\GeV $<$ \sqs  &  7.2  & 208.0 \\
 \hline
 4   &  Sector 6 off & all \sqs  & 60.0  & 206.5(*) \\
 \hline
\end{tabular}
\caption[.]{
\label{tab:ENERGY2000}
Definition of the  \sqs\  regions used to analyse the data collected in 2000.
The three energy bins for the period in which the detector
was fully operational are referred to in the text by their approximate 
average centre-of-mass energies: 205, 207 and 208 GeV. The ``Sector 6 off'' data are referred to as 206.5(*). 
}
\end{center}
\end{table}


To evaluate the signal efficiencies and background contaminations, 
simulated events were generated using several different programs. 

The background process \eeto~\qqbar$(n\gamma)$ was generated with
{\tt PYTHIA 6.125}~\cite{pytjet}.
For $\mu^+\mu^-(\gamma)$ and $\tau^+\tau^-(\gamma)$,
{\tt DYMU3}~\cite{dymu3} and {\tt KORALZ 4.2}~\cite{koralz} were used,
respectively. The {\tt BHWIDE} generator~\cite{bhwide} was used for 
Bhabha events. Simulation of four-fermion final states was performed using 
{\tt EXCALIBUR}~\cite{excalibur} and {\tt grc4f}~\cite{grc4f}.

Two-photon interactions giving hadronic final states were generated using
{\tt TWOGAM}~\cite{twogam}, {\tt PHOJET}~\cite{phojet} and 
{\tt PYTHIA 6.143}~\cite{pytjet}, while leptonic final states were generated using the 
generator of~\cite{bdk}, including radiative corrections for the \ee\mumu\ and
\ee\tautau\ final states.

{\tt SUSYGEN 2.2004}~\cite{susygen} was used to generate chargino, neutralino, 
slepton, and sbottom signal events and to calculate cross-sections and branching ratios
for these channels.
For the nearly mass-degenerate case, the chargino decays were modelled with 
the results of the computations of reference~\cite{cdg}.
Stop events were generated according to the expected differential cross-sections
using the {\tt BASES} and {\tt SPRING} program packages and taking special
care in the modelling of the stop hadronisation~\cite{stopgen}.

In all cases except for stop generation, {\tt JETSET}
7.4~\cite{pytjet}, tuned to LEP1 data~\cite{tune}, was used for 
quark fragmentation.

The generated signal and background events were passed through the
detailed simulation of the detector, DELSIM~\cite{delphi}, and then processed
with the same reconstruction and analysis programs as the real data.
The  faster DELPHI simulation code SGV~\cite{sgv} was also used for 
signal efficiency studies in some analyses. 
SGV is a model of the detector response which has been tuned to the
data independently from DELSIM. For charged particles the perigee parameters resulting
from the physical process generation were modified to take into account
parametrized effects of multiple scattering, detector resolutions and acceptance.
For neutral particles geometric acceptance and parametrized calorimetric
resolutions were taken into account. Bremsstrahlung and photon conversion in the
tracking system were also simulated. 

The numbers of simulated events from different background processes were
several times the numbers in the real data.




\section{Descriptions of the analyses}
\label{sec:analysis}

The analyses described below can be divided into two stages. The first stage
was very similar for all searches and consisted of the selection of charged
and neutral particles followed by an event preselection. These
are described in sections \ref{sec:trksel} and \ref{sec:presel}.
In the subsequent stage, the analyses differed according to the 
characteristics of the various signals. This is described in sections
\ref{sec:slepana} to \ref{sec:ananeu}.
In most searches several different topologies were considered, accounting
for the different possible final states. Particle identification and
reconstruction algorithms common to several searches are described
in section \ref{sec:id&jet}.
Different analysis techniques were chosen for the various searches:
some analyses were based on successive requirements on individual 
event variables (``sequential cut analyses''), others used multidimensional
techniques based on likelihood ratios or neural networks. These techniques
are briefly discussed in section \ref{subsubsec:analysistech}.

As discussed above, the sensitivity of the searches depends on 
the mass difference between the produced sparticles and the LSP,
which determines the visible energy released in the process.
Typically, for small \DM\ the signatures of sparticle
production are similar to those of two-photon interactions (\gamgam\ events). 
For high \DM\ sparticle production events resemble four-fermion final 
states such as \WW\ and \ZZ. For intermediate \DM\ values, 
the background is composed of several SM processes (in particular
two-fermion ones). 
The coverage of all the relevant \DM\ regions often requires the combination of different 
searches or the optimisation of the selection criteria separately in
each \DM\ interval.
Sparticle searches for very low values of  \DM\ are particularly challenging and 
required different preselections.

\subsection{Basic selections and techniques}
\label{sec:basic}

\subsubsection{Particle selection}
\label{sec:trksel}

The following quality requirements were applied to the charged and neutral particles
observed in the detector. 

Charged particles were required to have momentum $p$ above 100~MeV/$c$ 
and below $0.75\sqrt{s}$, a relative momentum error less than 100\%,
and to extrapolate
back to within 5 cm of the main vertex in the transverse ($r\phi$) plane and 
10 cm/$\sin \theta$ in the longitudinal ($z$) direction.
Similar but more stringent criteria were applied to particles
whose tracks extrapolated to the TPC, but which gave no signal in the TPC.
Whenever tracks reconstructed with TPC information were required in the
subsequent analyses, 
at least five of the 16 pad rows had to contribute hits. 

In the stau analysis  (see section \ref{sec:stauana}) charged particle 
tracks were required to have TPC information, or all three of the
detectors 
VD, ID and OD used in the reconstruction of the track.
In addition, only tracks with polar angle  $\theta >$ 15\dgree\ 
were kept.

In the nearly mass-degenerate chargino search (see section \ref{sec:chadegsel})
there was no lower bound on the momentum 
for tracks at polar angle above $25^\circ$, while $p>150$~MeV/$c$ was required otherwise. 
In addition, different impact parameter requirements were 
applied (6 cm and 12 cm in the transverse and logitudinal directions,
respectively).

Energy clusters in the calorimeters were taken as neutral particles 
if not associated to a charged particle and if above an energy threshold 
which was 900 MeV for deposits in the hadron calorimeter
and ranged from 300 to
500 MeV for deposits in the electromagnetic calorimeters (depending on the
region of the detector). 
Cuts removing clusters created by radioactivity in the lead of the HPC
or by particles from cosmic ray showers were also applied.
\subsubsection{General event preselection}
\label{sec:presel}

In the general preselection, events were kept if there were 
at least two charged particles, at least one of them had  a 
transverse momentum 
above 1.5~\GeVc, and the transverse energy of the 
event~\footnote{The transverse energy is defined as the sum of the absolute values of the 
transverse momenta of all particles in the event: $c \sum_i \pT_i$.} exceeded 4~GeV.
This rejected mostly
two-photon interactions (for which most of the energy is deposited 
in the forward regions of the detector), 
zero- or one-prong final states (like \eeto~\gamgam,
e$\gamma \to$ e$\gamma$) and beam-related backgrounds (such as
beam-gas interactions).

For chargino searches in nearly mass-degenerate scenarios a 
different preselection was used:
at least two charged particles were required, as well as one isolated electromagnetic
cluster with transverse energy above 
a \rs-dependent threshold close to 5~GeV
and a mass recoiling against it above 90~\GeVcc.
Two-photon and beam-gas backgrounds were reduced by rejecting events with a large
fraction of the detected energy in the forward region of the 
detector.

\subsubsection{Particle identification and reconstruction algorithms}
\label{sec:id&jet}

The following criteria for particle and event classification were common to the 
different searches. 

Particle jets were reconstructed using three different approaches:

\begin{itemize}
\item
The {\tt DURHAM}~\cite{durham} algorithm was used to cluster the particles into a fixed 
number of jets: two or four.
\item
The {\tt LUCLUS}~\cite{pytjet} algorithm was applied with the critical distance set to 
$d_{\mathrm join}=10$~\GeVc\ or  $d_{\mathrm join}=2.5$~\GeVc. The final number of jets is, thus, variable
(and lower in the first case).

\item
A specific algorithm optimised for the low multiplicity jets resulting from $\tau$
decays was used for $\tilde{\tau}$ pair-production searches.
This method considered all possible ways of
clustering the charged particles in the event into groups,
always requiring the invariant mass to be below 2~\GeVcc. 
Clearly identified leptons were considered as a single group, 
except for pairs of oppositely charged, well identified electrons
close together which were allowed to be grouped with other particles,
since they could come from a converted photon.

If possible, the event was clustered into two groups with invariant mass
below 2~\GeVcc. If no such combination existed, the one with the smallest
number of groups was kept. When more than one way of obtaining two groups
both with invariant mass below 2~\GeVcc\ was found,
the grouping yielding the lowest sum of masses was retained.
Once the best grouping of the charged tracks was found,
it was attempted to associate the neutrals in the event 
to the particle groups. Also in this step, the grouping yielding the lowest
sum of masses was chosen. However, as very few long-lived neutral hadrons are expected in
$\tau$-decays, neutral hadronic clusters were
not included in the groups, but treated as isolated neutrals.
Also electromagnetic clusters which 
could not be merged into any of
the groups without the invariant mass exceeding 2~\GeVcc,
were left as isolated neutrals.
In addition, a special procedure was applied to 
identify and correct for
neutral clusters that were likely to be either
bremsstrahlung photons or
a shower induced by an electron that was not correctly
assigned to the track by the reconstruction program.

The charge of each cluster was taken as the sum of the charges of all its particles. 
 
\end{itemize}

The thrust variable used in several
analyses throughout this paper  was computed
using all the particles meeting the particle selection requirements.


Isolated leptons or photons in the event are often very important
in distinguishing signal and background.
In the present searches, 
the isolation criteria depended 
on the multiplicity of the event:
\begin{itemize}
\item
In low multiplicity searches, charged particles were classified as
isolated if the
total charged energy, excluding the energy of the particle itself, within
$10^\circ$ of the track 
direction was below 2 \GeV . Slightly  tighter cuts in the impact parameters (1~cm and 5~cm in $r\phi$ and $z$, respectively) and 
in the momentum error were also applied. 
\item
In high multiplicity searches, a photon  was considered isolated if its angular separation from 
any neutral or charged particle was greater than 15\dgree. A lepton was tagged 
as isolated if its angular separation from all the jets (computed without the lepton
using the LUCLUS algorithm with $d_{join}=40$~\GeVc) was greater than 20\dgree.
\end{itemize}

The identification of a track as a muon, electron, or hadron
was ``tight'',``loose'', or ``veto'' (or none at all). 
Tight identification was unambiguous.
A particle could simultaneously be loosely identified as several
different species. Excluded particle species were vetoed.

The identification of muons was provided primarily by the DELPHI standard 
algorithm described in ~\cite{delphi}, which relies on the association of charged 
particles to signals in the muon chambers and the HCAL.

Electron and photon identification was performed by the algorithm described in
~\cite{remclu} which combines deposits in the EM calorimeters with 
tracking information and takes possible radiation and interaction effects into account 
by a clustering procedure in an angular region around the main shower.
In the $\tilde{\tau}$ analysis (see section \ref{sec:stauana}), 
the clustering procedure was not used for
tracks in the barrel region of the detector, since it tended to
treat charged pions from $\tau$ decays wrongly,
if they were accompanied  closely by neutral pions.
In low multiplicity topologies a very loose electron identification 
based on the ratio $E/p$ between the energy deposited in the EM calorimeter
and the momentum of the associated charged track was also used.

Below, whenever the identification level is not specified, 
it is implied to be ``tight'' for electrons and ``loose'' for muons.
In the case of electrons, ``tight'' identification basically 
adds some isolation requirements to the identification ones.

e/$\gamma$ separation inside the acceptance of the STIC luminometer 
was performed on a statistical basis, using the veto information of the two
planes of the scintillator counters placed in front of it.

In the stau analysis
a particle was considered as a tight hadron if it was not classified
as a muon or electron and had an associated 
energy in the  hadron calorimeter exceeding 50\% of its momentum,
or else  was considered as
a loose hadron if it had hadronic energy associated and
it was not tightly identified as a muon or electron.
If both the electromagnetic and hadronic
energies were small (less than 1 and 6 \GeV, respectively),
and the difference between the hadronic energy and the
track momentum was above 10 \GeV, the particle was
assigned both the loose muon and loose hadron
code.

Decays of b-quarks were tagged using a probabilistic method based on
the impact parameter of tracks with respect to the main vertex~\cite{btag}.
${\mathcal P}^+_E$ stands for the
corresponding probability estimator for tracks with positive impact parameter,
the sign of the impact parameter being defined by the jet direction.
The combined probability ${\mathcal P}_{com b}$ included additional
contributions from properties of reconstructed secondary
vertices.

All searches made use of the information from the hermeticity 
taggers~\cite{tag} to
reject events with photons  in the
otherwise insensitive region at polar angles around 40\dgree.
If there were active taggers
not associated to reconstructed jets,
the event was rejected if the tagger was located 
in the direction of the missing momentum.
In the stau analysis, where neutrinos from tau decays made the
estimation of the direction of the missing momentum
unreliable, events containing active and isolated
taggers were rejected irrespective of the 
direction of the missing momentum.

\subsubsection{Analysis techniques}
\label{subsubsec:analysistech}
\vspace{0.35cm}
\noindent
\underline{Likelihood ratio method}
\vspace{0.35cm}

In the likelihood ratio method used, several discriminating variables
are combined into one on the basis of their one-dimensional probability
density functions (pdf's). 
If the variables used are independent, this gives
the best possible background suppression for a given signal 
efficiency~\cite{ander}.
For a set of variables $\left\{ x_{i}\right\} $, the pdf's
of these variables are estimated by normalised 
frequency distributions for the signal and the background samples.
We denote the pdf's of these variables $f_{i}^{S}(x_{i})$ for
the signal events and $f_{i}^{B}(x_{i})$ for the background events submitted 
to the same selection criteria.
The likelihood ratio function is defined as $\mathcal{L}_{R}$ $ = \prod\limits_{i=1}^{n}\frac{%
f_{i}^{S}(x_{i})}{f_{i}^{B}(x_{i})}$. Events with $\mathcal{L}_{R}>\mathcal{L}_{R_{CUT}}$\ are selected
as candidate signal events.
The choice of variables and the value of $\mathcal{L}_{R_{CUT}}$ were optimised
using simulated event samples 
by minimising the signal cross-section that was expected to be excluded
at 95\%\ confidence level in the absence of a signal.

\vspace{0.35cm}
\noindent
\underline{Neural networks}
\vspace{0.35cm}

A neural network provides a different way of defining 
one discriminating variable from multidimensional distributions of
event variables given as inputs. 
The neural network used below (see section~\ref{sec:squarkana})  contains
three layers of nodes: the input layer where each neuron corresponds to a 
discriminating variable, the hidden layer, and the output layer which is the
response of the neural network. The program used in the squark analysis was 
{\tt SNNS}~\cite{snns}. A ``feed-forward'' architecture is implemented and
the ``back-propagation'' algorithm is used to train 
the network with simulated events. 
An independent validation sample was also used not to overtrain the network. 
A way of enhancing the efficiency of the network without increasing too much the
number  of its parameters is to define a separate output node for 
each type of event that the neural network should separate.
More details are given in section~\ref{sec:squarkana}.

\subsection{Slepton searches}
\label{sec:slepana}


Supersymmetric partners of electrons, muons
and taus were searched for. 
In this paper, data collected at 
centre-of-mass energies between 189 and 208 GeV
were analysed, and were combined with previous 
results \cite{slep}.

The track selection and the general event preselection described in 
sections \ref{sec:trksel} and  \ref{sec:presel}, respectively, were used.
The analyses were then performed in two stages. Firstly, a loose
selection was used to obtain a sample of low multiplicity events. 
Events with less than ten charged particles
and
a visible invariant mass above 4.5 \GeVcc\
were retained for further analysis.
Different selections were then applied in each of the three channels.

\subsubsection{Selectron searches}
\label{sec:selana}

To search for selectrons, the general topology required was two
acoplanar electrons and missing energy. 
All candidates with exactly two well reconstructed and isolated
particles (according to the definition described in section~\ref{sec:id&jet}), 
oppositely charged and with momentum above 1 \GeVc,
were first selected. 
One of the two charged
particles was required to be tightly identified as an electron, and 
the event was rejected 
if the other was identified as a muon.

\begin{figure}[htbp]
\begin{center}
\vspace{1.cm}
\includegraphics[width=7cm]{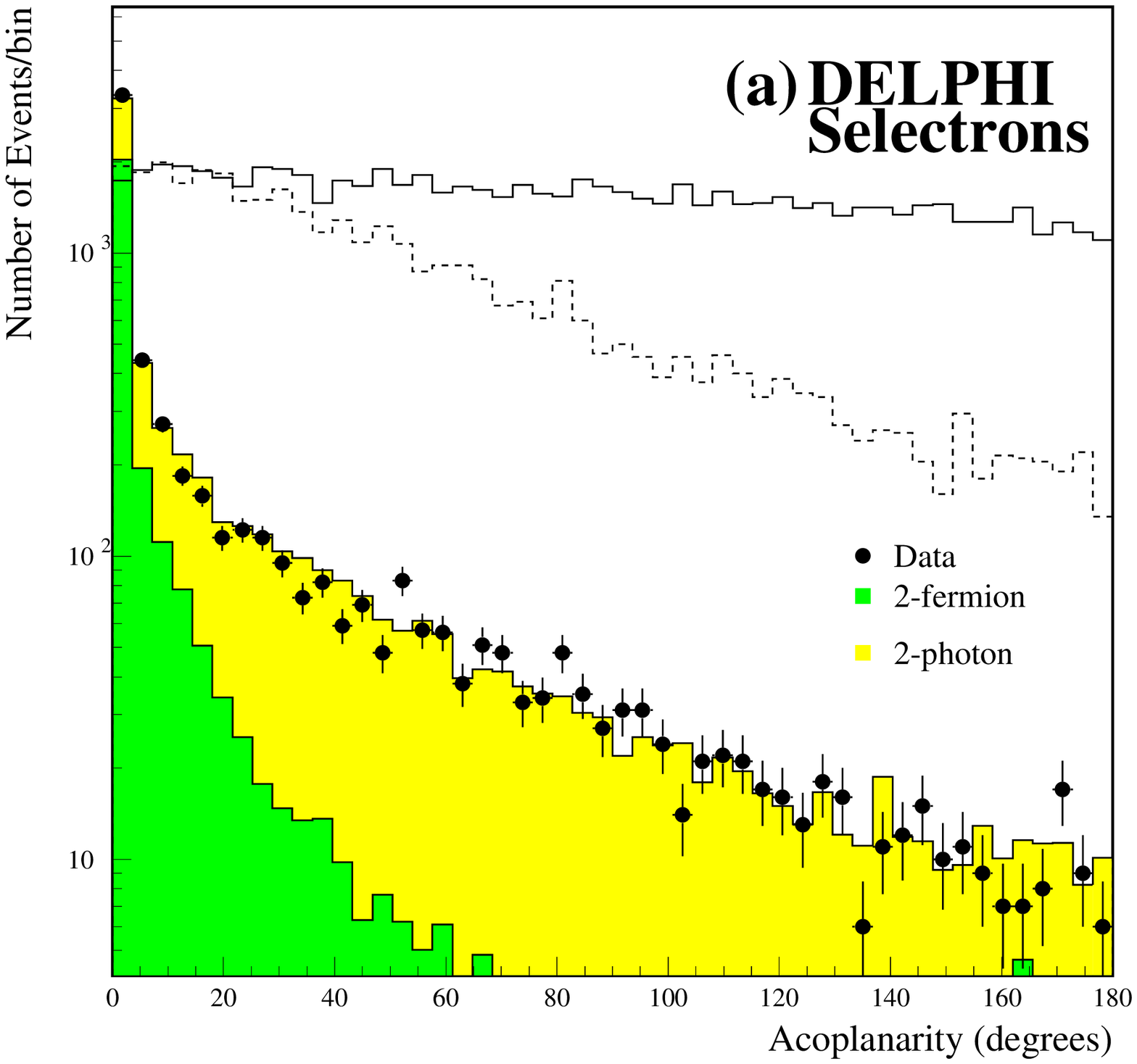}
\includegraphics[width=7cm]{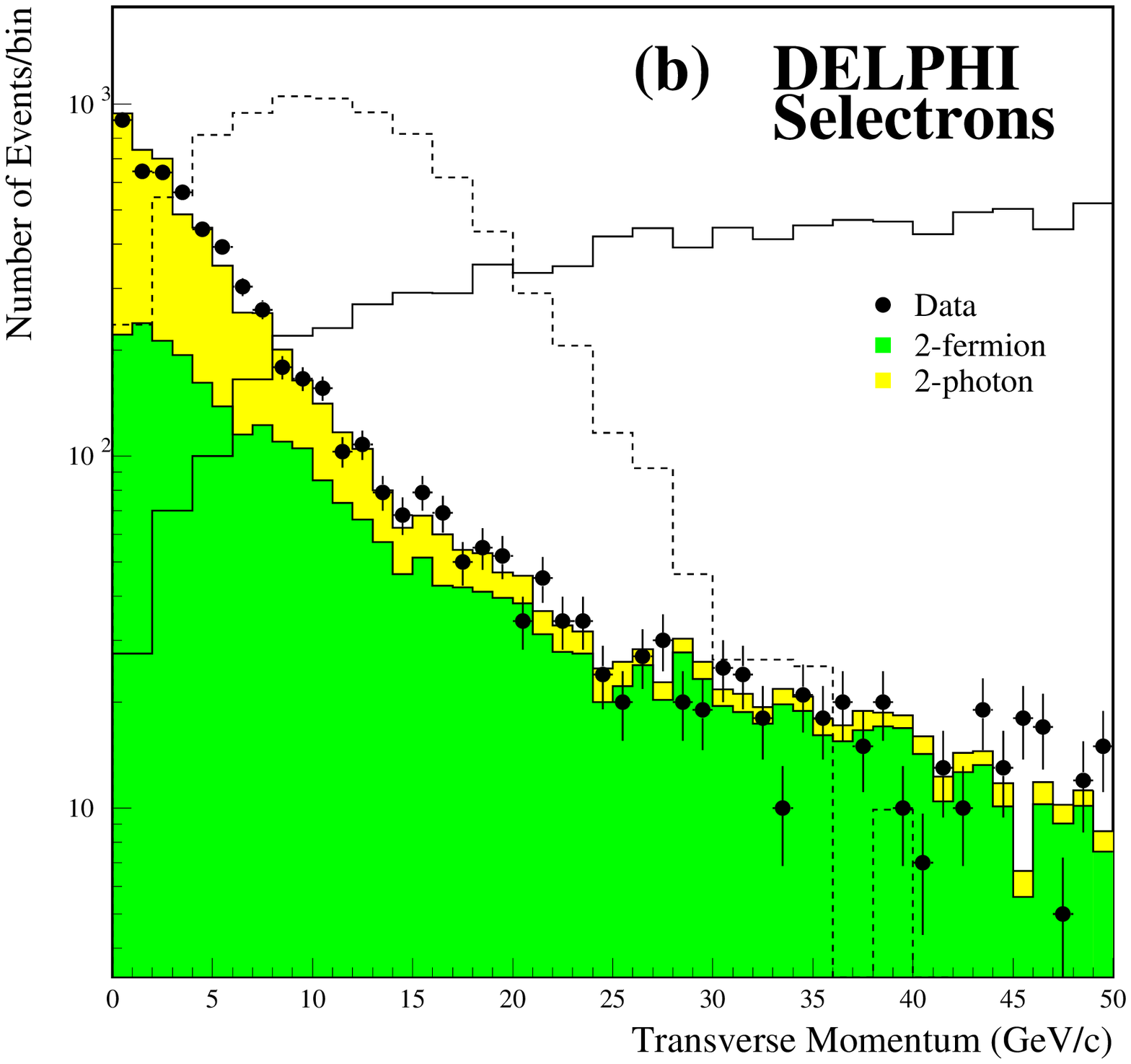}
\includegraphics[width=7cm]{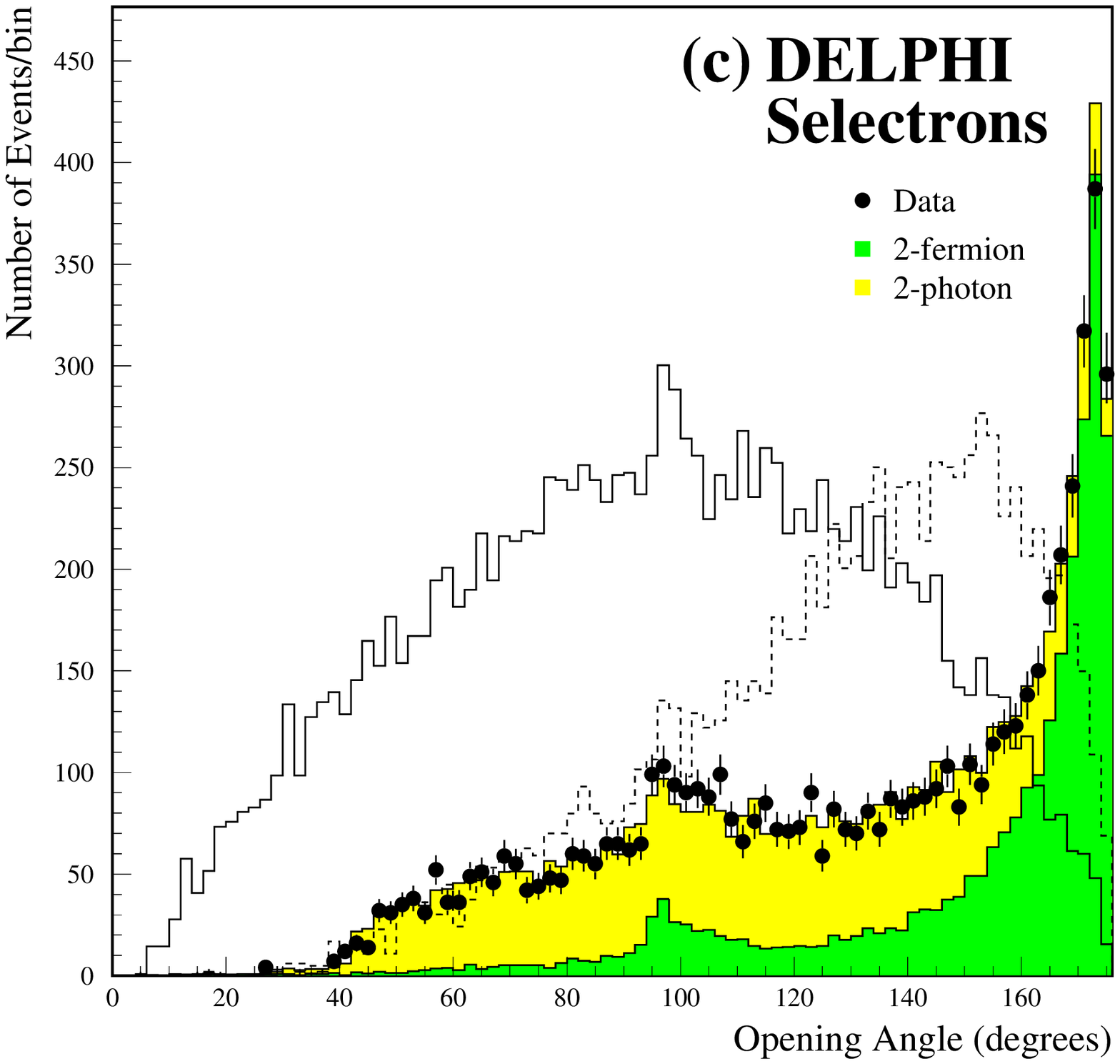}
\includegraphics[width=7cm]{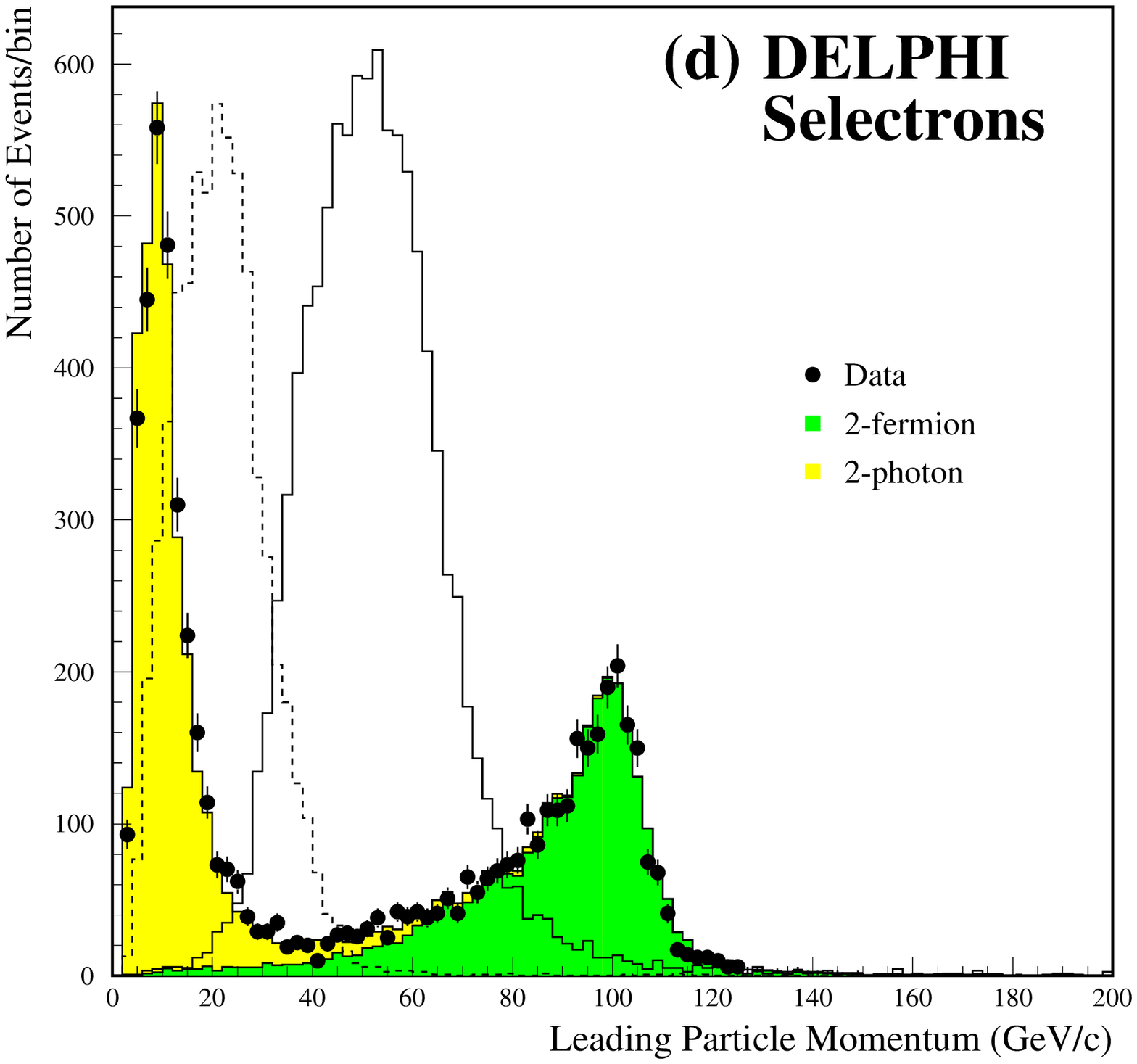}
\caption{Comparison of data and simulation in the selectron channel at preselection level.
The dots with  error bars show the data, 
 shaded histograms show the simulation. Plots include data taken in the year 2000 when the DELPHI detector was fully operational. 
 The plots show: (a) the acoplanarity, (b) the transverse momentum, (c)
the opening angle, (d) the momentum of the leading charged particle.
Possible signals corresponding to the mass combinations
\msel=90~\GeVcc,\MXN1=10~\GeVcc\ (solid) and
\msel=50~\GeVcc,\MXN1=40~\GeVcc\ (dashed) are
shown by the superimposed open histograms.
The signal normalisation is arbitrary.
} 
\label{fig:slep1}
\end{center}
\end{figure}

 At this stage of the selection
the sample consisted mainly of Bhabha and two-photon events.
Satisfactory agreement was observed between the data and simulated
background, as shown in 
figure~\ref{fig:slep1}~\footnote{In order to show the different background
contributions, the largely dominant Bhabha background was suppressed 
in these plots by demanding
that the opening angle between the two tracks was below 176$^\circ$.
}.

A series of tighter cuts reduced the SM background further.
As two-photon events are predominantly at low polar angles and with low
momentum, the visible energy was required to be greater than 
15~GeV, the energy deposited in the low angle STIC calorimeter
less than 4~GeV and the
total transverse momentum with respect to the beam axis greater than 5~\GeVc.
To reduce the number of Bhabha events an upper limit on the visible
energy of 100~GeV was imposed, while also requiring that the neutral
energy not associated to the charged particle tracks be less than 30~GeV.
 Events were also rejected if there were more than four neutral clusters
in total, each with energy above 500~MeV. Bhabha events are coplanar
and with a large opening angle, hence it was demanded that the 
opening angle between the two tracks be lower than 165$^\circ$
and the acoplanarity
be greater than 15$^\circ$.
Constraints were also imposed on the momenta of the two particles,
requiring that both tracks had momentum greater than 2 \GeVc. 

\subsubsection{Smuon searches}
\label{sec:smuana}

\begin{figure}[htbp]
\begin{center}
\vspace{1.cm}
\includegraphics[width=7cm]{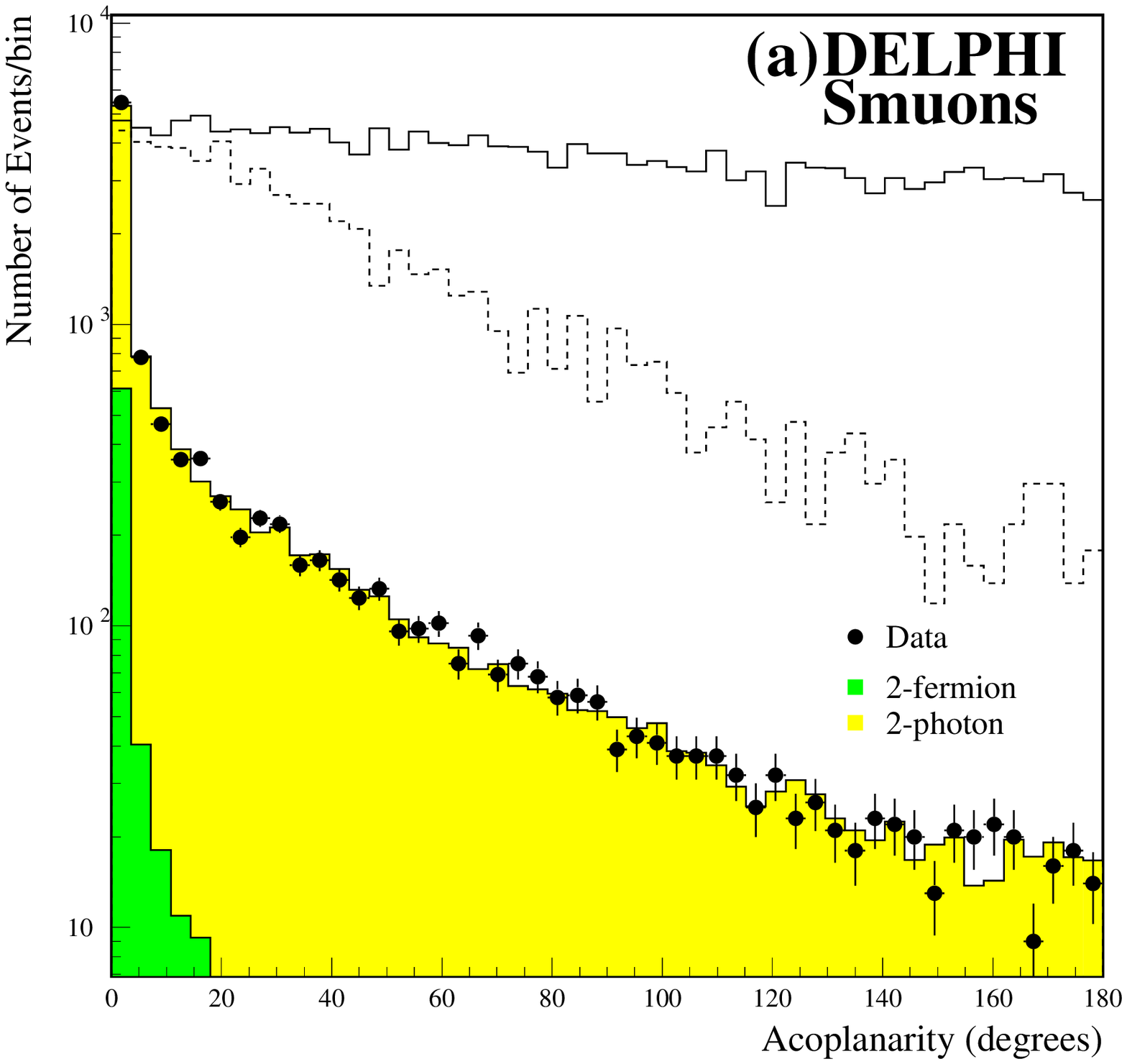}
\includegraphics[width=7cm]{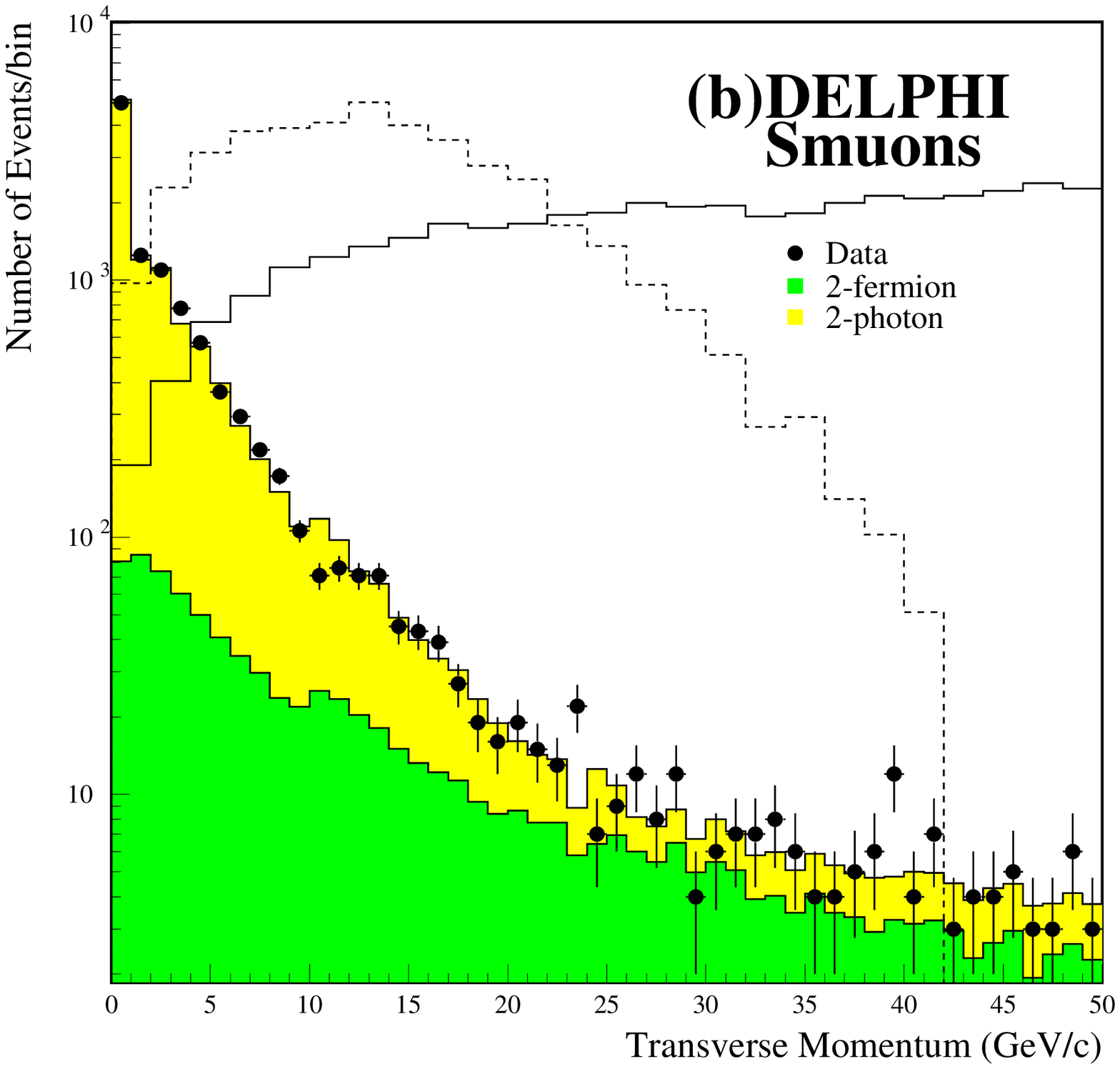}
\includegraphics[width=7cm]{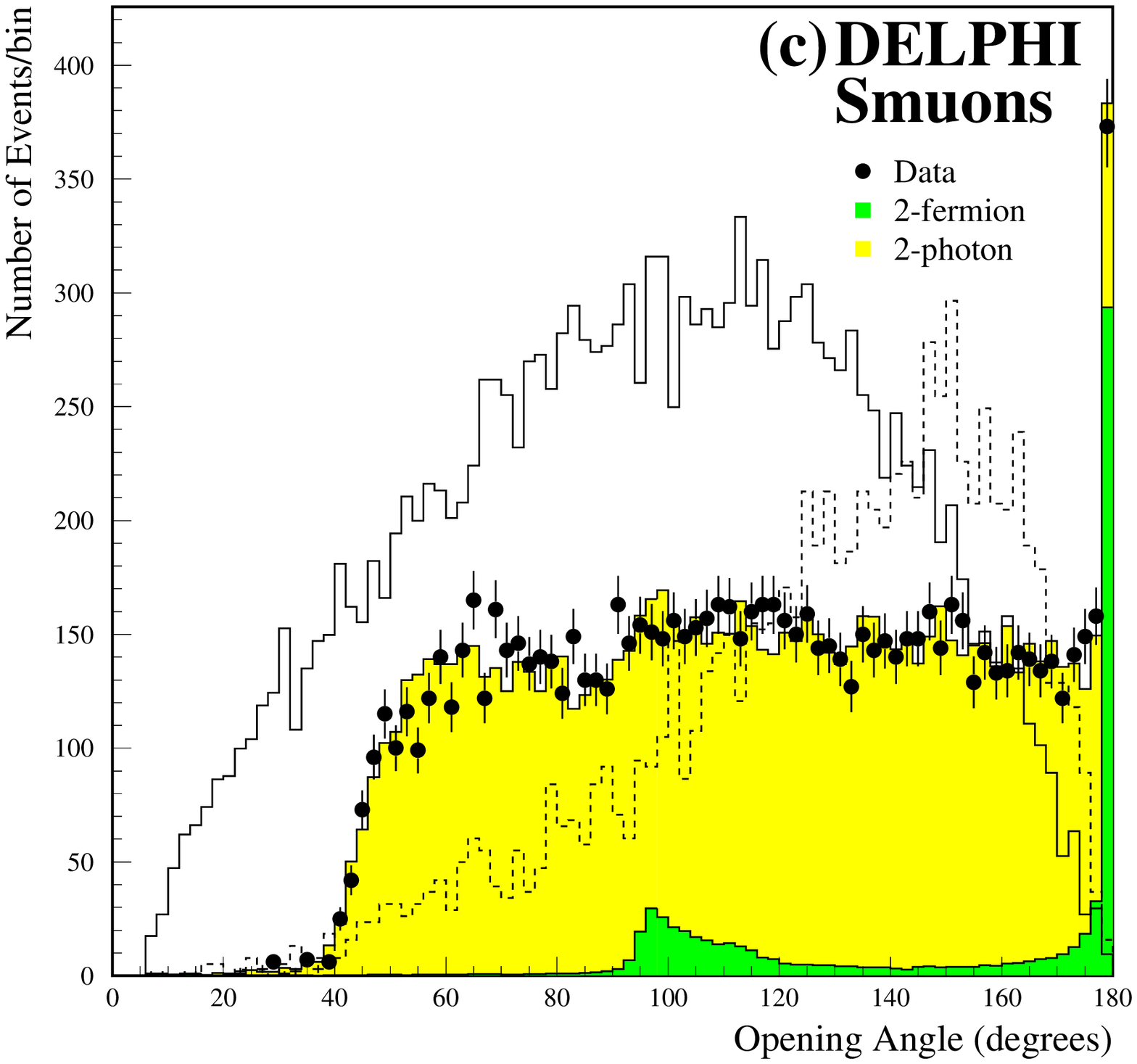}
\includegraphics[width=7cm]{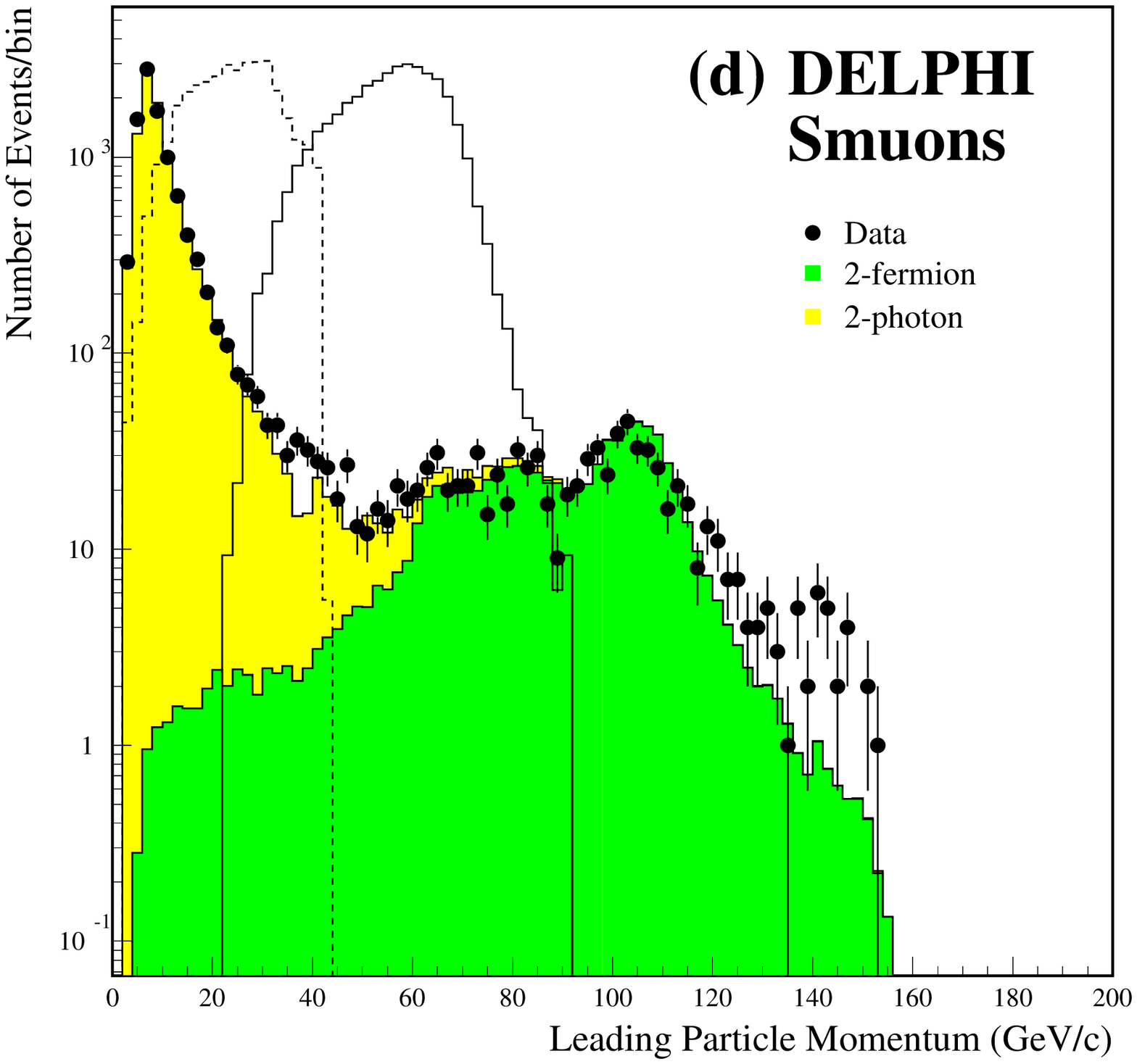}
\vspace{0.5cm}
\caption{Comparison of data and simulation in the smuon channel at preselection level.
The dots with  error bars show the data, 
 shaded histograms show the simulation. Plots include data taken in the year 2000 when the DELPHI detector was fully operational 
. The plots show: (a) the  acoplanarity, (b) the transverse momentum, (c)
the opening angle, (d) the momentum of the leading charged particle.
Possible signals corresponding to the mass combinations
\msmu=90~\GeVcc,\MXN1=10~\GeVcc\ (solid) and
\msmu=50~\GeVcc,\MXN1=40~\GeVcc\ (dashed) are
shown by the superimposed open histograms.
The signal normalisation is arbitrary.} 
\label{fig:slep2}
\end{center}
\end{figure}

   Smuon pair-production with decays to muon plus neutralino is expected
   to give acoplanar muons and missing energy.
All candidates with exactly two well reconstructed and isolated
particles (according to the definition given in section~\ref{sec:id&jet}), 
oppositely charged and with momentum above 1 \GeVc,
were first selected. 
 At least one
   of the particles had to be loosely identified as a muon. It was further
required
   that neither particle be identified as an electron. The selected
   sample consisted mainly of two-photon events and fair agreement 
   between real data and simulated background was observed 
   (see figure~\ref{fig:slep2}; 
   the slight discrepancies visible in figure~\ref{fig:slep2}(d), in the
tail of the two-photon distribution at about 40~\GeVc, and for high momenta,
are in kinematic regions rejected in the next step of the analysis).

   To reduce the SM background further, a series of tighter cuts
   were applied. To remove two-photon events, the visible energy was
   required to be greater than 10~GeV. Also, the energy in the STIC
   had to be less than 1~GeV. It was also demanded that the transverse
   momentum be greater than 5~\GeVc.
   To remove  \eeto $\mu^+\mu^-$ events, an upper limit of 120 GeV on the
   visible energy was imposed, whilst also requiring the unassociated
   neutral energy to be less than 10~GeV, with no more than two neutral
   clusters. This background was further suppressed by accepting only
   events in which the opening angle between the tracks was less than
   165$^\circ$ and the acoplanarity was greater than 15$^\circ$. 
   To reduce $t$-channel W pair contamination, events were rejected if the 
   positively
   charged muon was within 40$^\circ$ of the e$^+$ beam direction, or the 
   negatively charged muon was within 40$^\circ$ degrees of the e$^-$ beam 
direction.

\subsubsection{Stau searches}
\label{sec:stauana}
\begin{figure}[htbp]
\begin{center}
\mbox{\epsfysize=18.0cm \epsffile{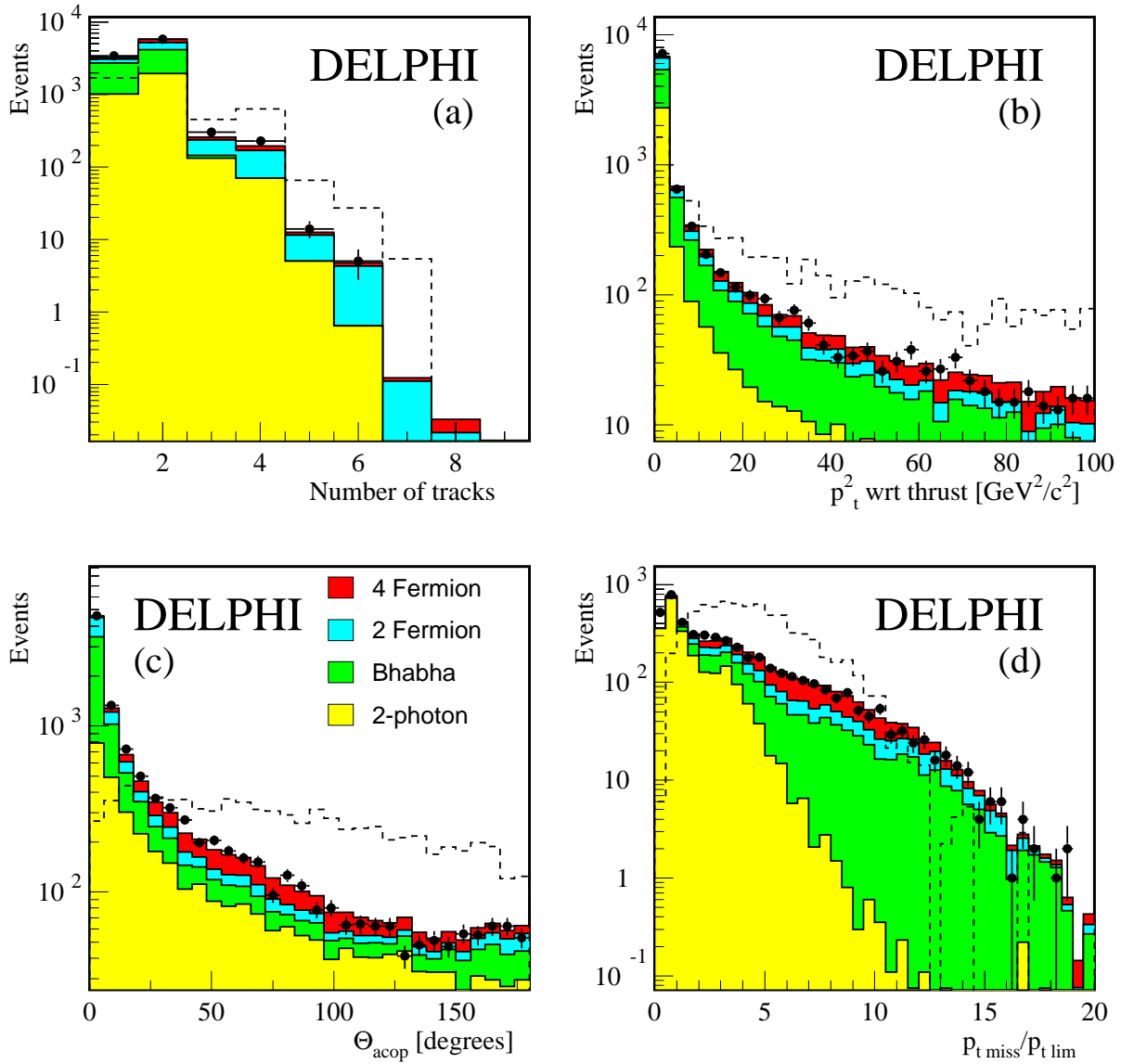}}
\vspace{0.5cm}
\caption{A preselection comparison of data and simulation in the stau
  analysis. The plots show: (a) the number of charged particles,
  (b) the square of the transverse momentum 
with respect to the
thrust axis,  
(c) the acoplanarity, 
(d) the missing transverse momentum divided by the maximum missing
transverse momentum
in two-photon events with no beam-remnant electrons
in the detector acceptance (ie. in ``no-tag'' events).
The dots with
  error bars show the data, 
  while the simulation is shown shaded.
A typical signal (\mstau = 83 \GeVcc ,  $M_{\mathrm LSP}$ = 0 \GeVcc) is shown by 
the superimposed open histogram, with arbitrary normalisation. 
} 
\label{fig:stau:datamc}
\end{center}
\end{figure}

Events with two acoplanar taus and high
missing energy can be the signature of stau pair-production.
Due to the scalar nature of the stau,
the two taus are produced centrally in the detector. 
To select this topology, 
the particles in the events were grouped into clusters
according to the algorithm described in section~\ref{sec:id&jet}.
Events with exactly two particle clusters
(possibly accompanied by isolated neutral particles)
were considered further if
they contained at least one charged particle 
with 
momentum above 1~\GeVc\ and a relative error less than
30\%. This particle had to be
 isolated 
(no neutral or charged particles in a cone of 
20\dgree\ half-angle around it),
or 
above 30\dgree\ in polar angle,
and its calorimetric energy could not exceed 
the momentum by more than three times the expected error
on the calorimetric measurement.
A comparison of data and simulated SM 
background at this
stage~\footnote{
In order to show the contribution of all 
classes of background,
the largely dominant Bhabha background was suppressed
in these plots 
by demanding that the missing transverse momentum was above 4 \GeVc,
and that the acoplanarity was above 1\dgree.}
is shown in figure~\ref{fig:stau:datamc}.
It was further required that
there were no more than six charged particles in the event and
that the total charge was 0 or $\pm$1. 

Tight and loose electrons, muons and hadrons were defined
as described in section~\ref{sec:id&jet}.
In addition, if a particle had a loose identification
for one species, and was vetoed for the other two,
it was considered as tightly identified.
If a track pointed to a gap in the electromagnetic
calorimetry, it was considered as a loose
electron.
Tracks with no identification information were
treated as loose electrons.
Neutral clusters passing the criteria of section~\ref{sec:trksel} were
used if their angle to the beam  was above 15\dgree.
Furthermore, no identified hadronic secondary interactions
inside the tracking system were allowed.

Beyond this point, the analysis differed depending on
whether a stau with mass above or below \MZ/2 was searched for.

~\\
\noindent
{\underline{\it Search for staus with large mass}}\\

\noindent
To suppress the 
two-photon background, it was required that
the total transverse momentum imbalance 
exceeded 4~\GeVc,
the total calorimetric energy below $30\dgree$ in polar angle
did not exceed 
20\% of the beam momentum and the total momentum of
the event was within the region 
$\theta (\Sigma \vec p)\ >\ 30\dgree$.

To reduce the background from radiative return to the \Zn,
none of the clusters was allowed to have a total momentum 
($p^{\rm JET}$) above 70\% of the beam momentum,
the momentum of isolated photons had to be
less than 10\% of the beam momentum,
and the acoplanarity was required to be above $12\dgree$.
To reject \eeto \hspace{0.1cm}$\Zn/\gamma \rightarrow $ \tautau\
events where the decay of one $\tau$ yielded visible products with large
momentum, while the decay of the other $\tau$ yielded soft products,
the momentum transverse to the thrust axis was required to
exceed 0.7 \GeVc.

This selection was supplemented by cuts that depended on the region of
the (\mstau,$M_{\mathrm LSP}$) plane considered.
At any given point in the plane,
it was demanded that the accepted events in that point
were kinematically compatible with the corresponding signal:
the maximal momentum the
$\tau$ can have in the lab frame, neglecting  m$_{\tau}$, is
$
p_{max} = 
  \frac{\sqrt{s}}{4} 
    \left ( 1 - \left [ {M_{\mathrm LSP}}/{\mstau} \right ] ^2  \right )
    \left ( 1 + \sqrt{ 1 - 4{\mstau}^2/{s} } \right )
$,
which is also the end-point of the spectrum of the visible
$\tau$ - jet momentum. 
Hence $p^{\rm JET}_{high}$ (the larger of
the two jet momenta) was required to be less than $p_{max}$.   

For large \DM = $\mstau - M_{\mathrm LSP}$,
the remaining background from two-photon events can be
removed by 
requiring large
transverse momentum imbalance.
At smaller values of \DM\, such a cut would greatly reduce the
signal detection efficiency.
Therefore, if \DM\ was below (above) 20 \GeVcc\ the
total missing transverse momentum
(\Ptmis)
had to exceed 0.8 (1.2) times the maximum
transverse momentum a $\gamma \gamma$ event could have
without one of the beam-remnant electrons being deflected into
the STIC ($i.e.$ by an angle greater than $\theta_{max}$ = $1.82\dgree$).
This limit depended on the centre-of-mass energy:
$p_T^{\rm lim} =  \sqrt{s} \sin \theta_{max} / ( 1 + \sin \theta_{max} ) =
 0.031 \ \sqrt{s} $~\footnote{
The missing transverse momentum was estimated in four different 
ways: from the transverse momentum of the two jets,
as that of all particles passing the quality cuts,
as that of all reconstructed particles except those
identified as bremsstrahlung photons,
and as that of all reconstructed particles.
The cut was applied on the smallest of these.
This gave
stability against possible errors in reconstruction 
and against the presence of noise or cosmics.
}.
In addition, in the region with \DM\ below 20 \GeVcc\ 
there should be no calorimetric energy 
below $30\dgree$, 
and $\theta (\Sigma \vec p)$ had to exceed   $45\dgree$.

At this stage of the analysis,
the dominant remaining background at
large \DM\
were \WW\ events.
Only events with both W bosons decaying leptonically
were still present.
Electrons and muons in \WW\ events might come either directly from
the W-decay, or indirectly from $\tau$-decays.
In the former case, which is dominant, 
the momentum of the detected lepton tends to
be higher than in the signal, where all electrons and muons
would be indirect. In the latter case, the momentum spectrum of
background and signal events is similar.

In order to reduce the background from \WW\ events with leptons from direct W decays,
it was
demanded that the highest momentum of any 
tightly or loosely identified lepton in the
event was less than $(0.1 \DM\ +  0.6) P^{lept}_{W min}$,
where  $P^{lept}_{W min}$ is the lowest momentum a lepton
in the decay $W \rightarrow \ell \nu$ can obtain in the lab-frame if the
W is on-shell.
In the region of large \DM, it was also demanded that
there be no more than one
tightly identified electron or muon in the event.

To suppress further the \WW\ background,
and in particular the component with leptons from indirect decays, 
the events were analysed as if
they were indeed \WW\ events.
The $\theta$ angle of the
W yielding the cluster with positive charge was estimated ($\theta_{W^+}$)
and selections in
the 
($\theta_{W^+}$,$p^{\rm JET}_{high}$)-plane  
were used to discriminate
the signal from this background~\footnote{At this 
stage of the analysis about  99 \% of the events 
had a cluster with charge equal to +1 (this selection
was not aplied to events where there was no cluster with positive charge).}.
The $\theta_{W^+}$ was estimated as follows.
If neither W decayed to a tau,
$\theta_{W^+}$ can be calculated exactly
(albeit usually with two-fold ambiguity).
The W decay to a muon
or an electron
could be distinguished from the decay to the $\tau$  by the presence of
a single track cluster with 
at least a loose lepton identification
and a momentum above the lowest possible
momentum for the charged lepton from the
decay of an on-shell W.
Decays that did not fulfil these requirements
were assumed to be decays to $\tau$.
If one of the  W decayed to a $\tau$
the solution for $\theta_{W^+}$ is approximate, 
since  the momentum of the $\tau$ is unknown.
In this case, the momentum of the initial $\tau$
was estimated as the average momentum of taus from
W decay, calculated for tau momenta above the
measured jet momentum.
The direction of the momentum
of the tau was taken
as the measured jet direction.

The W pair-production
process tends to have higher
$p^{\rm JET}_{high}$ and
$\theta_{W^+}$.
In addition, $p^{\rm JET}_{high}$ and $\theta_{W^+}$ are correlated 
in such processes, while they are independent in \stau\ production.
Thus the signal selection criteria in 
the 
($\theta_{W^+}$,$p^{\rm JET}_{high}$)-plane were:  
$$
\theta_{W^+} < 
\left \{ 
  \begin{array} { l l }
    \mbox{$ -3.0 (x - 0.325) + A $} & \mbox{${\rm if\ }  x < 0.325 $} \\
    \mbox{$ A $}            & \mbox{${\rm if\ } 0.325 \le x < 0.52 $} \\
    \mbox{$-2.1 (x - 0.52) + A $}   & \mbox{${\rm if\ } x \ge 0.52$} \\
  \end{array} 
\right.
$$
where $\theta_{W^+}$ is in radians and $x = p^{\rm JET}_{high} / p_{beam}$, 
and $A$ is a constant
chosen to be 1.6 (2.1) for \DM\ below (above) 20 \GeVcc.
The boundary of the selected region  
in the 
($\theta_{W^+}$,$p^{\rm JET}_{high}$)-plane 
thus defined 
closely follows a curve of
constant ratio between the probability density functions for signal and
background (the likelihood ratio).

~\\
\noindent
{\underline{\it Search for light staus without coupling to the \Zn\ }}\\

\noindent
A light stau 
can be excluded 
to a large extent 
using LEP1 results,
as discussed in section~\ref{sec:lep1}. 
This is however not possible 
when 
the coupling to the \Zn\ vanishes 
(the stau mixing angle gives the minimum cross-section). 
The large mass
analysis described in the previous section looses its efficiency for stau
masses below 15~\GeVcc. This is mainly due to the
fact that the staus are highly boosted at such small masses, 
failing the acoplanarity cut.


Therefore a specific search was required for small \mstau\ 
at the minimal cross-section mixing angle.    
Two search regions were identified:
one optimised for very small masses, \mstau\ below 10 \GeVcc,
referred to as the ``very small mass analysis'',
and one for larger masses, optimized
for  
\mstau\ above  10 \GeVcc and 
\DM\ between $m_{\tau}$ and  4 \GeVcc\ referred to as
the ``small mass analysis''
(the large mass analysis shows good sensitivity for
\DM\ above 4 \GeVcc\ down to \mstau\ = 15 \GeVcc). 

The signal events in the relevant kinematic regions are characterised
by containing two taus at large angles to the beam,
and being softer than two-fermion events but slightly
harder than $\gamma \gamma$ events. Due to the sizable
boost of the staus, the two jets tend to be rather
back-to-back. 
Two-tau events were selected as described in the previous section,
with the additional requirement that the topology was either 
1-prong and 3-prong, or two 1-prongs. In the latter case,
there should not be two tightly identified leptons of the same flavour
in the event.
To select central events, the polar angle of the most energetic particle 
in each detector hemisphere had to be
above 50\dgree,
and the sine of the polar angle of each jet had to be above 0.8.
The acollinearity was required to be 
above 0.4\dgree,
$p^{\rm JET}_{max}/p_{beam}$ had to be below
90\%, and
the total reconstructed mass of the visible system had to be above 4.5 \GeVcc. 
There had to be no energy in a 30\dgree\ cone around the beam axis.

In the small mass region,
$\theta (\Sigma \vec p)$ had to be above 55\dgree\ in polar angle,
the visible mass had to be below
$\left [ 15 (\DM - m_{\tau})-(\mstau - 25)/3-(\DM - m_{\tau})(\mstau - 25)/2+15 \right ]$  \GeVcc,
and 
\Ptmis\ 
had to exceed max(0.05, (\DM $-$ $m_{\tau}$)/4) $p_T^{\rm lim}$.

In the very small mass region,
it was required that 
\Ptmis\ 
was above 0.01$p_T^{\rm lim}$,
$p^{\rm JET}_{max}/p_{beam}$ was above
15\%, and $\theta (\Sigma \vec p)$ was above 15\dgree\ to
the beam.
The acollinearity had to be below 15\dgree.

\subsection{Squark searches}
\label{sec:squarkana}


Supersymmetric partners of top and bottom quarks were searched for.
The data 
collected at centre-of-mass energies from 189 to 208~\GeV\ were analysed. 
The dominant 
decays of the stop and sbottom squarks are assumed to be 
$\stqone \to \mathrm{c}\XN{1}$ and $\sbqone \to \mathrm{b}\XN{1}$, 
respectively, and the final topology is two
acoplanar jets and missing energy. In the non-degenerate scenario (\DM\ $>$ 10~\GeVcc), 
the neural network analysis has already been presented in~\cite{squa}. 
This analysis has been extended down to \DM\ = 5~\GeVcc.
In addition, a new analysis based on a sequential cut approach has been developed 
to search for stops nearly degenerate in mass with the LSP, investigating \DM\ values between
2 and 10 \GeVcc. Moreover, this analysis has been extended to \DM\ values up to 20 \GeVcc\
in order to cross-check the non-degenerate analysis.

The track selection and the general event preselection described in 
sections \ref{sec:trksel} and  \ref{sec:presel}, respectively, were used.


\subsubsection{Non-degenerate scenarios}
\label{sec:squarkana1}

To select hadronic events,
the number of charged particles reconstructed with TPC information
was required to be
greater than three, and  the energy in the STIC to be 
less than 70\% of the detected
energy. The polar angle of the thrust axis had to be above
20$^\circ$.
The following event quality cuts were then applied.
The percentage of good tracks, the ratio of
the number of charged particle tracks after 
the particle selection 
to the number before, 
had to be greater than 35\%. In addition, the scalar sum of 
charged particle momenta reconstructed with TPC information 
was required to be 
greater than 55\% of the total energy in the event, and the total
number of charged particles to be greater than six. 

To remove radiative return events,
the energy of the most energetic neutral particle was required to be less 
than 40~\GeV.
Additional cuts were then applied to restrict the selection to events with missing
energy. The transverse missing momentum had to be greater than 4~\GeVc, the polar 
angle of the missing momentum had to 
be above 20$^\circ$  
and the energy in a $40^\circ$ cone around the z axis
was required to be less than 40\% of the total detected event energy. 
Finally, the visible mass
of the events was required to be less than 95~\GeVcc. 

\begin{figure}[p!]
\begin{center}
\begin{tabular}{cc}
(a) & (b) \\
\includegraphics[width=7cm]{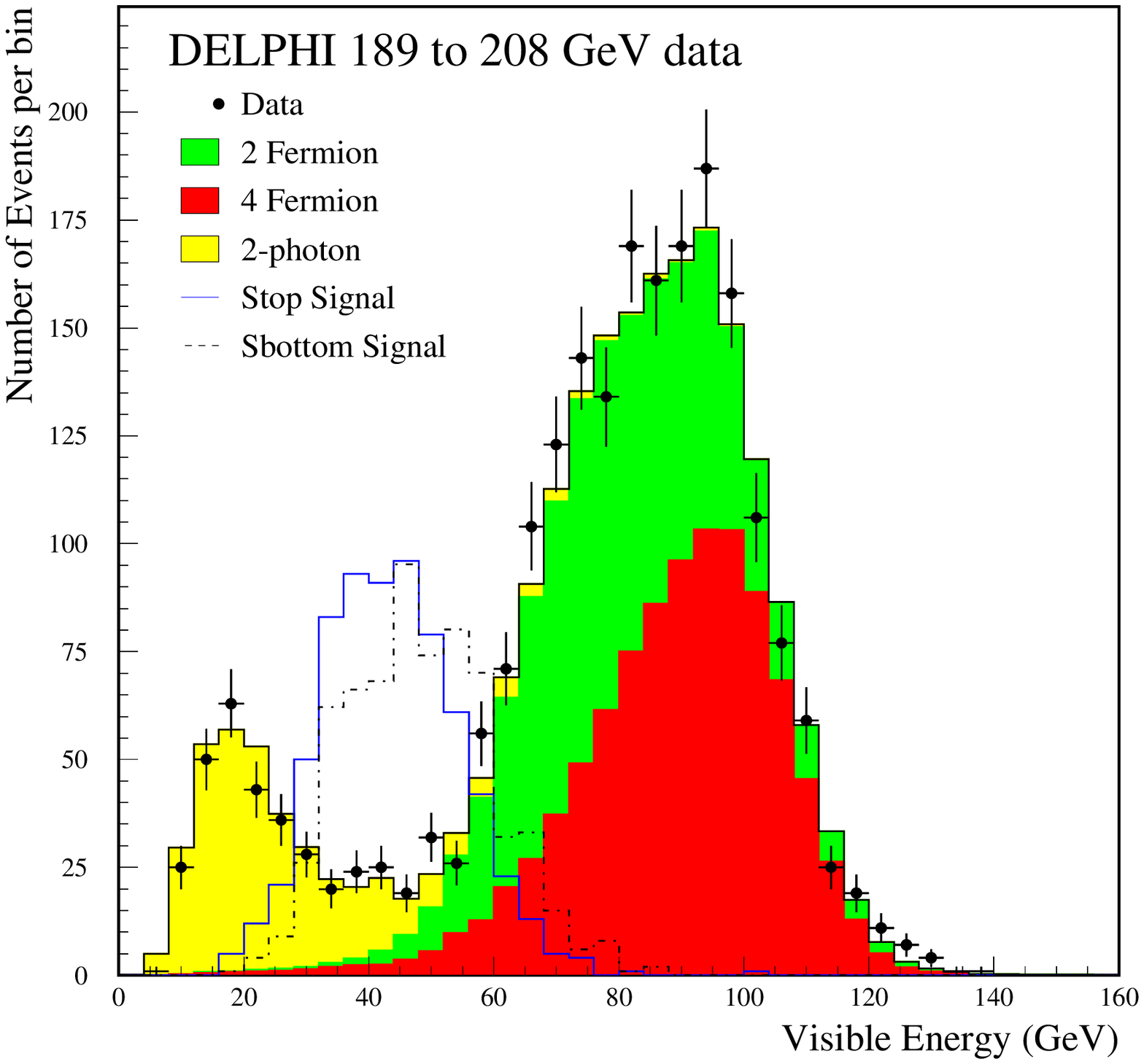} &
\includegraphics[width=7cm]{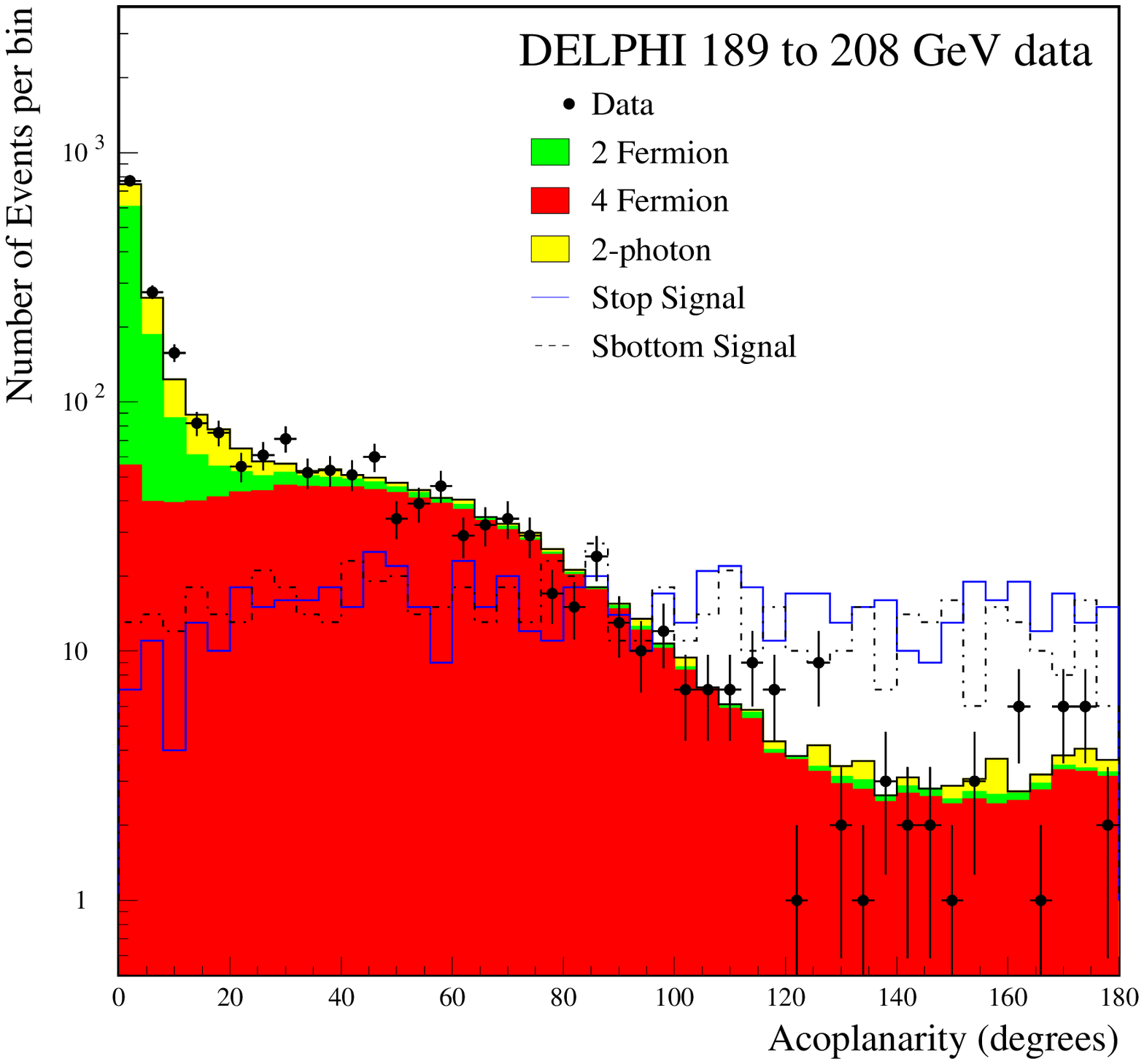} \\
(c) & (d) \\
\includegraphics[width=7cm]{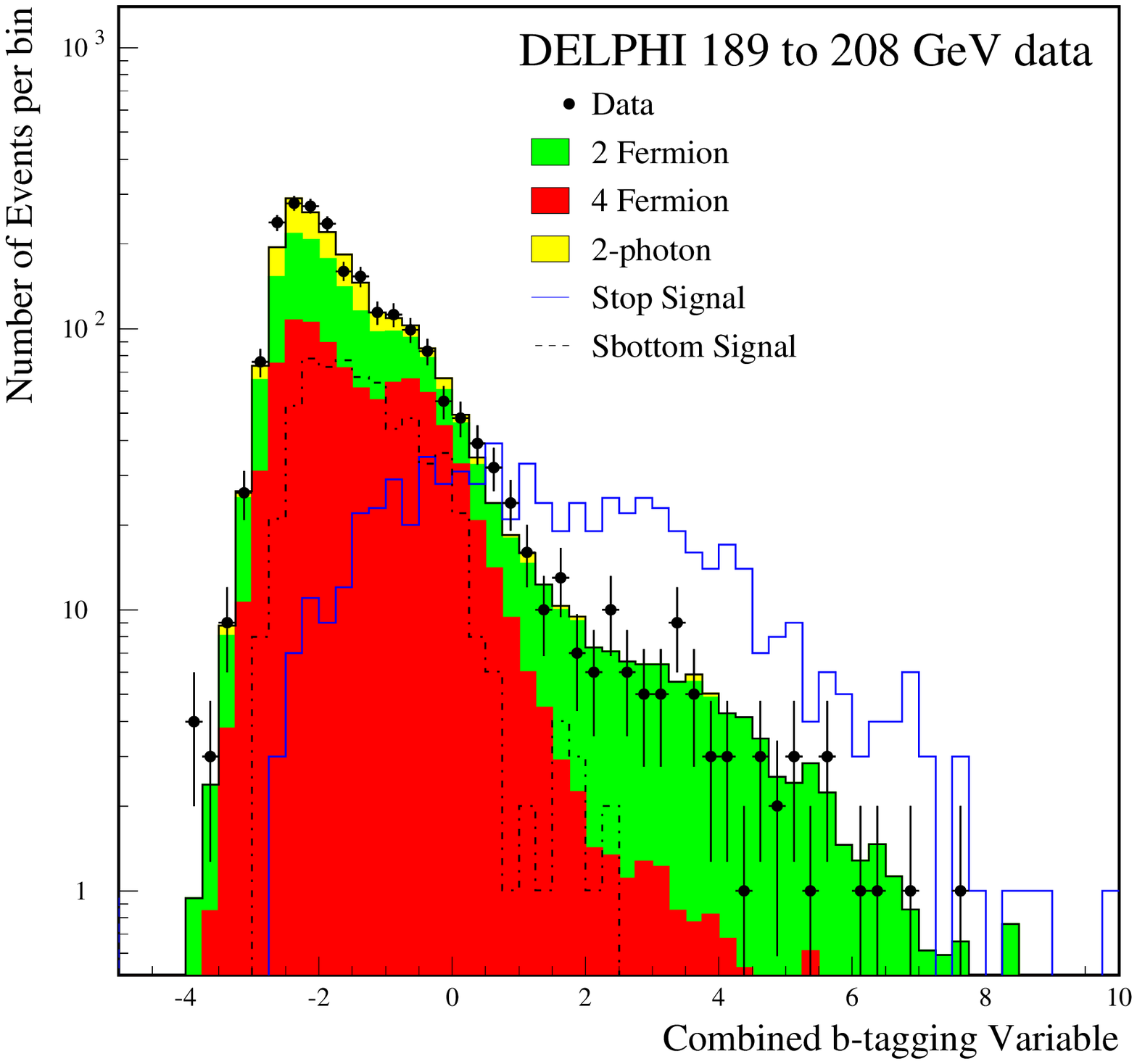} &
\includegraphics[width=7cm]{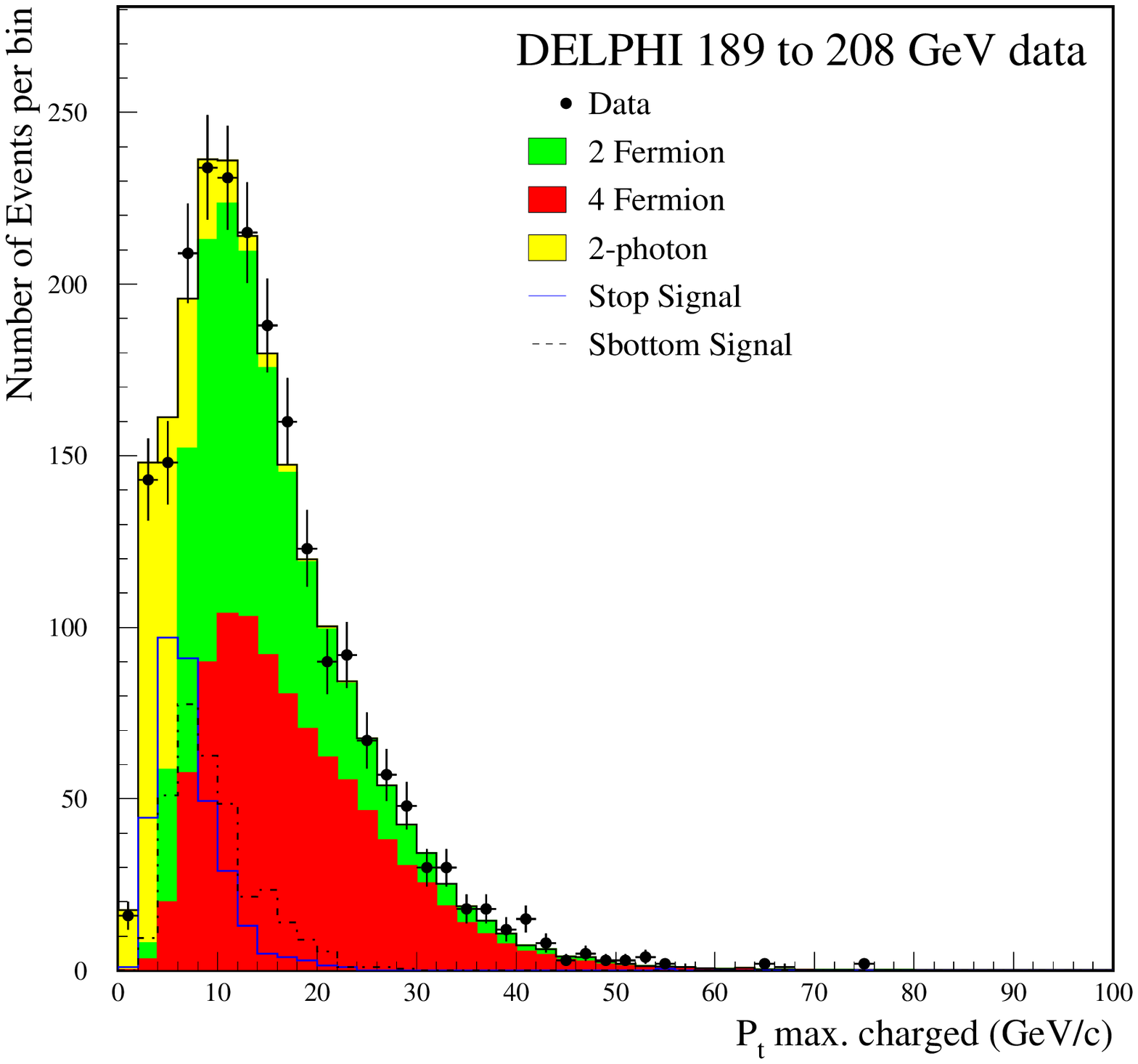} \\
\end{tabular}
\caption{Comparison of data and simulation at the preselection level in the
non-degenerate squark analysis. Plots include all DELPHI data from 189 to 208
GeV: (a) visible energy, (b) acoplanarity, (c) combined b-tagging variable, (d)
maximal transverse momentum of a charged particle. The expected
signal distributions at $\sqrt{s}=200~\GeV$ are shown for one possible
stop and sbottom signal 
($ M_{\tilde {\mathrm q}}$=90~\GeVcc, \MXN1\ = 60~\GeVcc), with arbitrary
normalisation.}
\label{fi:mcsquark}
\end{center}
\end{figure}

The number of 
events selected by this preselection 
was 2178 for 2143$\pm$8 expected (combined data from 
$\sqrt{s}=$189 to 208~\GeV).
Figure~\ref{fi:mcsquark} shows a comparison between data and simulated events.
At this level, for \DM\ $>$ 10~\GeVcc, stop signal efficiencies
ranged from 20\% to 
70\% depending on the mass difference between the stop and the neutralino.
Sbottom efficiencies were quite similar except at low \DM\ where, for
example, the efficiency for \msbqone=90~\GeVcc\ and \MXN{1}=85~\GeVcc\
was close to zero,
because the $b$ quarks are produced almost  
at rest.

The final selection of events was performed using neural network techniques
(see section \ref{subsubsec:analysistech}).
Separate searches were made for two different ranges of \DM: \DM\ $>$ 20~\GeVcc\
and 5 $<$ \DM\ $\leq$ 20~\GeVcc. 
Events were forced into 
two jets using the Durham algorithm.
The neural network structure was as follows.
There were ten input nodes (variables), 
ten hidden nodes (in one layer) and three output nodes.
The ten input variables were:
the ratio between the transverse missing momentum and the visible energy, the transverse
energy, the visible mass, the softness defined as 
$M_{\mathrm jet1}/E_{\mathrm jet1}+M_{\mathrm jet2}/E_{\mathrm jet2}$, the acollinearity, the quadratic sum of
the transverse momenta of the jets 
$\sqrt{(P_{t}^{\mathrm jet1})^2+(P_t^{\mathrm jet2})^2}$, the acoplanarity, 
the sum of the first and third Fox-Wolfram moments, the polar angle of the 
missing momentum and finally the combined b-tagging event probability. 
For each \DM\ window a neural network 
with three ouput nodes was trained to discriminate 
the signal from the combined two-fermion and four-fermion backgrounds, and 
from the $\gamma\gamma$ interactions leading to hadronic final states.

\begin{figure}[p!]
\begin{center}
\begin{tabular}{cc}
 (a) & (b) \\
\includegraphics[width=7cm]{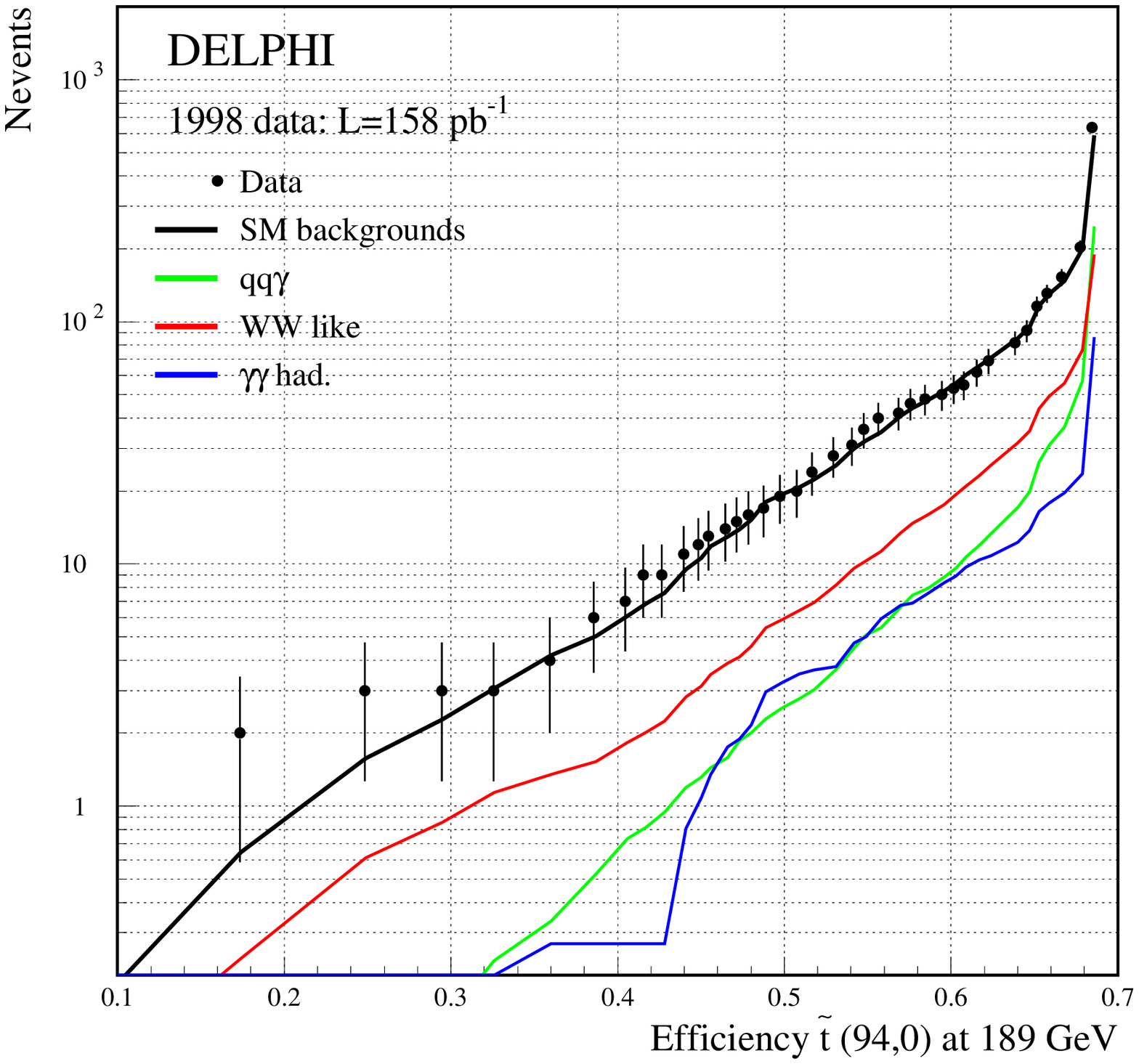} &
\includegraphics[width=7cm]{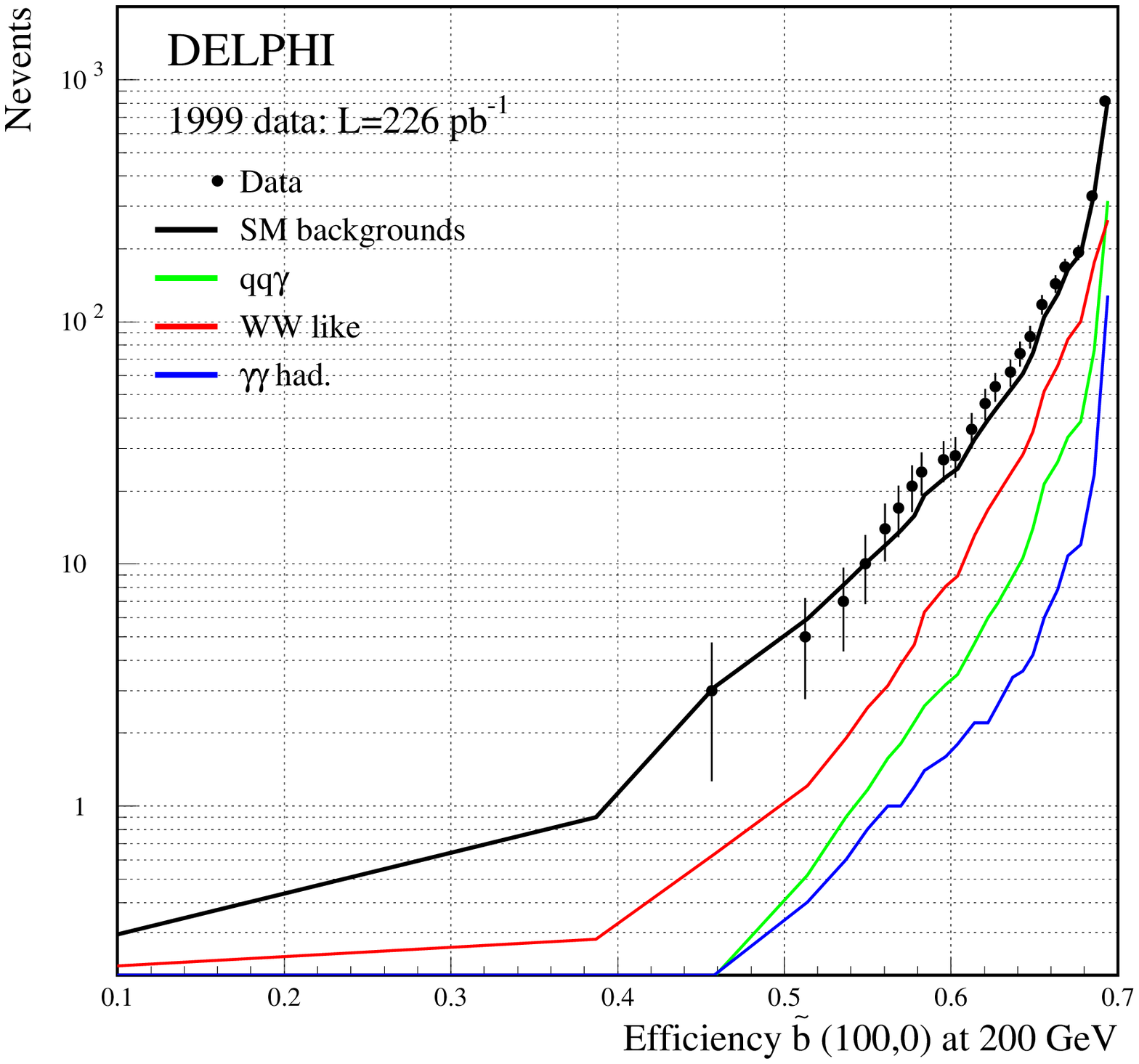} \\
 (c) & (d) \\
\includegraphics[width=7cm]{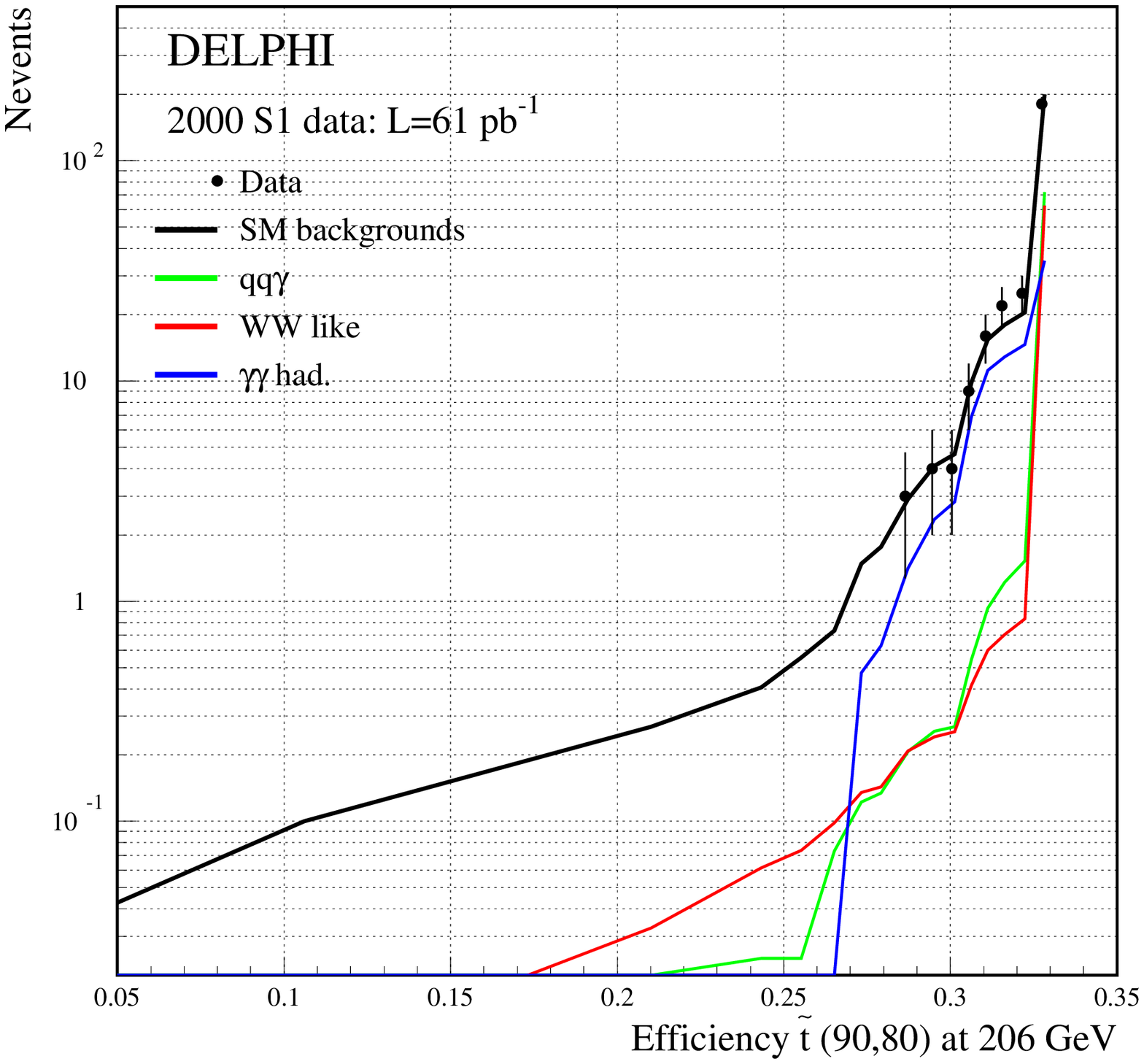} &
\includegraphics[width=7cm]{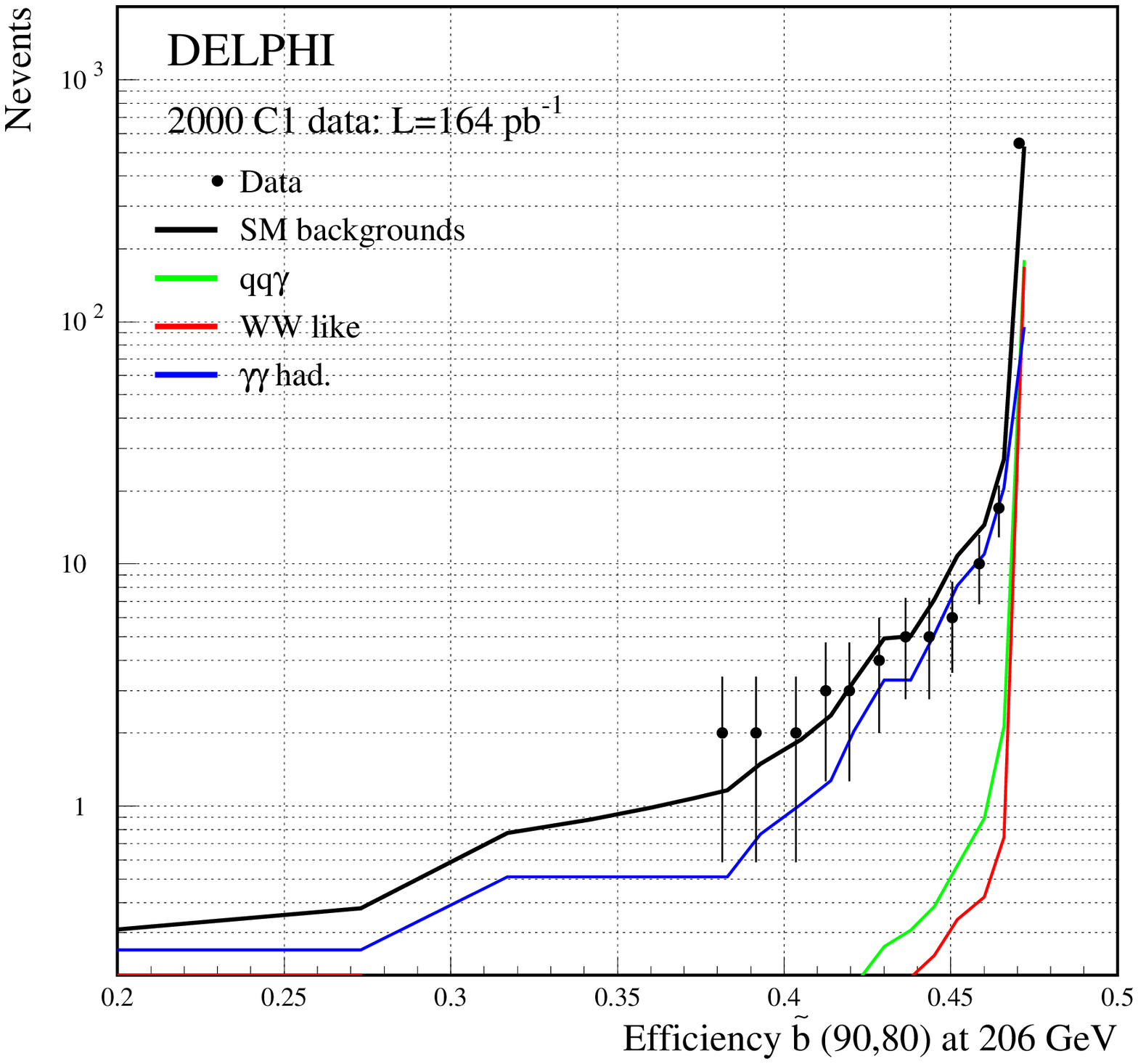} \\
\end{tabular}
\caption{Number of events as a function of the signal detection efficiencies
in the non-degenerate squark analysis:
(a) 1998 data at 189~\GeV: stop analysis for $\DM >20~\GeVcc$,
(b) 1999 data from 192 to 202~\GeV: sbottom analysis for $\DM >20~\GeVcc$,
(c) 2000 data with TPC sector 6 off: 
stop analysis for $5 \leq \DM \leq 20~\GeVcc$,
(d) 2000 data with TPC sector 6 on: sbottom analysis for $5 \leq \DM \leq 20~\GeVcc$. The efficiencies are for a given combination of  squark and LSP masses,
indicated in the parentheses. }
\label{fi:sqnneffi}
\end{center}
\end{figure}

Although the three output nodes proved useful in training the network,
the selection was made according to the output of the signal node only.
Figure~\ref{fi:sqnneffi} shows the 
number of events as a function of the signal efficiency for the two mass 
analysis windows of the stop and the sbottom searches. The number of 
events in the data is in agreement with the SM 
background predictions over the full range of neural network outputs.
The optimisation of the final cuts was performed by 
minimising the 
confidence level of the signal hypothesis expected in the absence of a signal.
~\cite{alexread}.


\subsubsection{Nearly mass degenerate scenarios}
\label{sec:squarkana2}

%
%

Due to the large Yukawa coupling (see section \ref{sec:phenomenology}),
the stop ($\tilde{t_1}$) and the sbottom  ($\tilde{b_1}$) can be light
and nearly degenerate in mass with the LSP.
The effective coupling of the stop to charm and neutralino results
from loops and is thus  small. In addition,
the width of the decay $\stq \to c\XN{1}$ is proportional to 
$\mstq (1 - \MXN1 ^2/ \mstq ^2)^2$,     
and therefore proportional to \DM. So
if \DM\ gets small enough, 
the stop acquires a sizable lifetime 
and may form a quasi-stable (decaying inside the tracking volume) or 
even stable
stop hadron (see~\cite{stablest} for this case). 
The current analysis focusses on a stop decaying promptly into a charm 
particle and the LSP. 

%
%
%




The event preselection required, in addition to the criteria described in section~\ref{sec:presel},
that not more than 30\% of the total visible
energy was carried by particles with tracks seen in the VD and ID only.
To eliminate Bhabhas and leptonic $\gamma\gamma$ backgrounds, the charged multiplicity 
was required to be 
greater than five.
The $\gamma\gamma$ background was further suppressed by requiring the 
energy in a forward cone of $30^\circ$ around the beam direction 
to be at most 40\% of the total visible energy and smaller than 2 GeV, 
and that no energy was deposited in the STIC calorimeter.
To avoid the relatively low hadronic energy region, where the $\gamma\gamma$ background
is  not well reproduced by the MC, 
the total transverse charged energy
was required to  be greater than 7 GeV, the total transverse energy of tracks
reconstructed with TPC information 
and the total transverse momentum had to be greater than 4 GeV and 
3.5~\GeVc, 
respectively, and
the number of tracks with TPC information had to be at least four. 



\begin{figure}[p!]
\begin{center}
\begin{tabular}{cc}
(a) & (b) \\ 
\includegraphics[width=6.5cm]{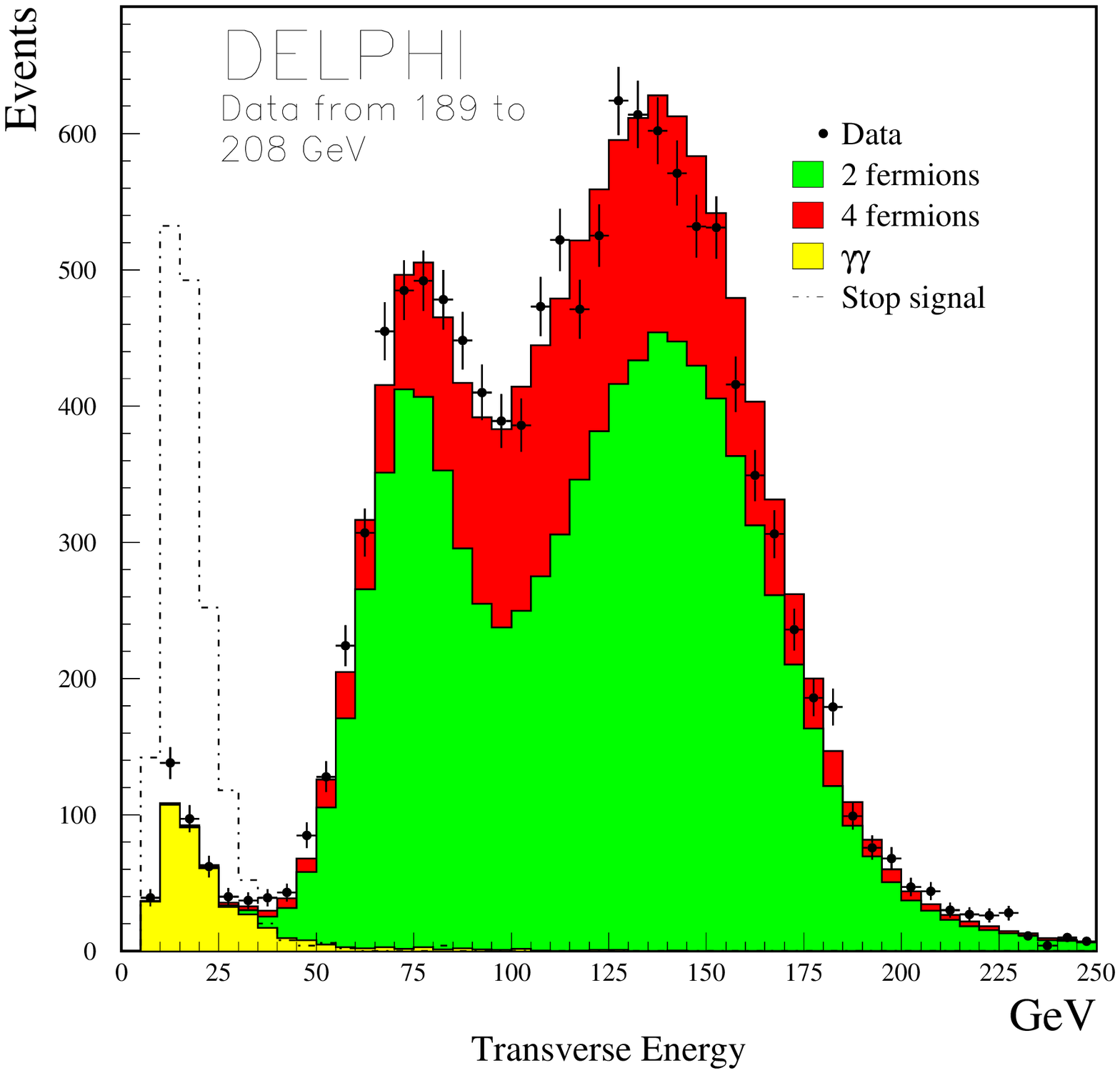} &
\includegraphics[width=6.5cm]{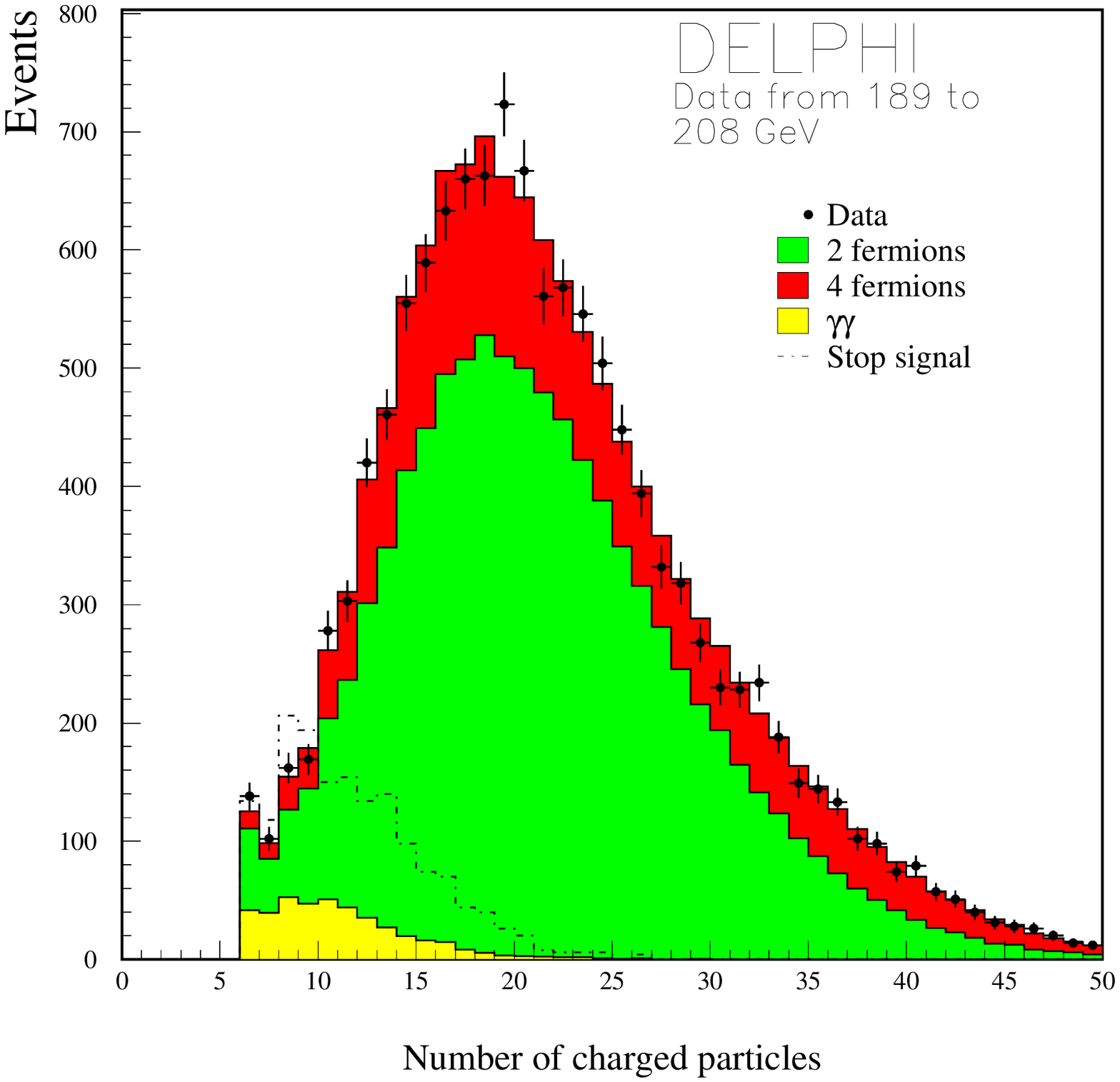} \\
(c) & (d) \\
\includegraphics[width=6.5cm]{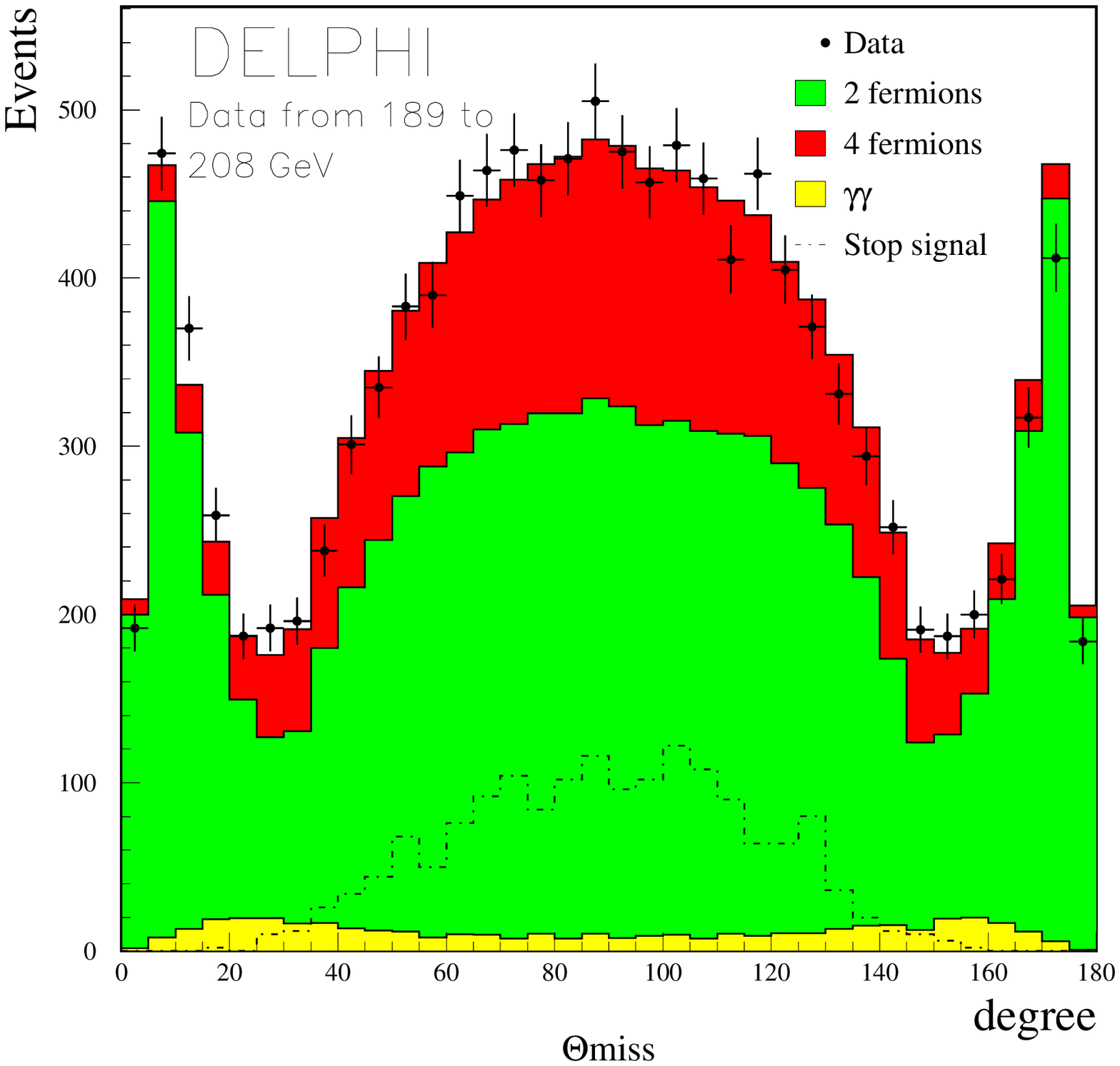} &
\includegraphics[width=6.5cm]{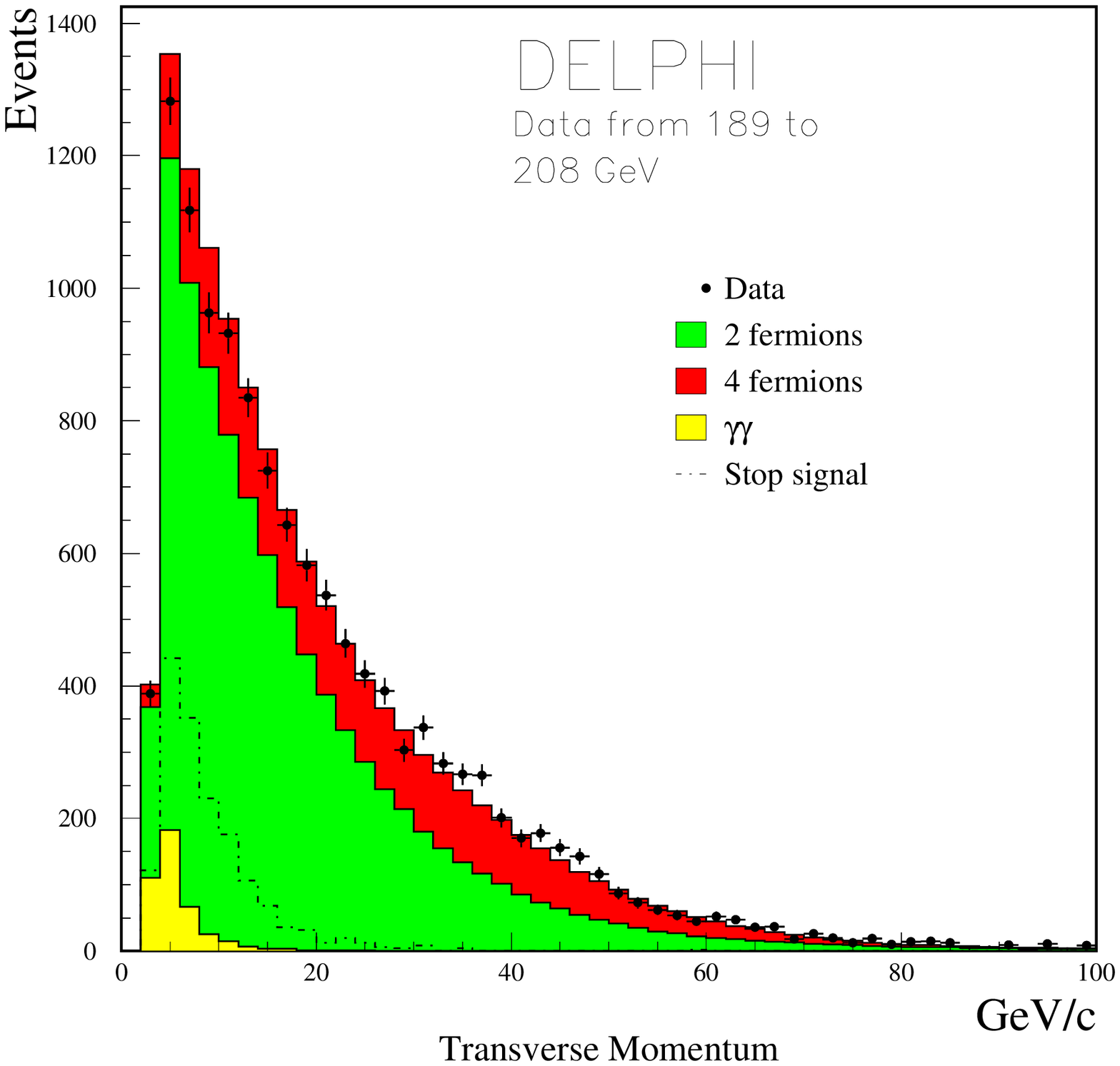} \\
(e) & (f) \\
\includegraphics[width=6.5cm]{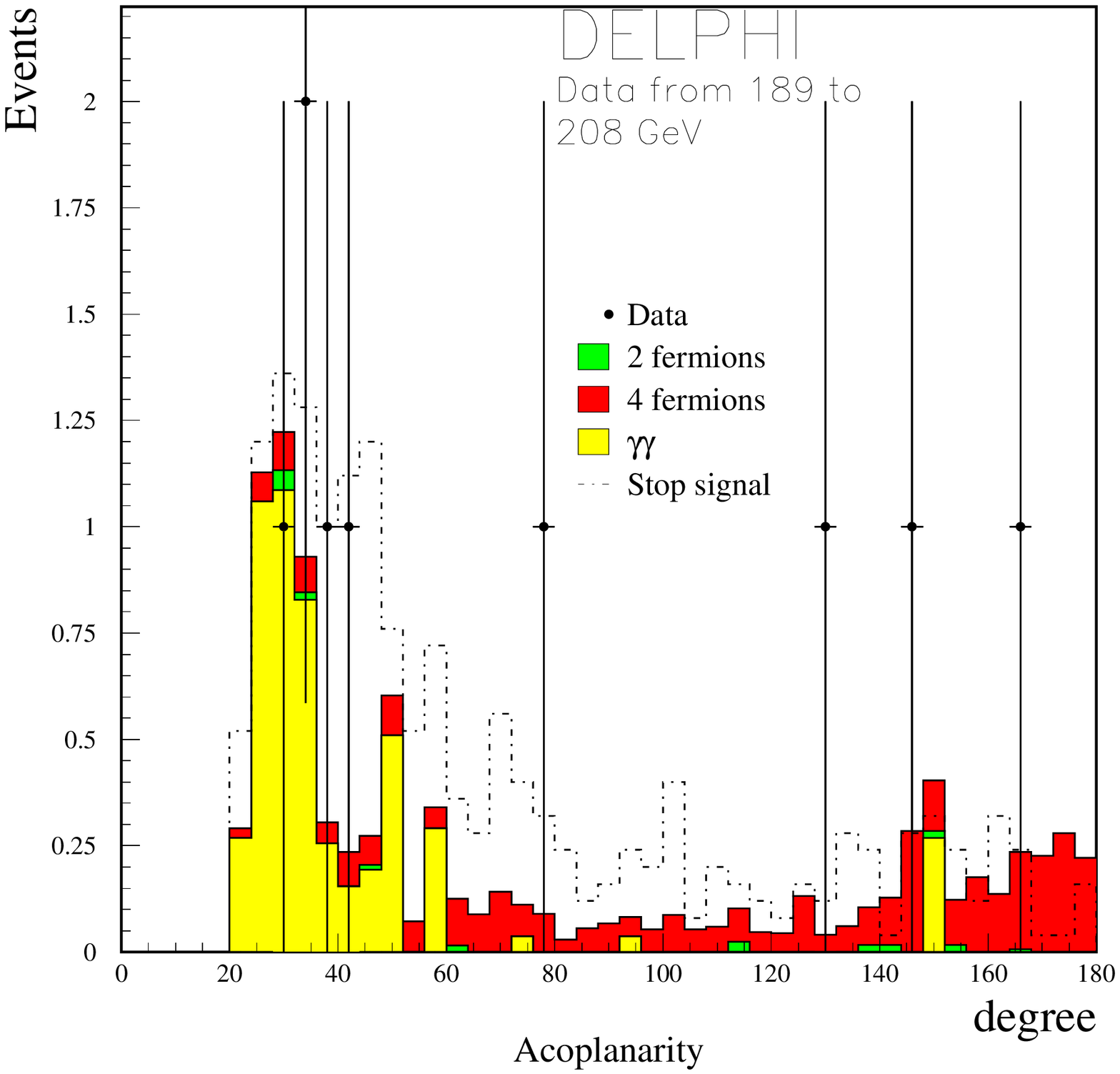} &
\includegraphics[width=6.5cm]{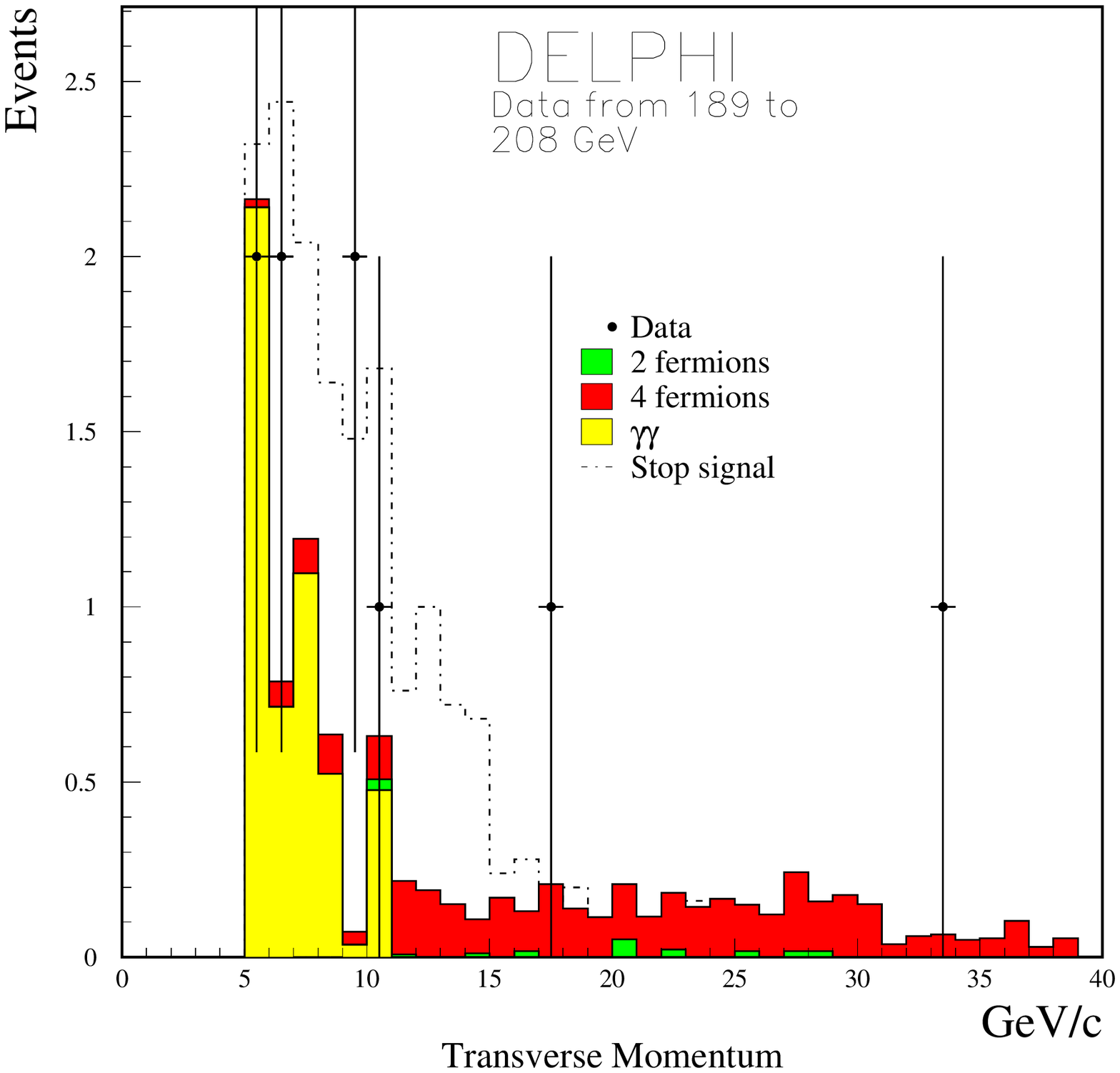} \\
\end{tabular}
\caption{Comparison of data and simulation in the 
nearly degenerate squark analysis at preselection and final selection 
level. The distributions of (a) the total 
transverse energy, (b) total number of charged particles, 
(c) angle between the missing momentum and the beam axis and 
(d) transverse momentum 
are shown at preselection level. At the final selection
level the distributions of (e) the acoplanarity and (f) the transverse
momentum are shown. 
All DELPHI data from 189 to 208~GeV is included.
A signal corresponding to 
\mstq\ =60~\GeVcc\ and  \DM\ =4~\GeVcc\ is also shown
(with arbitrary normalisation at preselection level and normalised to
a cross-section of 0.16 pb in (e) and (f)).
}
\label{fi:datamc-dg}
\end{center}
\end{figure}

The agreement between data and MC after this preselection is shown in 
figure~\ref{fi:datamc-dg}(a)~to~\ref{fi:datamc-dg}(d).
Figure~\ref{fi:datamc-dg} also demonstrates that the two-fermion and
four-fermion backgrounds dominate at this stage of the selection. 
The smaller contribution from two-photon interactions 
can  be reduced at this level using for example the different  
polar angle distribution with respect to the signal.

A further selection was performed in order to reduce the remaining backgrounds.
Events having mainly barrel activity were selected. This was achieved 
by requiring that the energy within a cone of $60^\circ$ around the beam direction
was less than 10 GeV and 
that the polar angle of the missing momentum
was above 45\dgree.
Most of the remaining two- and four-fermion
background was rejected 
by demanding that the transverse momentum of
the most energetic particle was less than 10 GeV/c and that the total
transverse energy was less than 40 GeV.
Finally, the total transverse momentum was required to be greater than 5 GeV/c and 
the scaled acoplanarity~\footnote{The scaled acoplanarity is the acoplanarity of the two jets 
multiplied by the sine of the minimum angle between a jet and the beam axis.}
greater than 20\dgree. This cut removed 
most of the remaining background from two-photon processes.     
The agreement between data and simulation after this selection is shown in 
figure~\ref{fi:datamc-dg}(e,f).


\subsection{Chargino searches}
\subsubsection{Non-degenerate scenarios}
\label{sub:CHASEL}

The search for charginos in the non-degenerate scenarios
covers the case when the mass difference  
\DM = \MXC{1}$-$ \MXN{1}\ is above 3 \GeVcc .
In order to take all possible signatures of chargino decays into account,
events were 
divided into four mutually exclusive topologies:
\begin{itemize}
\item{the \ll\ topology, with no more than five charged particles and no isolated photons;}
\item{the \jjl\ topology, with more than five charged particles, at least one isolated lepton and no 
isolated photons;}
\item{the \jjjj\ topology, with more than five charged particles and no isolated photons or leptons;}
\item{the \rad\ topology, with at least one isolated photon.}
\end{itemize}
The signal events selected in a given topology are mostly events from the corresponding decay
channel, but events from other channels may also contribute. For instance, for low \DM\
(and thus low visible energy) 
some events with hadronic decays are selected in the \ll\ topology, and some mixed decay  events 
with the isolated lepton unidentified enter into the {\it jets} topology. This migration effect tends to 
disappear as \DM\ increases. This effect was taken into account in the final efficiency and limit 
computations.

The signal events were simulated using 132 combinations of \XPM{1} and \XN{1} masses for nine 
chargino mass values 
(\MXC{1}~$\approx$~103, 102, 100, 98, 94, 85, 70, 50  and 45~\GeVcc) and with \DM\ ranging 
from 3~\GeVcc\ to 80~\GeVcc. 
A total of 264000 chargino events (2000 per mass combination) was generated. The kinematic 
observables (acoplanarity\footnote{To compute the  acoplanarity and acollinearity the particles 
were forced into two jets by the {\tt DURHAM} algorithm.}, \Evis, \Ptmis, etc.) of the signal events were studied 
in terms of their mean value and standard deviation, and six \DM\ regions were defined, each 
containing signal events with similar properties (see table  \ref{tab:REGION}). 

In each of these 24 windows (four topologies, six \DM\ regions), a likelihood ratio function 
($\mathcal{L}_{R}$, see section~\ref{subsubsec:analysistech}) was defined.
The variables $\left\{ x_{i}\right\} $ used to build 
the $\mathcal{L}_{R}$ functions in the present analysis 
were \cite{char}: the visible energy (\Evis), visible mass (\Mvis), missing transverse momentum (\Ptmis),
polar angle of the missing momentum, number of charged particles, total number of particles, 
acoplanarity, acollinearity, ratio of electromagnetic energy to total detected energy, 
percentage of total energy 
within 30$^\circ$ of the beam axis, kinematic information concerning the isolated photons and leptons and 
the two most energetic charged particles, and finally the jet characteristics.\\ 

\begin{figure}[ht]
\begin{center}
\begin{tabular}{cc}
\hspace*{-1.3cm}
\begin{minipage}[c]{8.0cm}
\epsfxsize=9.3cm  
\epsffile{plotb2061jj_108_108_n_151_p.eps_v5}
\end{minipage}
&
\hspace*{0.4cm}
\begin{minipage}[c]{8.0cm}
\epsfxsize=9.3cm  
\epsffile{plotb2061ll_103_103_n_3_p.eps_v5}
\end{minipage}
\\
\end{tabular}
\vspace{0.5cm}\\
\begin{tabular}{cc}
\hspace*{-0.9cm}
\begin{minipage}[c]{8.0cm}
\hspace*{-0.5cm}
\epsfxsize=9.3cm  
\epsffile{plotb2061gg_108_108_n_21_p.eps_v5}
\end{minipage}
&
\hspace*{0.01cm}
\begin{minipage}[c]{8.0cm}
\epsfxsize=9.3cm  
\epsffile{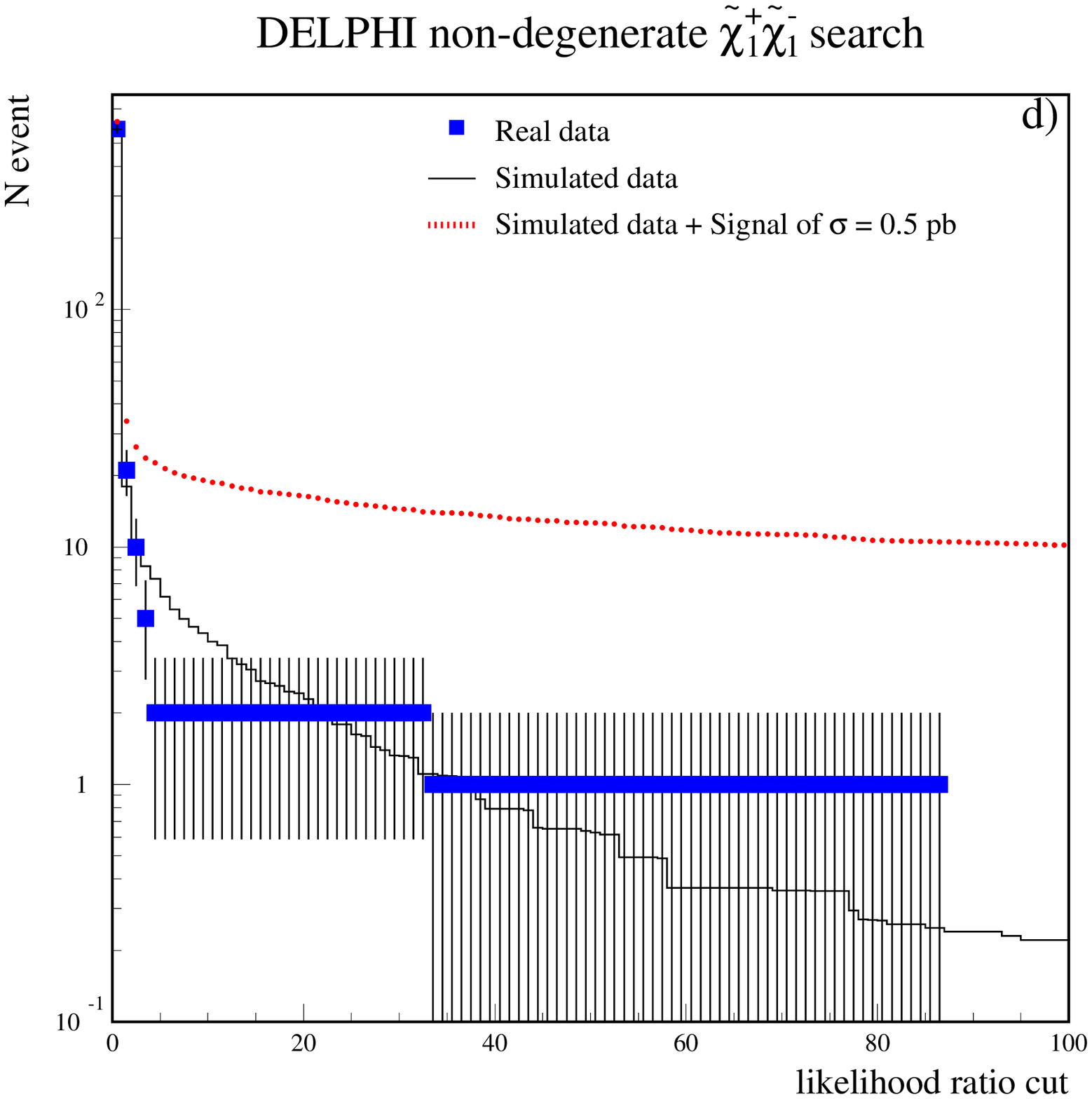}
\end{minipage}
\\
\end{tabular}
\vspace{0.5cm}
\caption[]{ a), b) and c) show comparisons between real data (dots) and simulated background events 
(histogram) for the $jets$, $\ell\ell$ 
and $rad$ topologies respectively, using a logical  {\tt OR} of the 6 preselection cut 
functions of the corresponding topology. The dashed lines indicate how a characteristic 
chargino signal would appear (arbitrary normalisation). d) shows the number of events selected by the 
standard chargino analysis as a function of the $\mathcal{L}_{R_{CUT}}$  cut in the \jjl\ topology for 25 $\leq$ \DM\ $<$ 35~\GeVcc.
 The squares are the data and the solid line is the background simulation. The dotted curve shows
a possible signal, \MXC{1}=102.8~\GeVcc\ \MXN{1}=73~\GeVcc, of 0.5 pb. In all cases the data collected in the year 2000 with the TPC sector 6 
on are shown.} 
\label{fig:DATAMC}
\end{center}
\end{figure}

\begin{figure}[ht]
\begin{center}
\begin{tabular}{cc}
\hspace*{-1.3cm}
\begin{minipage}[c]{8.0cm}
\epsfxsize=9.3cm  
\epsffile{plotb2071jj_108_108_n_151_p.eps_v5}
\end{minipage}
&
\hspace*{0.4cm}
\begin{minipage}[c]{8.0cm}
\epsfxsize=9.3cm  
\epsffile{plotb2071ll_103_103_n_3_p.eps_v5}
\end{minipage}
\\
\end{tabular}
\vspace{0.5cm}\\
\begin{tabular}{cc}
\hspace*{-0.9cm}
\begin{minipage}[c]{8.0cm}
\hspace*{-0.5cm}
\epsfxsize=9.3cm  
\epsffile{plotb2071gg_108_108_n_21_p.eps_v5}
\end{minipage}
&
\hspace*{0.01cm}
\begin{minipage}[c]{8.0cm}
\epsfxsize=9.3cm  
\epsffile{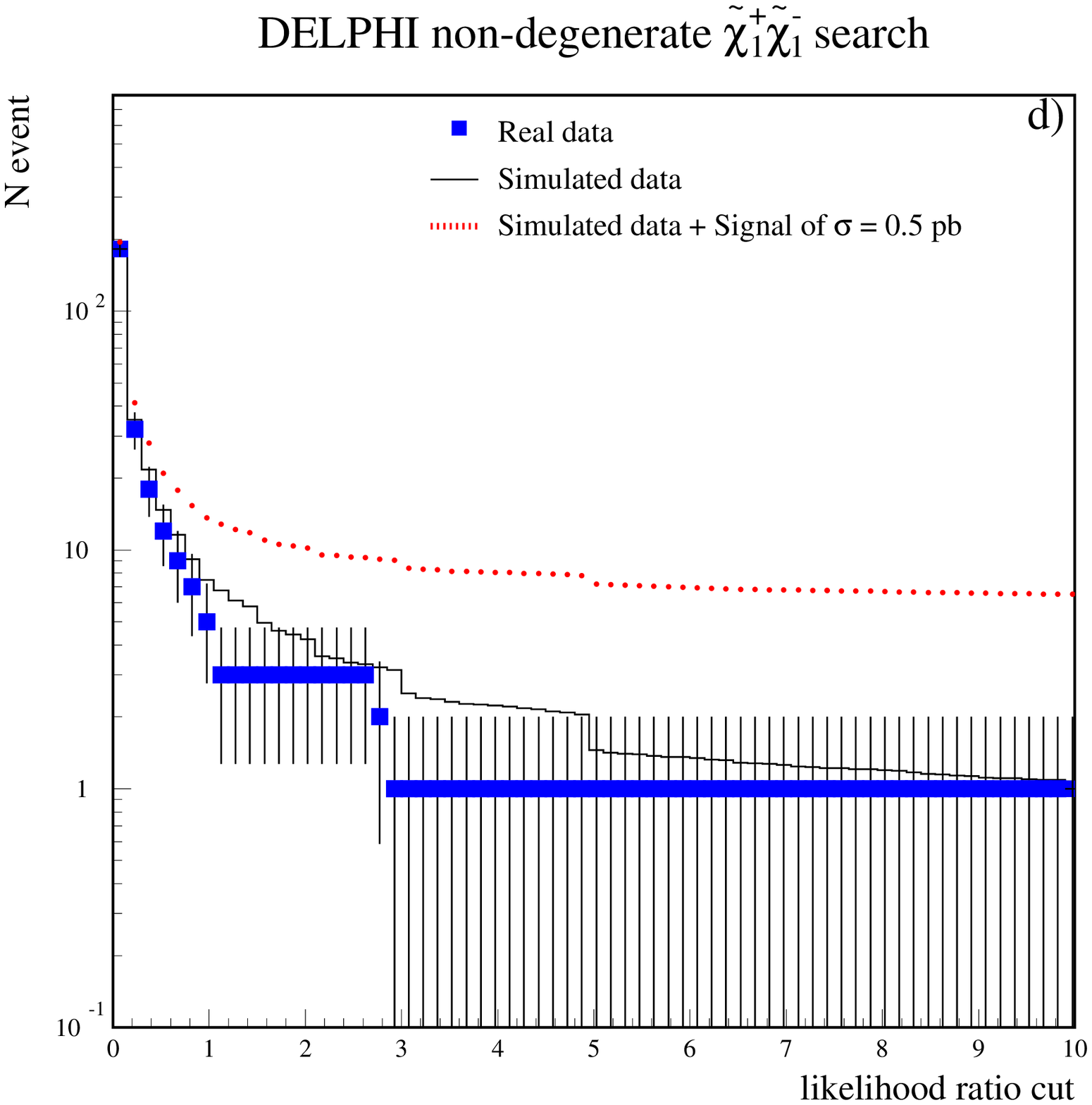}
\end{minipage}
\\
\end{tabular}
\vspace{0.5cm}
\caption[]{ a), b) and c) show comparisons between real data (dots) and simulated background events 
(histogram) for the $jets$, $\ell\ell$ 
and $rad$ topologies respectively, using a logical  {\tt OR} of the 6 preselection cut 
functions of the corresponding topology. The dashed lines indicate how a characteristic 
chargino signal would appear (arbitrary normalisation). d) shows the number of events selected by the 
standard chargino analysis as a function of the $\mathcal{L}_{R_{CUT}}$  cut in the \jjl\ topology for 25 $\leq$ \DM\ $<$ 35~\GeVcc.
 The squares are the data and the solid line is the background simulation. The dotted curve shows
a possible signal, \MXC{1}=102.8~\GeVcc\ \MXN{1}=73~\GeVcc, of 0.5 pb. In all cases the data collected in the year 2000 with the TPC sector 6 off  are shown.} 
\label{fig:DATAMCS1}
\end{center}
\end{figure}

The generation of these 24 likelihood ratio functions was performed as follows:

\begin{itemize}
\item{The signal distributions of all the variables  $\left\{ x_{i}\right\} $  were built 
with signal events generated with parameter sets giving rise to charginos and 
neutralinos with masses in the 
corresponding \DM\ region. For each \DM\ region the events were classified according to the
above topological cuts. The background distributions were built with background events passing the same topological cuts.}
\item{Preselection cuts\cite{chapre}, different for each \DM\ region, were applied in order to reduce the 
backgrounds with largest cross-section 
(two-photon interactions and Bhabha events) and to generate the pdf's.
The total background was reduced to 5\% of the one
passing the general event preselection (section \ref{sec:presel}).  
The pdf's were then generated as mentioned in section~\ref{subsubsec:analysistech}.} 
Figure~\ref{fig:DATAMC}(\ref{fig:DATAMCS1}) shows the distributions of some event variables
for the \jjjj, \ll\ and \rad\ topologies 
for the 2000 data and simulation with the TPC sector 6 on (off). 
\item{To reduce statistical fluctuations, a smoothing was performed 
by passing the 24 sets of pdf's for signal and background through a triangular filter~\cite{FILTER}.}
\item{In each window all the combinations of the pdf's were tested, starting from a minimal set of four variables. 
Every combination defined an $\mathcal{L}_{R}$  function 
and 
an $\mathcal{L}_{R_{CUT}}$,
as described in section~\ref{subsubsec:analysistech},
using 
the single channel formula~\cite{pdg96}. The parameters entering this computation were the 
number of expected background events and the window efficiency of the chargino selection,
defined  as the mean efficiency of the chargino-neutralino mass sets belonging to the investigated window~\footnote{The efficiency of one chargino-neutralino 
mass set is defined as the number of events satisfying $\mathcal{L}_{R}\! >\! \mathcal{L}_{R_{CUT}}$  divided by the total number of chargino 
events satisfying the topological cuts.}. Figures~\ref{fig:DATAMC}(d) and 
\ref{fig:DATAMCS1}(d) show the good agreement obtained between real and simulated 
events as a function of the likelihood ratio cut,
for $25 \leq \DM < 35$~\GeVcc\ in the \jjl\ topology.} 
\item{The combination of variables corresponding to the lowest excluded cross-section defined the $\mathcal{L}_{R}$ 
function and the $\mathcal{L}_{R_{CUT}}$ of each window.}
\end{itemize}


   


\begin{table}[ht]
 \begin{center}
   \begin{tabular}{|c|c|}
 \hline
\multicolumn{2}{|c|}{\DM\ regions} \\
 \hline
 1   &  3$\leq$\DM $<$ 5~\GeVcc \\
 \hline
 2   &  5$\leq$\DM $<$ 10~\GeVcc \\
 \hline
 3   &  10$\leq$\DM $<$ 25~\GeVcc \\
 \hline
 4   &   25$\leq$\DM $<$ 35~\GeVcc \\
 \hline
 5   &  35$\leq$\DM $<$ 50~\GeVcc \\
 \hline
 6   &   50~\GeVcc$\leq$\DM  \\
 \hline
\end{tabular}
\caption[.]{
\label{tab:REGION}
Definition of the \DM\ (mass difference between the chargino and 
lightest neutralino) regions for the chargino search in non-degenerate scenarios.}
\end{center}
\end{table}

\subsubsection{Nearly mass-degenerate scenarios}
\label{sec:chadegsel}

The search for charginos in the nearly mass-degenerate scenarios uses several different
techniques, depending on the lifetime of the chargino, which in turn depends on the 
mass difference \DM\ between the chargino and the lightest neutralino (this is the
only relevant dependence, at least in the heavy slepton hypothesis).
When \DM\ is below the mass of the pion, the chargino lifetime is usually long enough
to let it pass through the entire detector before decaying. This range of \DM\ can be
covered by the search for long-lived heavy charged particles. 
For \DM\ of a  few hundred \MeVcc\, the chargino can decay inside the main tracking
devices. Therefore, a search for secondary vertices or kinks can be used to
cover this region. 
As the mass difference increases, the mean lifetime shortens until the position
of the $\tilde\chi_1^{\pm}$ decay can hardly be distinguished from the main event
vertex. In this case, the tagging of an energetic ISR photon
can help in exploring the \DM\ region between a few hundred \MeVcc\ and 3~\GeVcc.
The selection criteria
are similar to the ones used in the analysis of previous data,
which have been described in~\cite{isr}. 

\vspace{0.35cm}
\noindent
\underline{Search for long lived charginos}
\vspace{0.35cm}

Long lived charginos can either be ``quasi-stable'' (decay outside  the tracking
system) or decay ``visibly'' inside  the  tracking devices.

The search for heavy stable charged particles is described in~\cite{GMSB-contribution}.
The method used to identify heavy stable particles relied on the ionisation loss 
measurements in the TPC and on the absence of signal in the DELPHI Cherenkov 
radiation detectors (RICH). 
Heavy stable charged particles crossing the detector would be seen in the tracking
system and have as distinctive signature the absence of  Cherenkov radiation and an
anomalous energy loss in the TPC.
Three different search windows were used in the search for heavy stable charginos:
\begin{itemize}
\item the charged particle had momentum above 15~\GeVc, and no photons in 
any of the two radiators of the RICH (liquid, refractive index $n=1.28$, and
gas, $n=1.0015$) were associated to the track;
\item the charged particle had momentum above 5~\GeVc, high ionisation loss in the TPC,
 and no signal in the gas RICH;
\item the charged particle had momentum above 15~\GeVc, a TPC ionisation loss not exceeding
 70\% of the expectation for a proton, and no signal in the gas RICH.
\end{itemize}
A fourth search window considered in \cite{GMSB-contribution} was not included,
in order to treat the two hemispheres of the event independently. 

If a heavy charged particle decays inside the central tracking devices 
(at a radius between 10~cm and 1~m) then both the incoming and the outgoing track
can be reconstructed, and the angle between the tracks can be calculated. 
Such a search for kinks was originally designed to search for long-lived staus
in the Gauge Mediated SUSY Breaking scenario~\cite{GMSB-contribution}. A similar
technique was applied to search for mass-degenerate charginos, with some specific
features needed because the visible decay products carry very little momentum in
the nearly mass-degenerate case.
Details of the selection criteria can be found in \cite{isr}. 
Here only a
brief and qualitative summary of the most important selection cuts is given.

A set of rather loose general requirements was applied in order
to suppress the low energy background (beam-gas, beam-wall, etc), two-photon,
\ee\ and hadronic events. 
For each event passing the preselection cuts, all the charged particles
were grouped in clusters according to their measured point closest to the
interaction vertex.
A cluster with only one track with momentum above 20~\GeVc\ was considered
as a possible chargino candidate if it was compatible with a particle coming
from the interaction point.
For each single track cluster fulfilling the above conditions, a search was
made for a second cluster possibly formed by the decay products of the \XP{1}\
and defining a secondary vertex or kink with the chargino candidate.

Reconstructed secondary vertices could also be the result of particles
interacting in the detector material, or having a particle
trajectory reconstructed in two separate track segments.
Additional requirements rejected these backgrounds in the events with an
acceptable secondary vertex~\cite{GMSB-contribution}.
Finally, for an event to be accepted, at least one charged particle had to be
found in each hemisphere (defined by the plane containing the beam spot
and perpendicular to the line connecting the beam spot to the kink).

The search for events with tracks at large impact parameter described
in~\cite{GMSB-contribution} was not possible in this case:
events with only two extremely soft charged particles with large 
impact parameter are difficult both to trigger on and to 
discriminate from machine related noise.
Such events were however considered if a
high \pt\ ISR photon was present, as explained in the next section.

\vspace{0.35cm}
\noindent
\underline{Search for charginos with ISR photons}
\vspace{0.35cm}

The visible particles resulting from the decay of
a chargino nearly mass-degenerate
with the LSP have typically little energy and momentum. 
The trigger efficiency is low for such events, and there is a very large background 
from two-photon events.
The ISR photon tag improves detectability and, if the transverse energy of the photon
is above a threshold which depends on the minimal polar angle acceptance of the 
experiment, it rejects most of the two-photon background.

After the preselection, which was summarised in \ref{sec:presel}, the following 
requirements were applied to the data and simulation samples.

\begin{itemize}
\item{There had to be at least two and at most six good charged particles passing 
the quality criteria (see section~\ref{sec:trksel}), and no more than ten 
tracks in total.}
\item{The transverse energy of the ISR photon candidate was required to be greater
 than $(E_T^{\gamma})^{\min} \simeq 0.03\cdot\rs$.}
\item{The mass recoiling against the photon had to be above 
 $2 \MXC{1}-\delta M$, where the term $\delta M$ takes into account 
 the energy resolution in the electromagnetic calorimeters.}
\item{The photon had to be isolated by at least $30^\circ$ with respect
 to any other charged or neutral particle in the event.}
\item{The sum of the energies of the particles with polar angles within $30^\circ$
 of the beam axis ($E_{30}$) was required to be less than 25\% of the total visible
 energy. If the photon itself was below $30^\circ$, it was the ratio
 $(E_{30}-E_{\gamma})/(\Evis-E_{\gamma})$ that was required to be below $0.25$.}
\item{If the ISR photon candidate was detected in the very forward 
 calorimeter STIC, it must not  be correlated with a signal in the
 scintillators placed in front of the STIC.}
\item{$(\Evis-E_{\gamma})/ \rs$ had to be below a kinematic threshold
 which depended on \DM\ and on \MXC{1}\ (and in any case below 6\%).}
\item{The ratio of the absolute value of the missing transverse momentum over the
 total transverse energy had to be above $0.40 / c$ if $\DM>300$~\MeVcc, and 
 above $0.75 / c$ for smaller $\DM$.}
\item{If $\DM>1$~\GeVcc~, at least two charged particles in the event had to be
 consistent with coming from the beam interaction region.}
\end{itemize}

\begin{figure}[htb]
\centerline{
\epsfxsize=18cm\epsffile{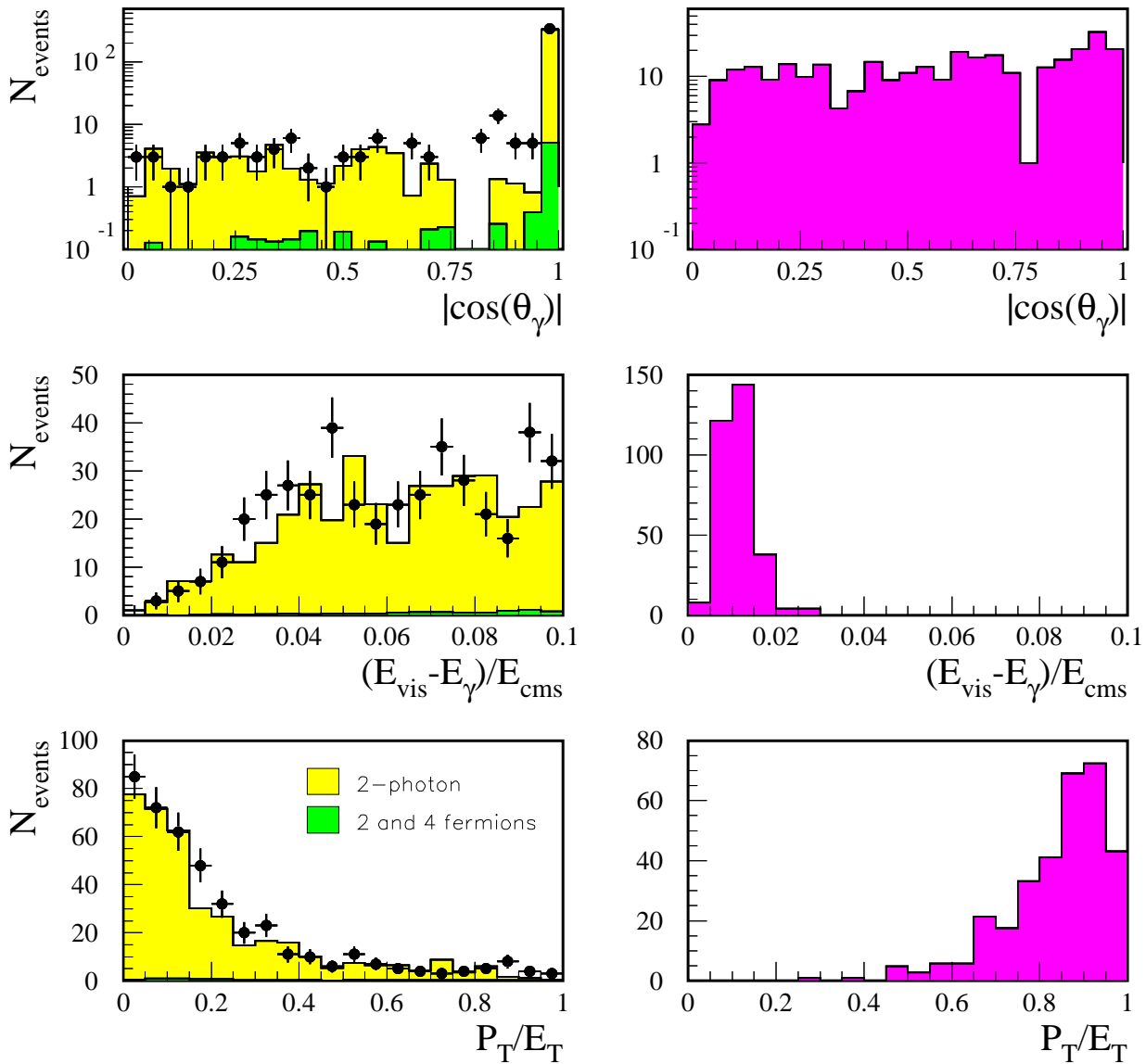}  }
\caption[]{
    Some of the variables used in the selection for mass-degenerate charginos with
    an ISR photon tag.
    In the plots on the left the data (dots) are compared with the SM expectations.
    On the right, as an example, the corresponding distributions (with arbitrary
    normalisation) are shown for the signal with $M(\tilde\chi_1^+)$ = 80~\GeVcc\
    and $\DM=1$~\GeVcc.
                     }

\label{fig:isrcomp}
\end{figure}

Distributions of some of the variables used in the final selection are shown in
figure~\ref{fig:isrcomp} for data, simulated SM background, and simulated signal events. 
Although there is a certain overall qualitative agreement of the various distributions, there is 
already an excess of data. On the other hand, the two-photon generators used in the 
simulation lack the events which have small $\gamma\gamma$ invariant mass, and in some
cases (namely, $\gamma\gamma \to \ee $) no ISR generation is implemented at all. 
Moreover, background processes such as beam-gas interactions are not included in the 
simulation. As in previous publications~\cite{isr}, the most likely explanation of
such disagreement is therefore a deficit of simulated background events rather than an
excess of data from possible new physics. As no attempt will be made in the following
to account for the backgrounds missing in the simulation, the limits that
will be obtained are conservative.


\subsection{Neutralino searches}
\label{sec:ananeu}

\newcommand{\mc}{\multicolumn}

The neutralino searches were designed to cover both 
$\XN{k} \XN{1}$ production 
with $\XN{k} \to \XN{1} + \ffbar$,
with a signature of acoplanar jets or leptons, and channels of
the type \XN{k}\XN{j} with $k$ or $j> 2$, which can lead to neutralino
cascade decays. To maximise the sensitivity several searches
were used, covering different topologies, namely:

\begin{itemize}
\item  a search for acoplanar jet events, as from $\XN{1}\XN{2}$
      with \XN{2}~$\to$~\XN{1}\qqbar; 
\item  a search for acoplanar lepton events, as from $\XN{1}\XN{2}$
      with \XN{2}~$\to$~\XN{1}\ee\ or \XN{2}~$\to$~\XN{1}\mumu; 
\item a search for multijet events, as from $\XN{i}\XN{j},i=1,2,j=3,4$
      with \XN{j}~$\to$~\XN{2}\qqbar\ and \XN{2}~ decaying to \XN{1}\qqbar\ or
      \XN{1}$\gamma$;
\item a search for multilepton events for the corresponding decays to 
      lepton pairs;
\item a search for cascade decays with tau leptons, e.g.
      \XN{2}\XN{1} production with \XN{2}~$\to$~\stau$\tau$ and
      \stau~$\to$~\XN{1}$\tau$;
\item a search for double cascade decays with tau leptons, e.g.
      \XN{2}\XN{2} production with the same \XN{2} decay chain 
      as above.
\end{itemize}

The different searches, briefly described below, were designed to be 
mutually exclusive in order to allow easy combination of the results. 
Thus events selected in the
likelihood-based searches for acoplanar leptons or jets of 
section \ref{sec:neu_2j2l} were explicitly rejected in the searches 
described in the subsequent sections. 
The track selection and the general event preselection described in 
sections \ref{sec:trksel} and  \ref{sec:presel}, respectively, were used.
The LUCLUS algorithm with $d_{\mathrm join}=10$~\GeVc\ 
was used for jet clustering in the 
analyses described below, with the exceptions of the
likelihood ratio acoplanar jets/leptons and double tau cascade searches,
as explicitly mentioned in the corresponding sections
(see section~\ref{sec:id&jet} for details on the jet clustering algorithms).

Data collected at 
centre-of-mass energies between 192 and 208 GeV
were analysed, and were combined with previous 
results \cite{neut189}. In the search for staus in \XN{2}\XN{2} production,
all data recorded at centre-of-mass energies between 189 and 208 GeV were
analysed.

\subsubsection{Acoplanar jets and acoplanar leptons searches}
\label{sec:neu_2j2l}

As mentioned above, the acoplanar jets and acoplanar
leptons topologies are dominant in most of the parameter space.
For these cases, a search based on the likelihood ratio method was performed 
and the sequential cut analyses described in~\cite{neut189} were used as a cross-check.

The characteristics of the neutralino decays are mainly determined by 
the value of \DM, here defined as the mass difference 
between the heavier of the produced neutralinos and the 
LSP (\XN{1}). 
The total energy of the visible final state particles, \Evis, was used
to distinguish between regions of different signal and background
characteristics in the optimisation of the selections.

\vspace{0.35cm}
\noindent
\underline{Likelihood Ratio analysis}


The first step of the analysis was to preselect the events dividing them into  
three mutually exclusive topologies: ee, $\mu\mu$, and \qqbar.
The ee and $\mu\mu$ topologies were defined as having exactly two isolated 
lepton candidates (see section \ref{sec:id&jet}). In the ee topology, at least one of
these had to be a tightly identified electron, and neither
identified as a muon. Similarly, the events in the $\mu\mu$
topology were required to contain at least one isolated loose muon
candidate and no isolated electron. 
The \qqbar\ topology was defined as events with more than five charged 
particles and no isolated photons or leptons.

In the second step, aimed at removing the dominant SM background processes,
events which fulfilled all of the following criteria were selected:

\begin{itemize}
\item the polar angles of the most energetic neutral and charged particles
were required to be above 10\dgree; 
\item the missing transverse momentum had to exceed 2~\GeVc, or 4~\GeVc\ 
      if the visible energy was less than 30~\GeV;
\item both the acoplanarity and acollinearity had to be greater than 3\dgree; 
\item the total visible energy had to be lower than 0.75\sqs.
\end{itemize}

The first two requirements remove the bulk of the \gamgam\ events and 
off-momentum beam electrons. The third and fourth reject two-fermion processes.
The last requirement removes mainly four-fermion events.

In the \qqbar\ topology an additional selection was applied
to suppress further the large \qqbar$(\gamma)$ 
background. This was based on the jets reconstructed using the Durham
algorithm and forcing the number of jets to two.
If the invariant mass of the event was within 40~\GeVcc\ of the 
Z mass, the acoplanarity of the two jets was required to be at 
least 10\dgree. 

 

For different values of visible energy \Evis\ (typical of different values of \DM), 
the kinematic properties of the signal were studied
in terms of the mean value and standard deviation of several event variables. 
Five 
\Evis\
regions were defined, each containing  
signal events with similar properties and SM background composition.
These regions are 
given in table  \ref{tab:neulike_dm}, together with the corresponding 
dominant SM background. 

\begin{table}[ht]
 \begin{center}
\begin{tabular}{|c|c|c|}
 \hline
\multicolumn{2}{|c|}{\Evis\ regions} & Main SM bkg.\\
 \hline
 \hline
 1   &  5$\leq$\Evis $<$ 20~\GeVcc & \gamgam\ \\
 \hline
 2   &  20$\leq$\Evis $<$ 50~\GeVcc & \gamgam, 2-fermions \\
 \hline
 3   &   50$\leq$\Evis $<$ 70~\GeVcc & 2-fermions\\
 \hline
 4   &  70$\leq$\Evis $<$ 110~\GeVcc & 2- and 4-fermions\\
 \hline
 5   &   110~\GeVcc$\leq$\Evis  & 4-fermions\\
 \hline
\end{tabular}
\caption[.]{
\label{tab:neulike_dm}
Definition of the visible energy regions of the neutralino search in the acoplanar leptons and jets 
topologies 
and corresponding dominant SM backgrounds.}
\end{center}
\end{table}

In the last step of the analysis,  for each of the 15 windows 
(three topologies, five \Evis\ regions) thus defined, a likelihood ratio  
function was computed. The variables used in the likelihood definition 
are listed below:
\begin{itemize}
\item global variables (all topologies):
visible energy, transverse energy, 
missing momentum, energy and direction of the most energetic charged and 
neutral particles, transverse momentum with respect to the thrust axis, 
polar angle of the missing momentum, thrust value, thrust direction and
acoplanarity;
\item variables specific to the \qqbar\ topology: jet directions, energies, 
widths, invariant mass and $y_{\rm cut}$ value;
\item variable specific to the ee and $\mu\mu$ topologies, namely
the invariant mass of the two charged particles.
\end{itemize}

For each variable used in the likelihood ratio function, the one-dimensional probability 
density function was defined according to the procedure described in 
sections \ref{subsubsec:analysistech} and \ref{sub:CHASEL}, and
events with $\mathcal{L}_{R}>\mathcal{L}_{R_{CUT}}$ were selected as 
candidate signal events. 

\begin{figure}[ht]
\begin{center}
   \vspace*{-1.cm}
    \begin{tabular}{cc}
   \vspace*{-0.5cm}
    \epsfxsize=8.cm\epsffile{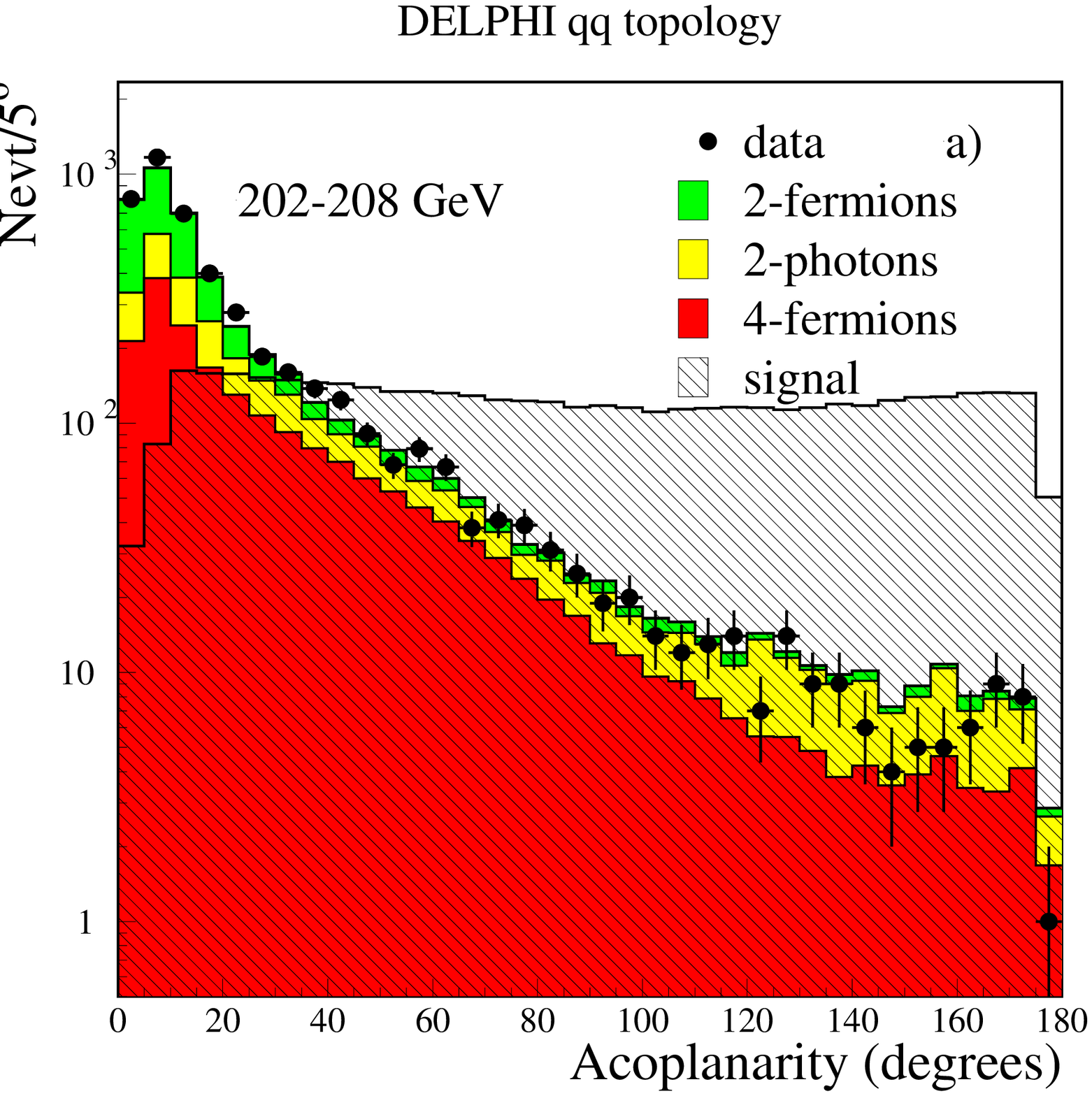} &
    \epsfxsize=8.cm\epsffile{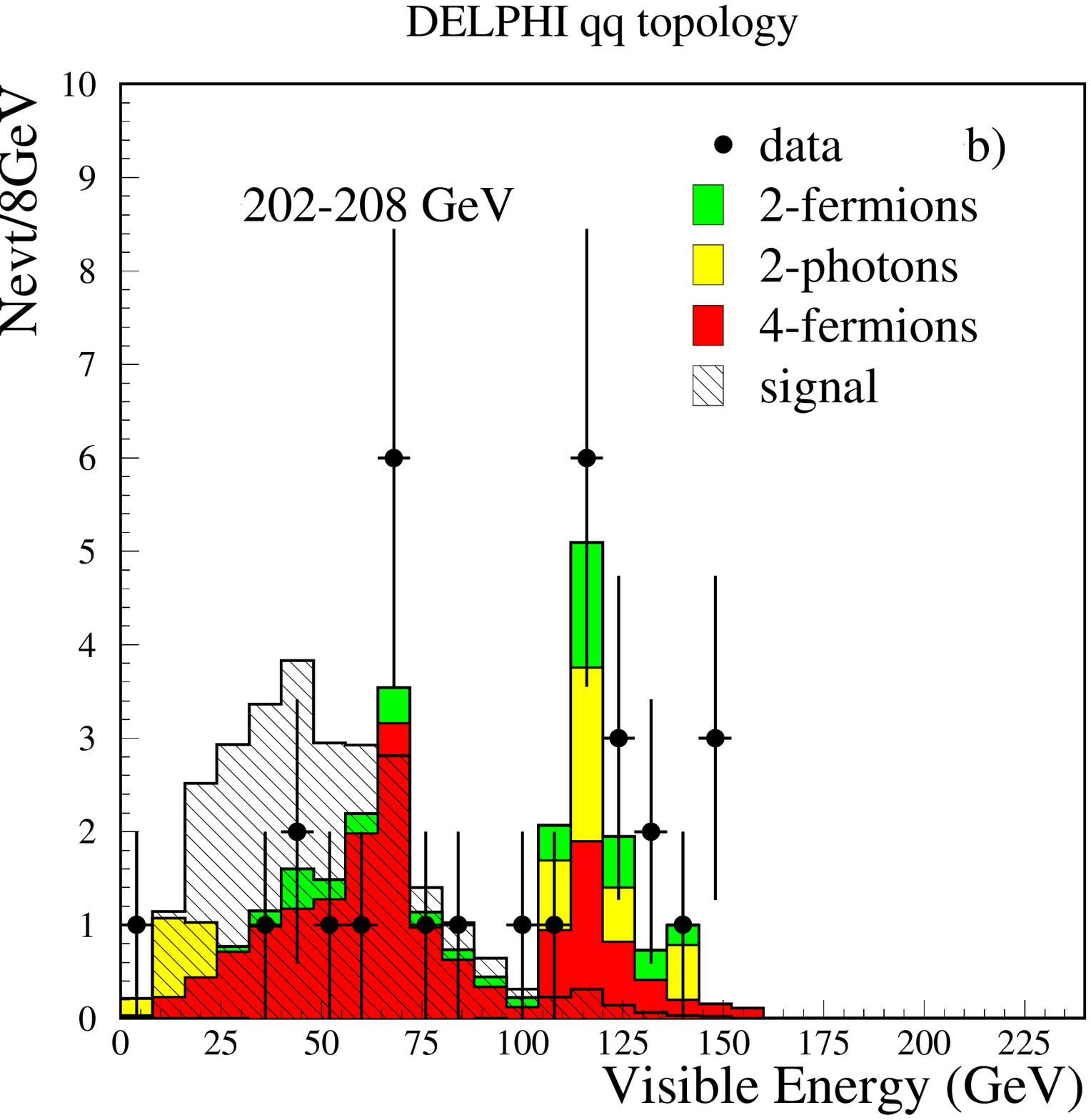} \\
    \vspace*{-0.5cm} 
    \epsfxsize=8.cm\epsffile{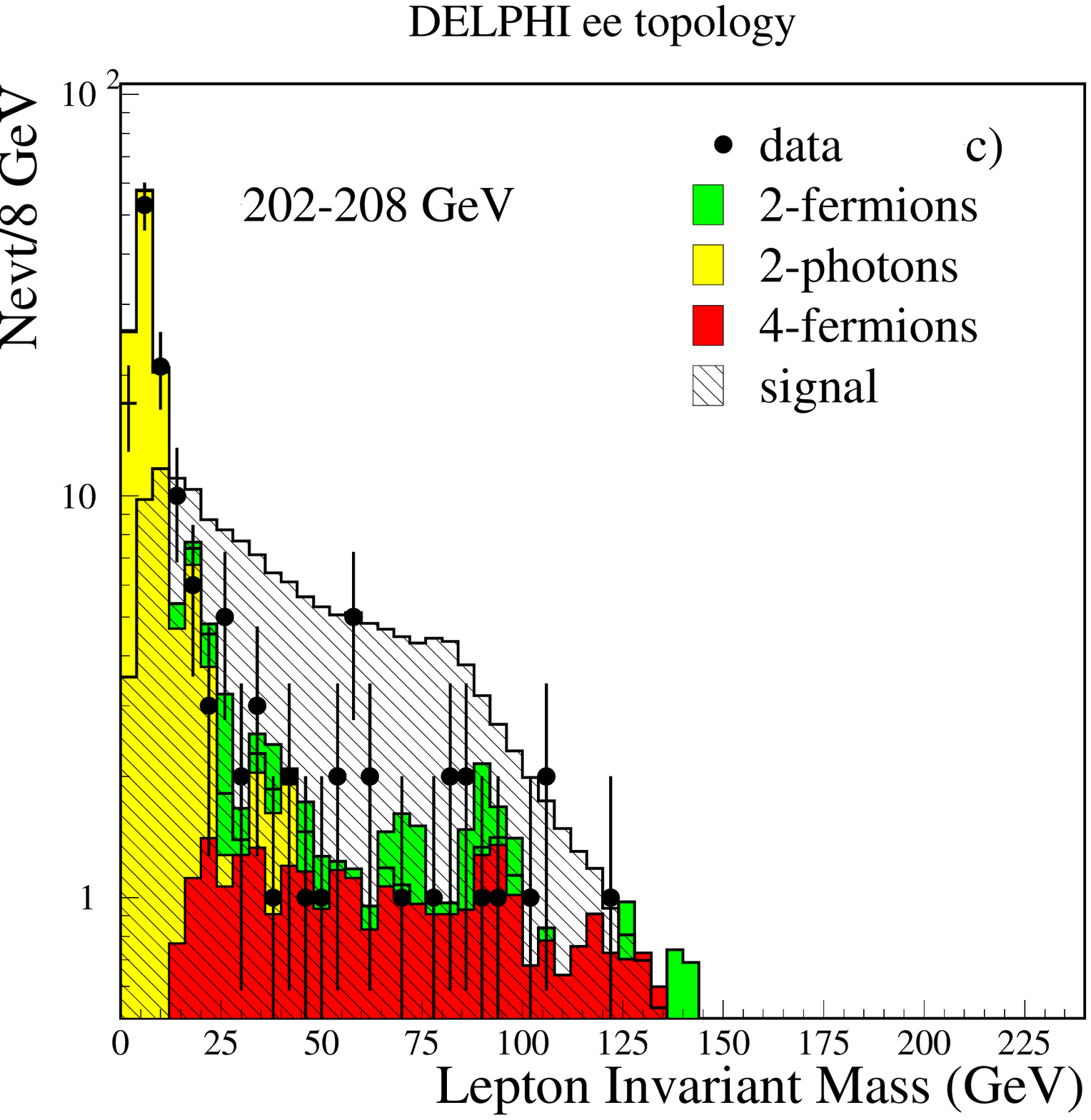} &
    \epsfxsize=8.cm\epsffile{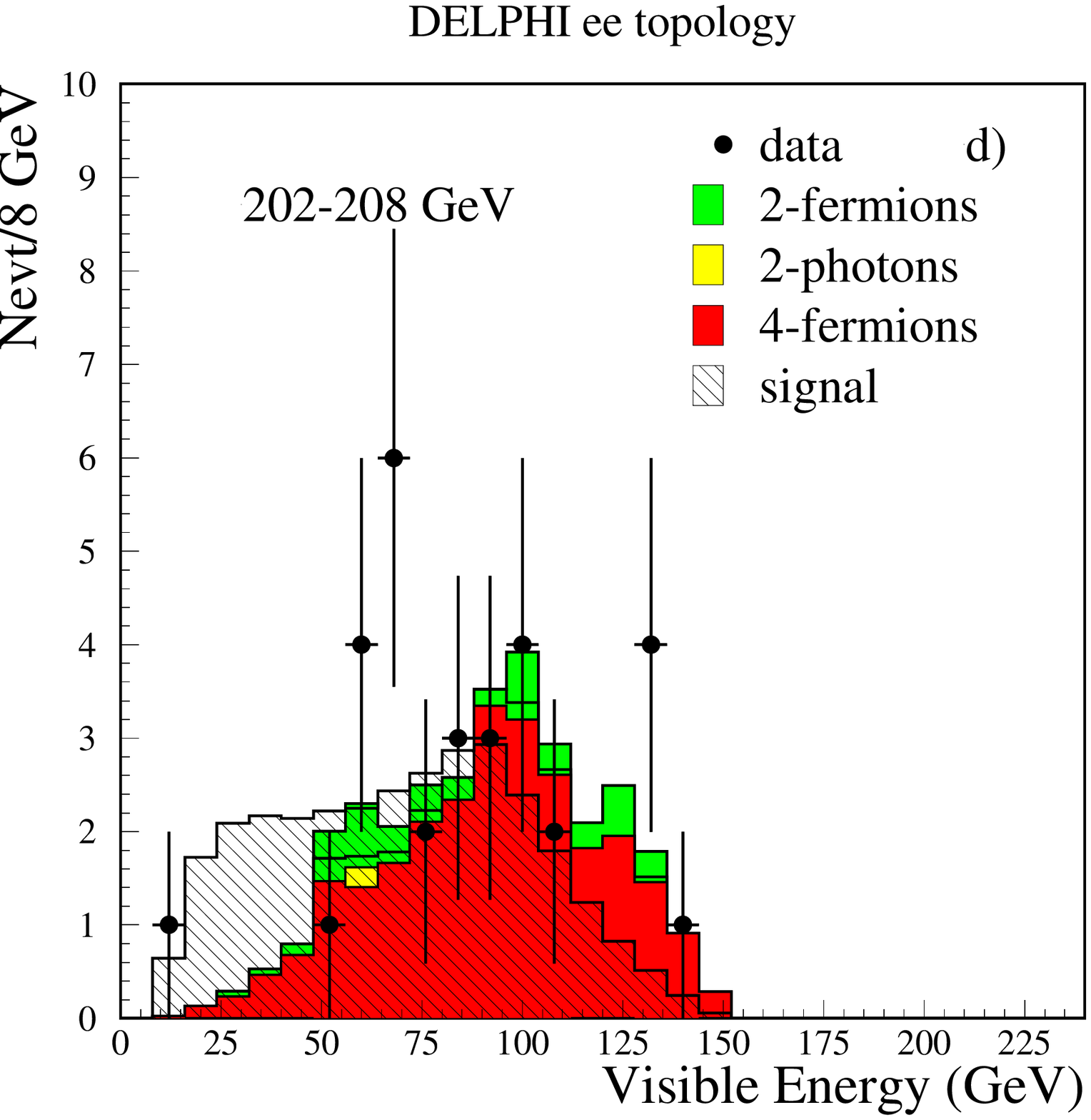} \\
    \vspace*{-0.5cm} 
    \epsfxsize=8.cm\epsffile{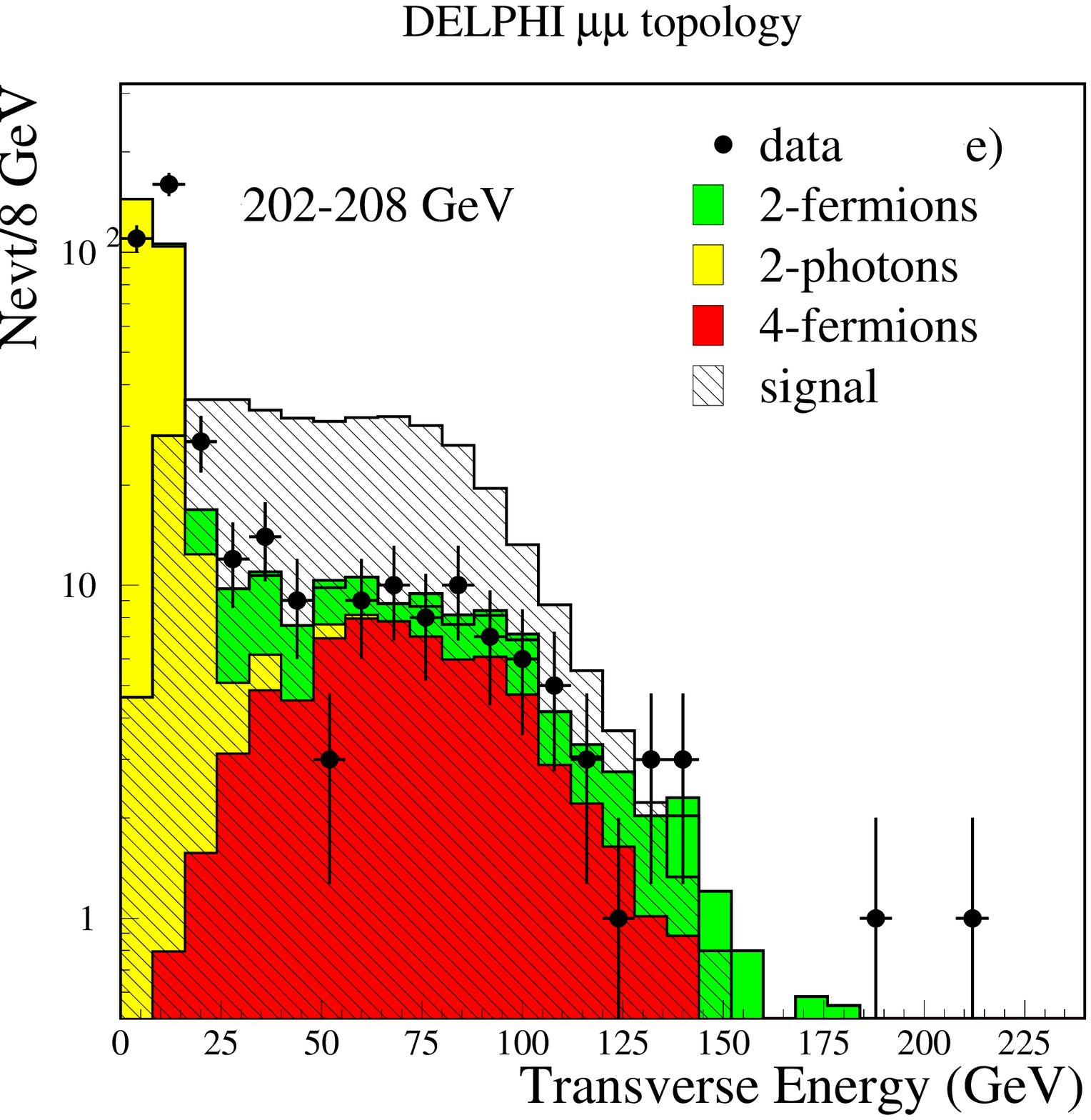} &
    \epsfxsize=8.cm\epsffile{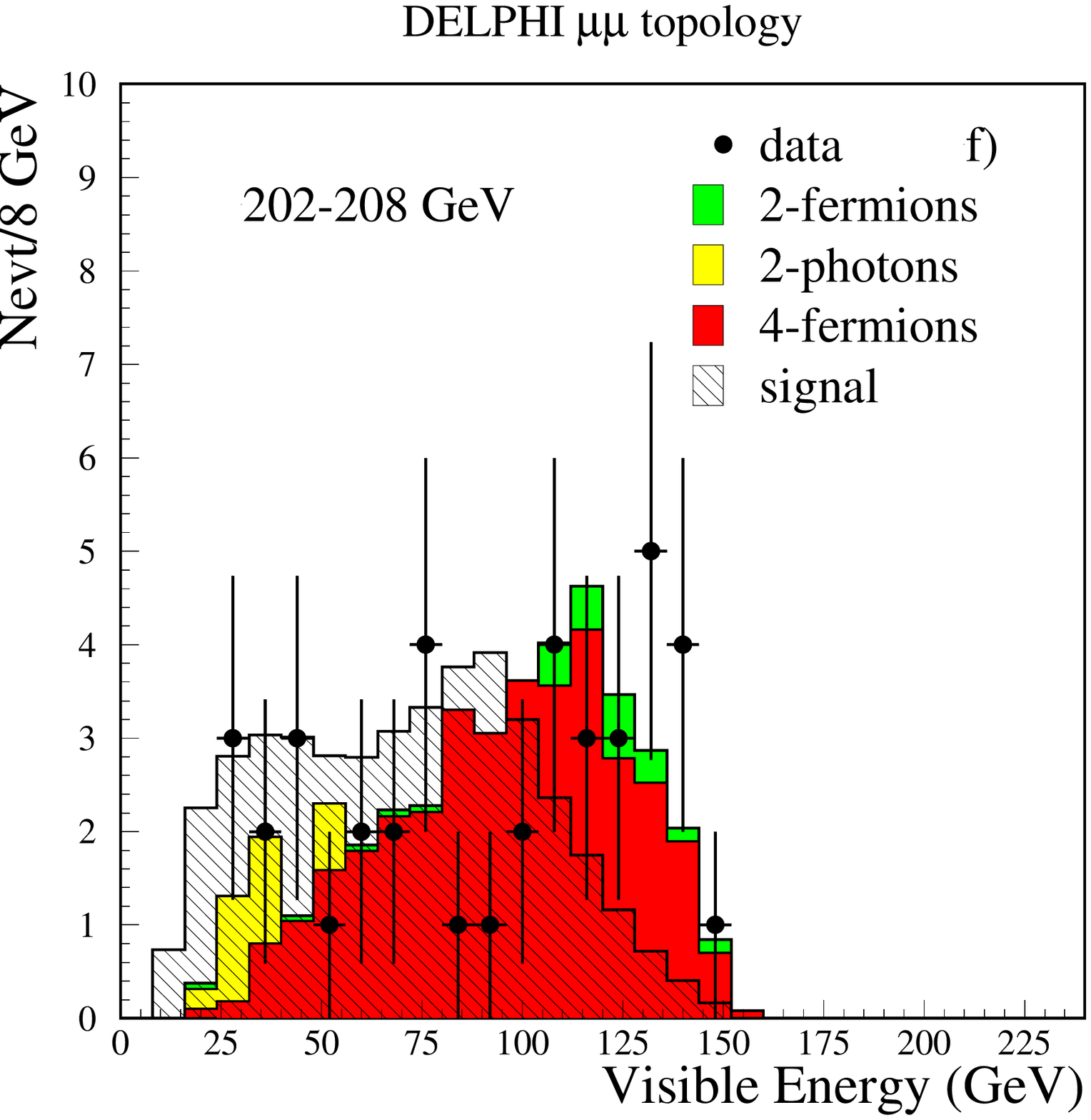} \\ 
    \end{tabular}
\end{center}
  \caption{On the left the comparison between the real and simulated 
  data is shown at preselection level for different variables for the 
  \qqbar, ee, and $\mu\mu$ topologies in the neutralino search.
  On the right the visible energy distribution is shown for 
  the events selected at the final stage, after the likelihood
  selection. The signal distribution contains the contribution of 
all generated mass combinations and is arbitrarily normalised.
 }
\label{fig:neulike_datamc}
\end{figure}

Some event variable distributions for real and simulated data 
before the likelihood selection are shown
in figure~\ref{fig:neulike_datamc}. 
A fair agreement between data and the SM expectation is found. 
A low visible energy event is present in the ee topology (see figure~\ref{fig:neulike_datamc}(d)),
in a region where low but non-zero background was expected from the 
SM (about 0.5 events with visible energy below 30 GeV expected from four-fermion processes). 
No significant disagreement between data and MC is  
found in any of the topologies.


\vspace{0.35cm}
\noindent
\underline{Sequential analysis}

The detailed selection criteria for the selection based on sequential
cuts are given in reference \cite{neut189} and have not been changed.
This analysis was used as a cross-check of the likelihood ratio
results. 
At the final selection level,  
criteria optimised for different 
\DM\ regions were designed. They were used as independent
selections in the derivation of the results.

\subsubsection{Multijet search}
\label{sec:neu_mj}

The multijet search was optimised for cascade decays of neutralinos with large
mass splittings, giving high energy jets. Events with energetic
photons, characteristic of the decay \XN{2}~$\to$~\XN{1}$\gamma$,
were treated separately.
Events with a photon signature were selected on the basis of reconstructed photons 
with a polar angle above 20\dgree, isolated by more than 20\dgree\ from the nearest charged particle. 
If there was only one such photon its energy was required to be between 10~GeV 
and 40~GeV; if more than one photon was present, at least two had to have energy greater than 10 GeV. 

This selection was  similar to the acoplanar jet selection of the sequential
analysis, but required
a rather large transverse energy and allowed any number of reconstructed
jets. Events selected by the searches for acoplanar leptons or acoplanar
jets were explicitly rejected.

The detailed selection criteria are similar to the ones described 
in~\cite{neut189}. Background studies based on data of 1998 and an  
improved energy-flow reconstruction motivated several changes in
the selection procedure. Three selection stages
(preselection, intermediate selection and final selection)
are defined and used in the figures and tables. 
In the following, the most important steps of this search are summarised. 

\begin{figure}[ht]
\begin{center}
\mbox{\epsfysize=17.0cm\epsffile{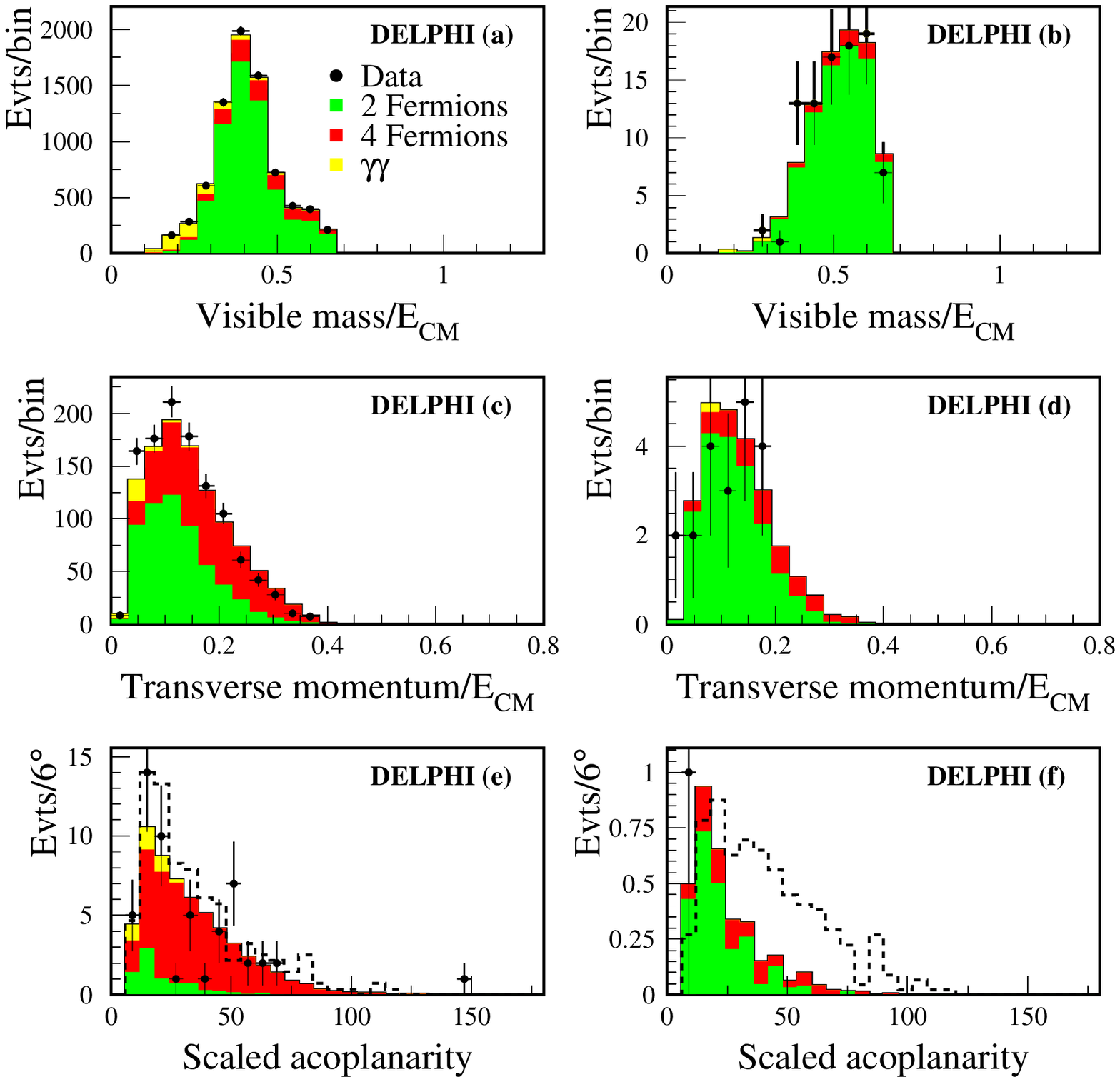}}
\caption[Data-MC comparison]{ {
Comparison of real data from 1999 (\sqs\ in the range
192--202~\GeV) and simulation 
for the neutralino multijet selection
without photons (a,c,e) and with photons (b,d,f), 
at the preselection (a,b),intermediate (c,d), and final selection(e,f)
stages. 
The distributions expected for \XN{3}\XN{2} production with 
\XN{3}~$\to$~\XN{2}\qqbar\ and \XN{2} decaying to \XN{1}\qqbar\ 
or \XN{1}$\gamma$ are shown as dashed histograms where 
\MXN{3}=112~\GeVcc, \MXN{2}=75~\GeVcc\ and \MXN{1}=41~\GeVcc.
The signals are normalised to cross-sections of 0.8~pb (e) and 0.1~pb (f).   
}}
\label{fig:neu_mjet99}
\end{center}
\end{figure}

\begin{figure}[ht]
\begin{center}
\mbox{\epsfysize=17.0cm\epsffile{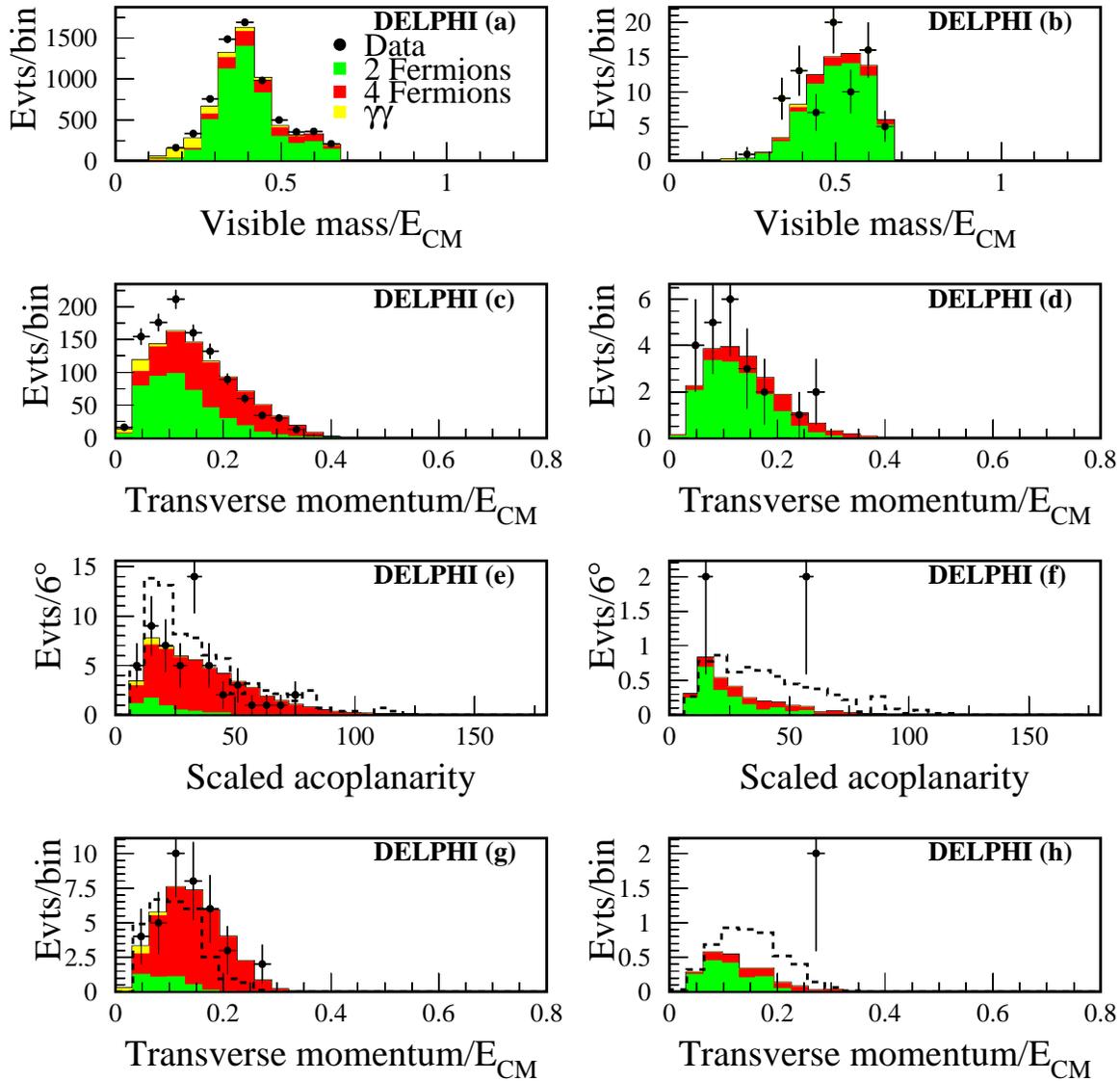}}
\caption[Data-MC comparison]{ {
Comparison of real data from 2000 (\sqs\ in the range
204--208~\GeV) and simulation 
for the neutralino multijet selection
without photons (a,c,e,g) and with photons (b,d,f,h),
at the preselection (a,b),intermediate (c,d), and final selection(e,f,g,h)
stages. In (g,h) only the data collected with the TPC sector 6 on are shown.
The distributions expected for \XN{3}\XN{2} production with 
\XN{3}~$\to$~\XN{2}\qqbar\ and \XN{2} decaying to \XN{1}\qqbar\ 
or \XN{1}$\gamma$ are shown as dashed histograms where 
\MXN{3}=112~\GeVcc, \MXN{2}=75~\GeVcc\ and \MXN{1}=41~\GeVcc.
The signals are normalised cross-sections of 0.8~pb (e), 
0.1~pb (f,h) and 0.4~pb (g).   
}}
\label{fig:neu_mjet}
\end{center}
\end{figure}

At preselection level,
at least five charged particles (passing the track selection described
in section~\ref{sec:trksel}) were required, at least
one of them with a transverse momentum exceeding 2.5~\GeVc.
The transverse energy of the event had to be greater than 25~GeV, the
visible energy was required to be less than 0.65\sqs, and the
missing momentum had to be less than 0.4\sqs/$c$. 
There were several requirements
aimed at selecting events with 
jets which were not dominated by
single particles with large reconstructed energy.
Figures~\ref{fig:neu_mjet99}(a,b) and \ref{fig:neu_mjet}(a,b) show the 
distributions of the visible mass divided by the centre-of-mass
energy for real data and simulated background events
passing the above selection.

At the intermediate selection level,  
radiative return to the \Zn, two-photon, and Bhabha background was reduced by 
excluding events with 
a neutral particle whose energy exceeded 60~GeV
or with more than 
40\%
of the visible energy within 30\dgree\ of the beam direction. Also,
the transverse momentum had to exceed 6~\GeVc. The total momentum and the
most energetic shower in the event were both required not to be close to
the beam direction. 
A comparison of the $c$\pT/$\sqrt{s}$ distributions for real data and simulated background
following the above selection is shown in figures~\ref{fig:neu_mjet99}(c,d) 
and \ref{fig:neu_mjet}(c,d). 
The excess in data visible in~\ref{fig:neu_mjet99}(c) 
and~\ref{fig:neu_mjet}(c) was found to be consistent with 
(low acoplanarity) radiative
return to the \Zn\ events with misreconstructed missing momentum.
In~\ref{fig:neu_mjet}(c) the excess is larger. 
The reason for this is that in the year 2000 the missing momentum 
reconstruction suffered from the problems with sector 6 of the TPC and
from the fact that the distortions in the TPC data had not yet been 
fully corrected for at the time of this analysis. The phase space
region in question was removed by criteria imposed later in the 
selection, and the excess has no impact on the analysis results.

At the final level of the selection
the acollinearity and scaled acoplanarity
(in a forced 2-jet configuration) 
had to be greater than 30\dgree\ and 10\dgree, respectively.
To reject \WW\ background it was required that there be no charged particle with
a momentum above 30~\GeVc, 
and no isolated lepton above 10~\GeVc\, or above 4~\GeVc\ with an isolation angle greater than 20\dgree.

For events with a photon signature, the mass recoiling against the
visible system was required to exceed 20\% of the centre-of-mass 
energy and the scalar sum of momenta reconstructed with TPC information
had to be less than 60\% of the visible energy. 

For the complementary sample, without a photon signature, several of the
criteria above were made stricter in order to reject \Zn$\gamma$ events.
The recoil mass was required to exceed 40\% of \sqs\ and the energy
deposited in electromagnetic calorimeters had to be less than 40~GeV.
No 
isolated neutrals with energy greater than
20~GeV
were allowed and the average momentum of particles with tracks reconstructed 
in the TPC had to be less than 4~\GeVc.

Figures~\ref{fig:neu_mjet99}(e,f) and \ref{fig:neu_mjet}(e,f) 
show the 
distribution of the scaled acoplanarity for real and simulated data 
after final selections, while figures~\ref{fig:neu_mjet}(g,h)
show $c$\pT/$\sqrt{s}$ for data with TPC sector 6 on.
A satisfactory agreement between data and SM simulation is found. 
Four events are observed while three were expected from SM simulation.

\subsubsection{Multilepton search}
\label{sub:neu_ml}

The multilepton search is sensitive to cascade decays involving leptons,
which can dominate if there are light sleptons. 
This search was described in~\cite{neut189} and is briefly summarised here. 

\begin{figure}[ht]
\begin{center}
\mbox{\epsfysize=17.0cm\epsffile{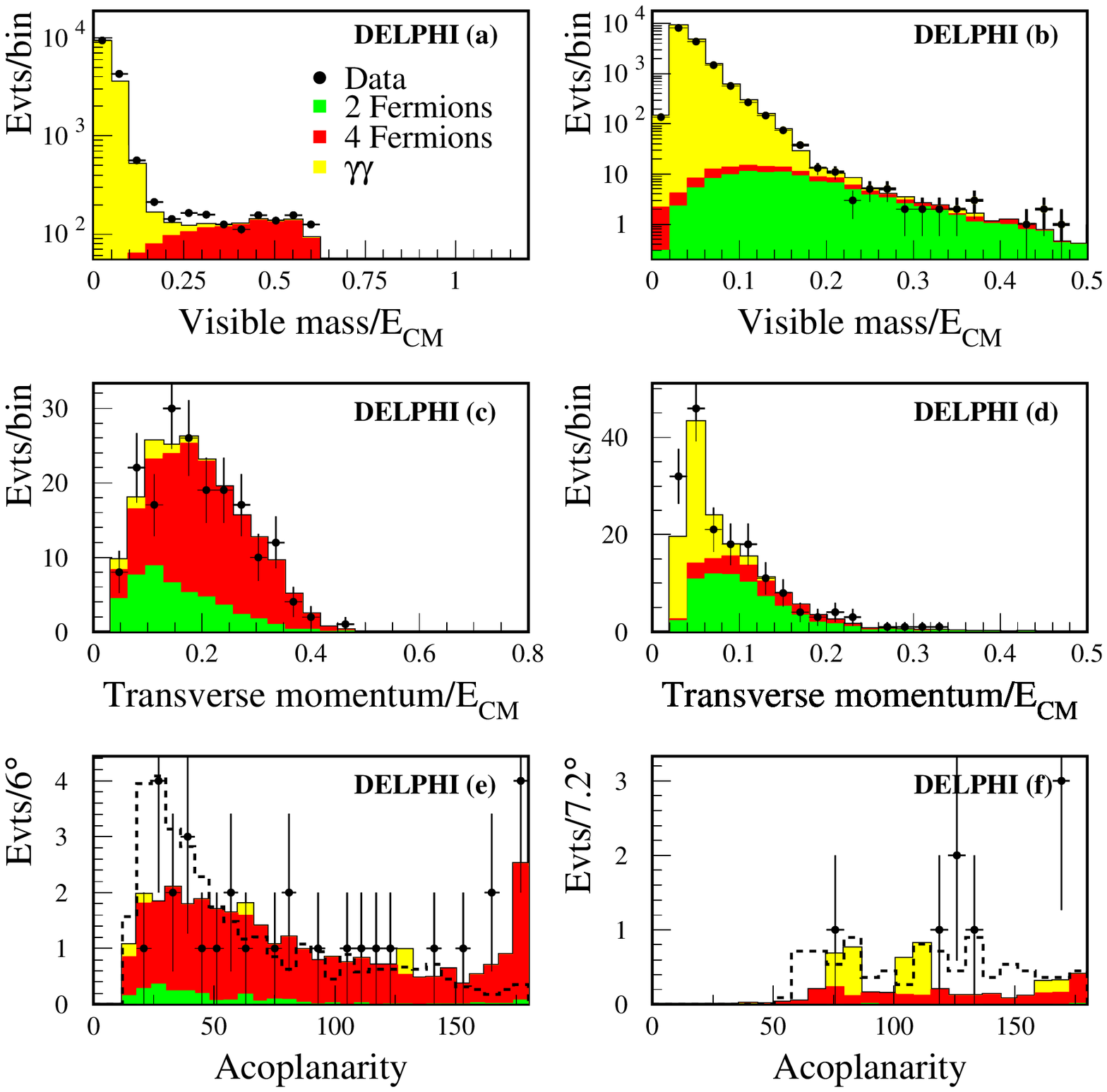}}
\caption[Data-MC comparison]{ {
Comparison of real data from 1999 (\sqs\ in the range
192--202~\GeV) and simulation 
for the neutralino multilepton search
(a,c,e) and asymmetric tau search (b,d,f), at 
the preselection (a,b), intermediate (c,d), and final selection (e,f)
stages. 
The dashed line in (e) shows the multilepton signal expected for
\XN{3}\XN{2} production with \XN{3}~$\to$~\XN{2}\ellell\ and 
\XN{2} decaying to \XN{1}\ellell\ where \MXN{3}=103~\GeVcc, 
\MXN{2}=51~\GeVcc\ and \MXN{1}=45~\GeVcc.
In (f) the dashed line shows the tau cascade signal for 
\XN{2}\XN{1} production with \XN{2}~$\to$~\stau$\tau$ and
\stau~$\to$~\XN{1}$\tau$ where \MXN{2}=75~\GeVcc, \mstau=44~\GeVcc\ 
and \MXN{1}=37.5~\GeVcc.
The signals are normalised to a cross-section of 0.4 pb.   
 }}
\label{fig:neu_mlxt99}
\end{center}
\end{figure}

\begin{figure}[ht]
\begin{center}
\mbox{\epsfysize=17.0cm\epsffile{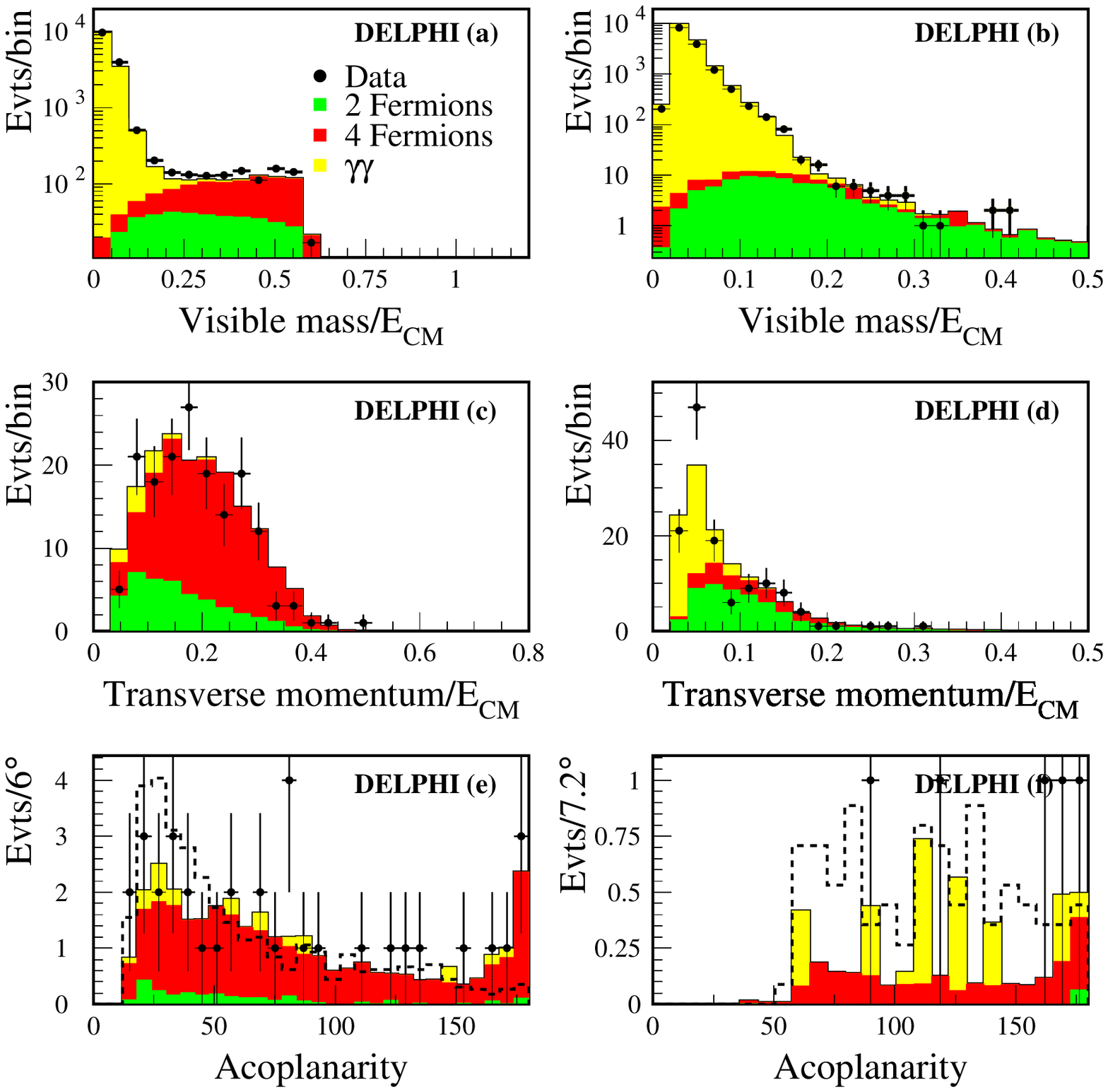}}
\caption[Data-MC comparison]{ {
Comparison of real data from 2000 (\sqs\ in the range
204--208~\GeV) and simulation 
for the neutralino multilepton search
(a,c,e) and asymmetric tau search (b,d,f), at 
the preselection (a,b), intermediate (c,d), and final selection (e,f)
stages. 
The dashed line in (e) shows the multilepton signal expected for
\XN{3}\XN{2} production with \XN{3}~$\to$~\XN{2}\ellell\ and 
\XN{2} decaying to \XN{1}\ellell\ where \MXN{3}=103~\GeVcc, 
\MXN{2}=51~\GeVcc\ and \MXN{1}=45~\GeVcc.
In (f) the dashed line shows the tau cascade signal for 
\XN{2}\XN{1} production with \XN{2}~$\to$~\stau$\tau$ and
\stau~$\to$~\XN{1}$\tau$ where \MXN{2}=75~\GeVcc, \mstau=44~\GeVcc\ 
and \MXN{1}=37.5~\GeVcc.
The signals are normalised to a cross-section of 0.4 pb.   
 }}
\label{fig:neu_mlxt}
\end{center}
\end{figure}

At the preselection level, well reconstructed low multiplicity 
events with missing energy and missing mass were selected. 
In particular, the total visible energy including badly reconstructed tracks
(not passing the track selection described in section~\ref{sec:trksel})
was required to be less than 140~GeV, the number of charged particles 
was required to be at least two and at most eight, and events with more 
than four neutral particles were rejected.
Figures~\ref{fig:neu_mlxt99}(a) and \ref{fig:neu_mlxt}(a) show a 
comparison between the visible mass
divided by the centre-of-mass energy
for real and simulated events passing the preselection.

The selection at the intermediate level served mainly to
reject \Zn$\gamma$, two-photon, and Bhabha events by requiring
significant transverse momentum and transverse energy
(\pT $>$ 8~\GeVc, \Etvis $>$ 25~GeV). 
The distributions of $c$\pT/\sqs\ for real and
simulated data, following the intermediate selection are compared in
figures~\ref{fig:neu_mlxt99}(c) and \ref{fig:neu_mlxt}(c).


At the final selection level events with two or more
charged particles were subjected to criteria
designed to reject Bhabha events. In addition, the charge asymmetry for
the two most energetic such particles was used to reject W pairs
decaying  leptonically. Events with four or more tracks were
clustered into jets, and those with exactly two
jets were rejected if their scaled acoplanarity
was less than 15\dgree.
Figures~\ref{fig:neu_mlxt99}(e) and \ref{fig:neu_mlxt}(e) show 
the acoplanarity distribution for real and simulated data at this
level.

\subsubsection{Asymmetric tau cascade search}
\label{sec:neu_xt}

The tau cascade search is sensitive to \XN{1}\XN{2} production
with $\XN{2}\! \to\! \stau\tau$ and $\stau\! \to\! \XN{1}\tau$,
where the second $\tau$ produced has very low energy. 
This search
was described in~\cite{neut189} and is briefly summarised
here. 

At the preselection level,
well reconstructed low multiplicity events with missing energy
and missing mass were selected. 
The selection was the same as for the multilepton
search (section \ref{sub:neu_ml}), with the additional
requirement of no more than two reconstructed jets.
At least two of the charged particles had also
to satisfy stricter criteria on reconstruction and impact parameters.
The distributions of the visible mass divided by the centre-of-mass energy
for real and simulated data at this
level are shown in 
figures~\ref{fig:neu_mlxt99}(b) and \ref{fig:neu_mlxt}(b).

At the intermediate selection level, 
the highest and second highest momenta of charged particles
were required to be below
40~\GeVc\ and 25~\GeVc, respectively, and at least one charged particle had to
have a transverse momentum above 2.5~\GeVc.
The criteria to reject \Zn$\gamma$, two-photon, and Bhabha events were
similar to those used in the multilepton search, 
except for 
the removal of the requirement on the transverse energy. 
Figures~\ref{fig:neu_mlxt99}(d) and \ref{fig:neu_mlxt}(d) show
the distributions of 
$c$\pT/\sqs\ for real and simulated data at the
intermediate selection level.

At the final selection level, 
events with 
two or more isolated
charged particles were required to have acollinearity and acoplanarity
above 60\dgree. The smaller of the two momenta 
had to be below 70\% of the greater one, and below 10~\GeVc.
For events with two reconstructed jets the scaled acoplanarity 
was required to be greater than 20\dgree, and the acoplanarity and
the acollinearity greater than 60\dgree.
The acoplanarity distributions for the resulting 
samples of real and simulated data events are shown in 
figures~\ref{fig:neu_mlxt99}(f) and \ref{fig:neu_mlxt}(f).

\subsubsection{Double tau cascade search}
\label{sec:neu_xtxt}

The final state of \XN{2}\XN{2} $\to$ $\stau \tau \stau \tau$
differs from that of stau pair-production only in the presence
of two soft taus from the stau decays.
The search criteria for stau pairs (see section~\ref{sec:stauana}) could 
therefore be applied with slight modifications, as follows.
Instead of exactly two  $\tau$-candidates, at least three were
required, but the momentum of the third most energetic
one had to be less than 4\% of the beam momentum.
The maximum number of tracks in the event was increased from 
seven to eight, and, as the \WW\ background is negligible, the
requirement on $p^{\rm JET}_{high}$ versus $\theta_{W^+}$ was removed.
Since the only SM processes yielding final states with four taus
are \ZZ, $\ZZ^*$ or $\Zn\gamma^*$,
the invariant mass of the two most energetic
$\tau$-candidates was required to be below 40 \GeVcc, 
$i.e.$ well below \MZ.
Variables specific
to the two-jet topology,
e.g. acoplanarity, were evaluated
using the two most energetic $\tau$ candidates only, and
the remaining requirements of the stau pair search were left unchanged.
In the cases where these requirements differed for high and low \DM,
the less stringent selection was applied here.



\section{Results and limits}
\label{sec:results}

In this section the results of the
event selections and 
the estimated signal efficiencies are presented for each channel. 
No evidence of a significant excess with respect to the SM expectation was found in any of the 
channels, and limits on masses and cross-sections were set.
In a relatively model-independent approach, 
cross-section upper limits are derived. 
Lower limits on the sparticle masses are also obtained, 
under assumptions 
which depend on the channel and will be specified case by case. All limits quoted are at 95\%
confidence level (CL).

\subsection{Limit computation}
\label{sec:limcomp}


Depending on the searches, limits were derived using the multichannel
Bayesian method~\cite{obraztsov} or the modified frequentist likelihood ratio 
method~\cite{alexread}. 
Both these approaches allow search results in multiple channels to be combined,
taking into account efficiency, expected background, number of candidates 
and centre-of-mass energy in each channel. 

In the Bayesian multichannel approach\cite{obraztsov}, the a posteriori
probability density of the mean total number of signal events 
(assuming a uniform prior) is used
to compute a 95\% CL upper limit and compare it to the mean predicted
for different signal assumptions.
In the likelihood ratio approach, global multi-channel likelihood
functions for the signal plus background 
and background-only hypotheses are evaluated for the 
experimental outcome \cite{alexread}.
One then defines the confidence level in the signal plus background
hypothesis as the probability, under this hypothesis, to obtain a lower
likelihood ratio than experimentally observed. 
In the modified frequentist approach used here this confidence 
level is renormalised by the probability to obtain a value 
below the observed one 
in the absence of a signal. All points where the resulting
confidence in the signal, CL$_s$, is less than 5\% are considered excluded.
More details are given below channel
by channel.

The two methods typically give similar results.
Both methods take a posteriori knowledge about the background into account, 
and thus give physically reasonable and conservative 
results in the case of downward fluctuations of the background. Background 
fluctuations could nevertheless significantly affect the range of exclusion, 
and for this reason limits expected in the absence of a signal are also 
given below.


\subsection{Systematic uncertainties}
\label{sec:sysgen}
The excluded cross-section and mass ranges can be affected by the 
systematic uncertainties related to the SM background rate and
 signal efficiency determinations.
These are mainly due to imperfections in:
\begin{itemize}
\item  the description of the differential cross-sections;
\item  the modelling of fragmentation and hard gluon radiation;
\item  the modelling of the detector response.
\end{itemize}

The inaccuracy of the differential cross-section description for 
the various
SM processes was studied extensively in the context of their measurement
at LEP and is typically small, except for the hadronic $\gamma\gamma$ 
background. 
Modelling of fragmentation and hard gluon radiation was studied both
in the context of background and
efficiency. It mainly affects
the searches for the stop quark, due to the uncertainties
in the  modelling of the stop fragmentation.
However, the largest contribution to the systematics on both
background rate and efficiency determination arises from the modelling 
of the detector response. This was addressed in  several ways,  
depending on the type of the final state searched for.
Below, a review of all of these systematic effects is given. The methods used 
to address them
are then discussed. Finally, the methods used to propagate
those uncertainties in the limit computation are briefly presented.
   
\subsubsection{Review of the main systematic sources}
\label{sec:sysrev}

{\underline{ \it Description of the differential cross-sections:}} \\

\noindent
The uncertainty on the differential cross-sections within the acceptance
region of the most important
 SM processes is~\cite{stopgen,MCgene}
less than  1\%\ for \WW\ production, around 0.5\% for $Z\gamma$, 2\% for \ZZ\ and
 about 5\%~\footnote{
This number is valid in the phase-space regions where these processes are not 
$\gamma\gamma$-like.}
for \Wev\ and \Zee . 
The uncertainty of the description of the 
$Z\gamma^*$ process is typically  5\%, reaching 20\% in the 
region of very small masses of $\gamma^*$. 
While 
the uncertainties are small for
two-photon interactions giving leptons  
(less than 5\%),
the description of the hadronic $\gamma\gamma$ processes is 
subject to  
large uncertainties
in some regions of the phase  space. 
The effect was studied comparing various 
two-photon
generators~\cite{MCgene}.
In the squark analysis (see section \ref{sec:squarkana})
at the preselection level {\tt PYTHIA}~\cite{pytjet} results
were cross-checked with  {\tt TWOGAM}~\cite{twogam}  and differences 
up to 15\% were observed, with  {\tt PYTHIA} giving a higher background.
It has been shown~\cite{notegg} that the {\tt PYTHIA} 
generator represents a more correct background estimation. 
A systematic uncertainty of 15\% on the hadronic $\gamma\gamma$\ background was 
assumed.
The impact
of these uncertainties on the final background rate determination  
will depend on the 
importance of the two-photon background in the different cases. 
In most of the analyses presented in this paper
the contribution of the hadronic $\gamma\gamma$\
process at the final selection level is negligible.
The search for
multi-jet
final states 
without isolated photons
from heavier neutralino decays 
(see section~\ref{sec:ananeu}) may serve as an example. In this case,
40\% of the background originates from 
the \WW\ process,
 40\% from  \Wev\ and \ZZ\ 
processes,
and  15\% from 
$Z\gamma$.
If the presence of photon(s) is required, 80\% of the background 
originates from 
the $Z\gamma$ process. 
In summary, 
the net uncertainties arising from the description of the above processes
are of the order of a few percent at most.   

~\\
\noindent
{\underline{\it Fragmentation and hard gluon radiation modelling:}}\\

\noindent
The description of hard gluon radiation
leads to uncertainties of at most 10\% in the relevant parameter space, 
in particular for the rate of b-tagged four-jet events from the $Z\gamma$ process.
This is
unimportant for most of analyses presented here. 
Fragmentation uncertainty related effects are typically of the order of 1\%.
However in the searches for the stop (see section~\ref{sec:squarkana}), the stop hadronisation scheme
is important and a dedicated  generator was used for these studies. 
The stop hadronisation
 was performed non-perturbatively and 
the $\epsilon$ parameter 
of the Peterson function, which regulates the stop
 fragmentation and hadronisation, was varied. 
The analysis was applied to the different signal samples thus obtained 
and the results were compared. 
The relative systematic error on the efficiency as a consequence of imperfect
signal simulation was taken
to be 7\% for $ \DM\! > 10~\GeVcc$ 
and 10\% for $\DM\! \leq 10~\GeVcc$. For the sbottom analysis,
 7\% was used 
for all \DM.

~\\
\noindent
{\underline{\it Modelling of the detector response:}}\\

\noindent
The modelling of the detector response is the main source of
systematics in the searches for new particles.
The next section gives a review of the methods which were developed
to address primarily the influence of this  uncertainty on the 
SM background rate and efficiency determinations.





\subsubsection{Methods to evaluate detector response systematics}
\label{sec:sysmet}
{\underline{\it {The ``re-weighting'' method}}} \\ 

\noindent
This method estimates the systematic error
on
the number of selected background events from
imperfect simulation of the detector response, as well as  from
possible large uncertainties in the modelling of the differential 
cross-section for two-photon processes. 
It propagates the  difference between the 
data and the simulated background at preselection 
level  to the final selection level for
the variables relevant for the analysis.

The propagation is performed
as 
follows.
Each discriminating variable used in the event selection
is histogrammed both for data and simulated background
events at the preselection and final selection levels.
From the 
preselection
histograms a weight factor,
$N_{\mathrm Data}/N_{\mathrm Back}$,  
is calculated for each bin  
containing  at least 1 \% of the total number 
of background events and a non-zero number of data events.
For bins with zero data events, it was checked that the SM expectation
was in agreement with such an observation. 
These weight factors are then applied, bin by bin, to the
background histograms at the final selection level.
The contribution to the 
systematic  error  of each variable is then determined 
by subtracting the total number of selected background 
events from the total number of re-weighted background 
events at the final selection level.

The total positive (negative) systematic error of the
number of background events 
is then computed as a quadratic sum over positive
(negative) error
contributions from the different variables:

\begin{equation}
\nonumber
\delta^{\pm} = \sqrt{\sum_{N^{\mathrm RW}_{\mathrm Back}(i) \stackrel{>}{<} 
N_{\mathrm Back}} (N^{\mathrm RW}_{\mathrm Back}(i)-N_{\mathrm Back})^{2}}
\label{eq:wei}
\end{equation}

\noindent
where $N^{\mathrm RW}_{\mathrm Back}(i)$ is the total number of re-weighted 
background events from variable $i$ and $N_{\mathrm Back}$ is
the number of selected background events without re-weighting.

In this method the correlations between different variables
are not taken into account, but, as the quadratic sum is performed
over all error contributions, the total systematic error is
overestimated.
It was checked by the means of the Kolmogorov
test that rescaling the distribution of any of the variables
improves the agreement between the data and SM background
for the other variables as well.
Moreover, the statistical fluctuation in the bins
of the histograms used to calculate the weight factors lead to
an additional overestimation of the error.
Another drawback of this method is that the total systematic error
obtained grows with the number of studied variables.

The main advantages of this method are its simplicity and
conservative error estimation. 
Because of its conservative nature, it was
used in searches where
the expected background is small, and background fluctuations
have bigger effect on the expected limit than the systematic error.
For example, in the squark analysis the systematic
error on the background estimated with this method
was  15-60 \% (see section \ref{sec:squa_eff}). 
This
had only  a small effect on the final mass/cross-section limits.
%

~\\
\noindent
{\underline{\it {The ``shaking'' method}}} \\

\noindent
The ``shaking'' method \cite{shakeitZZ} attempts to correct 
at particle level the simulation description  
of the detector response and residual effects in the
fragmentation.
Changes in the particle multiplicity (by adding or removing particles)
are made in the simulation, in such a way that the multiplicities 
of neutral and charged particles in bins of momentum and polar angle 
are well described for the hadronic  data collected at the $Z$ peak.
The momentum and the polar angle of the added particles
are randomly changed with respect to the parent particle, while staying 
in the same phase 
space region. 

The method was tested in the study of the systematics of the \ZZ\ cross-section
measurement~\cite{shakeitZZ}.
The multiplicity adjustment is of the order of 1\%-2\%.
It leads to a better agreement between  the data taken
at energies above the $Z$ peak and the SM
background, typically predicting slightly higher observed missing energy
and momentum.

The method can be used to study the uncertainty on the background and
efficiency in hadronic high multiplicity topologies, with or without
isolated leptons (or photons).
In these topologies, the method gives an estimation of the systematic 
uncertainty 
including both the effect of ``shaking'' the variables and the consequences
of additional reconstructed tracks which lie inside the cone for the isolated 
lepton or photon. 


This method was used to study the systematic error of background and efficiency
in 
hadronic neutralino
topologies, and the systematic error 
on the chargino (non-degenerate scenario)
detection efficiency in the
hadronic, semileptonic and radiative topologies.
Tables \ref{deteffsyst} and \ref{neuteffsyst} show,
for chargino and neutralino searches respectively
(sections \ref{sub:chasdt_res} and \ref{sec:neuresult}), 
the relative difference between 
the efficiencies
obtained with and without ``shaking''. 
The values
in table \ref{deteffsyst}
correspond to high multiplicity hadronic and semi-leptonic
final states and are averaged 
over several points with model parameters corresponding to
the specific \DM\   region (see section~\ref{sub:CHASEL} for
the  \DM\ region definition).
The values shown 
in table \ref{neuteffsyst} 
correspond 
to the different high multiplicity topologies in the neutralino search
and they are averaged over the
interesting range of masses.   

\begin{table}[htb]
\begin{center}
\footnotesize
\begin{tabular}{|c|c|c|c|c|c|c|} 
\hline
 \DM\ & 1 & 2 & 3 & 4 & 5 & 6 \\ \hline
 \jjjj  & 6.8 \% & 8.8 \% & 3.1 \% & 0.5 \% & 1.0 \% & 3.4 \% \\
 \jjl  & 12.9 \% & 4.1 \% & 0.5 \% & 0.4 \% & 0.0 \% & 0.9 \% \\
\hline
\end{tabular}
\end{center}
\caption{\small The relative difference in the chargino detection efficiency
obtained with and without ``shaking'' in the year 2000 simulation 
with the  
detector fully operational. The values shown correspond 
to the different \DM\ regions of the hadronic (\jjjj)
and semi-leptonic (\jjl) topologies in the chargino production
(see section \protect{\ref{sub:chasdt_res}}).  
\label{deteffsyst}}
\end{table}

\begin{table}[htb]
\begin{center}
\footnotesize
\begin{tabular}{|c|c|} 
\hline
 topology & efficiency change  \\ \hline
 multijets without $\gamma$  & 2.9 \%  \\
 multijets with $\gamma$  & 1.3 \%  \\
 acoplanar jets  & 1.3 \% \\
\hline
\end{tabular}
\end{center}
\caption{\small The relative difference in the neutralino detection efficiency
obtained with and without ``shaking'' in the year 2000 simulation 
with the  detector fully operational. 
The values shown correspond 
to the different high multiplicity topologies in the neutralino production
(see section \protect{\ref{sec:neuresult}}) and they are averaged over the
interesting range of masses.   
\label{neuteffsyst}}
\end{table}

Except for the
lowest \DM\  regions in the chargino analysis,
 where systematic uncertainties
 due to tracking and neutral energy
reconstruction have the largest effect, 
the variation in  the detection
efficiency is very small. The efficiencies for the  ``shaken'' signal
are typically larger than for the ``unshaken'' signal. 

In hadronic neutralino topologies, 
the  background estimated with the shaking
method was typically 10\% higher than the ``unshaken'' background. 
The variation of both background and efficiency is consistent with ``shaking''
predicting slightly higher observed missing energy. 

~\\
\noindent
{\underline{\it {Methods dedicated to low multiplicity topologies ("smearing" technique)}}} \\

\noindent
The uncertainty on the  efficiency of the muon, electron and photon identifications is expected
to be a dominant effect in the low multiplicity topologies requiring identified particles.
Studies on back-to-back di-muon events and back-to-back di-electron events and photons from
radiative return to the
$Z$ peak
point to 
background errors 
and relative efficiency errors
of the order of 3\% each. 
 The uncertainty on the track momentum reconstruction arising from the modelling of the detector
response was also studied  and 
a small effect was found, both on efficiency and background estimates.
These estimates are relevant for  
neutralino and chargino leptonic topologies,
and for all slepton searches.

\subsubsection{Methods to propagate statistical and systematic uncertainties in the mass/cross-section limits}
\label{sec:syslim}

Systematic and statistical uncertainties on the  parameters involved in the calculation 
of the limits (mainly, the uncertainties related to the SM background rate and signal 
efficiency determinations) were propagated into the final results using two methods. The analyses
that rely on the modified likelihood ratio method adopted the procedure explained in~\cite{alexread}.
For the analyses that use the multichannel method of~\cite{obraztsov}, a different
procedure based on the same Bayesian approach was chosen~\cite{appap}.
A probability distribution function (pdf) was assumed for the efficiency and the background
for every signal channel. Such a pdf was assumed to be either gaussian or binomial,
depending on the statistical accuracy of the estimations of the signal or background,
and on the methods used to evaluate the uncertainty of the parameter.  

The effect of including systematics in the computation of the limits can become relevant
whenever there are large (order of 10\% or more) relative errors on the efficiency and, in
particular, on the expected background. 
The degradation of the limits is much less significant, however, than would be indicated by the
simplistic and over-conservative approach of reducing all efficiencies and increasing all 
backgrounds by one standard deviation.




\subsection{LEP1 limits \label{sec:lep1}}

In this section limits on the masses of SUSY particles from LEP1
data \cite{LEP1LIMITS} are briefly reviewed. 
In most cases the LEP1 limits have been superseded by LEP2 results, such
as those presented in the present paper. However,
for certain situations they are still relevant. This is
particularly true for limits deriving from comparisons of the
measured \Zn\ decay widths to SM expectations.
%
%
Such limits are relatively insensitive 
to the details of the decays of SUSY particles, although they depend on 
the coupling of the sparticles to the \Zn\ (which is affected by the
sparticle field composition). 


From 1990 to 1995 LEP was run at centre-of-mass energies near the \Zn\ resonance.
Model independent fits to all the lineshape and asymmetry data have been
carried out, giving accurate values of the resonance parameters~\cite{DELPHI00}.
The total decay width of the \Zn\ boson was measured with a precision of 
about 2.5 \MeVcc. 
Decay channels of the \Zn\ opened by new physics would increase the \Zn\ width.
Thus the difference between the measured width and the SM value may be used to
constrain SUSY models. If the new particles decay invisibly, limits can be 
derived in a straightforward
manner from the comparison of the measured invisible width to the SM prediction.
The combined LEP result gives 
$\Gamma_{\mathrm inv}^{\mathrm new}<2.0$~\MeVcc\ at 
95\%~confidence level~\cite{PDG2000}. 
Whether the new particles are visible or invisible, 
they will contribute to
the measured values of the total width $\Gamma_Z$. 
Confronting the measured \Zn\ width with the SM
expectation an upper limit on the extra partial width 
$\Gamma^{\mathrm new}<3.2$~\MeVcc\ was 
obtained~\cite{PDG2000}~\footnote{In~\cite{Kmoenig} a method 
designed to be more model-independent gives 
$\Gamma^{\mathrm new}<6.3$~\MeVcc\ using older data.}.

From the limit on 
$\Gamma_{\mathrm inv}^{\mathrm new}$,
a limit on the sneutrino mass of 43.7 \GeVcc\ may be 
obtained~\cite{PDG2000}. 
A lower mass limit for the lightest chargino of approximately 45~\GeVcc,
independent of the field composition and of the decay modes, has been derived
from the analysis of the \Zn\ width and decays.   
Limits for other sparticles depend both on masses and couplings.
To a large extent left-handed sleptons below 40 \GeVcc\ can also be 
excluded using the agreement of the \Zn\ decay width 
with the SM prediction~\cite{PDG2000}.

The composition of the neutralinos affects their production cross-sections 
and the light states may decouple from the \Zn. The production then 
proceeds through $t$-channel selectron exchange, giving a small cross-section 
if the selectrons are heavy. 
Hence no general limit on the LSP mass can be derived from LEP1.

For the third family, the production rates may be 
affected by potential large mixing of the weak eigenstates. 
For the stau mixing angle
giving the minimal cross-section, the coupling to the \Zn\ vanishes and
no exclusion is possible using this method~\footnote{Selectron 
mass limits of the order of 50 \GeVcc\ 
have been derived from single photon searches
at early colliders, for a nearly massless
photino LSP~\cite{Hearty}.}. 
Also for squarks, the LEP1 limits depend on the mixing angle.
Left squarks below 45 \GeVcc\ are excluded by the \Zn\ invisible 
width if they are nearly mass-degenerate with the LSP,
decaying invisibly. If the decay is visible, the limit from the total
width should be applied instead.
Squarks with non-zero mixing cannot be excluded by this method,
as the coupling to the \Zn\ varies and can vanish~\footnote{A sufficiently 
light stop would contribute  through loop corrections
to the partial width $\Gamma_{bb}$.}.

Direct searches at LEP1 set mass limits above 40~\GeVcc\ for sfermions
in the case of a decay into a fermion and a neutralino, 
with \DM\ $>$ 5~\GeVcc\ and provided the production cross-section is
not supressed due to $t$-channel contributions (in the case of 
selectrons) or to \Zn\ decoupling (in the case of third family 
sleptons and squarks).



\subsection{Slepton searches}
\label{sec:slepresult}

\subsubsection{Smuon and selectron searches}
\label{sec:slep_res1}

\vspace{0.35cm}
\noindent
\underline{Efficiencies and selected events}
\vspace{0.35cm}

   The efficiency for signal detection depends on the masses of 
   the slepton and neutralino. 
   The cuts used to reject the backgrounds resulted in typical signal
   efficiencies of 50\% both for the selectron and the smuon channels.  

   The number of events selected at each energy in the data, together
   with the estimate of the background, is shown in tables~\ref{tab:slep1} and \ref{tab:slep1b}
   for the selectron and smuon analyses.
   It can be seen that the principal background arises from leptonic
   decays of W pairs.

The effect of systematic uncertainties on background and efficiency
evaluation was studied with the ``smearing'' method
(see section \ref{sec:sysgen}). 
The variations  both in  the detection
efficiency and in the background were found to be  of the
order of 3\% on average.

\begin{table}[hbt]
\begin{center}
\begin{tabular}{|l||c|c|c||c|c|c|} \hline
     &    \multicolumn{3}{|c||}{Selectrons} &  \multicolumn{3}{|c|}{Smuons}
\\ 
\hline
\hline
year &  1998  & 1999  & 2000   &  1998  & 1999  & 2000 
\\ 
\hline
\hline
2-fermion events &  5.6  &  8.4  &  4.2    &  2.1  &   2.2  &  1.4 \\   
4-fermion events & 29.7  & 40.9  & 34.4    &  18.8  &   25.5   &  20.8 \\
$\gamma\gamma$ events&  1.7 & 2.5  & 2.2   &  0.5  &  0.6  &  1.6 \\
\hline
Total  & 37.0  &  51.8  &  40.8   &  21.4  &  28.3  &  23.8  \\
\hline
Data   &  40   &  52    & 49      &   19  &   23  &  28  \\
\hline
\end{tabular}
\end{center}
\caption[.]{\small 
Selectron and smuon candidates in the different data sets,
together with the number of background events expected.
The systematic uncertainties on the background were estimated to be of
the order of 3\% (see section \protect{\ref{sec:sysgen}}).
\label{tab:slep1}}
\end{table}

\begin{table}[htb]
\begin{center}
{\footnotesize
\begin{tabular}{|c|c|r@{ $\pm$ }c|c|r@{ $\pm$ }c|c|r@{ $\pm$ }c|}
\hline
$\sqrt{s}$ & \mc{9}{c|}{Selectron search} \\
\cline{2-10}
(GeV)     & \mc{3}{c|}{Preselection} &  \mc{3}{c|}{Intermediate} &  
\mc{3}{c|}{Final} \\
\cline{2-10}
&  Data & \mc{2}{c|}{MC} & Data & \mc{2}{c|}{MC} & Data & \mc{2}{c|}{MC}  \\
\hline
189   &29460  & 28558  &7  & 1925  &1937.0 &2.1 & 40  & 37.0  &0.4  \\
192   & 4797  & 4649   &7  & 315  & 319.7 &2.1 &  6  &  7.0  &0.4  \\
196   & 13705 & 14046  &9  & 812 & 847.2 &2.8 & 21 &  17.8 &0.6  \\
200   & 14709 & 14653  &8  & 993 &1001.0 &2.5 & 14 &  18.5 &0.6  \\
202   &  7125 & 7154   &4  & 474 & 491.5 &1.4 &11  &   8.6 &0.3  \\
205   &13229  & 12687  &15 & 832 & 774.8 &4.7 & 22 &  14.7 &0.3  \\ 
207   &12735  & 13315  &6  & 810  &813.2  &2.0 & 13 &  15.4 &0.4  \\ 
208   & 1202  & 1231   &1  &  78 &  75.2 &0.2 & 3  &  1.4  &0.04 \\
206.5(*) & 8963  & 9655  &6  & 524 & 495.6 &1.8 & 11 &   9.4 &0.5  \\

\hline

All &105295 &105948 &23 &6763  & 6755.2&7 & 141 & 129.8&1.3 \\ 

\hline

$\sqrt{s}$ & \mc{9}{c|}{Smuon search} \\
\cline{2-10}
(GeV)     & \mc{3}{c|}{Preselection} &  \mc{3}{c|}{Intermediate} &  
\mc{3}{c|}{Final}  \\
\cline{2-10}
&  Data & \mc{2}{c|}{MC} & Data & \mc{2}{c|}{MC} & Data & \mc{2}{c|}{MC}  \\
\hline
189   &18759  & 18591 &7  &5378  &5501.2 &2.1 & 19  & 21.4  &0.4  \\
192   & 2985  & 3058  &7  &854   & 860.2 &2.1 &  3  &  3.4  &0.4  \\
196   & 8908  & 8874  &9  &2659 &2617.3 &2.8 &  8 &  10.4 &0.6  \\
200   &9974   &10253  &8  &2883 &3017.8 &2.5 & 6  &   9.9 &0.6  \\
202   &  4768 & 4676  &4  &1439 &1377.8 &1.4 & 6  &   4.6 &0.3  \\
205   & 8773  & 8695  &15 &2659 &2609.5 &4.7 & 7  &   8.5 &0.3  \\ 
207   & 9084  &9024   &6  &2805 &2854.7 &2.0 & 11 &   8.9 &0.4  \\ 
208   &  820  & 835   &1  & 253 & 247.4 &0.2 & 0  &   0.8 &0.04 \\
206.5(*) & 6411  & 6549  &6  &1696 &1621.6 &1.8 & 10 &   5.6 &0.5  \\

\hline
All   &70482 &70555  &2 &20626 &20707.5&2  & 70  &73.4 &0.4  \\ 
\hline

\end{tabular}

\caption{\small
Results of the selectron and smuon searches 
at the different selection levels and centre-of-mass energies. The number of events selected 
in data and
expected from the SM simulation are given. Simulation errors are statistical;
(*) indicates the 2000 data taken with the sector 6 of the TPC off.}
\label{tab:slep1b}
}
\end{center}
\end{table}

\vspace{0.35cm}
\noindent
\underline{Limits}
\vspace{0.35cm}

   The results
   are presented in terms of excluded regions in the slepton-neutralino
   mass plane, obtained using all the analysed data sets 
   combined with data taken 
   previously at lower energies~\cite{slep}. The method in \cite{alexread}
   was used. 

   Limits on slepton masses can be derived using several assumptions.
   In the MSSM, right-handed sleptons are expected to have lower
   masses and lower cross-sections for a given mass. 
   Hence the assumption was made that only right-handed 
   selectrons and smuons are sufficiently low in mass to be 
   pair produced at LEP. This leads to conservative mass limits.
   
   Excluded regions in the slepton-neutralino mass plane
   were obtained taking into
   account the signal efficiencies for each slepton-neutralino mass
   point, the cross-section for right-handed slepton production and
   the 
   branching ratios
   squared for the direct decay 
   $\tilde{\ell}  \rightarrow  \ell \chi^0_1$,
   together with the number of data and background events kinematically
   compatible with the mass combination under test.
   The estimate of the SUSY cross-section and 
   branching ratios 
   for each mass point were
   determined with the SUSY parameters tan~$\beta$=1.5 and 
   $\mu\! =\! -200\!$~\GeVcc.
   In addition, for the smuons a further exclusion curve was derived setting the
   branching ratio of $\tilde{\mu}  \rightarrow  \mu \chi^0_1$ to 1.

\begin{figure}[htbp]
\begin{center}
\epsfysize=15.0cm\epsffile{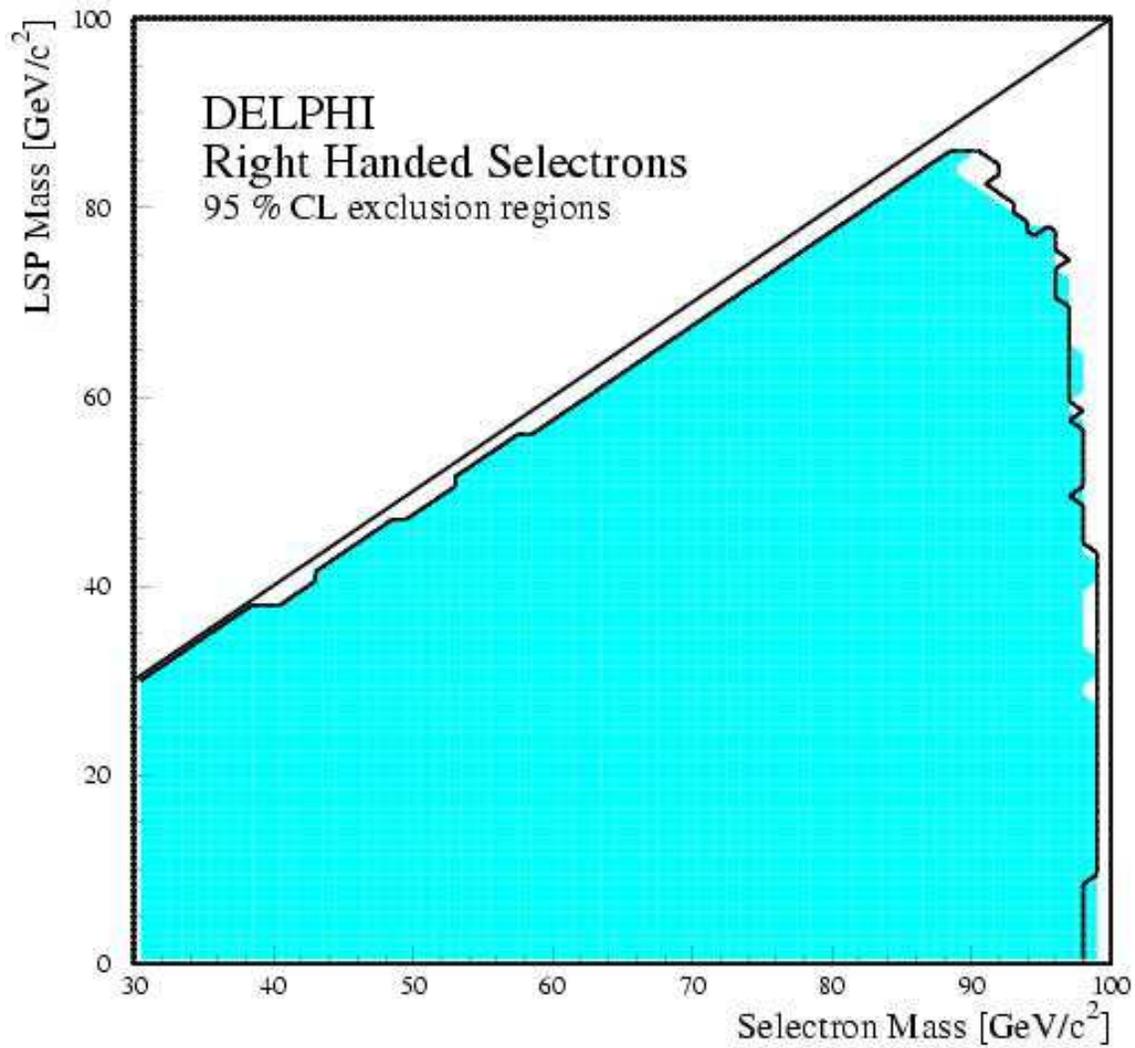}
\vspace{0.5cm}
\caption{
Excluded region
in the (\mselr\, $M_{LSP}$) plane.
The solid line shows the expected limit and
the shaded region shows the obtained limit.
The cross-section and the branching ratios for each mass point were
determined with the SUSY parameters 
tan~$\beta$=1.5 and $\mu\! =\! -200\!$~\GeVcc\
(for a discussion on the low selectron mass region see 
section~\ref{sec:lep1}).
}
\label{fig:sel_xcl}
\end{center}
\vspace{-1.0cm}
\end{figure}

   Figure~\ref{fig:sel_xcl} shows the
   excluded region for \selr\selr\
   production.
   For a mass difference 
   between the selectron and the neutralino above 
   5~\GeVcc, right-handed selectrons are excluded
   up to masses of 98~\GeVcc, for a neutralino mass up to 
   60~\GeVcc, beyond which the limit is weaker. 
   For \DM\ $\ge$ 15~\GeVcc,
   the excluded mass range is up to 94~\GeVcc.

\begin{figure}[htbp]
\begin{center}
\epsfysize=15.0cm\epsffile{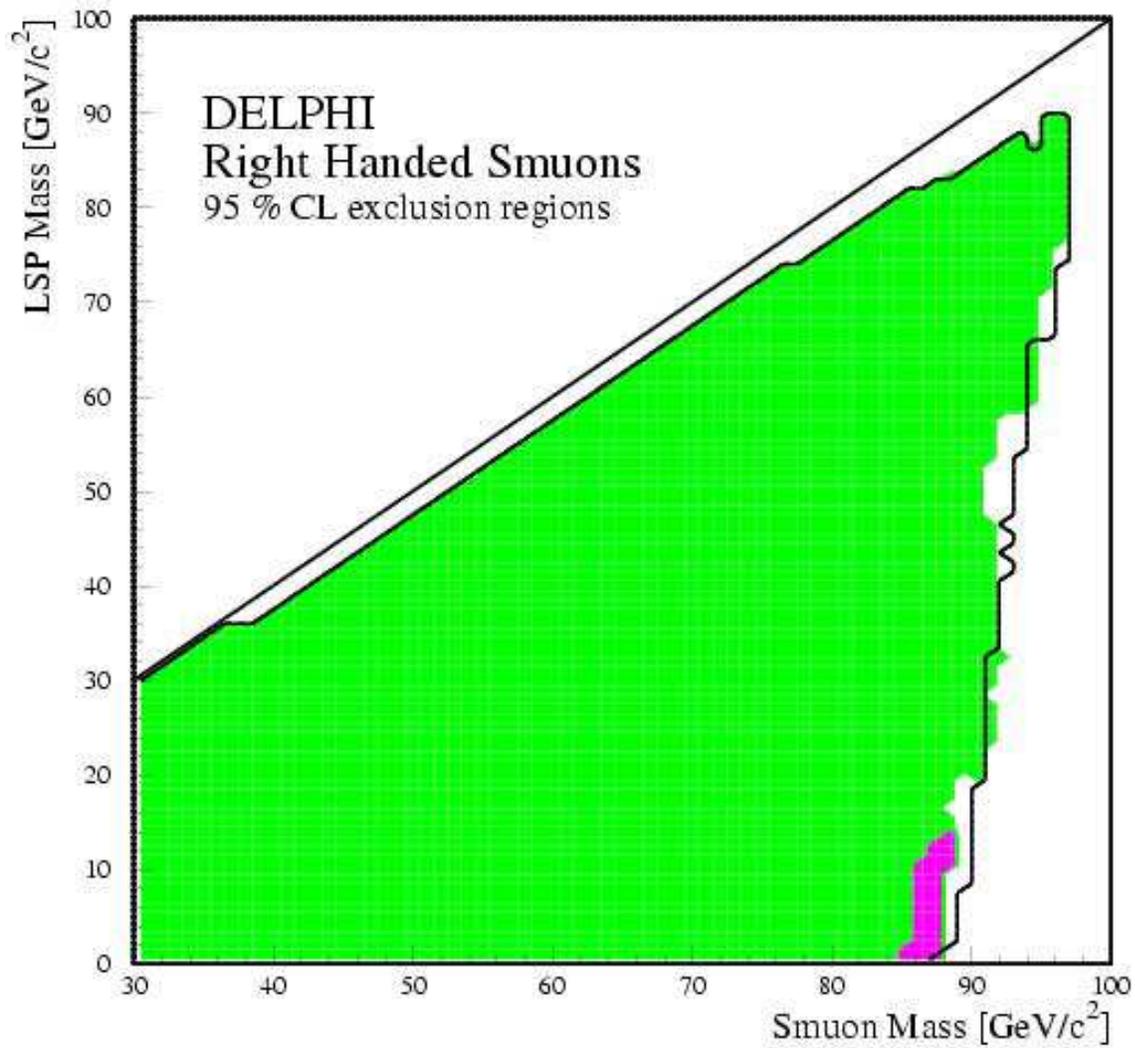}
\vspace{0.5cm}
\caption{
Excluded region
in the (\msmur\, $M_{LSP}$) plane.
The shaded  region shows the obtained 
limit. 
The lighter shaded region is excluded taking  
the branching ratios for each mass point with the SUSY parameters 
tan~$\beta$=1.5 and $\mu\! =\! -200\!$~\GeVcc. 
The solid line shows the corresponding expected limit. 
If the branching ratio
of  $\tilde{\mu}  \rightarrow  \mu \chi^0_1$ is set to 1, the 
darker shaded region is also excluded 
(for a discussion on the low smuon mass region see section~\ref{sec:lep1}).
}
\label{fig:smu_xcl}
\end{center}
\end{figure}

   Figure~\ref{fig:smu_xcl} shows the excluded regions for 
   \smur\smur. 
  The excluded region is shown both 
  taking the branching ratios for each mass point with the SUSY parameters 
tan~$\beta$=1.5 and $\mu\! =\! -200\!$~\GeVcc\ (lighter shaded region)
and  setting the  branching ratio
of  $\tilde{\mu}  \rightarrow  \mu \chi^0_1$ to 1 (darker shaded region).
For the smuons the limit is determined at small neutralino masses,
provided the mass difference between the smuon and the neutralino is above
5~\GeVcc.
Assuming the branching ratio
BR($\tilde{\mu}  \rightarrow  \mu \chi^0_1$)=1, masses up to 88~\GeVcc\ are excluded.   

The effect of the systematic and statistical errors of background and efficiency on
the mass limits was evaluated in a conservative
way by  changing the background and the efficiency by $\pm$ 3\%. 
The resulting change in the limit was less than 1 \GeVcc.

\subsubsection{Stau search}
\label{sec:slep_res2}

\vspace{0.35cm}
\noindent
\underline{Efficiencies and selected events}
\vspace{0.35cm}

\begin{figure}[htbp]
\begin{center}
\mbox{\epsfysize=18.0cm \epsffile{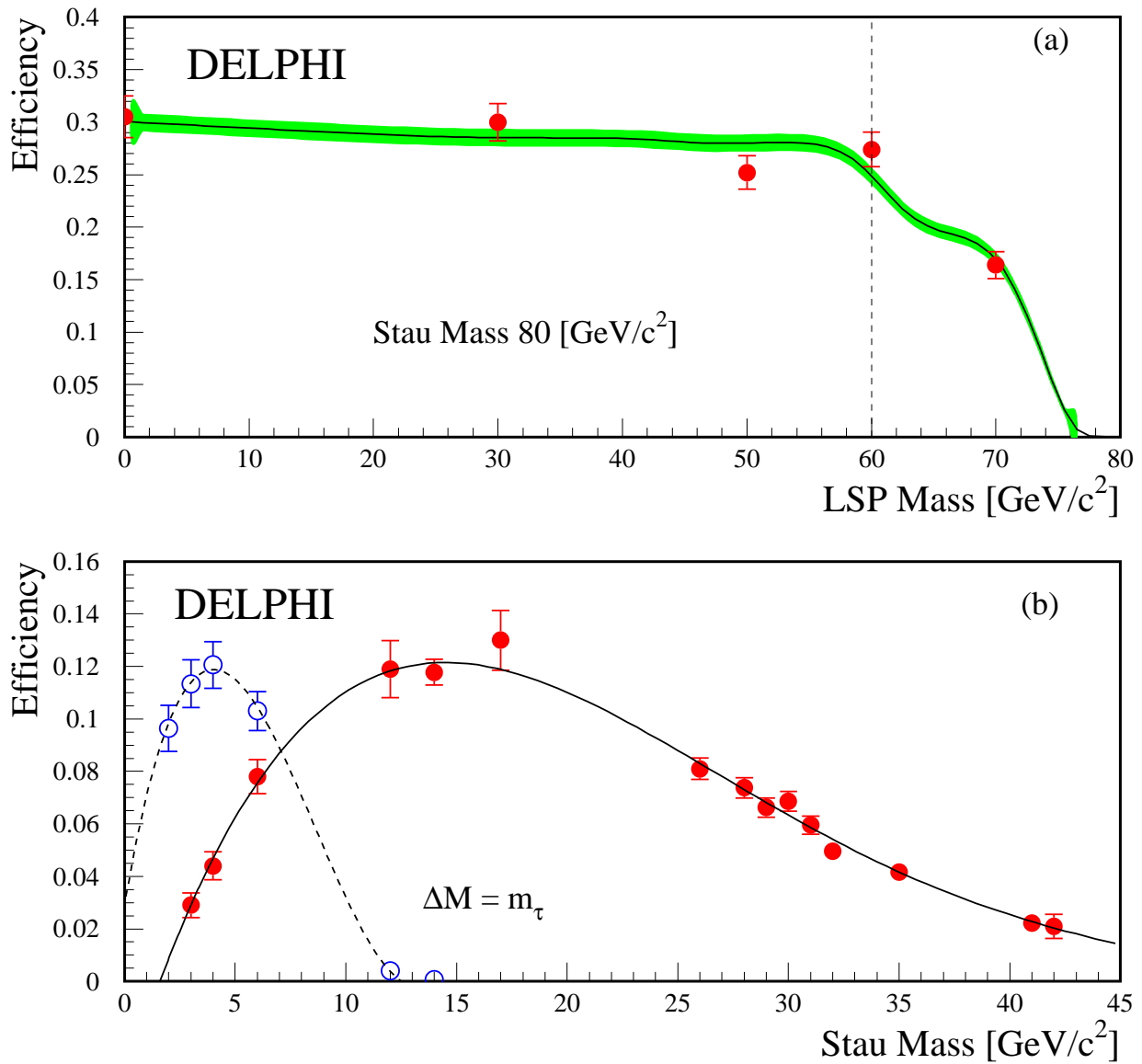}}
\vspace{0.5cm}
\caption{The selection efficiency in the stau analysis.
  The plots show: (a) Efficiency as a function of $M_{LSP}$ at \mstau = 80 \GeVcc .
  The line shows the result of SGV, and the points with
  error bars show the result of
  the full DELSIM simulation. The shaded area indicates the statistical
  uncertainty of the estimate, and the vertical line shows the 
  position of the transition between the two \DM-regions.
  (b) Efficiency as a function of \mstau\ at \DM = $m_{\tau}$
for the low mass analysis
(filled circles) and the very low mass analysis (open circles). 
The points show the results of the full simulation
The lines are polynomial fits to the points. 
} 
\label{fig:stau:eff}
\end{center}
\end{figure}

The efficiencies of the stau search 
have been determined using 5000 events for each point
of a $1\:\GeVcc~\times~1\:\GeVcc$ grid in the (\mstau,$M_{\mathrm LSP}$) plane,
using SGV (see  section \ref{sec:samples}),
and 
range from 20\% to 30\% 
for \DM $>$ 20 \GeVcc.
The results have been verified with
the full DELSIM detector simulation and analysis chain
(figure~\ref{fig:stau:eff}(a)).
In the low mass search, 
the efficiency only needed to be evaluated at a limited number of points,
and in particular at \DM = $m_{\tau}$. 
Therefore,
the
full detector simulation could be used at \DM = $m_{\tau}$
for \mstau\ between 2 and 45 \GeVcc\ (figure~\ref{fig:stau:eff}(b)). 
It was verified with  SGV and, for a smaller number of \mstau\ 
values, with DELSIM
that the efficiencies were higher for
higher \DM. 

The systematic error on the efficiencies 
in the high \DM\ region  was obtained
by reversing the cuts designed to remove the $\tau$-pair
and \WW\ backgrounds, and comparing the number of selected events
with the SM expectation.
This method selected a sample which contained
97.6\% \WW\ 
and $\tau$-pair 
events, while a possible signal would not
exceed 1\%.
In terms of the most important kinematic distributions,
the events selected were nevertheless quite similar to a high \DM\ signal. 
The difference between
data and the simulated SM processes was 5$\pm$5\%,
and  5\% was taken as the estimate of the systematic
uncertainty.
For low \DM, the uncertainty was estimated by the
maximal scatter of the efficiency obtained with
SGV with respect to the values obtained 
with DELSIM, and amounted to 15\%.

For the high mass stau-pair search, 
table~\ref{tab:stau:bkg} summarises the number of accepted events in
the data together with the expected
number of events from the different background channels. 
In all \DM\ regions, good agreement with the 
SM
expectation
was observed. 
Most of the selected SM events in the simulation
contained 
either one or two $\tau$'s (17\% and 67\%, respectively).
In 
two thirds
of the events with less than two $\tau$'s,
the lepton mistakenly taken as coming from
a $\tau$-decay had low momentum, $i.e.$ was indistinguishable
from a secondary lepton from a $\tau$-decay.
In the low mass analysis,
a total of 196 events kinematically
compatible with \mstau\ = 25 \GeVcc\ and \DM\ = $m_{\tau}$ were selected,
and the SM background was estimated to be 
196.1${ + 10.1 } \atop { - 4.1 }$.
The contribution from $\gamma \gamma \rightarrow \tau \tau$
to the background
was 91\%,
and the remainder was other $\gamma \gamma$ processes (8\%)
and $\tau$-pairs (1\%).
In the very low mass analysis,
59 events 
compatible with \mstau\ = 5.5 \GeVcc\ and \DM\ = $m_{\tau}$ were selected,
while the SM background was 
estimated to be 69.7 $\pm$ 1.8 events. The background was dominated by
$\tau$-pairs (89\%) 
and $\gamma \gamma$ processes (7\%). 
The
remaining 4\% came from four-fermion processes.

The systematic errors on the background estimates were obtained 
using  
the re-weighting method,
as described in section \ref{sec:sysmet}.

\begin{table}[hbt]
\begin{center}
{\footnotesize
\begin{tabular}{|c|c|c|c|c|} \hline
$\sqrt{s}$ & \multicolumn{2}{c|}{Preselection} &  
             \multicolumn{2}{c|}{Intermediate} \\  
(GeV)      
& \multicolumn{2}{c|}{ } & \multicolumn{2}{c|}{ } \\ 
\cline{2-5}
     & Data &  MC  & Data & MC  \\
\hline
189 & 2949   & 2916 $\pm$ 29   & 80  & 82 $\pm$ 2 \\   
192 & 473    & 444 $\pm$ 4     & 14  & 14 $\pm$ 0 \\  
196 & 1265   & 1333 $\pm$ 12   & 36  & 40 $\pm$ 1 \\  
200 & 1351   & 1264 $\pm$ 11   & 41  & 39 $\pm$ 1 \\  
202 & 716    & 652 $\pm$ 6     & 12  & 20 $\pm$ 1 \\  
205 & 1284   & 1228 $\pm$ 11   & 38  & 34 $\pm$ 1 \\  
207 & 1214   & 1230 $\pm$ 11   & 33  & 36 $\pm$ 1 \\  
208 & 132    & 116 $\pm$ 3     & 3   & 3 $\pm$ 0 \\  
206.5(*) & 842    & 857 $\pm$ 8     & 28  & 26 $\pm$ 1\\    

\hline
Total & 10226  & 10040 $\pm$ 38   & 285 & 294 $\pm$ 3 \\
\hline
Channel & \multicolumn{4}{c|}{Background composition}\\
\hline
  \WW\                & \multicolumn{2}{c|} { 8 \% }         & \multicolumn{2}{c|} { 85 \% }\\     
  4-fermion         & \multicolumn{2}{c|} { 0 \% }       & \multicolumn{2}{c|} { 2 \% } \\    
  2-fermion         & \multicolumn{2}{c|} { 16 \% }      & \multicolumn{2}{c|} { 1 \% } \\    
  bhabha            & \multicolumn{2}{c|} { 42 \% }  & \multicolumn{2}{c|} { 0 \% } \\    
  $\gamma \gamma$   & \multicolumn{2}{c|} { 34 \% }      & \multicolumn{2}{c|} { 12 \% }\\     

\hline
\end{tabular}
\begin{tabular}{|c|c|c|c|c|c|c|} \hline
$\sqrt{s}$ &  \multicolumn{6}{c|}{Final  } \\  
\cline{2-7}
(GeV)      & 
             \multicolumn{2}{c|}{High \DM\ } &  
             \multicolumn{2}{c|}{Medium \DM\ } &  
             \multicolumn{2}{c|}{Low \DM\ } \\  
\cline{2-7}
     & Data & MC & Data & MC & Data & MC  \\
\hline

189
& 16 & 17.8 $\pm$ 0.4  $ { + 1.7 }  \atop { -  2.8 }$    & 10& 10.4  $\pm$ 0.4  $ { + 1.0 } \atop { -  1.6 }$   & 0 & 1.6 $\pm$ 0.5  $ { + 0.2 } \atop { -  0.1 }$ \\
192
& 5 & 2.8   $\pm$ 0.0  $ { + 0.3 }  \atop { -  0.5 }$    & 2 & 1.6   $\pm$ 0.0  $ { + 0.1 } \atop { -  0.3 }$   & 0 & 0.1 $\pm$ 0.0  $ { + 0.0 } \atop { -  0.0 }$ \\
196
& 7 & 7.5   $\pm$ 0.1  $ { + 0.7 }  \atop { -  1.2 }$    & 3 & 4.2   $\pm$ 0.1  $ { + 0.4 } \atop { -  0.7 }$   & 0 & 0.3 $\pm$ 0.0  $ { + 0.0 } \atop { -  0.0 }$ \\
200
& 11 & 7.4  $\pm$ 0.1  $ { + 0.7 }  \atop { -  1.2 }$    & 9 & 4.5   $\pm$ 0.1  $ { + 0.4 } \atop { -  0.7 }$   & 0 & 0.6 $\pm$ 0.1  $ { + 0.1 } \atop { -  0.0 }$ \\
202
& 1 & 4.1   $\pm$ 0.1  $ { + 0.4 }  \atop { -  0.7 }$    & 1 & 2.6   $\pm$ 0.0  $ { + 0.2 } \atop { -  0.4 }$   & 0 & 0.3 $\pm$ 0.0  $ { + 0.0 } \atop { -  0.0 }$ \\
205
& 6 & 7.2   $\pm$ 0.1  $ { + 0.7 }  \atop { -  1.1 }$    & 4 & 4.7   $\pm$ 0.1  $ { + 0.4 } \atop { -  0.7 }$   & 1 & 0.4 $\pm$ 0.0  $ { + 0.1 } \atop { -  0.0 }$ \\
207
& 9 & 6.2   $\pm$ 0.1  $ { + 0.6 }  \atop { -  1.0 }$    & 5 & 3.9   $\pm$ 0.1  $ { + 0.3 } \atop { -  0.7 }$   & 2 & 0.5 $\pm$ 0.1  $ { + 0.1 } \atop { -  0.1 }$ \\
208
& 0 & 0.7   $\pm$ 0.0  $ { + 0.1 }  \atop { -  0.1 }$    & 0 & 0.4   $\pm$ 0.0  $ { + 0.0 } \atop { -  0.1 }$   & 0 & 0.2 $\pm$ 0.0  $ { + 0.0 } \atop { -  0.0 }$ \\
206.5(*)
& 5 & 4.8   $\pm$ 0.1  $ { + 0.5 }  \atop { -  0.7 }$    & 2 & 3.0   $\pm$ 0.1  $ { + 0.3 } \atop { -  0.5 }$   & 0 & 0.5 $\pm$ 0.0  $ { + 0.0 } \atop { -  0.0 }$ \\

\hline
Total 
& 60& 58.4  $\pm$ 0.5  $ { + 2.3 }  \atop { -  3.8 }$    & 36 & 35.3 $\pm$ 0.5  $ { + 1.3 } \atop { -  2.3 }$   & 3 & 4.5 $\pm$ 0.5  $ { + 0.2 } \atop { -  0.2 }$\\   
\hline
Channel & \multicolumn{6}{c|}{Background composition}\\
\hline
\WW\
& \multicolumn{2}{c|} { 82 \% }    & \multicolumn{2}{c|} { 77 \% }    & \multicolumn{2}{c|} { 54 \% }  \\  
  4-fermion   
& \multicolumn{2}{c|} { 3 \% }    & \multicolumn{2}{c|} { 4 \% }    & \multicolumn{2}{c|} { 2 \% }  \\  
  2-fermion
& \multicolumn{2}{c|} { 3 \% }    & \multicolumn{2}{c|} { 4 \% }    & \multicolumn{2}{c|} { 0 \% }  \\  
  bhabha   
& \multicolumn{2}{c|} { 0 \% }    & \multicolumn{2}{c|} { 0 \% }    & \multicolumn{2}{c|} { 0 \% }  \\  
  $\gamma \gamma $
& \multicolumn{2}{c|} { 12 \% }    & \multicolumn{2}{c|} { 15 \% }    & \multicolumn{2}{c|} { 44 \% }  \\

\hline
\end{tabular}
}
\end{center}
\caption[.]{\small 
Stau candidates,
together with the number of background events expected,
at the different selection levels and centre-of-mass energies.
The column labelled ``High \DM'' corresponds to the point
with $\mstau$=80 \GeVcc\ and $M_{\mathrm LSP}$=0 \GeVcc,
the one labelled ``Medium \DM'' to
$\mstau$=80 \GeVcc\ and $M_{\mathrm LSP}$=40 \GeVcc, and the one 
labelled ``Low \DM'' to
$\mstau$=65 \GeVcc\ and $M_{\mathrm LSP}$=60 \GeVcc.
The composition of the SM background for all centre-of-mass energies
summed is also given;
(*) indicates the 2000 data taken with the sector 6 of the TPC off.
\label{tab:stau:bkg}}
\end{table}



\vspace{0.35cm}
\noindent
\underline{Limits}
\vspace{0.35cm}

Excluded regions in the (\MXN{1},\mstau) plane were derived,
combining the analysed data with previous data sets~\cite{slep}.
For each 
\mstau -- $M_{LSP}$ 
mass combination, 
the predicted number of 
observed SUSY events was compared
with the observed number of kinematically
compatible 
events in data and simulated background.
The method presented in \cite{alexread}
was used and systematic errors were taken into account 
when calculating $CL_s$.
The largest effect of the systematic uncertainty on the limit was observed
at high \DM, but never exceeded 800~\MeVcc.

\begin{figure}[htbp]
\begin{center}
\mbox{\epsfysize=18.cm \epsffile{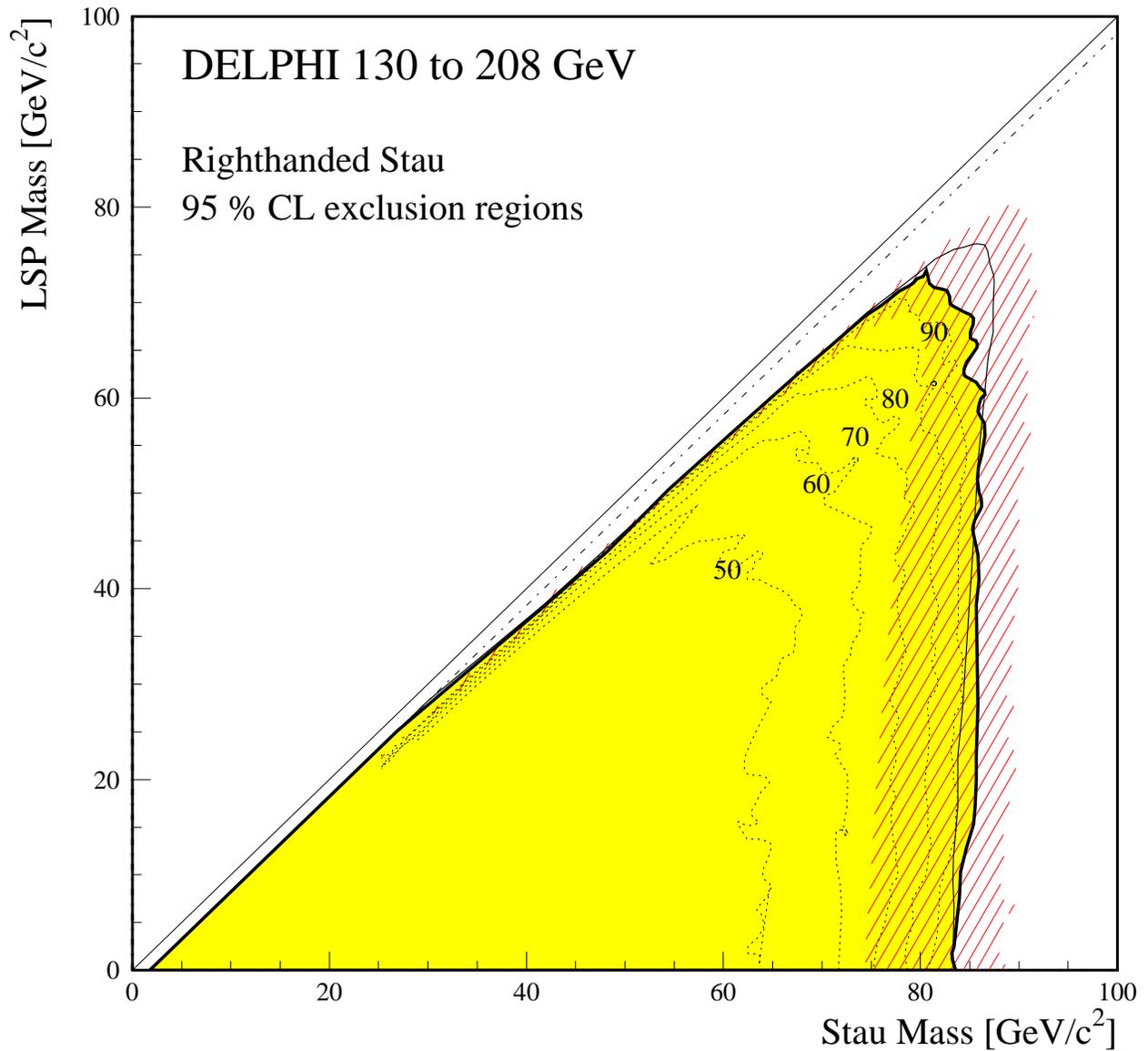}}
\vspace{0.5cm}
\caption{
Excluded region
in the ($M_{\stau_R}$ , $M_{LSP}$) plane.
The shaded 
region shows the region excluded
if \stau $\rightarrow \tau$ LSP is the only allowed channel, 
and the thin solid line shows the corresponding expected limit.
The dotted lines show the region excluded 
if the branching ratio of \stau $\rightarrow \tau$ LSP has the values
indicated in percent next to the lines, and it is assumed that the analysis has no sensitivity
to other decay-modes.
The dash-dotted line indicates \DM\ = $m_{\tau}$.
The observed limit is everywhere within the 95\% CL band for the expected limit
shown as the hatched area.}
\label{fig:staur:xcl}
\end{center}
\end{figure}

\begin{figure}[htbp]
\begin{center}
\mbox{\epsfysize=18.cm \epsffile{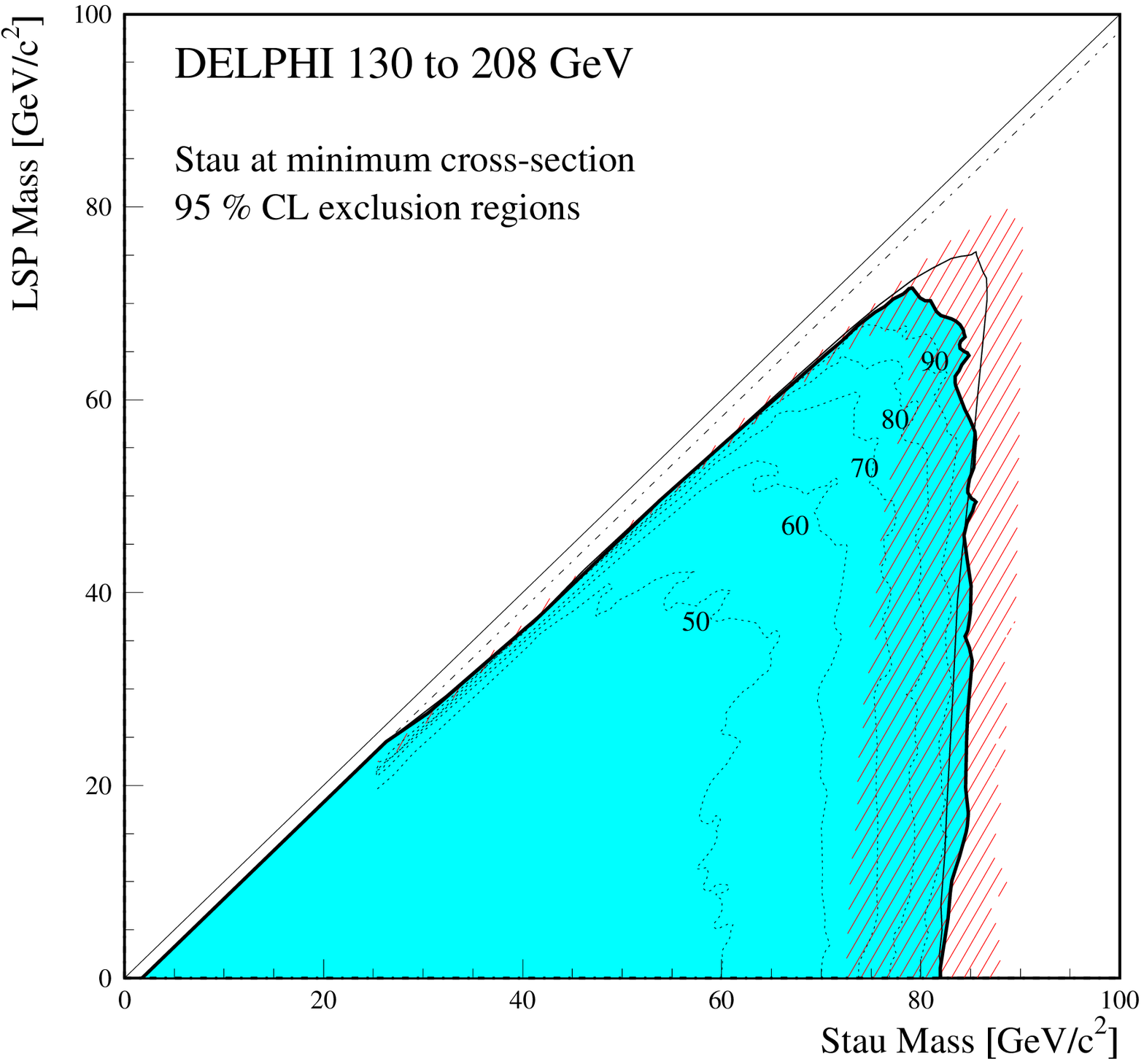}}
\vspace{0.5cm}
\caption{
Excluded region
in the (\mstau , $M_{LSP}$) plane obtained
for the minimal \stau\ pair-production cross-section.
The shaded 
region shows the region excluded
if \stau $\rightarrow \tau$ LSP is the only allowed channel, 
and the thin solid line shows the corresponding expected limit.
The dotted lines show the region excluded
if the branching ratio of \stau $\rightarrow \tau$ LSP has the values
indicated in percent next to the lines, and it is assumed that the analysis has no sensitivity
to other decay-modes.
The dash-dotted line indicates \DM\ = $m_{\tau}$.
The observed limit is everywhere within the 95\% CL band for the expected limit
shown as the hatched area.}

\label{fig:staumin:xcl}
\end{center}
\end{figure}

Figure~\ref{fig:staur:xcl} shows 
the $\stau_R$ excluded region.
Figure~\ref{fig:staumin:xcl} shows the excluded regions in the case of 
the mixing corresponding to the minimal production cross-section.
The cross-section has  a minimum at mixing
angle $52\dgree$ at LEP2 energies~\cite{bartl}.
In addition, the limit for $\stau_L$ was also evaluated,
and was found to be 84.7 \GeVcc\ (expected limit 84.9 \GeVcc) for
$M_{LSP}=0$ \GeVcc. 
The excess of candidates at low \DM\ for high stau-mass 
seen in both figures is
compatible with a statistical fluctuation: the observed limit was everywhere contained inside the
95\% confidence band for the expected limit, calculated 
from the Poisson
distributed background.

\begin{figure}[htbp]
\begin{center}
\mbox{\epsfysize=18.cm \epsffile{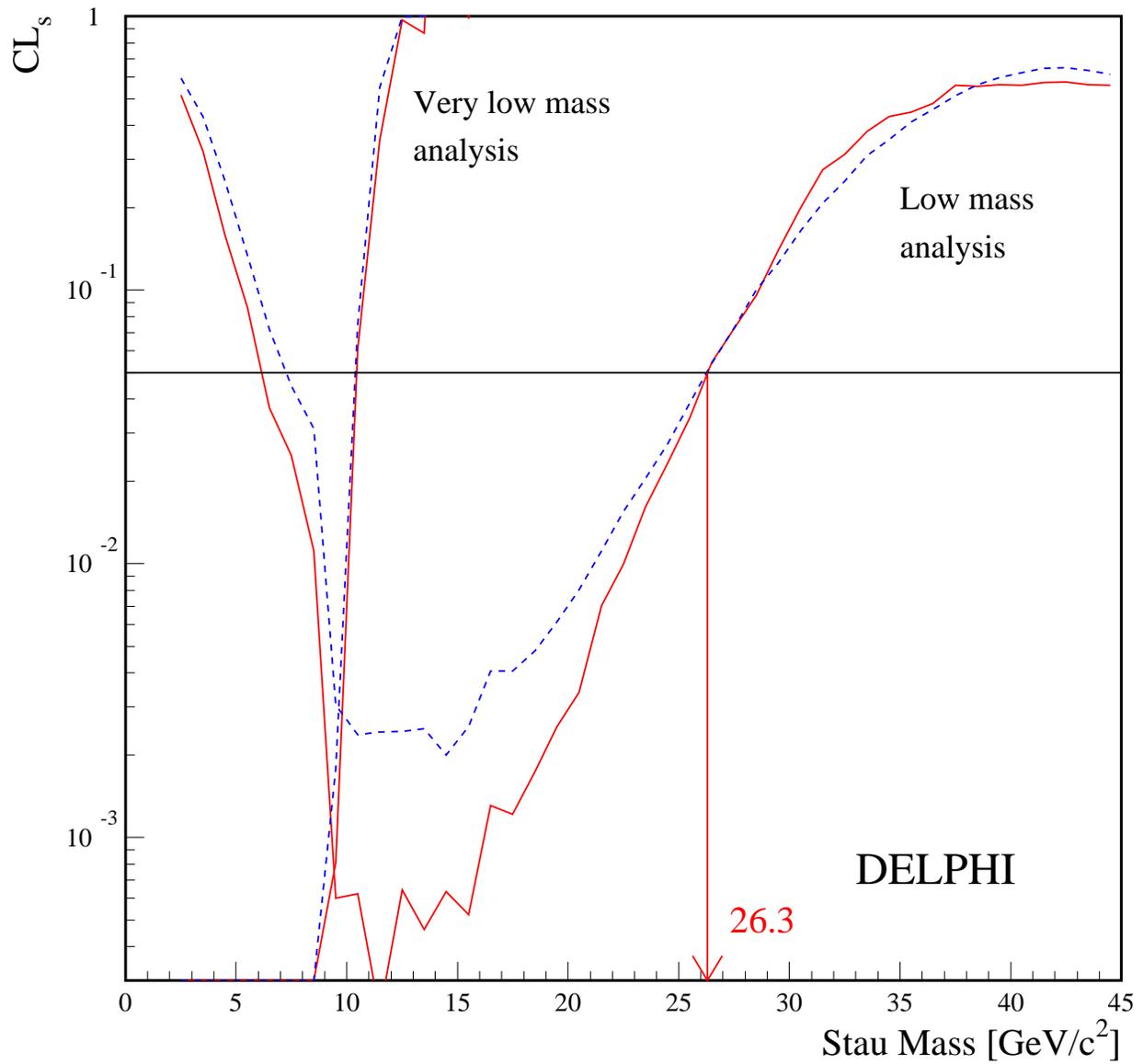}}
\vspace{0.5cm}
\caption{
The value of $CL_s$ versus
\mstau\ obtained
for the minimal \stau\ pair-production cross-section,
and \DM\ = $m_{\tau}$, which corresponds to the weakest
limit.
The solid
curve shows the obtained $CL_s$,
and the dashed line shows the corresponding expected value.
The two sets of curves correspond to the ``very low mass analysis'' and
the ``low mass analysis'', as indicated in the figure.
}

\label{fig:staumin:cls}
\end{center}
\end{figure}

From the low mass search, at the minimum cross-section, 
the lower limit on the stau mass was 26.3 \GeVcc\ for \DM\ = $m_{\tau}$,
31.7 \GeVcc\ for \DM\ = 3 \GeVcc, and 40.5 \GeVcc\ for \DM\ = 4 \GeVcc.
The corresponding expected limits were 26.3 \GeVcc, 35.9 \GeVcc, and
42.1 \GeVcc, respectively. 
The limit improves to 29.6 (31.1) \GeVcc\ 
for $\stau_R$ ($\stau_L$), with an expected limit of 30.0 (31.9) \GeVcc, for \DM\ = $m_{\tau}$. 
The low mass
search does not exclude \mstau\ below 6.3 \GeVcc\ for \DM\ below 3 \GeVcc.
As shown in figure \ref{fig:staumin:cls},
to cover this region combination with the very low \mstau\ search
is needed.
 


In summary, a stau mass
limit can be set at 
81.9 to  84.7 ~\GeVcc\ (depending on mixing) 
for mass differences between the
stau and the LSP above 15~\GeVcc.
The same limits hold for LSP masses below 68~\GeVcc\
and mass differences between the 
stau and the LSP above 6~\GeVcc .
The expected limit in the same region ranges from 
82.1 to  84.9 ~\GeVcc . 
The lowest stau mass allowed 
is
26.3 \GeVcc\ (any mixing-angle and any \DM $\ge$ m$_{\tau}$). 



\subsection{Squark searches}
\label{sec:sqresult}

\def\RES#1#2#3#4{$#1 \pm #2 ^{+ #3} _{- #4}$}

\subsubsection{Efficiencies and selected events}
\label{sec:squa_eff}

\begin{figure}[p!]
\begin{center}
\includegraphics[width=11cm]{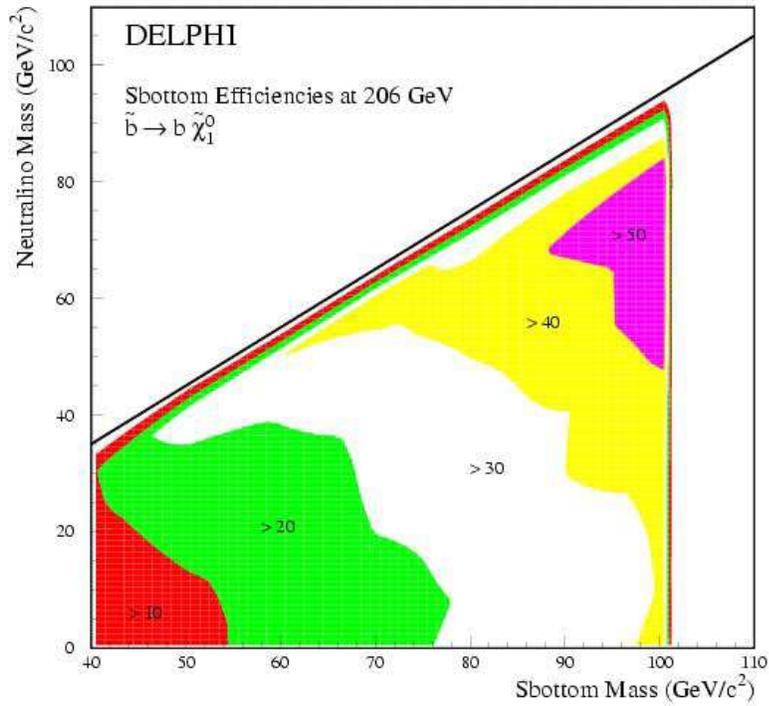}
\includegraphics[width=11cm]{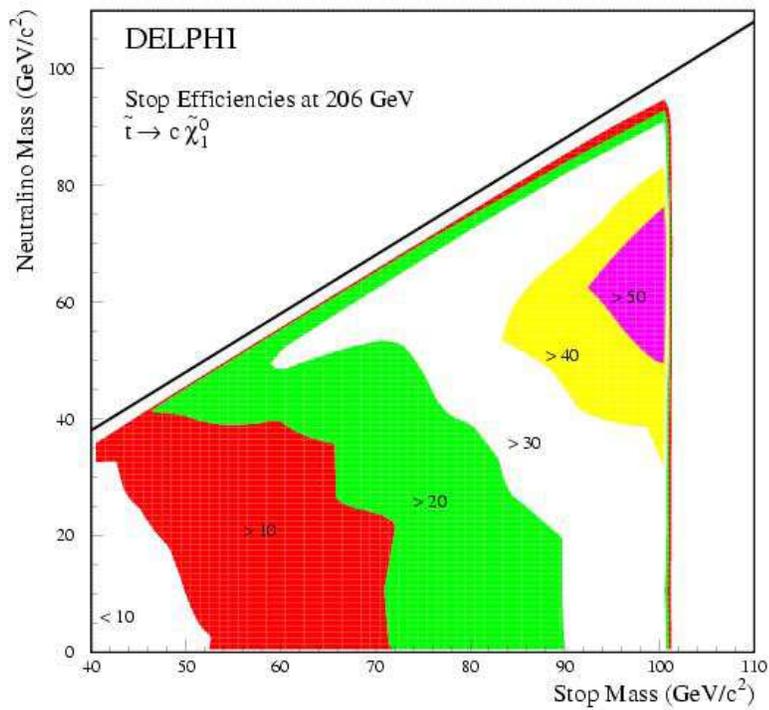}
\caption{Squark signal detection efficiencies (in \%) in
the plane  (M$_{{\tilde {\mathrm q}}_1}$,\MXN{1}) in the non-degenerate scenario.}
\label{fi:effisqu}
\end{center}
\end{figure}

The efficiencies of the 
stop and sbottom signal selection are summarised in figure~\ref{fi:effisqu}
for the non-degenerate scenario. 
Systematic uncertainties on the efficiency determination have been estimated
as explained in section~\ref{sec:sysrev}.
   
The number of events selected and the expected background 
at the final selection level are shown in table~\ref{tab:ressq}. The systematic uncertainties shown in the 
table have been estimated  using the ``re-weighting''  method described in 
section~\ref{sec:sysmet}. Moreover, in the mass limit computation, an additional relative uncertainty of 15\% related 
to the determination of the hadronic $\gamma\gamma$ contribution (see section~\ref{sec:sysrev}) has been added.

\begin{table}[h!]
\begin{center}
\begin{tabular}{|c||c|c||c|c|}
\hline
\multicolumn{5}{|c|}{Sbottom Analysis} \\
\hline
$\sqrt{s}$ & \multicolumn{2}{|c|}{\DM\ $\geq$ 20 \GeVcc} &
   \multicolumn{2}{|c|}{5 $<$ \DM\ $<$ 20 \GeVcc}\\
\cline{2-5}
(GeV)         & Data & MC & Data & MC \\
\hline
189      & 2 & \RES{0.43}{0.08}{0.12}{0.03}     & 1 & \RES{0.47}{0.20}{0.19}{0.05}  \\
192      & 0 & \RES{0.05}{0.01}{0.02}{0.02}     & 0 & \RES{0.06}{0.02}{0.01}{0.03}  \\ 
196      & 0 & \RES{0.17}{0.04}{0.03}{0.03}     & 1 & \RES{0.22}{0.08}{0.02}{0.07}  \\ 
200      & 0 & \RES{0.14}{0.03}{0.02}{0.03}     & 0 & \RES{0.27}{0.08}{0.05}{0.30}  \\ 
202      & 0 & \RES{0.07}{0.02}{0.03}{0.01}     & 1 & \RES{0.13}{0.04}{0.04}{0.03}  \\ 
205      & 0 & \RES{0.49}{0.05}{0.05}{0.06}     & 0 & \RES{0.43}{0.17}{0.05}{0.23}  \\ 
207      & 0 & \RES{0.36}{0.04}{0.07}{0.03}     & 0 & \RES{0.38}{0.17}{0.13}{0.16}  \\ 
208      & 0 & \RES{0.05}{0.01}{0.07}{0.00}     & 0 & \RES{0.04}{0.02}{0.06}{0.01}  \\
206.5(*) & 0 & \RES{0.33}{0.04}{0.01}{0.10}     & 0 & \RES{0.12}{0.03}{0.02}{0.05}  \\
\hline
Total    & 2 & \RES{2.10}{0.12}{0.17}{0.13}    & 3 & \RES{2.12}{0.34}{0.25}{0.42} \\
\hline
\multicolumn{5}{c}{ } \\
\hline
\multicolumn{5}{|c|}{Stop Analysis} \\
\hline
$\sqrt{s}$ & \multicolumn{2}{|c|}{\DM\ $\geq$ 20 \GeVcc} &
   \multicolumn{2}{|c|}{10 $<$ \DM\ $<$ 20 \GeVcc}\\
\cline{2-5}
(GeV)        & Data & MC & Data & MC \\
\hline
189    & 3 & \RES{2.28}{0.22}{0.78}{0.01} & 3 & \RES{0.87}{0.21}{0.28}{0.02}  \\ 
192    & 2 & \RES{0.92}{0.11}{0.17}{0.40} & 0 & \RES{0.27}{0.07}{0.01}{0.10}  \\
196    & 0 & \RES{2.35}{0.22}{0.34}{0.24} & 3 & \RES{0.78}{0.15}{0.08}{0.20}  \\
200    & 1 & \RES{2.14}{0.13}{0.07}{0.39} & 0 & \RES{0.91}{0.16}{0.13}{0.11}  \\
202    & 1 & \RES{1.16}{0.07}{0.49}{0.12} & 0 & \RES{0.49}{0.08}{0.18}{0.12}  \\
205      & 5 & \RES{2.00}{0.11}{0.34}{0.28} & 0 & \RES{0.75}{0.18}{0.07}{0.42}  \\
207      & 1 & \RES{2.32}{0.11}{0.43}{0.10} & 1 & \RES{0.78}{0.18}{0.16}{0.27}  \\
208      & 0 & \RES{0.19}{0.01}{0.26}{0.01} & 0 & \RES{0.08}{0.02}{0.14}{0.00}  \\
206.5(*) & 1 & \RES{2.67}{0.11}{0.01}{0.89} & 0 & \RES{0.41}{0.04}{0.03}{0.18}  \\
\hline
Total    &14 & \RES{16.03}{0.41}{1.2}{1.1} & 7 & \RES{5.34}{0.41}{0.43}{0.60}\\
\hline
\end{tabular}
\caption{Number of events selected by the squark analysis in the non-degenerate 
scenarios. The first errors are statistical and the second ones are systematic;
(*) indicates the 2000 data taken with the sector 6 of the TPC off.}
\label{tab:ressq}
\end{center}
\end{table}

\begin{table}[h!]
\begin{center}
\begin{tabular}{|c|c|c|c|c|}
\hline
\multicolumn{5}{|c|}{Stop (nearly degenerate) \DM\ $\leq$ 10 \GeVcc} \\
\hline 
\sqs         & \multicolumn{2}{|c|}{Preselection} &  \multicolumn{2}{|c|}{Final} \\
\cline{2-5}
(GeV)                      & Data & MC & Data & MC \\
\hline
189  & 3717 & 3717   & 3 & \RES{2.72}{0.34}{0.78}{0.33} \\
192   &  527 &  599   & 0 & \RES{0.33}{0.12}{0.10}{0.15} \\ 
196    & 1620 & 1623  & 2 & \RES{1.02}{0.21}{0.12}{0.17} \\ 
200  & 1667 & 1679   & 0 & \RES{1.12}{0.22}{0.27}{0.15} \\ 
202   &  867 &  793   & 0 & \RES{0.64}{0.16}{0.26}{0.05} \\ 
205  & 1469 & 1492     & 1 & \RES{1.32}{0.33}{0.14}{0.21} \\ 
207   & 1423 & 1468     & 2 & \RES{1.33}{0.33}{0.35}{0.21} \\ 
208    &  138 &  133   & 1 & \RES{0.12}{0.10}{0.17}{0.01} \\
206.5(*)& 1023 & 1133  & 0 & \RES{0.55}{0.19}{0.21}{0.16} \\
\hline
Total &12451 &   12637 & 9 & \RES{9.15}{0.72}{0.99}{0.55} \\
\hline
\end{tabular}
\caption{Number of events selected in the stop analysis in the nearly degenerate scenario.
The first errors are statistical and the second ones are systematic;
(*) indicates 2000 data taken with sector 6 of the TPC off.}
\label{tab:damcre_deg}
\end{center}
\end{table}

\begin{figure}[p!]
\begin{center}
\includegraphics[width=11cm]{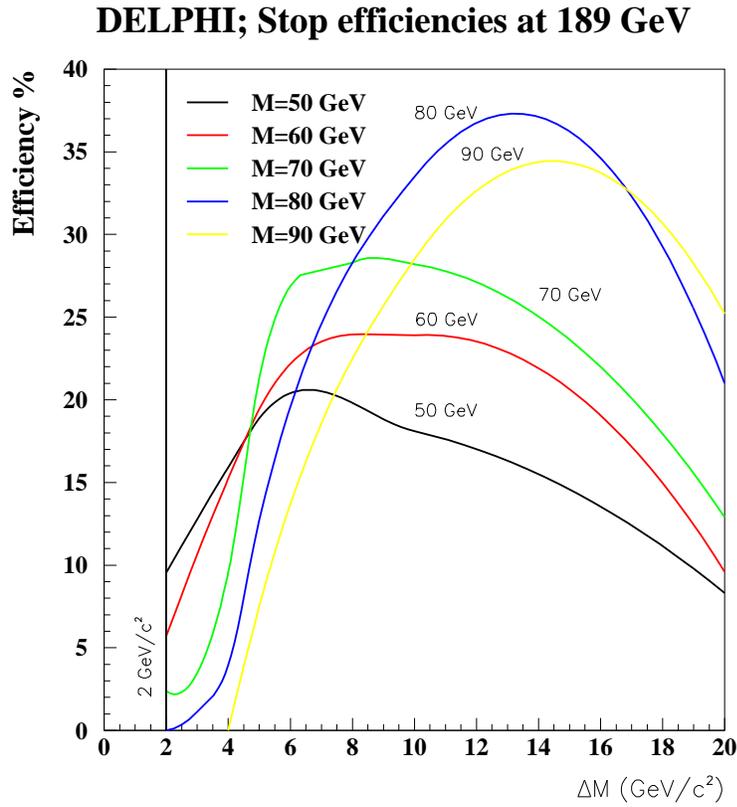}
\includegraphics[width=11cm]{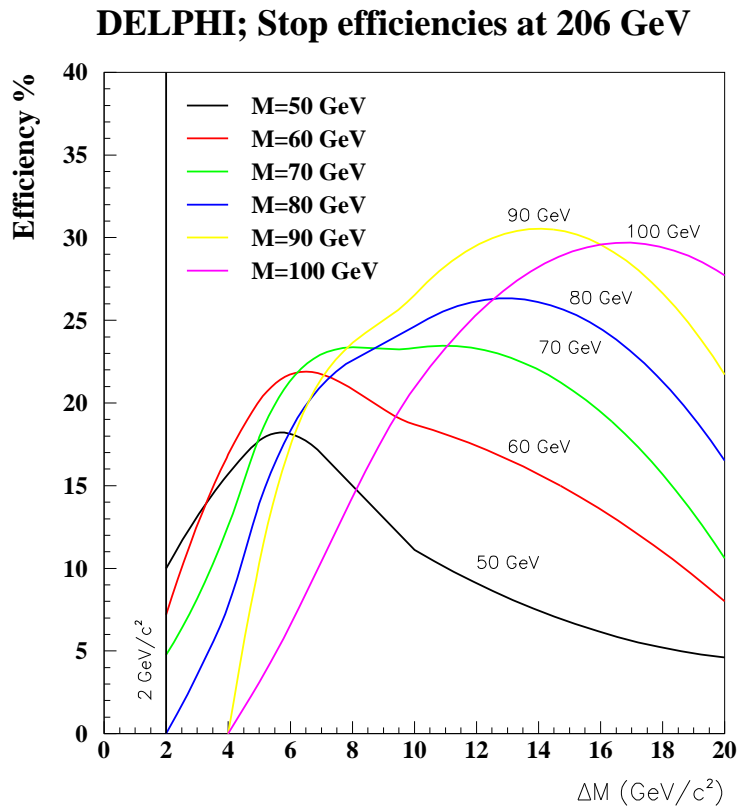}
\caption{Signal efficiency 
of the stop search in the nearly degenerate scenario for different mass hypothesis as a function of 
$\DM={\mathrm M}_{{\tilde{t}}_1}-{\MXN{1}}$,
for $\sqrt{s}$=189 and 206 GeV.}
\label{fi:eff_deg}
\end{center}
\end{figure}

In the nearly degenerate scenario, the efficiency for the stop signal is summarised 
in figure~\ref{fi:eff_deg}. The figure shows the variation of
the efficiency for signal selection as a function of \DM\ for different stop mass hypotheses, 
for centre-of-mass energies of 189 and 206 \GeV . 
Some examples of the detection efficiency obtained with this
analysis are quoted below: for a \DM\ of 2 GeV/c$^2$ and
a stop mass of 70 GeV/c$^2$, an efficiency of 2.4\% is obtained at
$\sqrt{s}=$ 189 GeV and of 4.8\% at $\sqrt{s}=$ 206 GeV. For a \DM\
of 4 GeV/c$^2$ and a stop mass of 80 GeV/c$^2$, an efficiency of 4.0\% is
achieved at $\sqrt{s}=$ 189 GeV and of 7.7\% at $\sqrt{s}=$ 206 GeV.

Table~\ref{tab:damcre_deg} gives the results for the nearly degenerate scenario in 
terms of number of events compared to the MC 
expectation after the preselection 
and at the final selection level. The systematic uncertainties were 
computed using the same method as for the non-degenerate case, 
and the same remark concerning the 
hadronic two-photon background applies.

\subsubsection{Limits}
\label{sec:squa_res}

Stop and sbottom cross-sections were calculated with the SUSYGEN 
program for
two squark mixing angles. For purely left-handed squarks
($\theta_{\tilde{q}}=0^\circ$), the cross-section is maximal. The squark mixing
angle which corresponds to the \Zn\ decoupling is $56^\circ$ for the stop and
$68^\circ$ for the sbottom, and it corresponds approximately to the minimal
cross-section. 
The program ALRMC~\cite{alexread} was used 
and 
systematic errors were taken into account in the definition of $CL_s$. 

\begin{figure}[p!]
\begin{center}
\includegraphics[width=11cm]{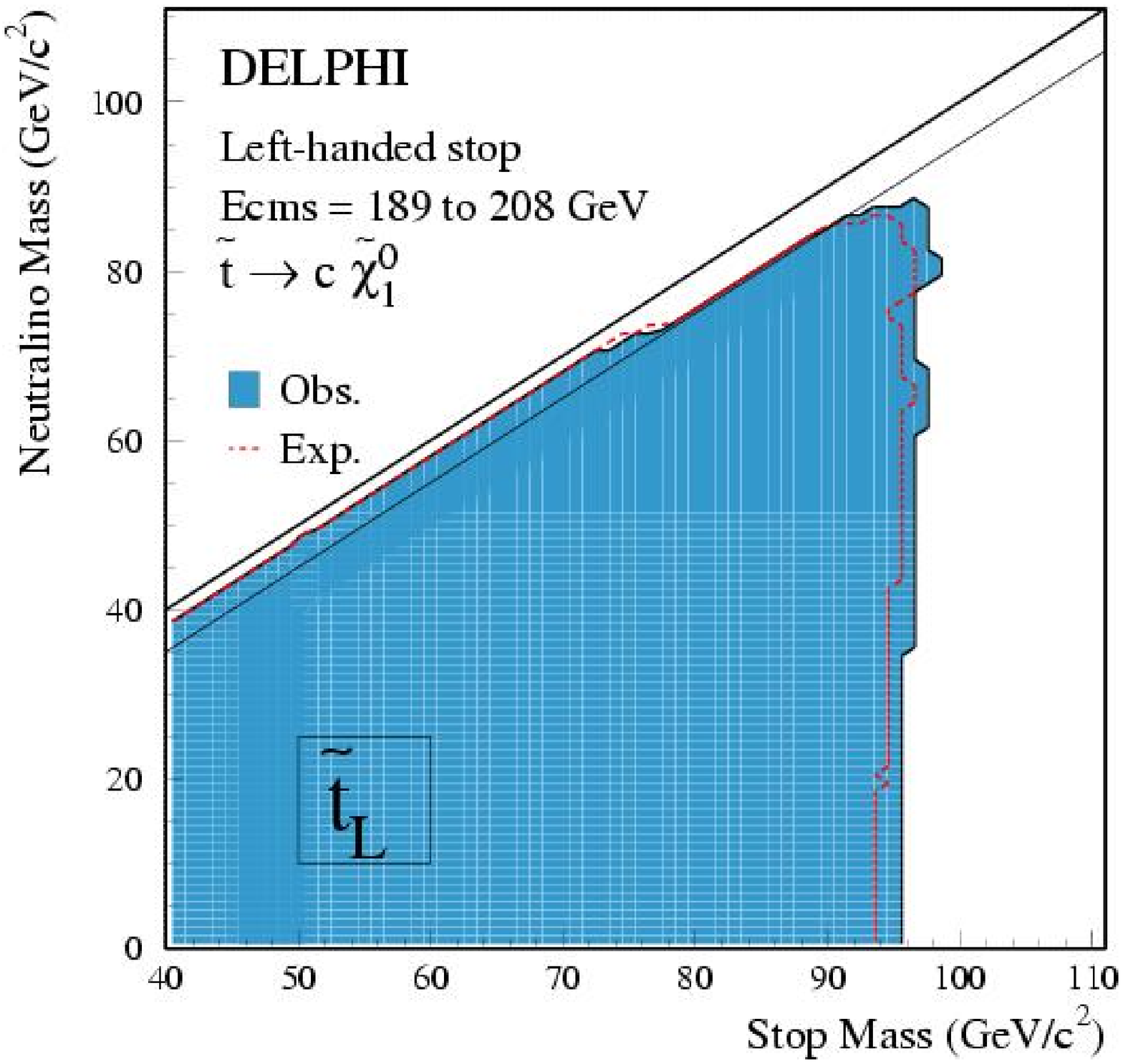}
\includegraphics[width=11cm]{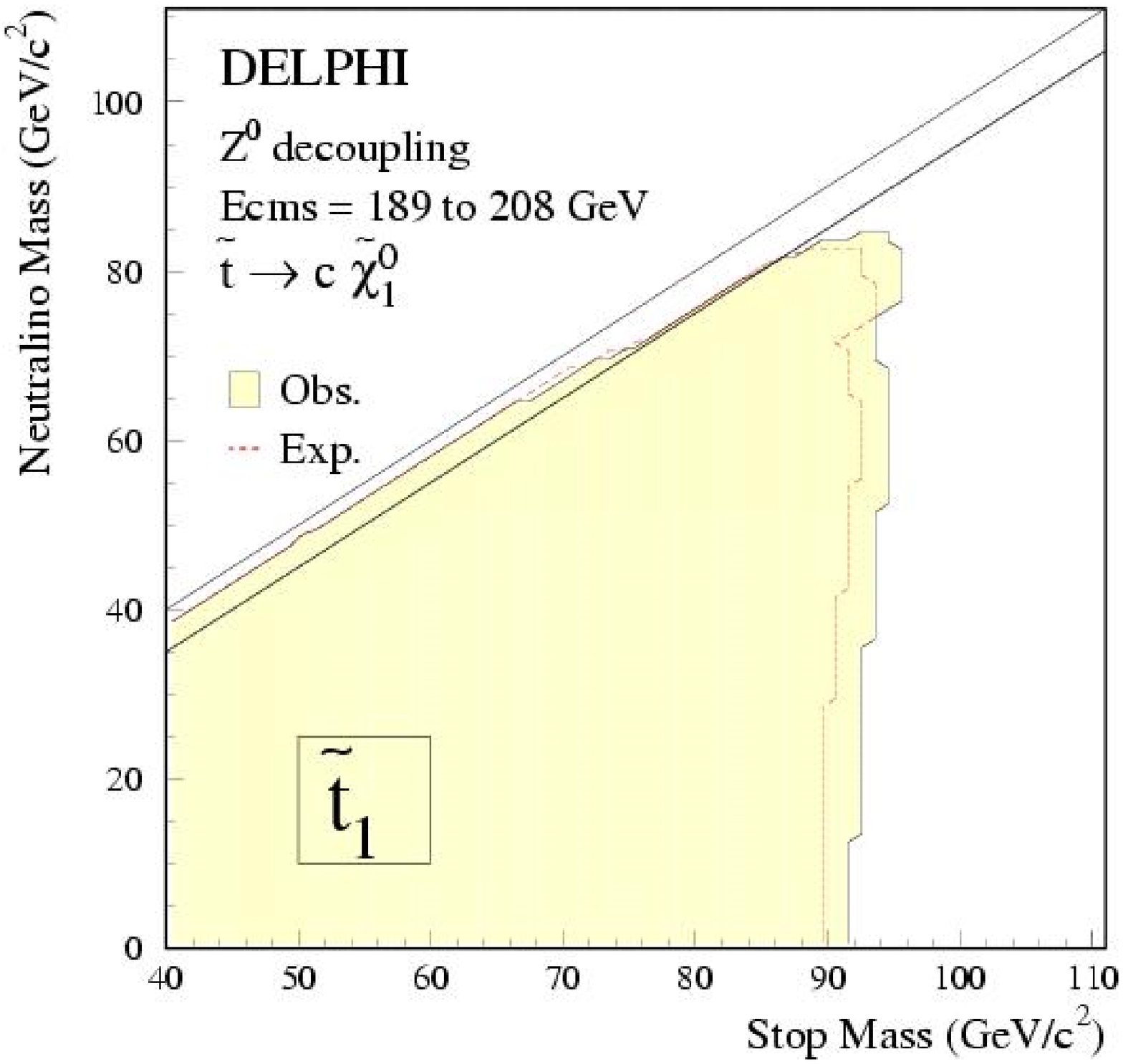} 
\caption{Excluded regions for the stop 
search in the plane (M$_{\tilde{t}_1}$,M$_{\tilde{\chi}^0_1}$) for purely left
handed stops (top) and for the states at the $Z$ decoupling (bottom). 
The shaded areas show the observed excluded regions and the lines correspond 
to the expected exclusions
(for a discussion on the low stop mass region see section~\ref{sec:lep1}).
}
\label{fi:stexclu}
\end{center}
\end{figure}
\begin{figure}[p!]
\begin{center}
\includegraphics[width=11cm]{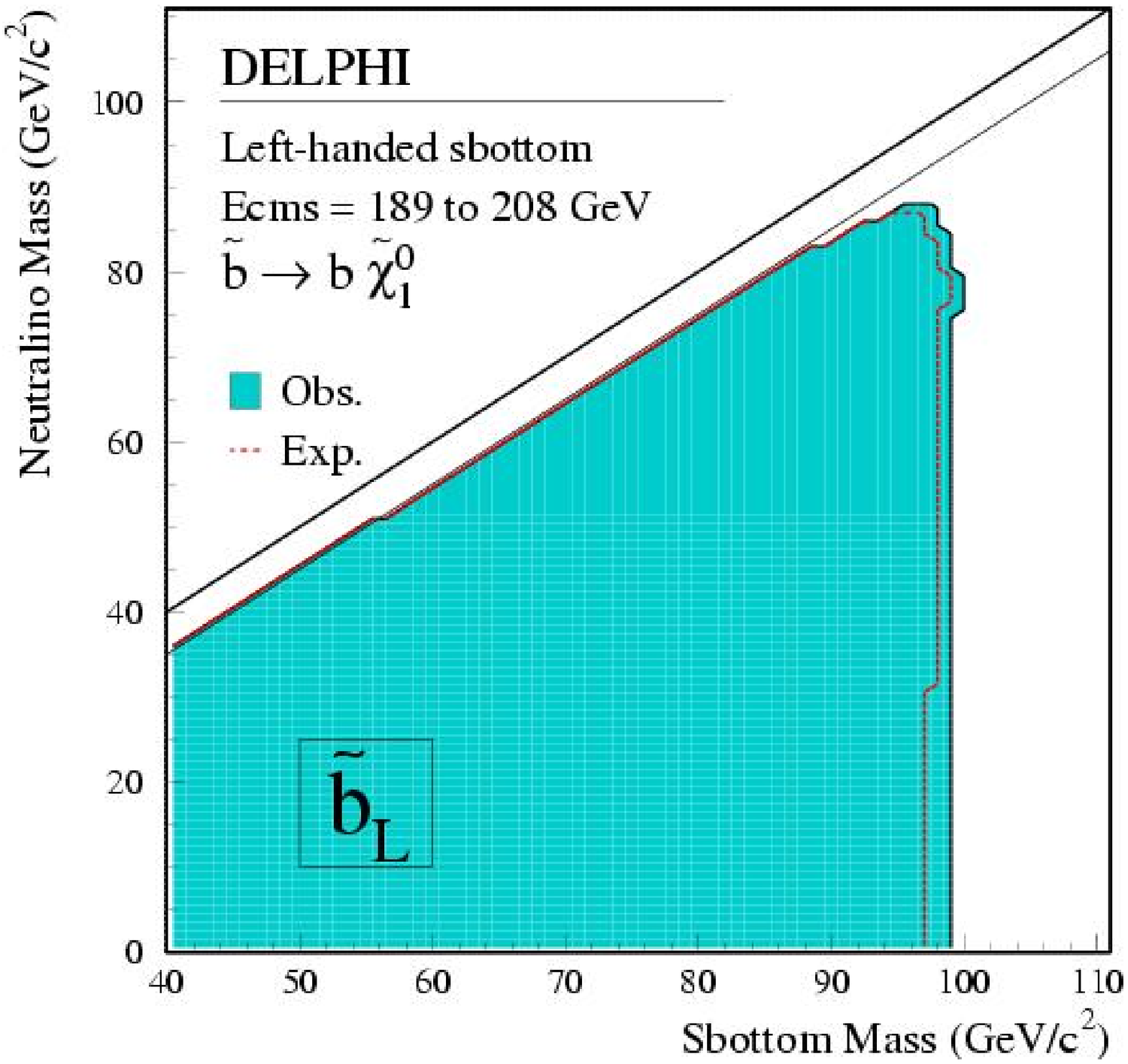}
\includegraphics[width=11cm]{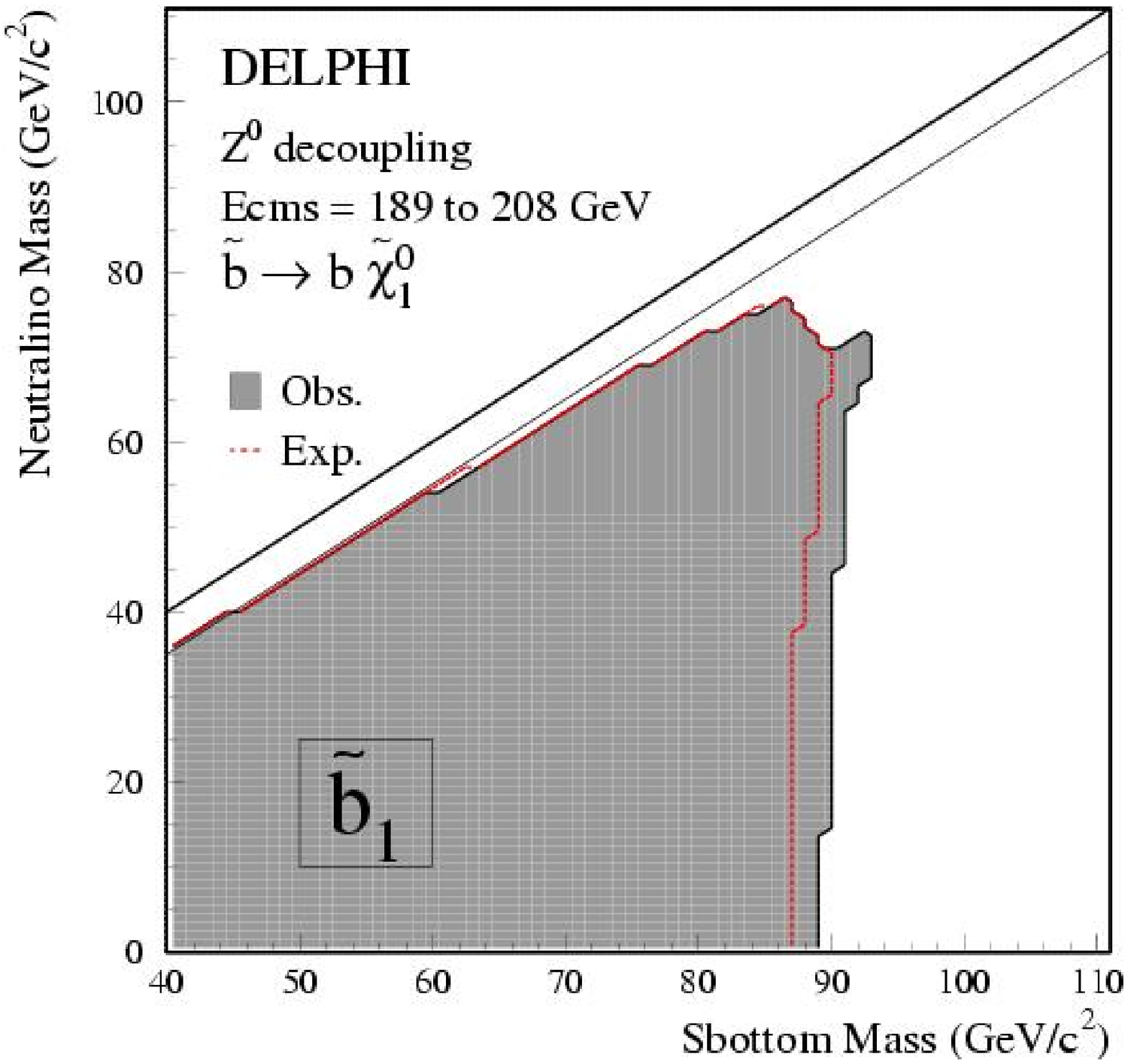}
\caption{Excluded regions at for the sbottom 
search in the plane (M$_{\tilde{b}_1}$,M$_{\tilde{\chi}^0_1}$) for purely left
handed sbottoms (top) and for the states at the $Z$ decoupling (bottom). 
The shaded areas show the observed excluded regions and the lines correspond 
to the expected exclusions
(for a discussion on the low sbottom mass region see section~\ref{sec:lep1}).
}
\label{fi:sbexclu}
\end{center}
\end{figure}
\begin{figure}[h!]
\begin{center}
\includegraphics[width=11cm]{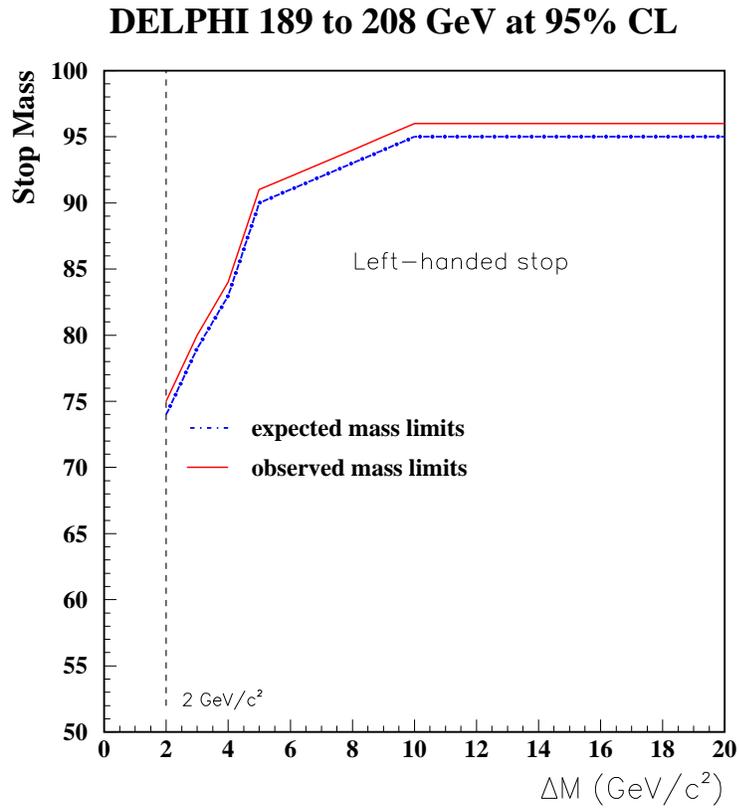}
\includegraphics[width=11cm]{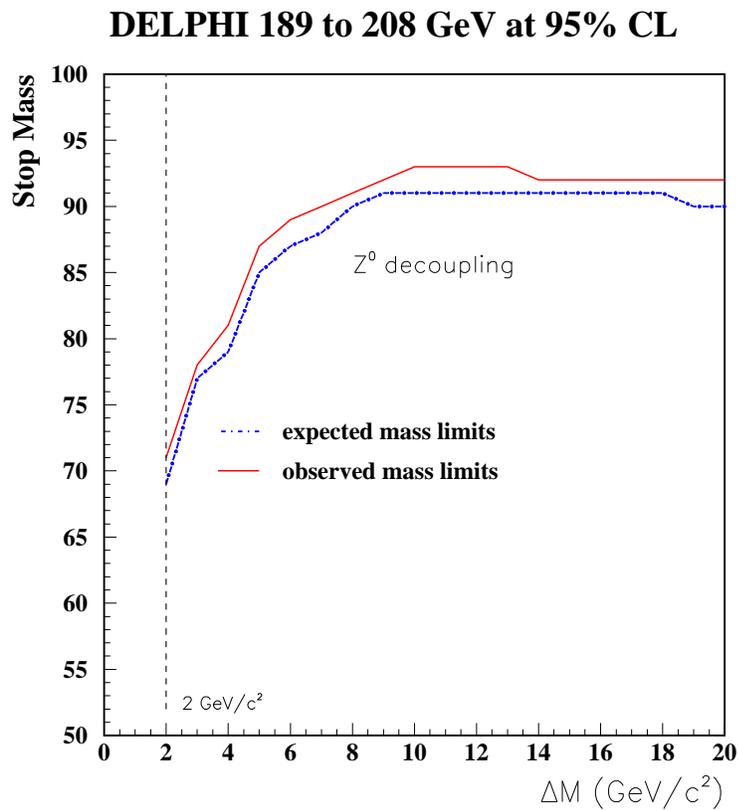}
\caption{Stop mass limit as a function of \DM\
for a purely left-handed stop 
and 
for the stop mixing angle which minimises the cross-section.}

\label{fi:excl_deg}
\end{center}
\end{figure}

Figures~\ref{fi:stexclu} and~\ref{fi:sbexclu} show the 
(M$_{\tilde{\mathrm q}}$,M$_{\tilde{\chi}^0_1}$)
regions excluded  by the search for
$\stq \to {\mathrm c} \tilde{\chi}^0_1 $  and 
$\sbq \to {\mathrm b} \tilde{\chi}^0_1 $  decays,
with the 100~\% branching ratio assumption,
both for purely left-handed states and for the states at the \Zn\ decoupling.
Figure~\ref{fi:excl_deg} shows the region excluded using only the nearly degenerate analysis for \DM\ values
between 2 and 20 \GeVcc .

Table~\ref{tab:sqmexclu} shows the limit on the squark masses as a function of
\DM\ obtained combining the two analyses. 
The introduction of the systematics in the confidence level
calculation has no effect on these numbers. Stop masses lower than 71~\GeVcc\ and 
sbottom masses lower than 76~\GeVcc\ are excluded if \DM~$\geq$~2~\GeVcc\ and 
\DM~$\geq$~7~\GeVcc, respectively,
for any squark mixing angle. These 
limits
become 75~\GeVcc\ and 93~\GeVcc\ for 
purely left-handed squarks. 

\begin{table}[h!]
\begin{center}
\begin{tabular}{|l||c|c||c|c|}
\hline
 & \multicolumn{2}{|c|}{Sbottom} & \multicolumn{2}{|c|}{Stop} \\
\hline
 & $\theta_{\tilde{b}}=$0 & $\theta_{\tilde{b}}=68^\circ$ & $\theta_{\tilde{t}}=$0 & $\theta_{\tilde{t}}=56^\circ$ \\
\hline
\DM~$\geq$~2~\GeVcc\  &  -  &  - & 75 & 71\\
\DM~$\geq$~3~\GeVcc\  &  -  &  - & 80 & 78\\
\DM~$\geq$~4~\GeVcc\  &  -  &  - & 84 & 81\\
\DM~$\geq$~5~\GeVcc\  &  -  &  - & 91 & 87\\
\DM~$\geq$~7~\GeVcc\  & 93  & 76 & 95 & 91\\
\DM~$\geq$~10~\GeVcc\ & 98  & 87 & 96 & 92\\
\DM~$\geq$~15~\GeVcc\ & 99  & 89 & 96 & 92\\
\hline
\end{tabular}
\end{center}
\caption{Lower limits on squark masses (in \GeVcc) as a function of \DM\ 
from the squark analysis in the non-degenerate and nearly degenerate scenarios.}
\label{tab:sqmexclu}
\end{table}


\subsection{Chargino searches}
\subsubsection{Non-degenerate scenarios}
 \label{sub:chasdt_res}

\vspace{0.35cm}
\noindent
\underline{Efficiencies and selected events}
\vspace{0.35cm}

\begin{figure}[hbt]
\begin{center}
    \begin{tabular}{c}
   \epsfysize=16.0cm\epsffile{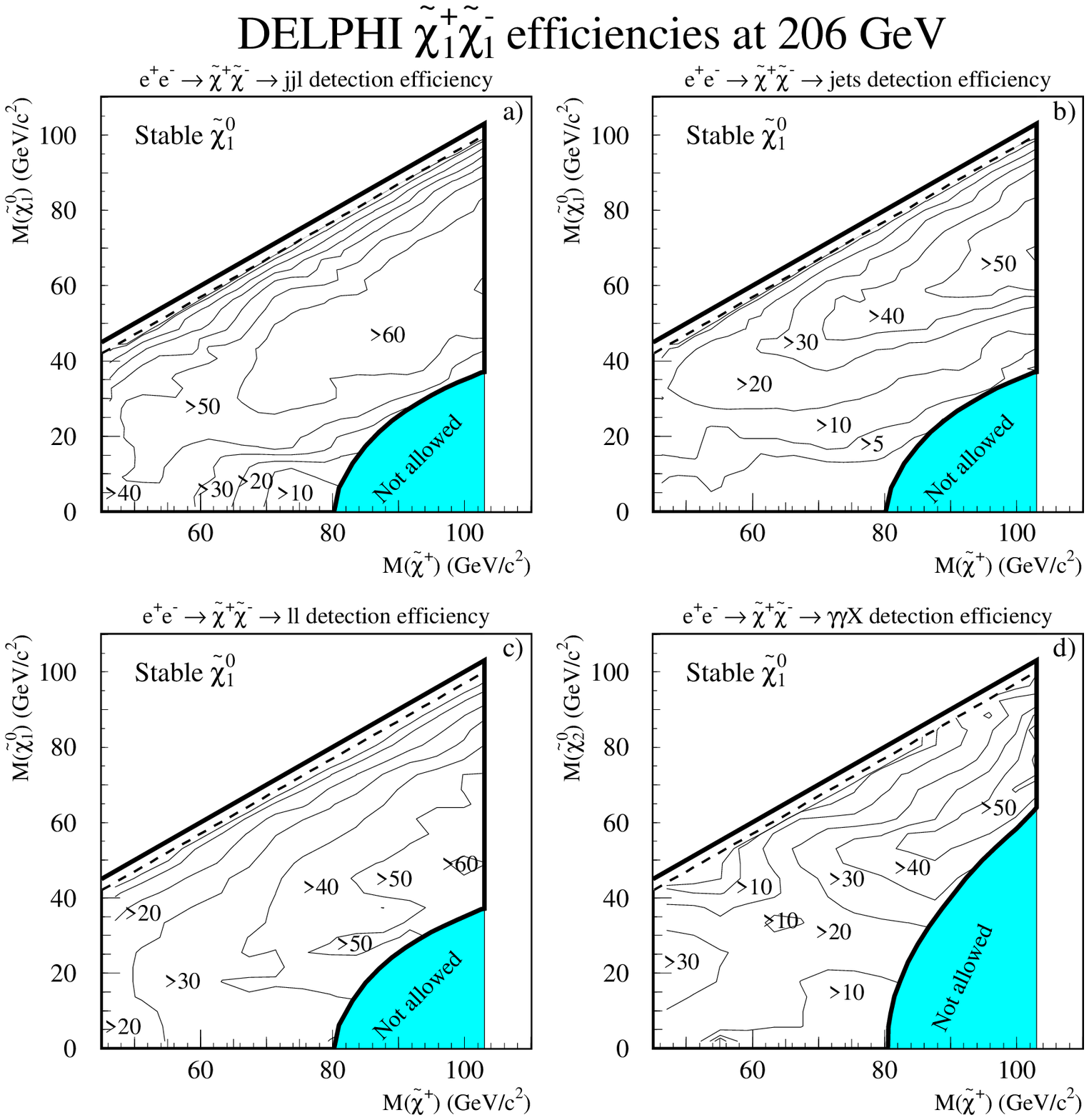} 
    \end{tabular}
\caption{Chargino pair-production detection efficiencies~(\%) for the four decay channels a) \jjl,
b) \jjjj, c) \ll\ and d) \rad, at 206 GeV in the ($\MXc,\MXn$) plane. TPC sector 6 is on and a
stable \XN{1}\ is assumed. The shaded areas are disallowed in the MSSM scheme.}
    \label{fig:CHAEFF}
\end{center}
\end{figure}
\begin{figure}[hbt]
\begin{center}
    \begin{tabular}{c}
   \epsfysize=16.0cm\epsffile{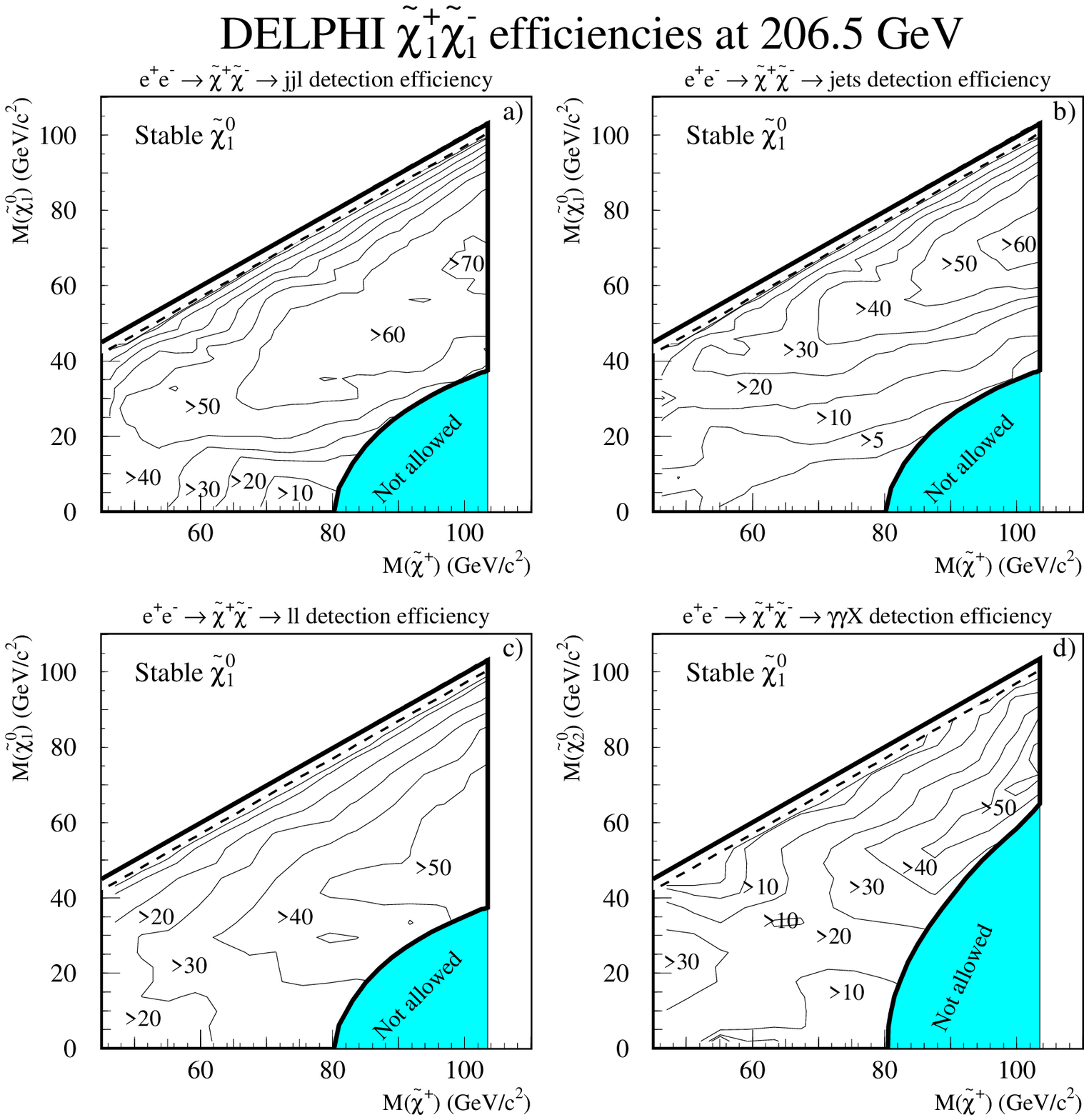} 
    \end{tabular}
\caption{Chargino pair-production detection efficiencies~(\%) for the four decay channels a) \jjl,
b) \jjjj, c) \ll\ and d) \rad, at 206.5~GeV in the ($\MXc,\MXn$) plane. TPC sector 6 is off and a
stable \XN{1}\ is assumed. The shaded areas are disallowed in the MSSM scheme.}
    \label{fig:CHAEFFS1}
\end{center}
\end{figure}

  The efficiencies of the chargino selection in the four topologies were computed separately for
 the 132 MSSM points using the $\mathcal{L}_{R}$ function and the $\mathcal{L}_{R_{CUT}}$ of the corresponding topology 
 and \DM\ region.
  To pass from the efficiencies of the chargino selection in the four topologies to the efficiencies in the four 
decay channels, all the migration effects were computed for all the generated points of the signal simulation. 
Then the efficiencies of the selection in the four decay channels were interpolated 
in the (\MXC{1},\MXN{1}) plane using the same method as in~\cite{char}. 
These efficiencies as functions of \MXC{1}\ and \MXN{1} are shown in 
figure~\ref{fig:CHAEFF}(\ref{fig:CHAEFFS1}) for
a mean centre-of-mass energy of 206(206.5)~GeV with the TPC sector 6 on(off).

Table \ref{tab:candchastand} gives the total 
number of events selected in data and expected from the SM simulation 
after the preselections and after the final selections,
for the different centre-of-mass energies.
The total number of background events expected in the different mass windows and
topologies is shown in tables \ref{tab:EXPEV1999}, \ref{tab:EXPEV2000} and 
\ref{tab:EXPEV2000S1}, together with the
number of events selected. 
In all the topologies, the number of selected events in the real data is compatible 
with the expectation from the 
background simulation.

The systematic errors shown in tables~\ref{tab:EXPEV1999}, \ref{tab:EXPEV2000} and 
\ref{tab:EXPEV2000S1} were obtained with the ``re-weighting'' method 
described in section \ref{sec:sysmet}.


\begin{table}[htb]
\begin{center}
{\footnotesize
\begin{tabular}{|c|c|c|c|c|c|}
\hline
$\sqrt{s}$ & $\int \cal L$ & \mc{2}{c|}{Preselection} &  \mc{2}{c|}{Final sel.} \\
\cline{2-6}
 (GeV)         &  (pb$^{-1}$)  & Data & MC  & Data & MC  \\
\hline
192       &  25.4 &  1966  &  $ 2012 \pm  11$  &  21  &  $23.7 \pm 0.9$  \\ 
196       &  76.2 &  5926  &  $ 5818 \pm  25$  &  85  &  $71.6 \pm 2.4$   \\ 
200       &  84.0 &  6433  &  $ 6331 \pm  22$  &  60  &  $72.3 \pm 1.7$ \\ 
202       &  40.4 &  3086  &  $ 2994 \pm  11$  &  26  &  $35.0 \pm 0.8$ \\ 
205       &  78.3 &  5796  &  $ 5734 \pm  21$  &  56  &  $54.6 \pm 1.3$  \\ 
207       &  78.8 &  5795  &  $ 5770 \pm  21$  &  63  &  $54.9 \pm 1.3$\\ 
208       &  7.2 &  530  &  $ 528 \pm  2$      &   3  &  $ 5.1 \pm 0.1$ \\ 
206.5(*) &  60.0 &  4119  &  $4356 \pm 16$  &  53  &  $58.7 \pm 1.8$ \\ 
\hline
all       &       & 33651  &  $33543 \pm 50$  & 367  & $375.9 \pm 4.0$ \\ 
\hline
\end{tabular}
}
\caption{\small
Numbers of events selected in data and expected from the SM simulation in the 
non-degenerate chargino search 
at the preselection and at the final  selection level, for the different centre-of-mass energies collected during the years 
1999 and 2000; (*) indicates the data collected with the TPC sector 6 off.
Simulation errors are statistical.}
\label{tab:candchastand}
\end{center}
\end{table}

\begin{table}[hbtp]
\begin{center}
\footnotesize
\begin{tabular}{|c||c|c|c|c||c|} 
\multicolumn{3}{l}{} &
\multicolumn{3}{c}{1999 data, ${\cal L}~=~227~pb^{-1}$} \\
\hline
  &  &  &  &  & \\
{  Topology:} & {  \jjl} & {  \ll} &  {  \jjjj} &  {  rad}  & {  Total} \\ 
  &  &  &  &  & \\
\hline \hline
 & \multicolumn{5}{|c|}{3 $\leq$ \DM $<$ 5~\GeVcc  } \\ \hline 
{ Obs. events:}  & 4 & 39 & 3 & 7 & 53 \\ 
{ Expect. events:} 
& 2.3$\pm 0.7 ^{+1.2} _{-0.1}$ 
& 49.2$\pm 2.3 ^{+2.6} _{-5.7}$ 
& 5.3$\pm 0.9 ^{+0.0} _{-2.7}$ 
& 5.3$\pm 0.7 ^{+0.9} _{-0.2}$ 
& 62.1$\pm 2.7 ^{+3.0} _{-6.3}$ \\ 
\hline 
 & \multicolumn{5}{|c|}{5 $\leq$ \DM $<$ 10~\GeVcc  } \\ \hline 
{ Obs. events:}  & 4 & 13 & 1 & 7 & 25 \\ 
{ Expect. events:} 
& 2.3$\pm 0.7 ^{+1.2} _{-0.1}$ 
& 11.9$\pm 1.1 ^{+0.5} _{-3.0}$ 
& 2.5$\pm 0.7 ^{+0.5} _{-0.7}$ 
& 5.3$\pm 0.7 ^{+0.9} _{-0.2}$ 
& 22.0$\pm 1.7 ^{+1.6} _{-3.1}$ \\ 
\hline 
 & \multicolumn{5}{|c|}{10 $\leq$ \DM $<$ 25~\GeVcc  } \\ \hline 
{ Obs. events:}  & 4 & 14 & 17 & 7 & 42 \\ 
{ Expect. events:} 
& 2.3$\pm 0.7 ^{+1.2} _{-0.1}$ 
& 14.3$\pm 0.9 ^{+1.7} _{-3.1}$ 
& 15.8$\pm 1.2 ^{+2.1} _{-1.4}$ 
& 5.3$\pm 0.7 ^{+0.9} _{-0.2}$ 
& 37.6$\pm 1.8 ^{+3.1} _{-3.4}$ \\ 
\hline 
 & \multicolumn{5}{|c|}{25 $\leq$ \DM $<$ 35~\GeVcc  } \\ \hline 
{ Obs. events:}  & 6 & 21 & 10 & 7 & 44 \\ 
{ Expect. events:} 
& 2.2$\pm 0.2 ^{+0.7} _{-0.0}$ 
& 25.1$\pm 1.0 ^{+4.2} _{-3.5}$ 
& 8.3$\pm 0.3 ^{+0.8} _{-0.0}$ 
& 5.3$\pm 0.7 ^{+0.9} _{-0.2}$ 
& 40.6$\pm 1.3 ^{+4.4} _{-3.5}$ \\ 
\hline 
 & \multicolumn{5}{|c|}{35 $\leq$ \DM $<$ 50~\GeVcc  } \\ \hline 
{ Obs. events:}  & 2 & 40 & 28 & 14 & 84 \\ 
{ Expect. events:} 
& 2.3$\pm 0.2 ^{+0.7} _{-0.1}$ 
& 45.1$\pm 1.1 ^{+3.1} _{-4.2}$ 
& 23.6$\pm 0.4 ^{+3.4} _{-0.4}$ 
& 12.9$\pm 0.8 ^{+1.7} _{-0.2}$ 
& 84.1$\pm 1.4 ^{+4.9} _{-4.2}$ \\ 
\hline 
 & \multicolumn{5}{|c|}{50 $\leq$ \DM                  } \\ \hline 
{ Obs. events:}  & 9 & 60 & 37 & 14 & 120 \\ 
{ Expect. events:} 
& 7.5$\pm 0.3 ^{+2.2} _{-0.3}$ 
& 68.2$\pm 1.2 ^{+4.4} _{-6.9}$ 
& 36.8$\pm 0.5 ^{+4.9} _{-0.4}$ 
& 12.9$\pm 0.8 ^{+1.7} _{-0.2}$ 
& 125.4$\pm 1.6 ^{+7.1} _{-6.9}$ \\ 
\hline 
 & \multicolumn{5}{|c|}{ TOTAL (logical .OR. between different \DM\ windows)} \\ \hline 
{ Obs. events:}  & 14 & 109 & 54 & 15 & 192 \\ 
{ Expect. events:} 
& 10.7$\pm 0.8 $ 
& 126.1$\pm 2.7 $ 
& 52.8$\pm 1.4 $ 
& 13.9$\pm 0.8 $ 
& 202.6$\pm 3.2 $ \\ 
\hline 
\end{tabular}
\end{center}
\caption[.]{
\label{tab:EXPEV1999}
\small Number of events observed in data and expected number of
background events in the different chargino search channels 
for all the data collected in 1999. The first errors are statistical and the second ones are systematic.
The ``re-weighting''  method  used to compute the systematics is described in section \ref{sec:sysrev}.}
\end{table}
\begin{table}[ht]
\begin{center}
\footnotesize
\begin{tabular}{|c||c|c|c|c||c|} 
\multicolumn{3}{l}{} &
\multicolumn{3}{c}{$<E_{cm}>$~=~206~GeV, ${\cal L}~=~164.4~pb^{-1}$} \\
\hline
  &  &  &  &  & \\
{ Topology:} & { \jjl} & { \ll} &  { \jjjj} &  { rad}  & { Total} \\ 
  &  &  &  &  & \\
\hline \hline
 & \multicolumn{5}{|c|}{3 $\leq$ \DM\ $<$ 5~\GeVcc  } \\ \hline
{ Obs. events:}  & 0 & 20 & 5 & 3 & 28 \\
{ Expect. events:} & 0.4 $ \pm 0.3 ^{+0.1} _{-0.0}$ & 20.6 $ \pm 1.9 ^{+0.3} _{-2.4}$ 
& 7.5 $\pm 0.9 ^{+1.3} _{-0.6}$ & 2.5 $\pm 0.5 ^{+1.1} _{-0.1}$ & 31.0 $\pm 2.2 ^{+1.7} _{-2.5}$ \\
\hline
 & \multicolumn{5}{|c|}{ 5 $\leq$ \DM\ $<$ 10~\GeVcc } \\ \hline
{ Obs. events:}  & 0 & 0 & 2 & 3 & 5 \\
{ Expect. events:} & 0.4 $\pm 0.3 ^{+0.1} _{-0.0}$ & 2.0 $\pm 0.5 ^{+0.1} _{-0.5}$ 
& 1.4 $\pm 0.3 ^{+0.5} _{-0.3}$ & 2.5 $\pm 0.5 ^{+1.1} _{-0.1}$ & 6.4 $\pm 0.8 ^{+1.2} _{-0.6}$ \\
\hline
 & \multicolumn{5}{|c|}{ 10 $\leq$ \DM\ $<$ 25~\GeVcc } \\ \hline
{ Obs. events:}  & 0 & 8 & 4 & 3 & 15 \\
{ Expect. events:} & 0.4 $\pm 0.3 ^{+0.1} _{-0.0}$ & 7.7 $\pm 0.9 ^{+0.9} _{-0.4}$ 
& 5.6 $\pm 0.7 ^{+1.3} _{-0.5}$ & 2.5 $\pm 0.5 ^{+1.1} _{-0.1}$ & 16.3 $\pm 1.3 ^{+1.9} _{-0.6}$ \\
\hline
 & \multicolumn{5}{|c|}{ 25 $\leq$ \DM\ $<$ 35~\GeVcc } \\ \hline
{ Obs. events:}  & 1 & 13 & 3 & 3 & 20 \\
{ Expect. events:} & 0.4 $\pm 0.1 ^{+0.0} _{-0.0}$ & 11.6 $\pm 0.9 ^{+0.7} _{-2.1}$ 
& 4.0 $\pm 0.3 ^{+1.0} _{-0.0}$ & 2.5 $\pm 0.5 ^{+1.1} _{-0.1}$ & 18.5 $\pm 1.1 ^{+1.6} _{-2.1}$ \\
\hline
 & \multicolumn{5}{|c|}{ 35 $\leq$ \DM\ $<$ 50~\GeVcc } \\ \hline
{ Obs. events:}  & 2 & 23 & 10 & 11 & 46 \\
{ Expect. events:} & 2.4 $\pm 0.4 ^{+0.1} _{-0.1}$ & 26.8 $\pm 1.0 ^{+1.2} _{-2.8}$ 
& 8.0 $\pm 0.4 ^{+2.0} _{-0.0}$ & 10.3 $\pm 0.5 ^{+1.6} _{-0.5}$ & 47.5 $\pm 1.2 ^{+2.8} _{-2.8}$ \\
\hline
 & \multicolumn{5}{|c|}{ 50~\GeVcc\ $\leq$ \DM } \\ \hline
{ Obs. events:}  & 3 & 40 & 22 & 11 & 76 \\
{ Expect. events:} & 3.8 $\pm 0.4 ^{+0.4} _{-0.3}$ & 38.9 $\pm 1.0 ^{+4.3} _{-1.1}$ 
& 18.7 $\pm 0.5 ^{+4.2} _{-0.0}$ & 10.3 $\pm 0.5 ^{+1.6} _{-0.5}$ & 71.7 $\pm 1.3 ^{+6.2} _{-1.2}$ \\
\hline
 & \multicolumn{5}{|c|}{ TOTAL (logical .OR. between different \DM\ windows)} \\ \hline
{ Obs. events:}  & 4 & 76 & 31 & 11 & 122 \\
{ Expect. events:} & 5.0 $\pm 0.5$ & 69.7 $\pm 2.2$ 
& 29.0 $\pm 1.3$ & 10.9 $\pm 0.7$ & 114.6 $\pm 2.7$ \\
\hline
\end{tabular}
\end{center}
\caption[.]{
\label{tab:EXPEV2000}
\small Number of events observed in data and expected number of
background events in the different chargino search channels 
for all the events recorded in 2000 with 
the detector fully operational ($<E_{cm}>$~=~206~GeV). The first errors are statistical and the second ones are systematic.
The ``re-weighting''  method  used to compute the systematics is described in section \ref{sec:sysrev}.}
\end{table}

\begin{table}[ht]
\begin{center}
\footnotesize
\begin{tabular}{|c||c|c|c|c||c|} 
\multicolumn{3}{l}{} &
\multicolumn{3}{c}{$E_{cm}$~=~206.5~GeV, ${\cal L}~=~60.0~pb^{-1}$} \\
\hline
  &  &  &  &  & \\
{Topology:} & { \jjl} & { \ll} &  { \jjjj} &  { rad}  & { Total} \\ 
  &  &  &  &  & \\
\hline \hline
 & \multicolumn{5}{|c|}{3 $\leq$ \DM\ $<$ 5~\GeVcc  } \\ \hline
{ Obs. events:}  & 0 & 10 & 2 & 2 & 14 \\
{ Expect. events:} & 0.2 $\pm 0.1 ^{+0.1} _{-0.0}$ & 18.8 $\pm 1.3 ^{+0.0} _{-4.0}$ 
& 2.1 $\pm 0.4 ^{+0.9} _{-0.0}$ & 1.3 $\pm 0.1 ^{+1.0} _{-0.0}$ & 22.5 $\pm 1.3 ^{+1.3} _{-4.0}$ \\
\hline
 & \multicolumn{5}{|c|}{ 5 $\leq$ \DM\ $<$ 10~\GeVcc } \\ \hline
{ Obs. events:}  & 0 & 4 & 4 & 2 & 10 \\
{ Expect. events:} & 0.2 $\pm 0.1 ^{+0.1} _{-0.0}$ & 3.6 $\pm 0.5 ^{+0.0} _{-0.5}$ 
& 3.9 $\pm 0.7 ^{+2.3} _{-0.1}$ & 1.3 $\pm 0.1 ^{+1.0} _{-0.0}$ & 9.1 $\pm 0.9 ^{+2.5} _{-0.5}$ \\
\hline
 & \multicolumn{5}{|c|}{ 10 $\leq$ \DM\ $<$ 25~\GeVcc } \\ \hline
{ Obs. events:}  & 0 & 3 & 7 & 2 & 12 \\
{ Expect. events:} & 0.2 $\pm 0.1 ^{+0.1} _{-0.0}$ & 2.31 $\pm 0.3 ^{+0.0} _{-0.4}$ 
& 5.4 $\pm 0.8 ^{+2.7} _{-0.0}$ & 1.3 $\pm 0.1 ^{+1.0} _{-0.0}$ & 9.3 $\pm 0.9 ^{+2.9} _{-0.4}$ \\
\hline
 & \multicolumn{5}{|c|}{ 25 $\leq$ \DM\ $<$ 35~\GeVcc } \\ \hline
{ Obs. events:}  & 0 & 3 & 4 & 2 & 9 \\
{ Expect. events:} & 0.4 $\pm 0.1 ^{+0.1} _{-0.1}$ & 5.3 $\pm 0.3 ^{+0.0} _{-2.2}$ 
& 3.2 $\pm 0.3 ^{+1.2} _{-0.0}$ & 1.3 $\pm 0.1 ^{+1.0} _{-0.0}$ & 10.3 $\pm 0.5 ^{+1.6} _{-2.2}$ \\
\hline
 & \multicolumn{5}{|c|}{ 35 $\leq$ \DM\ $<$ 50~\GeVcc } \\ \hline
{ Obs. events:}  & 0 & 10 & 6 & 5 & 21 \\
{ Expect. events:} & 0.7 $\pm 0.1 ^{+0.0} _{-0.1}$ & 11.7 $\pm 0.4 ^{+0.0} _{-5.2}$ 
& 5.0 $\pm 0.3 ^{+2.1} _{-0.0}$ & 4.3 $\pm 0.2 ^{+1.3} _{-0.0}$ & 21.6 $\pm 0.6 ^{+2.5} _{-5.2}$ \\
\hline
 & \multicolumn{5}{|c|}{ 50~\GeVcc\ $\leq$ \DM } \\ \hline
{ Obs. events:}  & 1 & 16 & 9 & 5 & 31 \\
{ Expect. events:} & 2.0 $\pm 0.1 ^{+0.0} _{-0.6}$ & 14.2 $\pm 0.4 ^{+4.1} _{-1.1}$ 
& 6.3 $\pm 0.3 ^{+2.4} _{-0.0}$ & 4.3 $\pm 0.2 ^{+1.3} _{-0.0}$ & 26.8 $\pm 0.5 ^{+4.9} _{-1.3}$ \\
\hline
 & \multicolumn{5}{|c|}{ TOTAL (logical .OR. between different \DM\ windows)} \\ \hline
{ Obs. events:}  & 1 & 31 & 16 & 5 & 53 \\
{ Expect. events:} & 2.3 $\pm 0.1 $ & 38.8 $\pm 1.4$ 
& 13.1 $\pm 1.0$ & 4.5 $\pm 0.2$ & 58.7 $\pm 1.8$ \\
\hline
\end{tabular}
\end{center}
\caption[.]{
\label{tab:EXPEV2000S1}
\small Number of events observed in data and expected number of
background events in the different chargino search channels 
for all the events recorded in 2000 with 
the TPC sector 6  off ($<E_{cm}>$~=~206.5~GeV). The first errors are statistical and the second ones are systematic.
The  ``re-weighting'' method used to compute the systematics is described in section \ref{sec:sysrev}.}
\end{table}


To study the systematic effect on the  detection efficiency
both the ``shaking'' method and the ``smearing'' method were used 
(see section \ref{sec:sysmet}).

For the high multiplicity topologies (\jjjj~\& \jjl) the 
``shaking'' method was used. The systematic uncertainty 
is larger in the two first \DM\ regions,
due to the more problematic event reconstruction. 
For these \DM\ regions the relative systematic uncertainty  on 
the detection efficiency determination is  between 4\% and 13\%, but at larger 
\DM\ the effect is less than 4\%. 
The relative difference for these two topologies can be seen 
in table \ref{deteffsyst}.
All the results from the ``shaking'' method for these topologies 
gave a higher efficiency, indicating that the ``unshaken'' 
efficiencies can be regarded as conservative.

For the leptonic (\ll) topology the ``smearing'' method was used.
The results from this study did not give consistently higher
efficiencies, but since the uncertainties are less than 1\%
for all the \DM\ regions it showed that this topology 
is much less sensitive to systematic effects.

In the case of the radiative topology, which consists of both high and low 
multiplicity events, both the ``shaking'' method and the ``smearing''
method were used to study the systematic uncertainty on
the efficiency. The two methods gave compatible results in
all the regions where they are both valid and the uncertainty
is less than 2\%.

\vspace{0.35cm}
\noindent
\underline{Limits}
\vspace{0.35cm}


  The simulated points were used to parameterise the efficiencies of the chargino 
selection criteria 
described in section \ref{sub:CHASEL} in terms of \DM\ and the mass of the chargino. 
The 
values of \DM, the chargino and neutralino masses, the cross-sections and the 
various decay branching 
ratios were then determined for a large number of points in the MSSM parameter space ($\mu $, $M_2$, \tanb). 
From these values and the appropriate efficiencies, the number of expected signal events can be calculated. 
Taking into account the expected background and the number of observed events, the corresponding point 
in the MSSM parameter space can be excluded, if the number of expected signal events is greater than the 
upper limit on the number of observed events of the corresponding \DM\ region, computed using 
the multichannel Bayesian formula (see section~\ref{sec:limcomp}). Systematic errors
were taken into account as described in section~\ref{sec:syslim}.

\begin{figure}[ht]
\begin{center}
\mbox{\epsfysize=21.0cm\epsffile{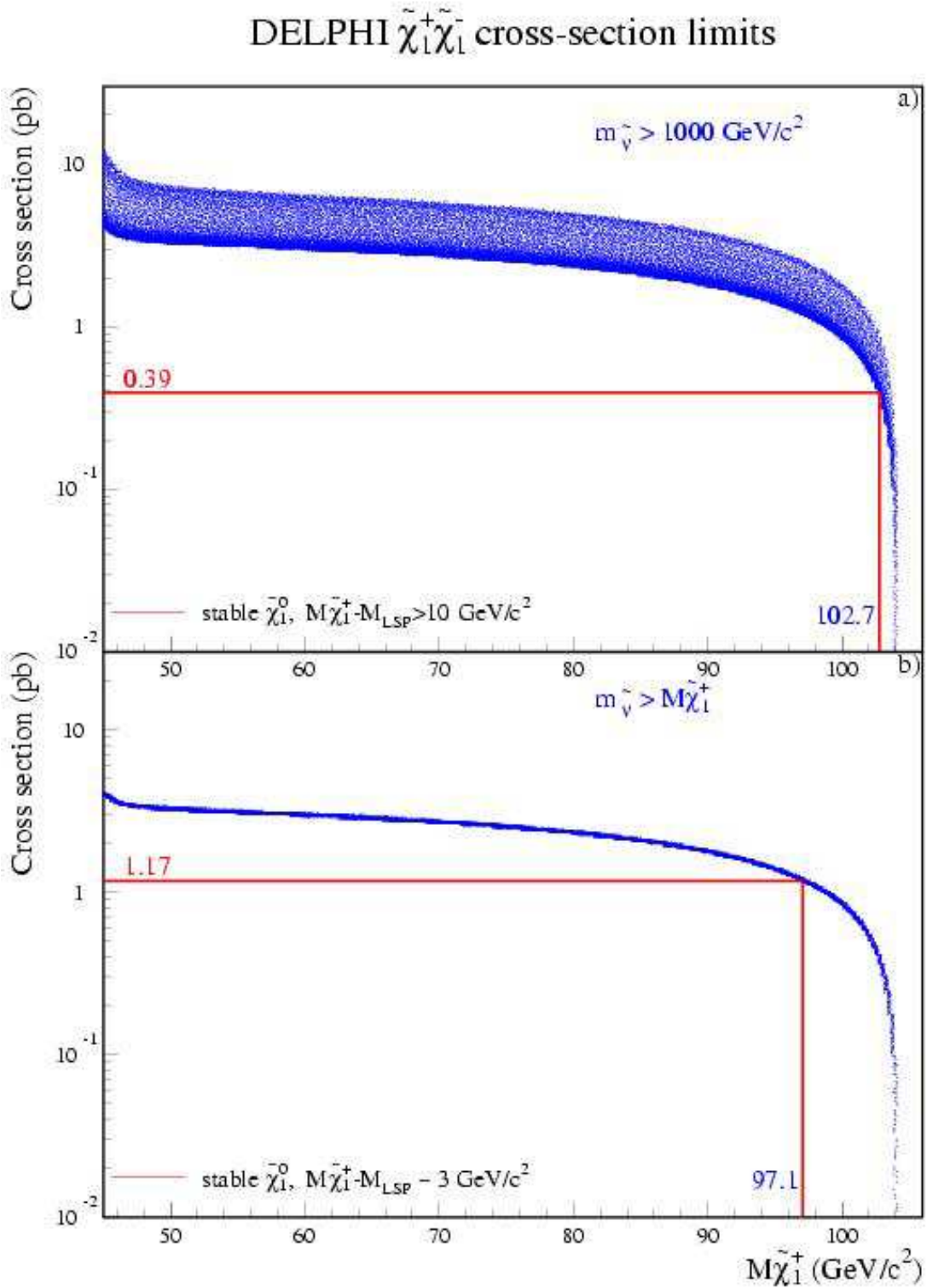}}
\vspace{-1.0cm}
\caption[]{Expected cross-sections in pb at 208 GeV (dots)
versus the chargino mass in a) for \DM $>$ 10~\GeVcc\ and b) for \DM $\sim$ 3~\GeVcc . The spread in the dots originates from the random scan over the
parameters $\mu$ and M$_2$.
A heavy sneutrino (\msneu $>$ 1000~\GeVcc) has been assumed in a) and $\msneu\! >\!\MXC{1}$ in b).
The minimal cross-sections below the mass limits are indicated by the horizontal lines.}
\label{fig:SMCHAL}
\end{center}
\end{figure}

  Figure~\ref{fig:SMCHAL} shows
the chargino production cross-sections as obtained in the MSSM
at \rs~=~208~\GeV\ for different chargino masses for the large \DM\  (\DM\ $>$ 10~\GeVcc)
and low \DM\   (\DM~=~3~\GeVcc) cases.
The parameters $M_2$ and $\mu $ were varied randomly in the ranges
0~\GeVcc\ $< M_2 <$ 3000~\GeVcc\ and $-$200~\GeVcc\ $ < \mu < $ 200~\GeVcc\
for three fixed different values of \tanb,
namely 1, 1.5 and 35. The random generation of the parameters led to 
an accuracy on the mass limit computation of the order
of 50~\MeVcc. Two different cases were considered for the sneutrino mass:
$\msnu\! >\! 1000~\GeVcc$ (for \DM~$>$~10~\GeVcc) and $\msnu\!>\!\MXC{1}$  
(for \DM\ $=$ 3~\GeVcc ). The
mass limits are valid also for $|\mu| > 200$~\GeVcc, since in this case
both the efficiency of the chargino search and the
chargino branching fractions are largely independent of $|\mu|$,
if  $\msnu\!>\!\MXC{1}$~\footnote{
This was investigated by a coarse scan of
random parameter sets with $|\mu| > 200$~\GeVcc, resulting in a mass limit 
greater than 103 \GeVcc.}.

\begin{figure}[ht]
\begin{center}
\mbox{\epsfysize=16.0cm\epsffile{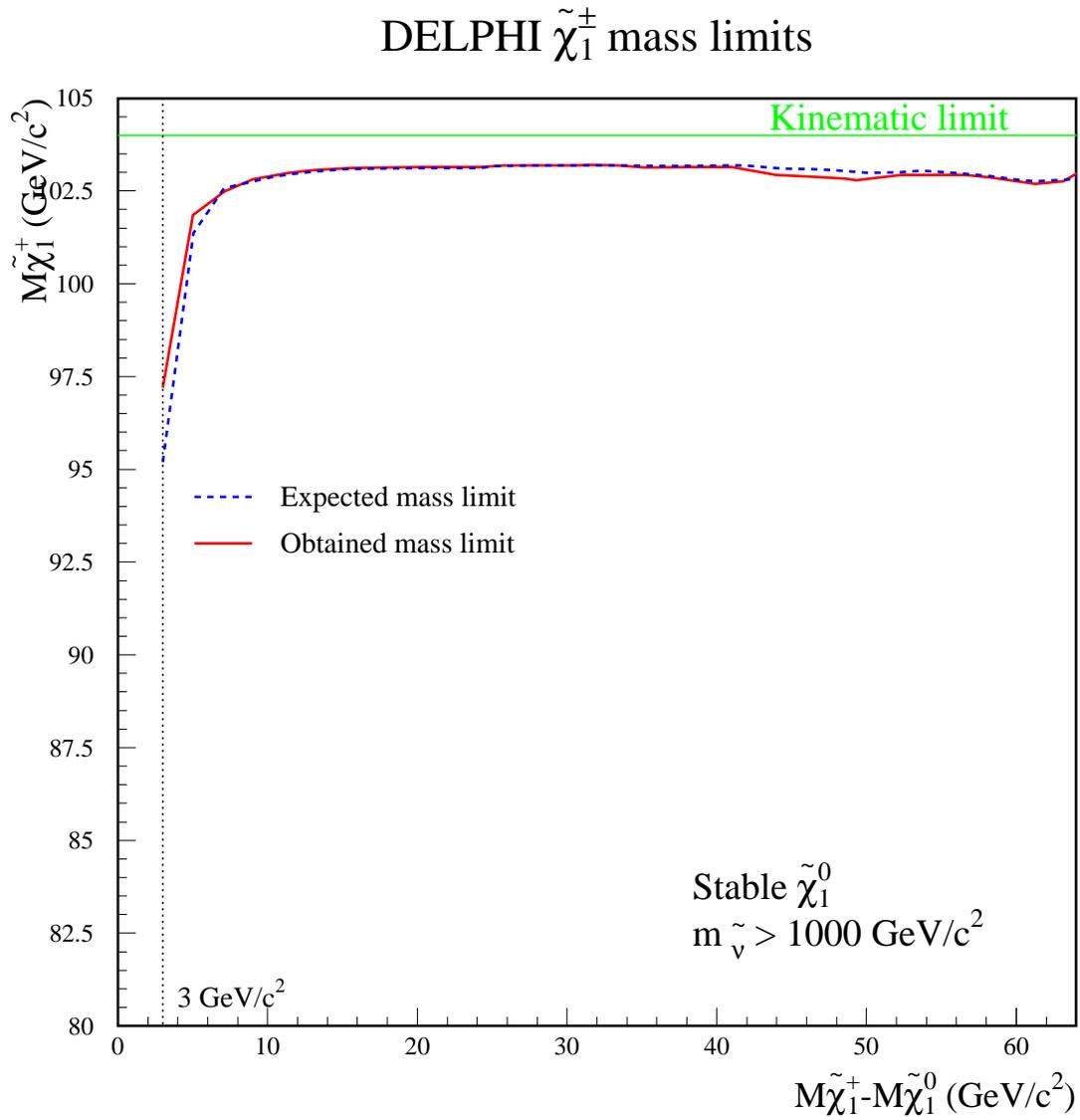}}
\vspace{0.5cm}
\caption[]{
The chargino mass limit as function of the \DM\ value
under the assumption of a heavy sneutrino.
The limit applies to the
case of a stable \XN{1}.
The straight horizontal line shows the kinematic limit.
}
\label{fig:MCHADM}
\end{center}
\end{figure}
\begin{figure}[ht]
\begin{center}
\mbox{\epsfysize=16.0cm\epsffile{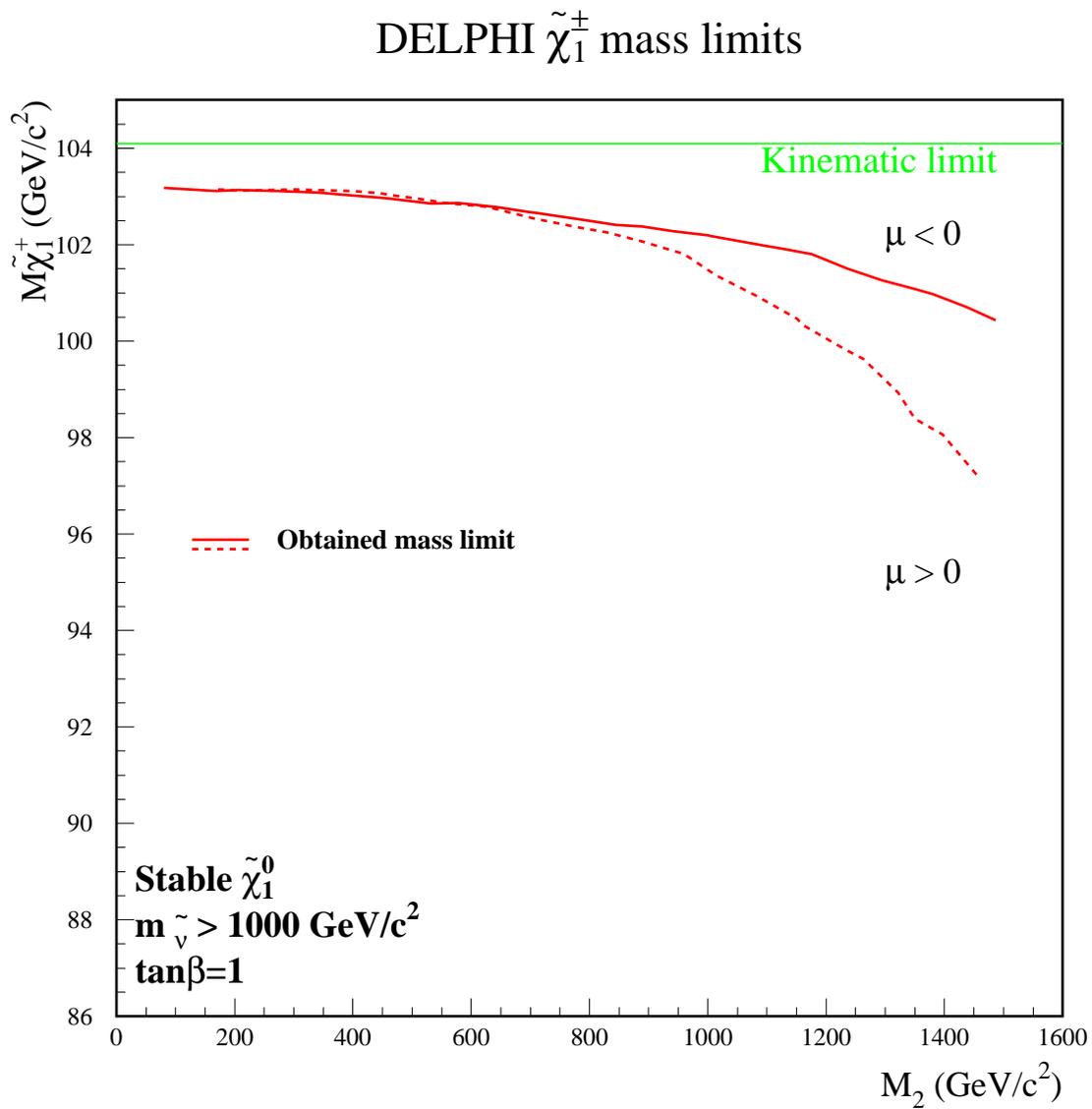}}
\vspace{14.23pt}
\caption[]{The chargino mass limit as function of $M_2$ for \tanb\ = 1, under the 
assumption of a heavy sneutrino. 
The straight horizontal line shows the kinematic limit of the chargino production. The limit 
applies in the case of a stable \XN{1}. 
}
\label{fig:MCHAM2}
\end{center}
\end{figure}

The chargino mass limits are summarised in table~\ref{tab:CHALIM}.
The dependence of the limit on
\DM\ and $M_2$ assuming a heavy sneutrino is 
shown in figures~\ref{fig:MCHADM} and \ref{fig:MCHAM2}.
The behaviour of the curve in figure~\ref{fig:MCHADM} depends 
very weakly on the relation between $M_1$ and $M_2$. Note that in figure~\ref{fig:MCHAM2}, for a fixed 
large 
value of $M_2$, the chargino mass limit is lower for positive $\mu$ than for negative  $\mu$. 
This is due to the
higher degree of mass degeneracy between the lightest chargino and the LSP
 found for positive $\mu$ compared to negative  
$\mu$, for a fixed value of  $M_2$.

 For \DM\ $>$ 10~\GeVcc\ with a large sneutrino mass ($>$~1000~\GeVcc),
the lower limit on the chargino mass ranges between 102.7~\GeVcc\ (for a mostly higgsino-like
chargino) and 103.4~\GeVcc\ (for a mostly wino-like chargino).

  For \DM~=~3~\GeVcc\ , the cross-section does not depend 
significantly on the sneutrino mass, since the region allowing small \DM\ 
values is located where the chargino is higgsino-like, due to the assumption of gaugino 
mass unification.
The lower limit for the chargino mass, shown in figure~\ref{fig:SMCHAL},
is 97.1~\GeVcc.


\begin{table}[ht]

\begin{center}
\footnotesize
\begin{tabular}{||c|c||c|c|c||} \hline \hline
           &          &                      &                &                  \\
{\bf Case} & $\msneu$ & $M^{min}_{\chi^\pm }$ & $\sigma^{208}_{min}$ & ${\rm N}_{95\%}$ \\
           &          &                      &                &                  \\
           & (\GeVcc) & (\GeVcc)             & (pb)           &                  \\ \hline \hline
\multicolumn{5}{||c||}{ } \\
\multicolumn{5}{||c||}{ $<E_{cm}>$~=~192-208~GeV} \\
\multicolumn{5}{||c||}{ } \\\hline 
                  &         &      &      &                \\
$\DM > 10$~\GeVcc & $>$~1000 & 102.7 & 0.39 & 13.8 \\
                  &         &      &      &       \\
$\DM = 5$~\GeVcc  & $>$~1000 & 101.7 & 0.57 & 7.3 \\
                  &         &      &      &       \\
$\DM = 3$~\GeVcc  & $>$\MXC{1}  & 97.1 & 1.17 & 18.1  \\
                  &         &      &      &                \\ \hline \hline 
\end{tabular}
\end{center}
\vspace{0.5 cm}
\caption{\small Lower limits for the chargino mass,  minimal
pair-production cross-sections at 208~GeV for chargino masses below the
limit and upper limits on the number of observed events.}
\label{tab:CHALIM}
\end{table}





\begin{figure}[ht]
\begin{center}
\mbox{\epsfysize=16.0cm\epsffile{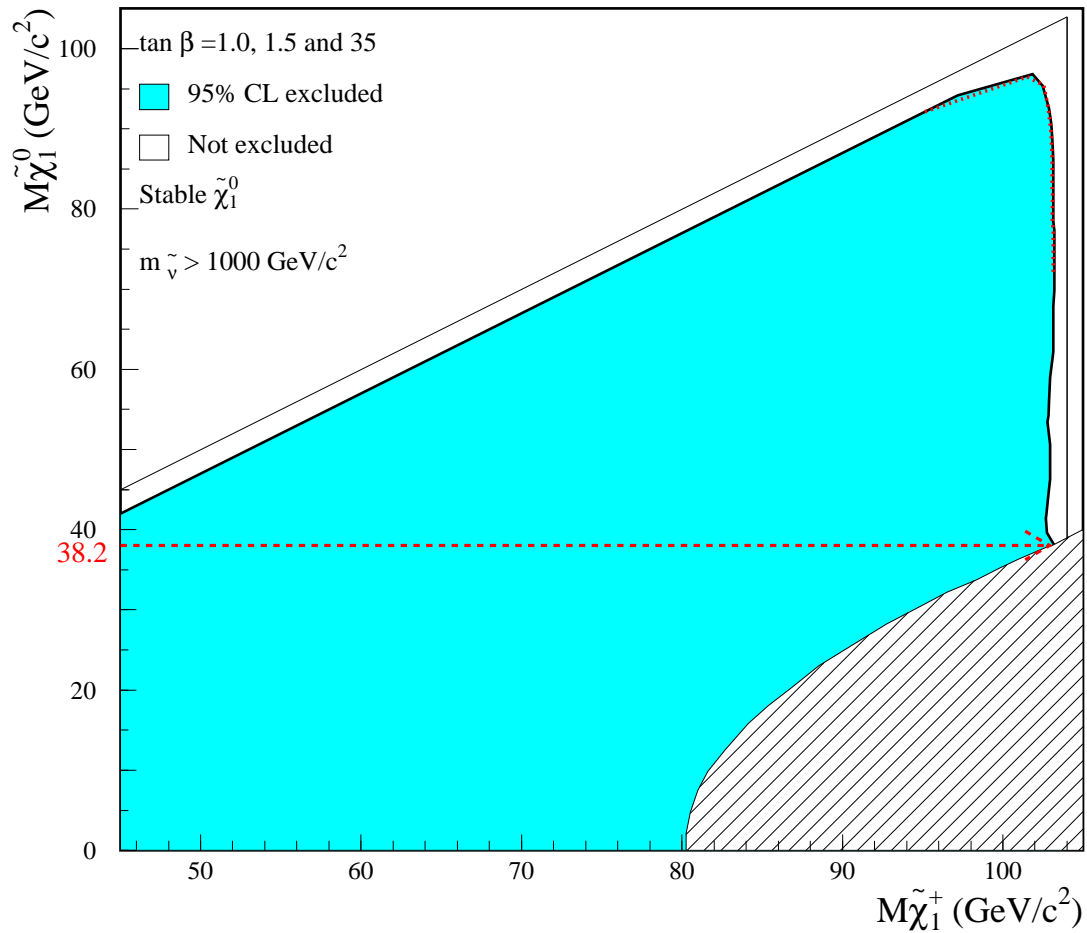}}
\vspace{14.23pt}
\caption[]{
Region excluded in the plane of the mass
of the lightest neutralino versus that of the lightest chargino
under the assumption of a heavy sneutrino, for \tanb~=~1.0, 1.5 and 35.
The thin lines show the kinematic limits in the production and
the decay. The dotted line (partly hidden by the shading) shows 
the expected exclusion. The dashed region is not
allowed in the MSSM. The limit applies in the
case of a stable \XN{1}. The mass limit on the lightest neutralino is 
indicated by the horizontal dashed line
(for a discussion on the low mass region see section~\ref{sec:lep1}).
}
\label{fig:NEUCHA}
\end{center}
\end{figure}

The results
can be translated into a limit on the mass of the lightest neutralino~\cite{char} also 
shown in the (\MXC{1},\MXN{1}) plane in figure~\ref{fig:NEUCHA}.
A lower limit of 38.2~\GeVcc\ on the lightest neutralino mass is obtained, valid for
\tanb\ $\ge$ 1 and a heavy sneutrino. This limit is reached for
\tanb~=~1, $\mu~=~-65.1$~\GeVcc, $M_2$~=~65.0~\GeVcc.

\subsubsection{Nearly mass-degenerate scenarios}
\label{sub:chadeg_res}

\vspace{0.35cm}
\noindent
\underline{Efficiencies and selected events}
\vspace{0.35cm}

\begin{figure}[htbp]
\centerline{
\epsfxsize=10.cm\epsffile{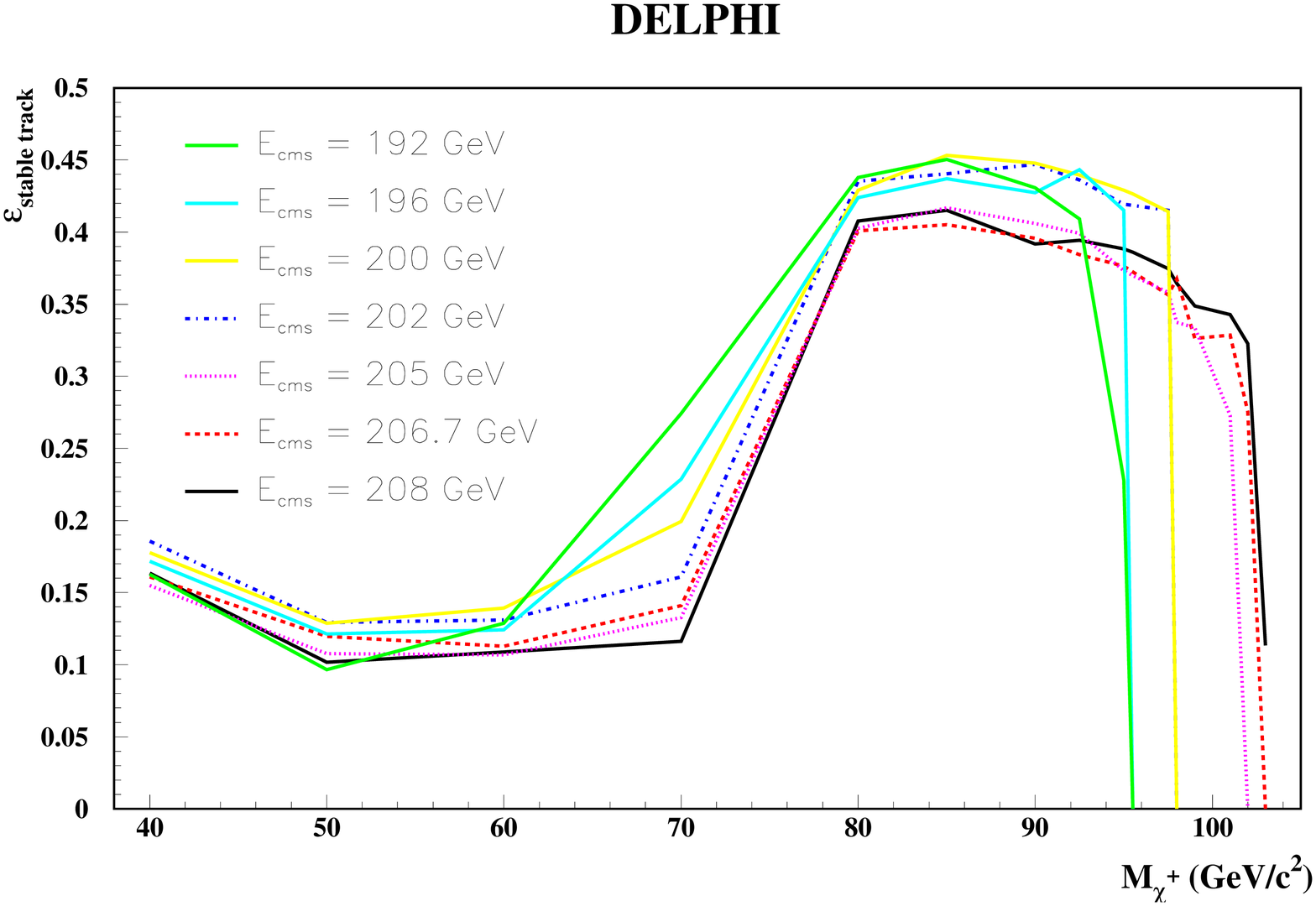} }
\caption[]{ Efficiency for selecting a single almost stable chargino, as function
    of its mass, at the centre-of-mass energies of the
    years 1999 and 2000.}
\label{fig:effstab}
\end{figure}

In the search for heavy stable charged particles, the three windows described in
section~\protect{\ref{sec:chadegsel}} were searched for mass-degenerate charginos.
No events were found in the 1999 and 2000 data, where $0.51 \pm 0.08$ and $0.15\pm 0.03$
events were expected, respectively.
The efficiency for selecting a single chargino track by using this technique is shown in 
figure~\ref{fig:effstab}, as a function of the mass of the chargino and at the various 
centre-of-mass energies.

\begin{figure}[htbp]
\centerline{
\epsfxsize=10.cm\epsffile{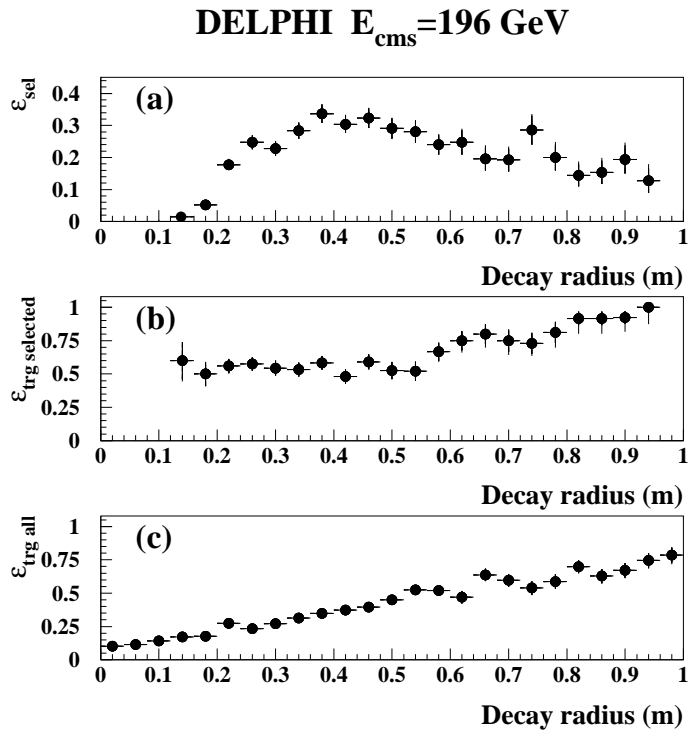} }
\caption[]{ (a) Efficiency for selecting a single 75~\GeVcc\ chargino in the search
    for displaced decay vertices (kinks) at the centre-of-mass energy of
    196~\GeV, as function of its decay radius.
    (b) Trigger efficiency for the selected charginos.
    (c) Trigger efficiency for all 75~\GeVcc\ charginos, whether or not they were selected.}
\label{fig:effkink}
\end{figure}

In the search for kinks, 42000 chargino events with masses between 60 and 
100~\GeVcc\ and mean decay length of 50~cm were generated at the centre-of-mass 
energies between 192 and 206~GeV. The events were used to map the selection and trigger
efficiency for the single chargino, as a function of its mass and decay position. As an 
example, the efficiencies for a 75~\GeVcc\ chargino at \rs=196~GeV are plotted in 
figure~\ref{fig:effkink}.
In the data of 1999 (2000 with full TPC, 2000 without the sector~6 of TPC), 5 (3, 1) events
passed the selection, while $3.7 \pm 1.0$ ($1.2 \pm 0.6$, $0.5 \pm 0.2$) were expected from 
the standard sources.

\begin{figure}[tbh]
\centerline{
\epsfxsize=12.0cm\epsffile{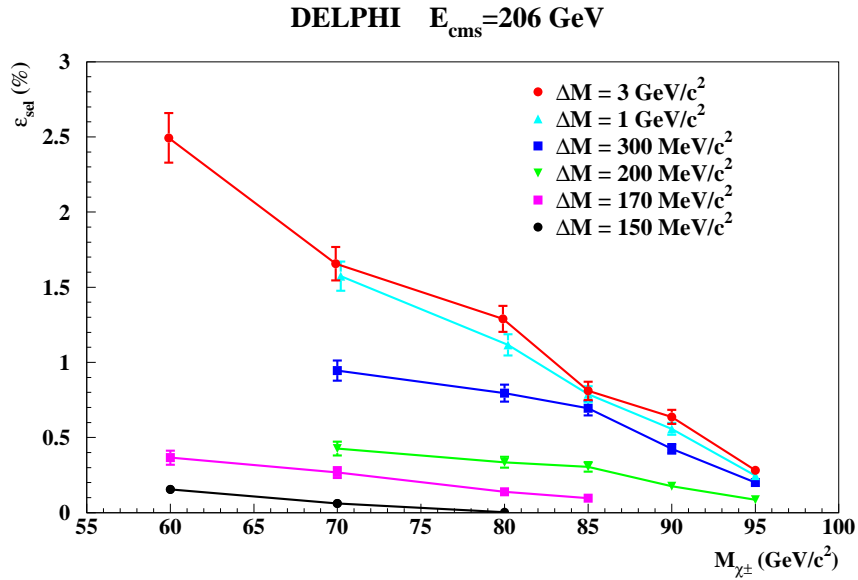} }
\caption[]{ Selection times trigger efficiency in the search for nearly mass-degenerate
  charginos in the search with an ISR photon. The efficiency for higgsinos at $\rs=206$~GeV
  at the different chargino masses and \DM\ values fully simulated is given as an example.      }
\label{fig:effisr}
\end{figure}

\begin{figure}[tbh]
\centerline{
\epsfxsize=12.0cm\epsffile{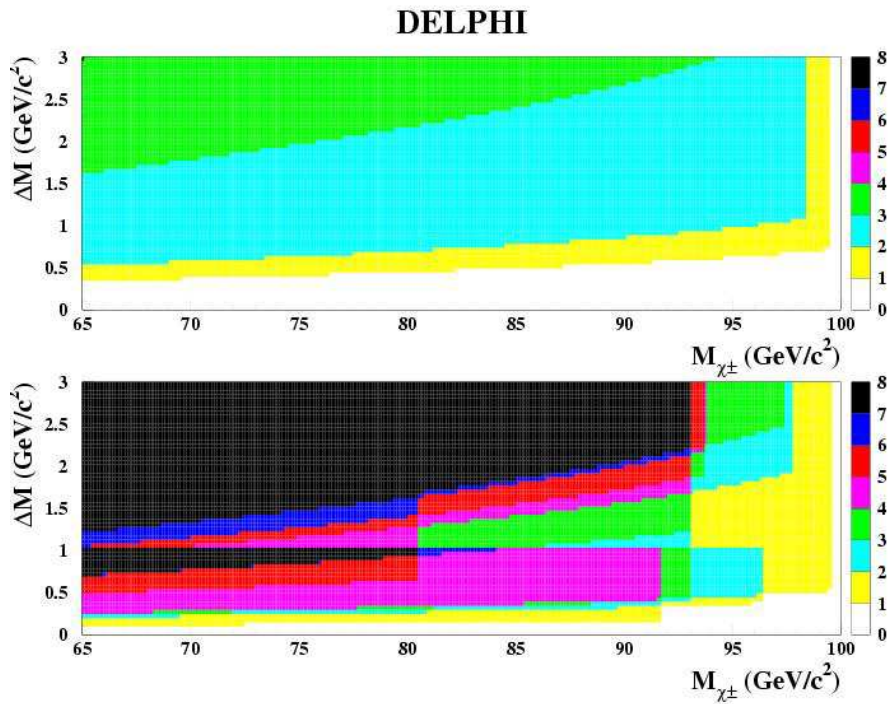} }
\caption[]{ Expected number of background  events (top) and number of selected events
  in the real data (bottom) after the final selection, for the year 2000 data with the TPC fully
  operational, as function of the points in the  plane $(\MXC{1},\DM)$ considered in the chargino
  search with an ISR photon.       }
\label{fig:cand}
\end{figure}

In the search with the ISR photon tag a total of almost 3 million $\XP{1} \XM{1}$ events
was generated at the centre-of-mass energies between 192 and 206~GeV; about 100~000 of
them had a high \pt\ photon within the detector acceptance and were passed 
through the full
detector simulation. The mass of the generated charginos ranged from 60 
to 95~\GeVcc, and 
$\DM$ from 150~\MeVcc\ to 3~\GeVcc. 
The selection efficiency was computed applying the
selection cuts to the samples of simulated events. 
The trigger efficiency was 
parameterised on the real data, separately for the isolated photon and for the system of
few low momentum particles. 
Compton and Bhabha events were used to assess the trigger efficiency for single photons,
as a function of the photon energy and polar angle. To estimate the trigger efficiency in
events with few soft charged particles, parameterised in terms of the transverse momentum
of the single particles and the total transverse energy, the redundancy of the trigger in
several classes of two-photon events was used instead. The overall trigger efficiency for
the whole event was finally considered as the convolution of the trigger efficiency of the
single photon and that of the system of low momentum charged particles.
The parameterisation was then applied on the simulation (both signal and background events).
Detection and trigger efficiency vary widely with $\MXC{1}$ and $\DM$; they also depend on
the field composition of the chargino, since the spectrum of the ISR radiation is different
for higgsinos and gauginos. 
The efficiency (including the trigger efficiency) for a signal (higgsino) at 206~GeV is shown in 
figure~\ref{fig:effisr}. The small probability of radiating a photon with transverse energy 
$E_T^{\gamma}$ above the chosen threshold is the main reason for having such a low efficiency.
As the selection cuts varied across the plane $(\MXC{1},\DM)$, so did the number
of candidate events remaining in the data and the amount of background expected from the 
SM simulations available. Figure~\ref{fig:cand} shows, as an example, the number of
events remaining in data and simulation for the year 2000 sample with fully operational TPC,
as a function of the position in the plane $(\MXC{1},\DM)$. In this sample, when \DM\ is below 
1~\GeVcc\ additional candidates are selected in the data but none in the simulated background.
The number of events remaining after the dedicated preselection and at the final 
step, after 
the logical OR of all selections, together with the integrated luminosity used for the 
analysis, are given in table \ref{tab:candisr}.
The excess of events seen in the data is of the same kind as the excess observed in the 
analyses done at lower energies \cite{isr} and, as in the past, it can be qualitatively
explained by the incomplete phase space coverage of the standard simulation used, and 
possibly some noise events like beam-gas collisions, as already discussed at the end
of section~\ref{sec:chadegsel}. The excess at the preselection level has been verified to
be almost entirely due to the cuts done at the generator level in the (hadronic) 
two-photon simulation used (in 1999 tighter cuts in the generation were applied to the
hadronic $\gamma\gamma$ samples). 
After the final selection, the distribution of the 17 candidates in the data is compatible 
with the integrated luminosities shown in table 17. 
Fewer candidates are observed in the SM simulation than in the data.
This is taken to indicate missing contributions in the modelling
of two-photon background, or beam-gas interactions.

\begin{table}[htb]
\begin{center}
{\footnotesize
\begin{tabular}{|c|c|c|c|c|c|}
\hline
$\sqrt{s}$ & $\int \cal L$ & \mc{2}{c|}{Preselection} &  \mc{2}{c|}{Final sel.} \\
\cline{2-6}
 (GeV)         &  (pb$^{-1}$)  & Data & MC  & Data & MC  \\
\hline
192         &  25.4 &   75  &  $ 40 \pm  2$  &  0  &  $0.6 \pm 0.1$  \\ 
196         &  76.2 &  162  &  $118 \pm  4$  &  4  &  $1.3 \pm 0.3$  \\ 
200         &  84.0 &  201  &  $123 \pm  4$  &  0  &  $1.4 \pm 0.3$  \\ 
202         &  40.4 &  109  &  $ 55 \pm  2$  &  0  &  $0.6 \pm 0.1$  \\ 
206         & 163.0 &  423  &  $393 \pm 18$  & 11  &  $3.5 \pm 1.3$  \\ 
206.5(*)    &  58.5 &  130  &  $113 \pm  4$  &  2  &  $1.0 \pm 0.4$  \\ 
\hline
All         &       & 1100  &  $842 \pm 19$  & 17  &  $8.4 \pm 1.5$  \\ 
\hline
\end{tabular}
}
\caption{\small
Number of events selected in data and expected from the SM simulation in the ISR photon
 search for
charginos nearly mass-degenerate with 
the lightest neutralino. Numbers are given at the preselection and at the final 
selection level, for the different centre-of-mass energies.
All year 2000 data were considered at the mean centre-of-mass energy of
206 GeV; (*) indicates the data collected with the TPC sector 6 off.
Simulation errors are statistical.}
\label{tab:candisr}
\end{center}
\end{table}


\noindent
\underline{Limits}
\vspace{0.35cm}

Having no evidence for a signal  in any of the three
searches for charginos nearly mass-degenerate with the lightest
neutralino, regions in the plane $(\MXC{1},\DM)$ can be excluded. First,
the two searches for long-lived charginos were combined, assuming that in a $\XP{1} \XM{1}$ 
event any of the two long-lived charginos can be tagged either as a kink or as a stable 
particle. Then the search with the ISR photon was considered 
for all events in which the chargino decay length was shorter than that required by the two
other methods. In all cases, the data  were 
combined with all previous data from the high energy phase of LEP \cite{isr}.

\begin{figure}[tbhp]
\centerline{
\epsfxsize=16.0cm\epsffile{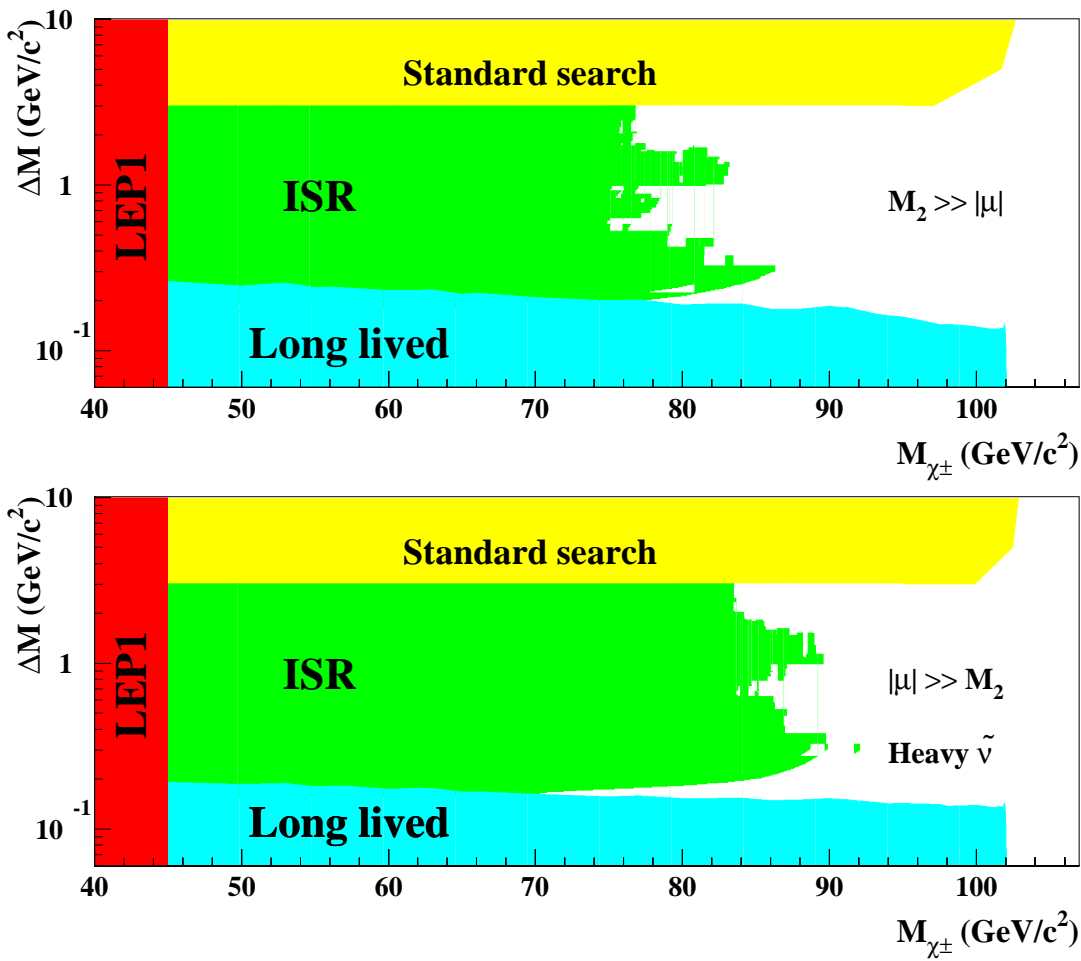} }
\caption[]{ Regions in the plane $(M_{\tilde\chi_1^+},\DM)$ excluded 
  using: the search for high $\DM$ charginos; the search for
  soft particles accompanied by ISR; and the search for long-lived charginos.
  The two scenarios are (upper plot) the one in which the lightest chargino is a higgsino
  and (lower plot) the one in which the lightest chargino is a gaugino. For the second
  scenario, the limits are valid in the heavy sfermion approximation, while for
  the higgsino scenario it is sufficient that $M_{\tilde f}>M_{\tilde\chi_1^+}$. }
\label{fig:limit}
\end{figure}

The limits obtained in this way are certainly model dependent: cross-sections, decay modes,
and the spectrum of the ISR radiation itself all depend on the gauge composition of the
chargino. They were obtained in the two SUSY scenarios in which a near mass-degeneracy
between the lightest chargino and neutralino is possible: the $M_2\gg\mu$ scenario, in which
the lightest chargino and neutralino are both almost pure higgsinos; and the $\mu \gg M_2$
scenario, in which the lightest chargino and neutralino are both almost pure gauginos.
In the gaugino scenario, the gaugino mass unification at large scale must be violated
in order to obtain low $\DM$; in the higgsino scenario this is not mandatory, therefore
this is the scenario to be taken into account in the constrained SUSY models.
These limits are shown in figure~\ref{fig:limit}, separately for the different techniques.
In the same figure, they are compared with the excluded region obtained in the search for larger
$\DM$ charginos.

With these new results, the $\DM$-independent lower limit on the mass 
of the chargino becomes $\MXC{1} > 75$~\GeVcc\ in the higgsino scenario 
and $\MXC{1} > 70$~\GeVcc\ in the gaugino scenario with heavy sneutrinos.
Both limits take into account a variation of $\tan \beta$ between $1$ and $50$, and a 
variation of $M_1$, $M_2$ and $\mu$ such that the mass difference between the chargino 
and the neutralino remains below 3~\GeVcc\ and $M_2 \le 2M_1 \le 10M_2$. 
In the higgsino scenario all sfermions are required to be heavier than the chargino,
while in the gaugino scenario they must be heavy enough not to modify significantly the 
cross-section (only the sneutrino was required to be heavier than 500~\GeVcc)
or the decay modes and widths.

Uncertainties in the selection efficiencies and in the expected background   were included
in the limits in a Bayesian way as described
in  section \ref{sec:syslim}, using the method in \cite{appap}.
The most important  contributions to the  systematic error is the uncertainty in
the background content: depending on the point, the excluded regions may vary as much as
2-3~\GeVcc. Such an effect comes only from the uncertainty in the simulation available,
since conservatively there was no attempt to increase the estimated
background to  take into account the regions of phase space not
included in the simulation. Other systematics, like the uncertainties in the
trigger efficiencies, were similarly taken into account by considering the configuration
which leads to the weakest limits.


\subsection{Neutralino searches}
\label{sec:neuresult}

\subsubsection{Efficiencies and selected events}
\label{sec:neu_eff}

More than 1.2 million \XN{1}\XN{2} events
were simulated
for different combinations of masses
with \MXN{1} and \MXN{2} ranging from 5~\GeVcc\ to 100~\GeVcc\ and
from 10~\GeVcc\ to 200~\GeVcc, respectively, and for
different \XN{2} decay modes (\qqbar\XN{1}, \mumu\XN{1}, \ee\XN{1},
\stau$\tau$ ).
The efficiencies were computed for each mass combination
and parameterised as functions of the two masses.
In addition, around 300\,000 \XN{2}\XNN{3}{4}\  events with cascade decays
were simulated. 
For cascade processes, the efficiencies were parameterised as functions
of  \MXN{1} and a second parameter, chosen to be either  \MXNN{3}{4}$-$\MXN{2} 
when considering the decay modes 
$\XN{2} \to \XN{1} \gamma $ and  $\XNN{3}{4} \to \qqbar \XN{2}$ ,
or 
 \MXNN{3}{4}+\MXN{2} when considering the decay modes 
$\XN{2} \to \qqbar \XN{1} $ and $\XNN{3}{4} \to \qqbar \XN{2}$.
The dependence on remaining parameters was found to be small in the 
relevant mass ranges and the efficiencies were averaged over them.

The efficiency for
\XN{2}\XN{2} $\to$ $\stau \tau \stau \tau$ 
was evaluated in a number of points with
\MXN{1} between 30 and 50 \GeVcc\ and mass differences
between the stau and the \XN{1} ranging from 2 to 5 \GeVcc.
The \XN{2} mass exceeded the \XN{1} mass by 30 to 45 \GeVcc.
These points were chosen because they correspond to a region
in the CMSSM parameter space which cannot be covered by
the chargino search  nor, due to the small mass difference,
by the stau search.
Efficiencies between 10 and 20\% were found, varying
little with the chosen masses.
These values were obtained with SGV and verified
with DELSIM in a sub-set of points.

The number of events selected at different centre-of-mass energies and
selection levels, together with the expected SM background, are given
in tables  \ref{tab:EVENTS_JJLL} to \ref{tab:EVENTS_XXTT}
for the different neutralino search topologies discussed in 
section \ref{sec:ananeu}.
The results of the sequential and likelihood ratio analyses for the
acoplanar leptons and acoplanar jets were found 
to be comparable. The sequential analysis 
performed less well
in the acoplanar jet search for large \DM\ values, whereas their 
results were very similar in the acoplanar lepton channels and in
the low \DM\ region in general.

Tables  ~\ref{tab:EVENTS_XXTT} and \ref{tab:neu_evnum} 
summarise the results at the final selection level of all neutralino searches.
Table \ref{tab:neu_searcheff} shows the main background sources
contributing in each channel and
the typical efficiency of each search for MSSM points where it
is relevant.

The effect of systematic uncertainties on background and efficiency
evaluation was studied with the ``shaking'' method for high multiplicity
topologies and with the ``smearing'' method for leptonic topologies
(see section \ref{sec:sysgen}). 
The variations in the detection
efficiencies were found to be small (the relative
change was below 3\% on average).
The efficiencies for the  ``shaken'' signal
are typically larger than for the ``unshaken'' signal.
The  background estimated with the shaking
method was typically 10\% higher than the ``unshaken'' background. 
The variation of both background and efficiency is 
mostly due to a slightly higher observed missing energy
in the ``shaken'' events.

\begin{table}[htb]
\begin{center}
{\footnotesize
\begin{tabular}{|c|c|r@{ $\pm$ }c|c|r@{ $\pm$ }c|}
\hline
$\sqrt{s}$ & \mc{6}{c|}{Acoplanar electrons search} \\
\cline{2-7}
(GeV)     & \mc{3}{c|}{Preselection} &  \mc{3}{c|}{Final} \\
\cline{2-7}
          &  Data & \mc{2}{c|} {MC}  & Data & \mc{2}{c|} {MC}   \\
\hline

192   &   65   &  60 &1   &  5  &  6.4 &0.1  \\ 
196   &  135   & 143 &3   & 19  & 15.7 &0.7  \\ 
200   &  192   & 180 &2   & 12  & 16.9 &1.0  \\ 
202   &   68   &  82 &1   &  8  &  9.7 &0.4  \\ 
205   &   64   &  77 &1   & 18  & 14.9 &0.7  \\ 
207   &   72   &  78 &1   &  9  & 15.0 &0.7  \\ 
208   &   10   & 6.5 &0.1 &  4  &  1.4 &0.1  \\ 
206.5(*) & 56  &  62 &1   & 16  & 12.6 &0.7  \\ 

\hline
$\sqrt{s}$ & \mc{6}{c|}{Acoplanar muons search} \\
\cline{2-7}
(GeV)     & \mc{3}{c|}{Preselection} &  \mc{3}{c|}{Final} \\
\cline{2-7}
          &  Data & \mc{2}{c|} {MC}  & Data & \mc{2}{c|} {MC}  \\
\hline

192   &   102   &  115 &3   & 13  &  7.7  &0.1 \\ 
196   &   298   &  289 &4   & 18  & 19.5  &0.4 \\ 
200   &   340   &  357 &3   & 15  & 21.0  &0.6  \\ 
202   &   200   &  160 &3   & 14  & 10.2  &0.2 \\ 
205   &   179   &  191 &2   & 18  & 19.5  &1.0  \\ 
207   &   180   &  193 &2   & 20  & 19.8  &1.1 \\ 
208   &    24   &   16 &0.4 &  3  &  1.9  &0.2 \\ 
206.5(*) & 179  &  191 &2   & 14  & 16.3  &0.3 \\ 

\hline
$\sqrt{s}$ & \mc{6}{c|}{Acoplanar jets search} \\
\cline{2-7}
(GeV)     & \mc{3}{c|}{Preselection} &  \mc{3}{c|}{Final} \\
\cline{2-7}
          &  Data & \mc{2}{c|} {MC} & Data & \mc{2}{c|} {MC}  \\
\hline

192   &    927   &  896 &6   &  3  &  3.1 &0.1 \\ 
196   &   2191   & 2218 &10  & 13  &  7.9 &0.3 \\ 
200   &   2886   & 2835 &12  &  9  & 10.4 &0.3  \\ 
202   &   1263   & 1250 &7   &  7  &  5.1 &0.2  \\ 
205   &   1458   & 1404 &8   & 14  & 12.7 &1.4 \\ 
207   &   1451   & 1416 &8   & 15  & 13.0 &1.4 \\ 
208   &    133   &  118 &2   &  2  &  1.2 &0.2 \\ 
206.5(*) & 1066  & 1014 &8   & 14  &  7.8 &0.4 \\ 

\hline
\end{tabular}
\caption{\small
Results of the likelihood ratio acoplanar jets and acoplanar leptons searches 
at the different selection levels and centre-of-mass energies. 
The number of events selected 
in data and
expected from the SM simulation are given. Simulation errors are statistical. 
The systematic uncertainties on the background were estimated to be of
the order of 3\% for the acoplanar leptons topologies and of the order of 10\%
for the acoplanar jets topologies on the final level (see section \protect{\ref{sec:sysgen}});
(*) indicates the 2000 data taken with the sector 6 of the TPC off.}
\label{tab:EVENTS_JJLL}
}
\end{center}
\end{table}

\begin{table}[htb]
\begin{center} 
{\footnotesize
\begin{tabular}{|c|c|r@{ $\pm$ }c|c|r@{ $\pm$ }c|c|r@{ $\pm$ }c|}

\hline
\sqs  & 
\mc{3}{c|}{ee selection} & \mc{3}{c|}{$\mu\mu$ selection} & \mc{3}{c|}{qq selection} \\

\cline{2-10}

(GeV) & Data & \mc{2}{c|}{MC} & Data & \mc{2}{c|}{MC} & Data & \mc{2}{c|}{MC} \\

\hline

192   &  0 &  2.4 &0.3  &  3 &  1.2 &0.2  &  1 &  3.3 &0.2  \\ 
196   & 11 &  7.5 &0.7  &  4 &  3.9 &0.5  &  9 & 10.1 &0.4  \\ 
200   &  6 &  9.3 &0.9  &  0 &  4.3 &0.6  &  5 & 11.6 &0.4  \\ 
202   &  5 &  4.3 &0.4  &  1 &  2.1 &0.3  &  5 &  5.5 &0.2  \\ 
205   &  5 &  6.3 &0.3  &  5 &  3.4 &0.2  &  7 & 11.5 &2.8  \\
207   &  6 &  7.4 &1.5  &  3 &  4.6 &1.2  & 12 & 10.9 &0.4  \\
208   &  2 &  0.7 &0.1  &  1 &  0.3 &0.1  &  4 &  0.9 &0.04 \\
206.5(*) &  6 &  4.4 &0.6  &  3 &  2.5 &0.4  & 10 &  8.0 &0.5  \\

\hline
All   & 41 & 42.3 &2.1  & 20 & 22.3 &1.5  & 53 & 61.8 &3.0  \\
\hline

\end{tabular}
}
\caption{\small
Results of the acoplanar lepton and acoplanar jet sequential searches at the final
selection level for the different flavours
and centre-of-mass energies. The number of events selected in data and
expected from the SM simulation are given. Simulation errors are statistical.
The systematic uncertainties on the background were estimated to be of
the order of 3\% for the acoplanar leptons topologies and of the order of 10\%
for the acoplanar jets topologies (see section \protect{\ref{sec:sysgen}});
(*) indicates the 2000 data taken with the sector 6 of the TPC off.}
\label{tab:neu_JJLLEVENTS}
\end{center}
\end{table}

\begin{table}[htb]
\begin{center}
{\footnotesize
\begin{tabular}{|c|c|r@{ $\pm$ }c|c|r@{ $\pm$ }c|c|r@{ $\pm$ }c|}
\hline
$\sqrt{s}$ & \mc{9}{c|}{Multijets without $\gamma$ search} \\
\cline{2-10}
(GeV)     & \mc{3}{c|}{Preselection} &  \mc{3}{c|}{Intermediate} &  
\mc{3}{c|}{Final} \\
\cline{2-10}
&  Data & \mc{2}{c|}{MC} & Data & \mc{2}{c|}{MC} & Data & \mc{2}{c|}{MC}  \\
\hline

192   & 873   & 916   &6  & 114 & 130.9 &2.0 & 3  &   6.7 &0.3  \\
196   & 2683  & 2664  &9  & 383 & 368.5 &2.7 & 23 &  19.3 &0.6  \\
200   & 2837  & 2733  &8  & 417 & 392.1 &2.5 & 20 &  21.4 &0.6  \\
202   & 1359  & 1323  &4  & 208 & 188.5 &1.4 & 8  &  10.0 &0.3  \\
205   & 2469  & 2412  &15 & 378 & 349.8 &4.7 & 15 &  17.5 &0.3  \\ 
207   & 2471  & 2415  &6  & 405 & 355.6 &1.9 & 20 &  17.9 &0.4  \\ 
208   & 213   & 212   &1  &  33 &  31.2 &0.2 & 3  &   1.6 &0.04 \\
206.5(*) & 1701  & 1248 &6  & 265 & 225.3 &1.8 & 17 & 12.9 &0.5  \\

\hline

All & 14606 & 13923 &22 & 2203 & 2042 &7 & 109 & 107.3 &1.2 \\ 

\hline

$\sqrt{s}$ & \mc{9}{c|}{Multijets with $\gamma$ search} \\
\cline{2-10}
(GeV)     & \mc{3}{c|}{Preselection} &  \mc{3}{c|}{Intermediate} &  
\mc{3}{c|}{Final}  \\
\cline{2-10}
& Data & \mc{2}{c|}{MC} & Data & \mc{2}{c|}{MC} & Data & \mc{2}{c|}{MC}  \\
\hline

192   & 11  &  10.5 &0.7 &  4 &  2.8 &0.4  & 0 & 0.4 &0.2  \\
196   & 29  &  31.1 &0.8 &  6 &  7.9 &0.4  & 0 & 1.0 &0.1  \\
200   & 33  &  31.6 &0.7 &  8 &  8.7 &0.4  & 1 & 1.3 &0.2  \\
202   & 17  &  15.5 &0.4 &  4 &  4.1 &0.2  & 0 & 0.6 &0.1  \\
205   & 28  &  29.2 &0.6 &  8 &  7.8 &0.3  & 2 & 1.0 &0.1  \\  
207   & 28  &  29.2 &0.7 &  7 &  7.8 &0.3  & 0 & 1.2 &0.1  \\  
208   &  1  &   2.5 &0.1 &  0 &  0.6 &0.02 & 0 & 0.1 &0.01 \\
206.5(*) &  24  &  14.2 &0.5 &  8 &  4.0 &0.2  & 2 & 0.7 &0.1  \\

\hline
All   & 171 & 164 &2 & 45 & 44 &1  & 5 & 6.3 &0.4  \\ 
\hline

\end{tabular}

\caption{\small
Results of the multijets without $\gamma$ and multijets with $\gamma$ searches 
at the different selection levels and centre-of-mass energies. The numbers of events selected 
in data and
expected from the SM simulation are given. Simulation errors are statistical.
The systematic uncertainties on the background were estimated to be of
the order of 10\% on the final level (see section \protect{\ref{sec:sysgen}});
(*) indicates the 2000 data taken with the sector 6 of the TPC off.}
\label{tab:EVENTS_MJ}
}
\end{center}
\end{table}

\begin{center}
\begin{table}[htb]
\begin{center}
{\footnotesize
\begin{tabular}{|c|c|r@{ $\pm$ }c|c|r@{ $\pm$ }c|c|r@{ $\pm$ }c|}
\hline
\sqs & \mc{9}{c|}{Multilepton search} \\
\cline{2-10}
(GeV)     & \mc{3}{c|}{Preselection} &  \mc{3}{c|}{Intermediate} &  
\mc{3}{c|}{Final} \\
\cline{2-10}
& Data & \mc{2}{c|}{MC} & Data & \mc{2}{c|}{MC} & Data & \mc{2}{c|}{MC}   \\
\hline

192   &  1710 &  1608 &12 &  26 &  22.6 &0.8 & 3  & 4.2  &0.4  \\
196   &  5158 &  4862 &35 &  64 &  69.3 &2.0 & 13 & 11.5 &1.0  \\
200   &  5923 &  5339 &37 &  61 &  68.3 &1.8 & 11 & 11.3 &0.5  \\
202   &  2900 &  2600 &18 &  36 &  33.0 &0.9 & 4  & 5.7  &0.3  \\
205   &  5532 &  5135 &48 &  57 &  62.0 &3.3 & 13 & 10.8 &1.0  \\
207   &  5553 &  5126 &34 &  55 &  62.5 &3.5 & 13 & 11.1 &1.1  \\
208   &   486 &   455 &3  &   7 &   5.6 &0.3 & 0  &  1.0 &0.1  \\
206.5(*) &  3911 &  4135 &26 &  46 &  44.8 &1.3 & 9  &  8.7 &0.7  \\

\hline
All   & 31173 & 29260 &85 & 352 & 368.1 &5.8 & 66 & 64.3 &2.1  \\
\hline

\sqs & \mc{9}{c|}{Asymmetric tau search} \\
\cline{2-10}
(GeV)     & \mc{3}{c|}{Preselection} &  \mc{3}{c|}{Intermediate} &  
\mc{3}{c|}{Final} \\
\cline{2-10}
& Data & \mc{2}{c|}{MC} & Data & \mc{2}{c|}{MC} & Data & \mc{2}{c|}{MC}  \\
\hline

192   &  1637 &  1828 &14 &  23 &  19.1 &1.0 &  1 &  0.6 &0.3  \\
196   &  5118 &  5536 &42 &  57 &  57.5 &2.7 &  4 &  1.7 &0.9  \\
200   &  5773 &  6139 &45 &  58 &  53.4 &2.3 &  1 &  2.1 &1.2  \\
202   &  2819 &  2994 &22 &  34 &  24.6 &1.1 &  2 &  0.9 &0.6  \\
205   &  5288 &  5804 &44 &  45 &  45.0 &3.9 &  0 &  1.1 &0.8  \\
207   &  5319 &  5873 &39 &  48 &  50.4 &3.0 &  4 &  1.9 &1.0  \\
208   &   491 &   521 &3  &   4 &   4.3 &0.3 &  0 &  0.2 &0.1  \\
206.5(*) &  3318 &  4613 &29 &  32 &  31.1 &1.6 &  1 &  1.5 &0.6  \\

\hline
All   & 29763 & 33308 &94 & 301 & 285.4 &6.5 & 13 & 10.0 &2.2 \\ 
\hline
\end{tabular}

\caption{\small
Results of the multilepton and asymmetric tau searches at the different selection levels and 
centre-of-mass energies. The numbers of events selected in data and
expected from the SM simulation are given. Simulation errors 
are statistical.
The systematic uncertainties on the background were estimated to be of
the order of 3\% on the final level (see section \protect{\ref{sec:sysgen}});
(*) indicates the 2000 data taken with the sector 6 of the TPC off.}
\label{tab:EVENTS_MLXT}
}
\end{center}
\end{table}
\end{center}

\begin{table}[hbt]
\begin{center}
\begin{tabular}{|l|c|c|} \hline
\sqs\ (\GeV) & Data & Background \\ \hline
\hline
 189             &  0   & 0.42 $\pm$ 0.17 \\ \hline
 192             &  0   & 0.02 $\pm$ 0.06 \\ \hline
 196             &  0   & 0.06 $\pm$ 0.02 \\ \hline
 200             &  0   & 0.25 $\pm$ 0.11 \\ \hline
 202             &  0   & 0.13 $\pm$ 0.05 \\ \hline
 205$-$208      &  0   & 0.21 $\pm$ 0.04 \\ \hline
\hline
All              &  0   & 1.10 $\pm$ 0.19 \\ \hline
\end{tabular}
\end{center}
\caption[.]{\small Background and candidates in the 
\XN{2}\XN{2} $\to$ $\stau \tau \stau \tau$ analysis.
This analysis did not include data from the period in 2000 when one 
sector of the TPC was off.
The total expected background over the
three years was 1.10 $\pm$ 0.19 events, 55 \% of which 
were
two-fermion
events, the 
remainder
being four-fermion events. 
The systematic uncertainties on the background were estimated to be of
the order of 5\% on the final level (see section 
\protect{\ref{sec:sysmet}}).

\label{tab:EVENTS_XXTT}}
\end{table}

\begin{table}[htb]
\begin{center}
{ \footnotesize
\begin{tabular}{|l|c|r@{ $\pm$ }c|c|r@{ $\pm$ }c|c|r@{ $\pm$ }c|c|r@{ $\pm$ }c|}
\hline
\hspace{-0.15cm} 1999
 &  \mc{3}{c|}{192 GeV}   &  \mc{3}{c|}{196 GeV} & \mc{3}{c|}{200 GeV} 
 &  \mc{3}{c|}{202 GeV} \\
 
\cline{2-13}

\hspace{-0.15cm} Search
& \hspace{-0.15cm} Data \hspace{-0.15cm} & \mc{2}{c|}{MC.} 
& \hspace{-0.15cm} Data \hspace{-0.15cm} & \mc{2}{c|}{MC.} 
& \hspace{-0.15cm} Data \hspace{-0.15cm} & \mc{2}{c|}{MC.} 
& \hspace{-0.15cm} Data \hspace{-0.15cm} & \mc{2}{c|}{MC.} \\
\hline
\hspace{-0.15cm} Acopl jets 
       & 3  & 3.1  &0.1 
       & 13 & 8.0  &0.3
       & 9  & 10.3 &0.3
       & 7  & 5.1  &0.2 \\
\hspace{-0.15cm} Acopl electrons 
       & 5  &  6.3 &0.1  
       & 19 & 16.1 &0.7  
       & 12 & 14.8 &0.8
       & 8  &  7.7 &0.3 \\
\hspace{-0.15cm} Acopl muons     
       & 13 &  7.7 &0.1 
       & 18 & 19.5 &0.4
       & 15 & 21.0 &0.5
       & 14 & 10.2 &0.2 \\
\hspace{-0.15cm} Multijets, $\gamma$'s   
       & 0  &  0.4 &0.2  
       & 0  &  1.0 &0.1  
       & 1  &  1.3 &0.2  
       & 0  &  0.6 &0.1 \\
\hspace{-0.15cm} Multijets, no $\gamma$  
       &  3 &  6.7 &0.3  
       & 23 & 19.3 &0.6  
       & 20 & 21.4 &0.6 
       &  8 & 10.0 &0.3 \\
\hspace{-0.15cm} Multileptons             
       &  3 &  4.2 &0.4  
       & 13 & 11.5 &1.0  
       & 11 & 11.3 &0.5  
       &  4 &  5.7 &0.3 \\
\hspace{-0.15cm} Asym tau cascades \hspace{-0.15cm}           
       &  1 &  0.6 &0.3  
       &  4 &  1.7 &0.9  
       &  1 &  2.1 &1.2  
       &  2 &  0.9 &0.6 \\
\hline
\hspace{-0.15cm} All   & 28 & 29 & 1 & 90 & 77 &2 & 69 & 82 &2 &43 & 40 &1 \\

\hline
\hspace{-0.15cm} 2000
 &  \mc{3}{c|}{205 GeV}   &  \mc{3}{c|}{207 GeV} & \mc{3}{c|}{208 GeV} 
 &  \mc{3}{c|}{206.5 GeV(*)} \\

\cline{2-13}
\hspace{-0.15cm} Search
& \hspace{-0.15cm} Data \hspace{-0.15cm} & \mc{2}{c|}{MC.}
& \hspace{-0.15cm} Data \hspace{-0.15cm} & \mc{2}{c|}{MC.}
& \hspace{-0.15cm} Data \hspace{-0.15cm} & \mc{2}{c|}{MC.} 
& \hspace{-0.15cm} Data \hspace{-0.15cm} & \mc{2}{c|}{MC.} \\
\hline

\hspace{-0.15cm} Acopl jets 
       & 14 & 12.7 &1.4 
       & 15 & 13.0 &1.4
       &  2 &  1.2 &0.2
       & 14 &  8.0 &0.4  \\
\hspace{-0.15cm} Acopl electrons 
       & 18 & 14.9 &0.7  
       &  9 & 15.0 &0.7  
       &  3 &  1.4 &0.1
       & 16 & 13.0 &0.7 \\
\hspace{-0.15cm} Acopl muons     
       & 18 & 19.5 &1.0  
       & 20 & 19.8 &1.1 
       &  3 &  1.8 &0.2
       & 15 & 15.4 &0.3  \\
\hspace{-0.15cm} Multijets, $\gamma$'s   
       &  2 &  1.0 &0.1  
       &  0 &  1.2 &0.1  
       &  0 &  0.1 &0.01  
       &  2 &  0.7 &0.1  \\
\hspace{-0.15cm} Multijets, no $\gamma$  
       & 15 & 17.5 &0.3  
       & 20 & 17.9 &0.4  
       &  3 &  1.6 &0.04 
       & 17 & 12.9 &0.5  \\
\hspace{-0.15cm} Multileptons             
       & 13 & 10.8 &1.0  
       & 13 & 11.1 &1.1  
       &  0 &  1.0 &0.1  
       &  9 &  8.7 &0.7  \\
\hspace{-0.15cm} Asym tau cascades \hspace{-0.15cm}         
       &  0 &  1.1 &0.8  
       &  4 &  1.9 &1.0  
       &  0 &  0.2 &0.1  
       &  1 &  1.5 &0.6  \\
\hline
\hspace{-0.15cm} All    & 80 & 78 &2 & 81 & 80 &2 & 11 & 7.3 &0.3 & 74  & 60 &1 \\  
\hline
\end{tabular}
}
\end{center}
\caption{\small
Results of the different neutralino searches.
For any given search, events are
explicitly rejected if accepted by one of the searches appearing
earlier in the table;
(*) indicates the 2000 data taken with the sector 6 of the TPC off.}
\label{tab:neu_evnum}
\end{table}

\begin{table}[htb]
\begin{center}
\begin{tabular}{|l|c|l|}
\hline
Search                 & Main SM bkg.                   & Eff. (\%)  \\ \hline
Acopl jets             & \WW,\ZZ                     &  10 -- 30  \\
Acopl electrons        & \WW,\gamgam $\to \ee$       &  10 -- 40  \\
Acopl muons            & \WW,\gamgam $\to \mumu$     &  10 -- 40  \\
Multijets, $\gamma$'s  & \Zg                         &  10 -- 20  \\ 
Multijets, no $\gamma$ & \WW,\Zg                     &  10 -- 40  \\  
Multileptons           & \WW                         &  30 -- 50  \\
Asym tau cascades      & \WW,\gamgam $\to \mumu$     &  13 -- 19  \\
Double tau cascades    & \ZZ,\Zn\Zstar,\Zn$\gamma^*$ &  10 -- 20  \\

\hline

\end{tabular}
\end{center}
\caption{\small
The main backgrounds and the typical efficiency of all of the neutralino searches for 
MSSM points where it is relevant is shown.
The efficiencies depend typically on the masses of the sparticles 
involved in the process.
}
\label{tab:neu_searcheff}
\end{table}

\subsubsection{Limits}
\label{sec:neu_res}

In the absence of a signal, upper limits on neutralino production cross-sections 
were derived, using the dependence of the calculated efficiencies 
on the masses involved. The results obtained in different search topologies
and at different centre-of-mass energies were combined using the Bayesian method \cite{obraztsov} (see section \ref{sec:limcomp}).

\begin{figure}[ht]
\begin{center}
\vskip 1cm
\begin{center}
\mbox{\epsfysize=7.5cm\epsffile{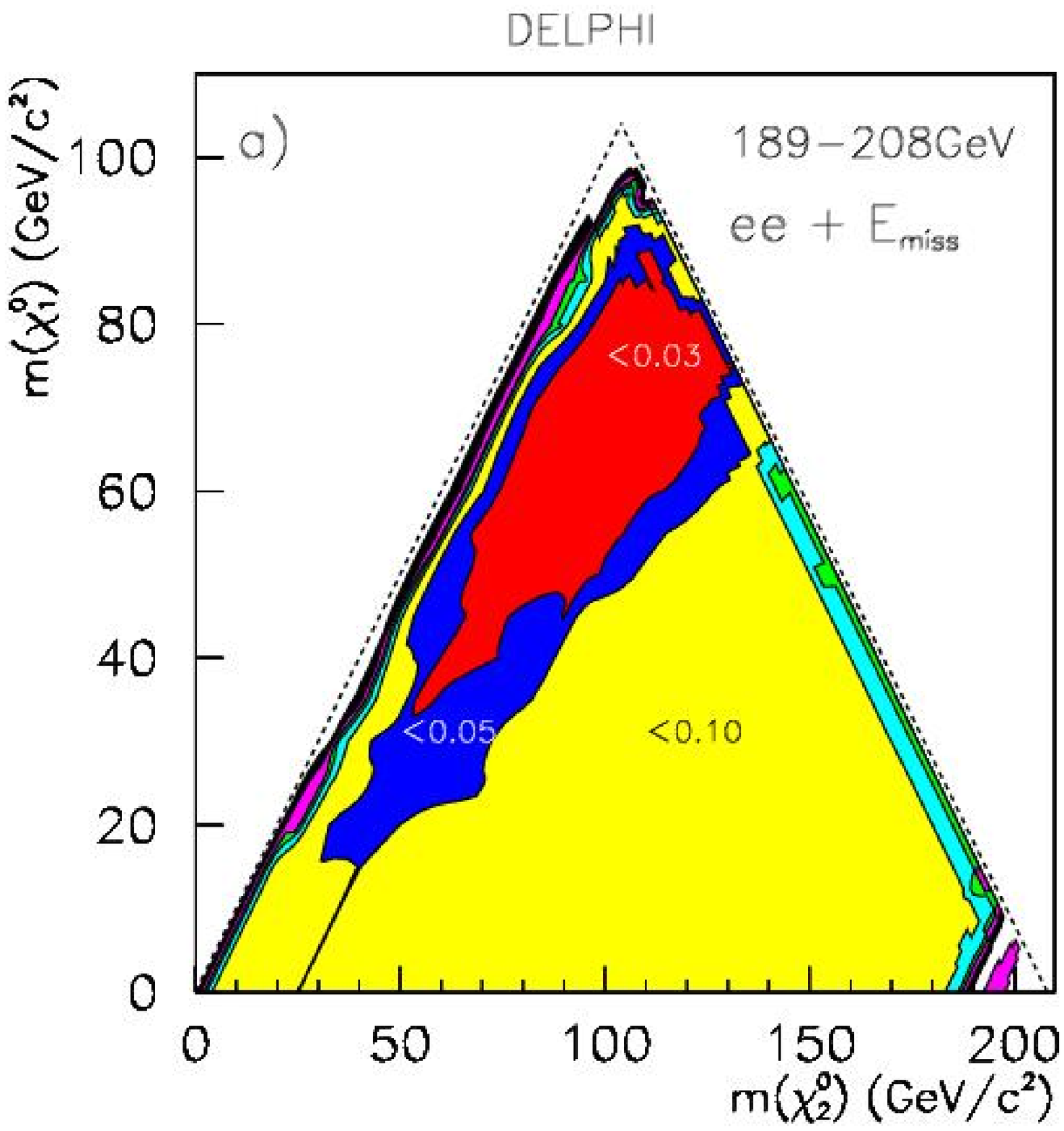}}
\mbox{\epsfysize=7.5cm\epsffile{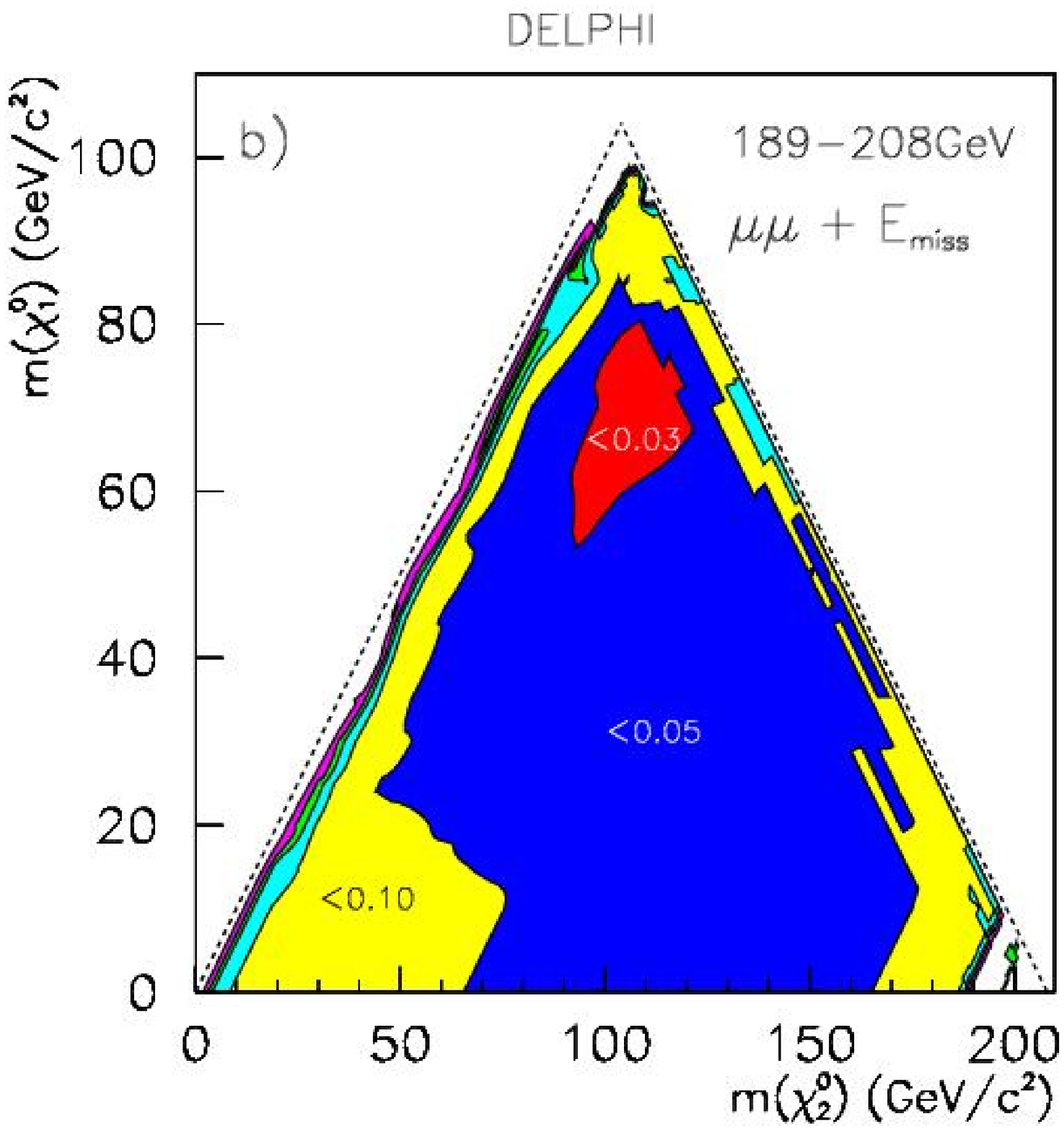}}
\mbox{\epsfysize=7.5cm\epsffile{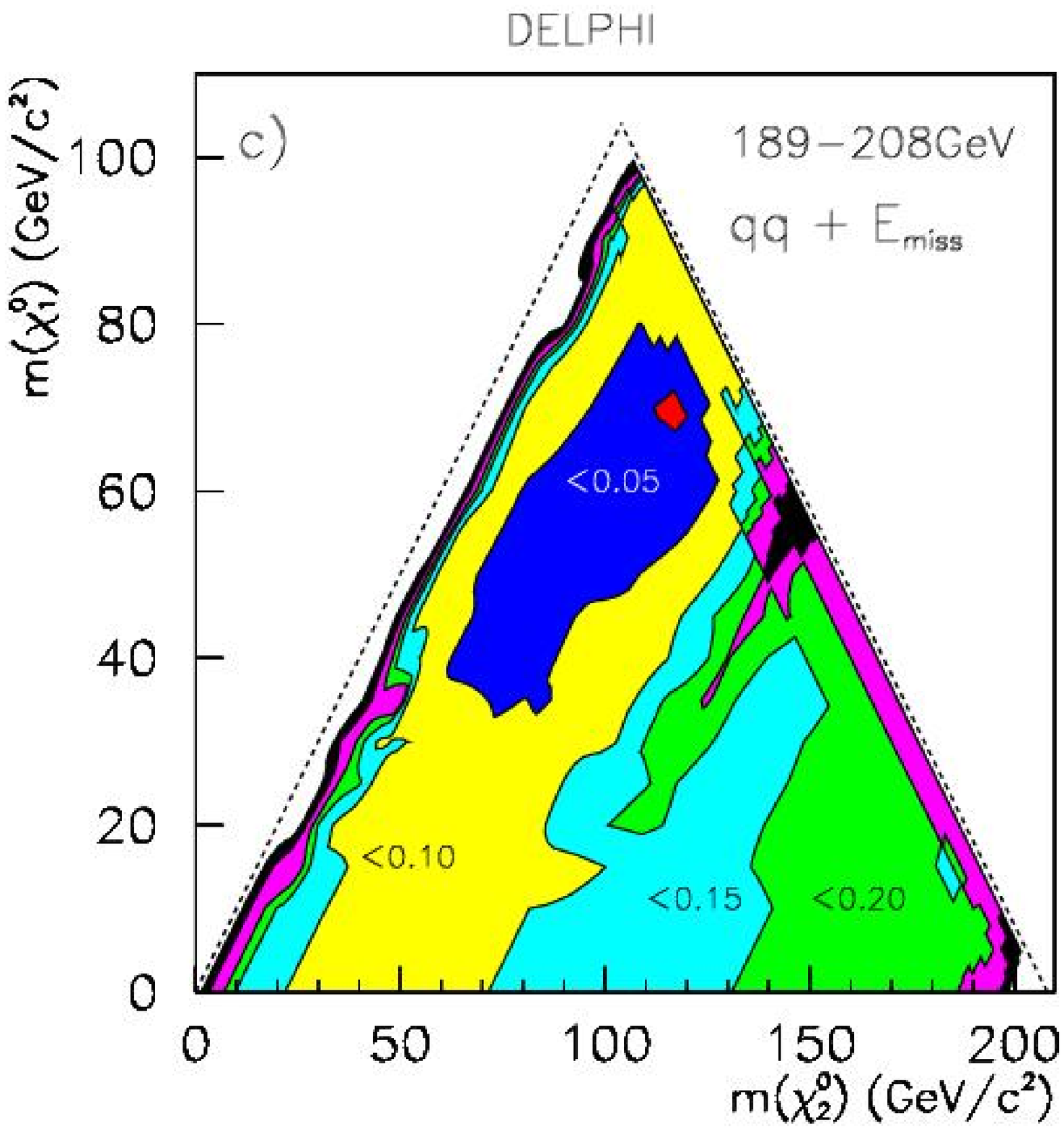}}
\mbox{\epsfysize=7.5cm\epsffile{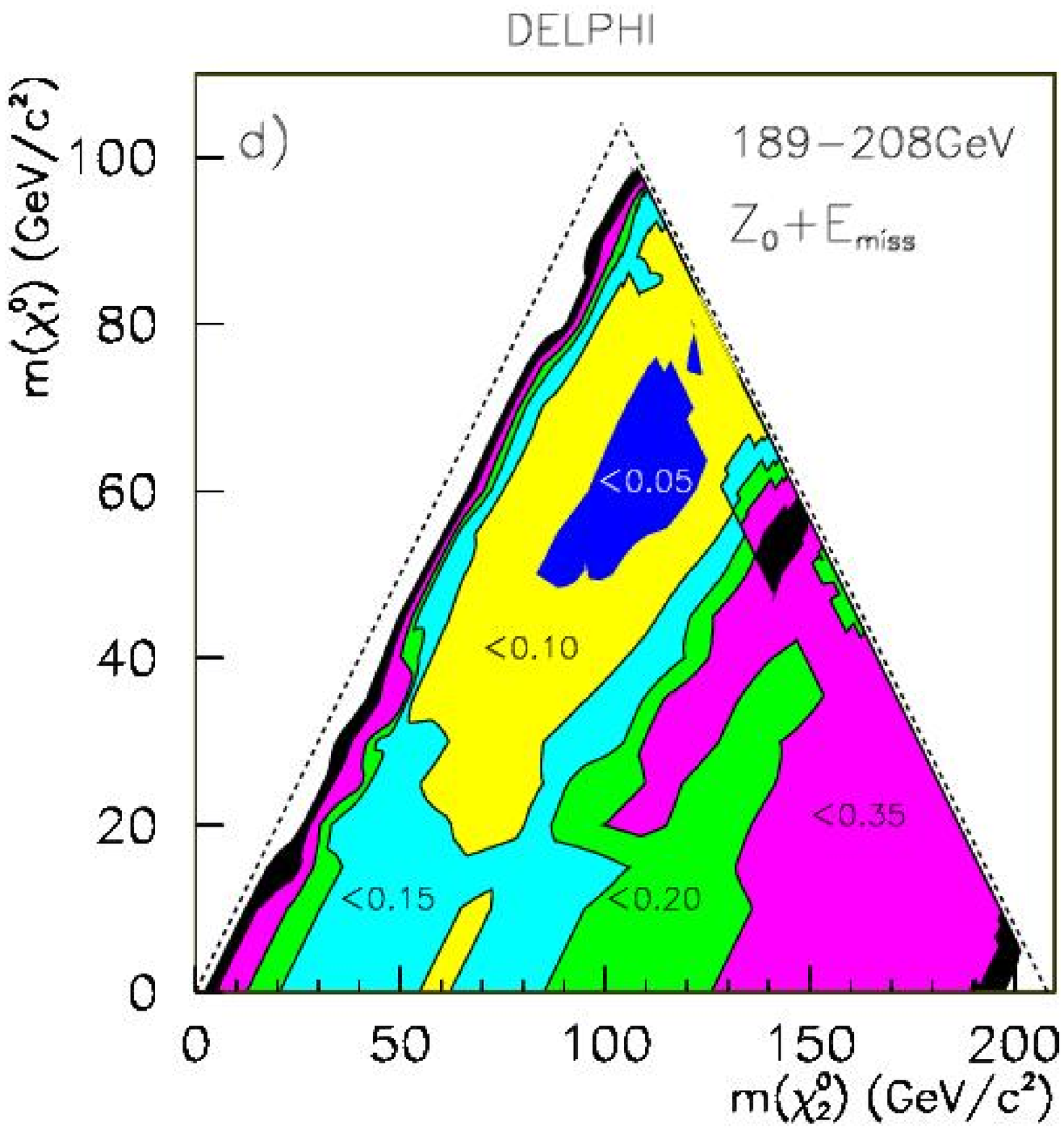}}
\end{center}
\vskip 1cm
\caption[Model independent cross-sections]{
Contour plots of the upper limits obtained on the cross-sections
for \XN{1}\XN{2} production at $\rs~=~206~\GeV$. 
The data at all
energies were used,
assuming the cross-section energy dependence expected at a chosen high
$m_0$ point in the higgsino region where the neutralino searches
play an important role
($m_0$=1~\TeVcc, $\mu$=-60~\GeVcc, $M_2$=200~\GeVcc). 
In each plot,
the different shadings correspond to regions where the cross-section limit in
picobarns is below the indicated number.
For figures a), b), c), \XN{2} decays into \XN{1} and a) \ee,
b) $\mu^+\mu^-$, and c) \qqbar, while in d) the branching
ratios of the \Zn ~were assumed, including invisible states.
The dotted lines indicate the kinematic limit and the
defining relation $\MXN{2}>\MXN{1}$.
}
\label{fig:M1M2LIM_ALL}
\end{center}
\end{figure}

The limits for \XN{1}\XN{2} production,
as obtained from the searches for acoplanar leptons and jets, are
shown in figure~\ref{fig:M1M2LIM_ALL} assuming different branching ratios.

\begin{figure}[ht]
\begin{center}
\mbox{\epsfysize=11.0cm\epsffile{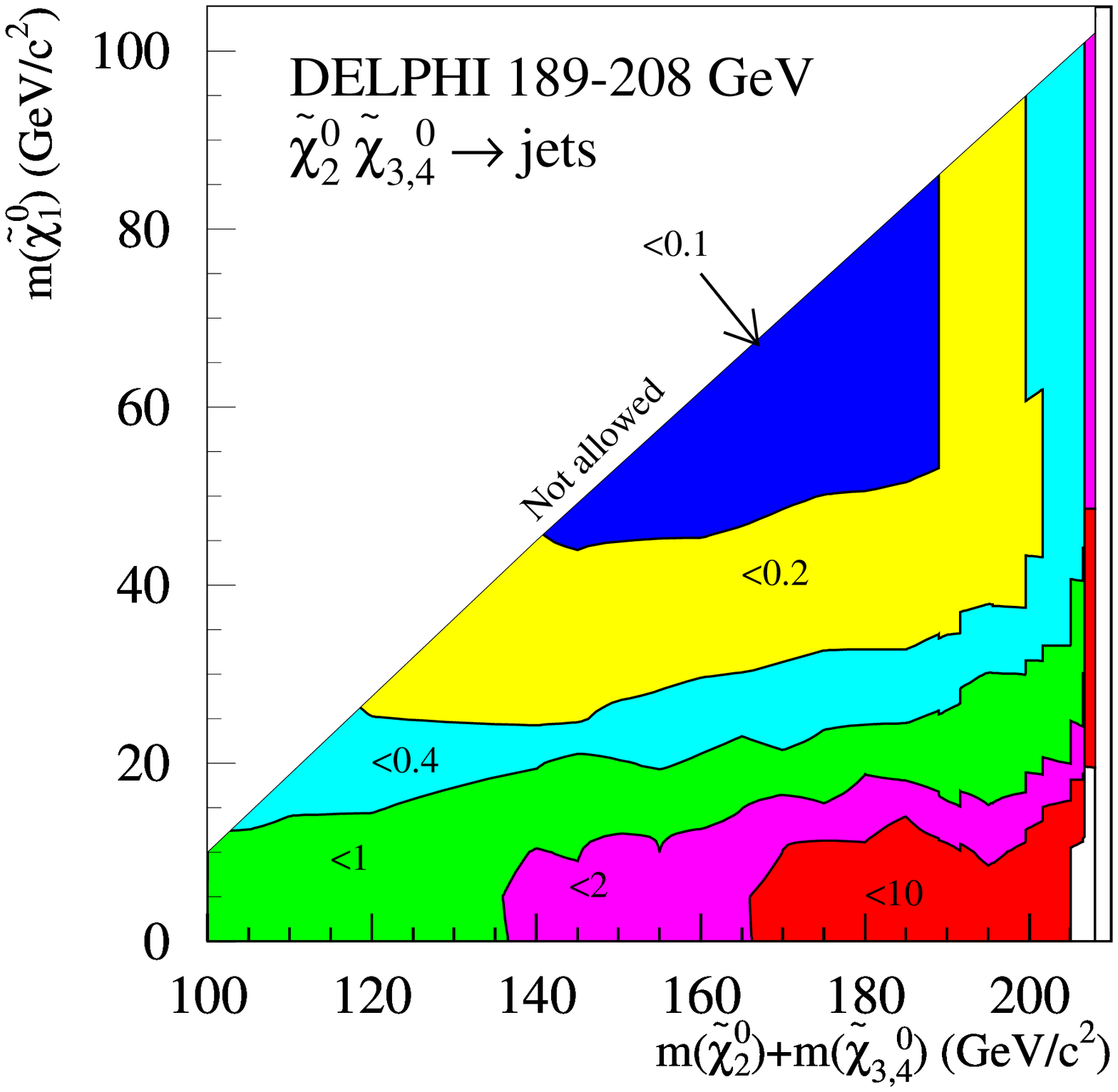}}
\vskip -0.5cm
\mbox{\epsfysize=11.0cm\epsffile{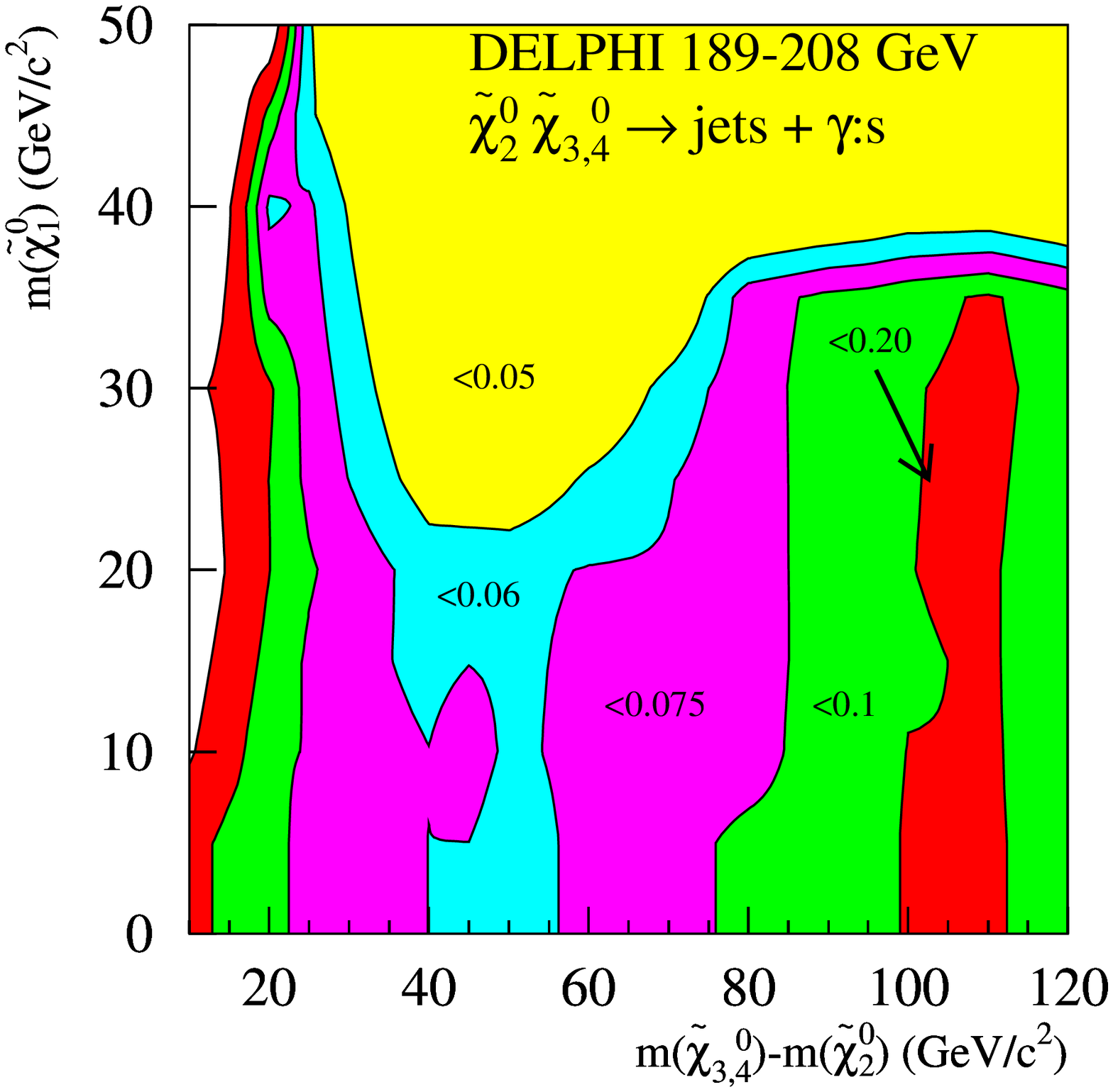}}
\vskip -0.5cm
\caption[Multijet cross-sections]{
Upper limits on the cross-sections
for \XN{2}\XN{i} production with 
\XN{i}$\to$\XN{2}\qqbar\ ($i$=3,4) including data up to $\rs~$=208~GeV. 
The different 
shadings correspond to regions where the cross-section limit in
picobarns is below the indicated number. The \XN{2}\ was assumed to
decay  100\% into \XN{1}\qqbar\ (left plot), and into \XN{1}$\gamma$ (right plot).
The limits in the left plot are based on the acoplanar jets and multijets
selections, while those in the right plot derive from the search for
multijets with photons.
}
\label{fig:MJXS}
\end{center}
\end{figure}

Similarly, figures~\ref{fig:MJXS}(a,b) show cross-section
limits for \XN{2}\XN{i} production ($i\! =$ 3 or 4). 
Figure~\ref{fig:MJXS}(a)
shows the limit obtained combining the results
of  the multijet and acoplanar jet searches in the case where
\XN{i}$\to$\XN{2}\qqbar\ and \XN{2}$\to$\XN{1}\qqbar.
Figure~\ref{fig:MJXS}(b) gives the corresponding
limits when \XN{2}$\to$\XN{1}$\gamma$,
as obtained from the search for
multijet events with a photon signature.

From the results of the search for 
\XN{2}\XN{2} $\to$ $\stau \tau \stau \tau$
described in section ~\ref{sec:stauana}~,
an equivalent production cross-section 
at $\sqrt{s}$ = 206.7~\GeV\ larger 
than 63~fb can be excluded. This limit is valid 
for 30~\GeVcc $ \le  \MXN{1} \le$ 50~\GeVcc\ and
\mstau$-$\MXN{1}$\le$ 5~\GeVcc\ in the gaugino region  $|\mu| >> M_2$, where
this channel is important for constraining the LSP mass.
To combine data at different centre-of-mass energies the cross-section  dependence on $\sqrt{s}$
given by CMSSM in the neutralino and stau mass ranges indicated above was assumed. 
The corresponding expected
limit was 68 fb.

The effect of the systematic and statistical errors of background and efficiency on
the cross-section limits was evaluated using the method described in  section
\ref{sec:syslim} and found to be very small.


\section{Combined  exclusions and mass limits in the  MSSM with gaugino and sfermion mass
unification}
\label{sec:combined}


\subsection{The method}

The method employed to set lower mass limits on the LSP mass and on the masses of 
other SUSY particles is to convert the negative results of the searches for charginos,
neutralinos, sleptons and squarks into excluded regions in the $(M_2,\mu)$ plane 
for different tan~$\beta$ values, and then to find the minimal allowed sparticle 
masses as a function of tan~$\beta$. It is described in detail in \cite{lsp,osakalsp}. 
The Higgs boson search in the 
maximal  $M_{h^0}$  scenario
was used to exclude   low values of  tan~$\beta$ \cite{higgs}. 

Unless stated otherwise, the
limits presented in the following are valid
for $M_2~\le$~1000~\GeVcc\ 
and for the $\mu$ region in which
the lightest neutralino is the LSP. The $\mu$
range depends on the values of \tanb\ and $m_0$,
and on the mixing parameters in the third family ($A_{\tau}, A_{t}, A_{b}$).
Unless stated otherwise, for  high values of  $m_0$ (above 500 \GeVcc) 
the $\mu$ range between 
$-$2000 and 2000 \GeVcc\ was scanned, but the scan range 
was increased if any limit point was found to be  close
to the scan boundary.

\subsubsection{Method of combining different searches}
\label{sec:comb}


In the scan of the SUSY parameter space
the efficiencies of the different searches, 
as obtained
in  
the previous sections,
were 
parameterised 
for the dominant channels, and used 
together with the  information on
the number of  events selected in data and the expected number
of background events. The excluded
regions obtained with the different searches were then simply superimposed.

In previous publications \cite{lsp,osakalsp} it was verified 
that these parameterisations led to  conservative results by  
comparing with  a parallel approach, 
where these searches were applied to samples produced combining
SGV
with SUSYGEN (see section \ref{sec:samples})
to simulate simultaneously all channels 
of chargino, neutralino, slepton
and squark production and decay.

The typical scan step size in $\mu$ and $M_2$ was 1~\GeVcc\, except
in the region of the LSP mass limit, where
the step size was decreased to 0.5~\GeVcc. The step
size in $m_0$ was variable, the density  of points being increased
in  regions of potentially difficult mass configurations. Special
care was taken to set up the scan logics in such a way that
no such configuration was overlooked. In particular, whenever  two nearby scan
points were excluded by  different searches, the scan was performed
with  smaller steps  between these points to check the
continuity of the exclusion.

\subsubsection{The influence of the \mbox{\boldmath${m_0}$}  
value and of the mixing in the third family}
\label{sec:influence}

The unification of sfermion masses to a common $m_0$ at the
GUT scale allows   sfermion masses at the Electroweak
Scale to be calculated as functions of $\tanb$, $M_2$ and $m_0$. 
In particular the masses of the  sneutrino (\snu), 
the left-handed selectron and smuon
 (\sell,$\smu_{L}$ ), and  the right-handed selectron and smuon 
($\selr$,~$\smu_{R}$)  can be expressed as
\footnote{It is worth noting that for $\tanb \ge 1$ ($\tanb < 1$) we have
$\cos 2\beta \le 0$ ($\cos 2\beta > 0$), so the \snu\ is never 
heavier (lighter)
than the \sell. }:

(1)   $\msnu^2 $ $=$ $m_0^2+ 0.77M_2^2 +0.5\MZ ^2 \cos 2\beta$

(2)   $M_L^2$ $=$ $m_0^2+ 0.77M_2^2  -0.27\MZ ^2 \cos 2\beta$

(3)   $M_R^2$ $=$ $m_0^2+ 0.22M_2^2 -0.23\MZ ^2 \cos 2\beta$

\vspace{0.2 cm}

{\underline{
In the large $m_0$ scenario, $m_0 = 1000$ \GeVcc\ was assumed}},
which implied
 sfermion masses of the same order.
In this case  only charginos, neutralinos and the Higgs boson could be
produced at LEP~\footnote{If $|\mu|$ and/or mixing terms for the third family sfermions are
sufficiently large, they can be light for large $m_0$ as well, this case will be discussed
further on.} and the limits arise from a combination of the searches 
for these particles described in the previous sections of this paper and
in~\cite{higgs}.

For large $m_0$, the chargino pair-production cross-section 
is large and the chargino
is excluded nearly up to the  kinematic limit,  provided
$M_2<200$ \GeVcc. 

It may also be remarked that at 
small $M_2$, 
\DM=\MXC{1}~$-$~\MXN{1}\
is large, resulting in increased background from $W^+ W^-$ production.
However, if $|\mu|$ is small as well, the chargino  tends to decay via 
$\XPM{1} \to \XN{2} W^*$ to  the 
 next-to-lightest neutralino \XN{2}, which
then decays by  $\XN{2} \to \XN{1} \gamma $ or 
$\XN{2} \to \XN{1} \Zn^* $ . 
For setting the mass limits, it is therefore important that
the chargino search includes topologies with photons stemming from
the decays $\XPM{1} \to \XN{2} W^* \to \XN{1} \gamma W^*$, since
the search for topologies with photons  does not suffer
from $W^+W^-$ background and is effective for large \DM\ (close to \MW).

\begin{figure}[ht]

\vspace{2.0 cm}
\mbox{\epsfysize=16.0cm\epsfxsize=14cm\epsffile{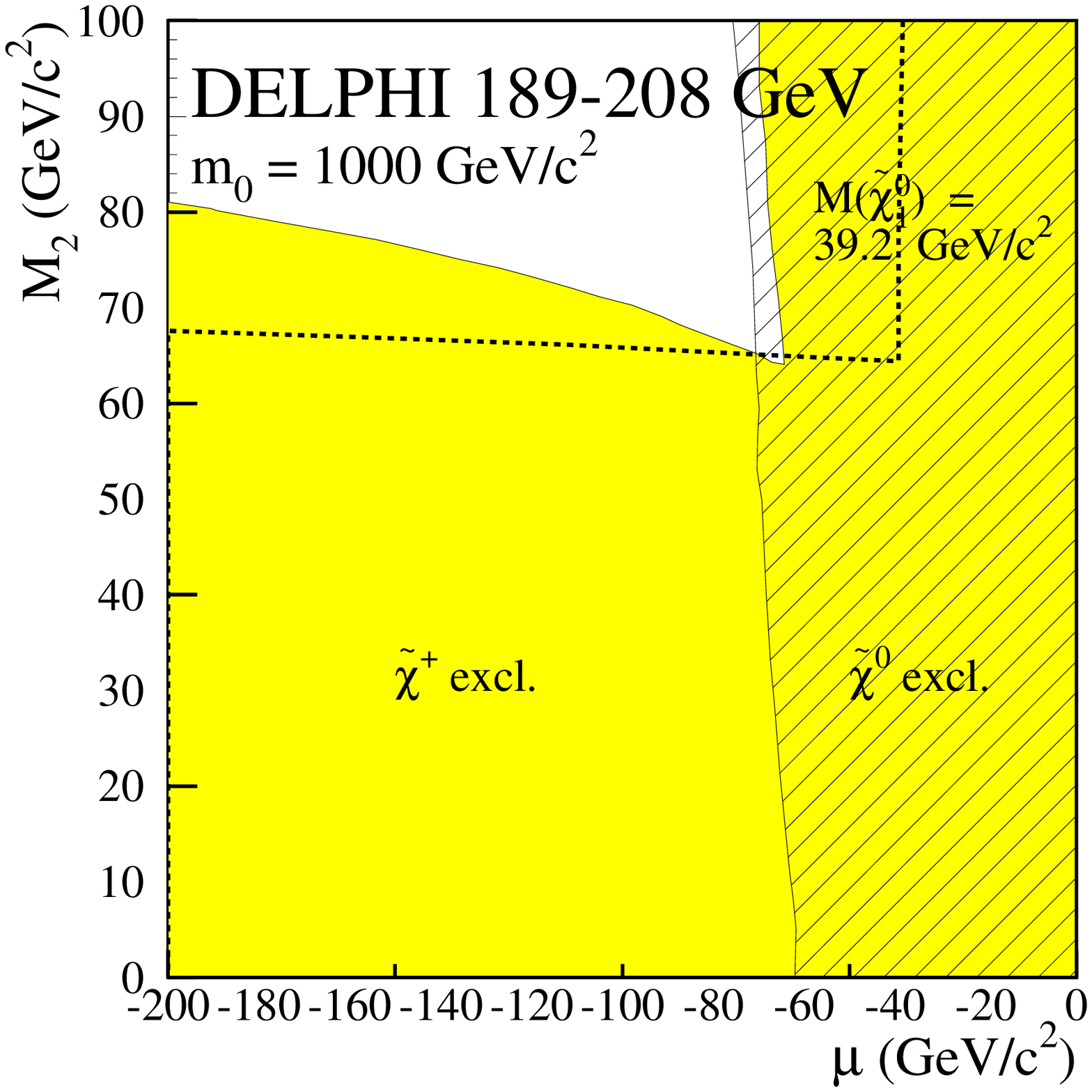}}
\vskip -0.1 cm
\begin{center}
\caption[MSSM limits in ($\mu$,$M_2$) plane]{
Excluded regions in the ($\mu$,$M_2$) plane for $\tanb\ = 1$ for $m_0$~=~1000~\GeVcc .
The shaded areas 
show  regions excluded by searches for charginos and the hatched areas
 show regions excluded by searches for neutralinos. 
The thick dashed curve shows the isomass contour for  
$\MXN{1}=$~39.2 \GeVcc,  the lower limit
on the LSP mass  obtained at \tanb=1. The chargino exclusion
is close to the isomass contour
for $\MXC{1}$ at the kinematic limit.
From chargino searches alone the lower limit on $\MXN{1}$
is $\MXN{1}=$~38.2 \GeVcc.}
\label{fig:mumlsp}
\end{center}
\end{figure}

Of the detectable
neutralino production channels
({\em i.e.} excluding \XN{1}\XN{1}), 
the \XN{1}\XN{2} and \XN{1}\XN{3} channels are dominant for large regions in 
the parameter space, but in order to cover 
as much as possible of the parameter space channels
like \XN{2}\XN{3} and \XN{2}\XN{4} must also be considered, giving cascade
decays with multiple jets or
leptons in the final state. At large $m_0$ the production cross-section
for all these neutralino
production channels drops to very small values for
$|\mu| \geqsim$ 75 \GeVcc. This
is because the two lightest neutralinos
have very small higgsino component (photino $\XN{1}$ and 
zino $\XN{2}$ ) and
their {\it s}-channel pair-production is therefore suppressed,
while pair-production
of heavier neutralinos is not  kinematically
accessible.
Nevertheless,  for $\tanb < 1.5$  and
$M_2>68$ \GeVcc\   the neutralino exclusion reaches beyond the kinematic
limit for chargino production at negative $\mu$  
(see figure~\ref{fig:mumlsp}).
This region of  $\tanb$  is now also excluded
by the searches for the production of the lightest
Higgs boson \cite{higgs}.\\

{\underline{ For medium $m_0$, 100~$\GeVcc \leqsim m_0 \leqsim$~1000~$\GeVcc$}}, 
the $\XN{1}\XN{2}$  production
cross-section  in the gaugino-region ($|\mu| \geqsim$~75 \GeVcc) grows quickly
as $m_0$ falls, due to  the  rapidly rising contribution from the 
selectron {\it t}-channel exchange. Meanwhile,
the chargino production cross-section in
the gaugino region
drops slowly, but it  remains large enough to allow chargino
exclusion  nearly up to the kinematic limit for  $m_0 \geqsim$~200~\GeVcc.
For lower $m_0$ ($\sim~100$~\GeVcc), 
the chargino production cross-section  
in the gaugino region is close to  its minimum, while
the neutralino
production cross-section is very much enhanced.
Consequently,   the region of the ($\mu,M_2$) parameter 
space excluded by the searches
for  neutralino production at small $m_0$ is larger than the one excluded
by the search for  chargino and neutralino production at large $m_0$, as shown 
in  \cite{neut189,lsp,osakalsp}. \\

\underline{For  small $m_0$, $m_0 \leqsim$~100~\GeVcc,  and
small $M_2$, $M_2 \leqsim$~200~\GeVcc}, the situation is much
more complicated because
light sfermions affect not only the production cross-sections 
but also  the decay patterns
of charginos and neutralinos. Sleptons can also be searched for in direct
pair-production. Excluded regions at small $m_0$ arise from
the combination of searches for chargino, neutralino and slepton
production.\\

For small $m_0$ and $M_2$ the  sneutrino is light, and for  
\MXC{1} $>$ \msneu\
the chargino decay mode   \XPM{1}~$\to$~\snu$\ell$
is dominant,  leading to an experimentally undetectable
final state if \MXC{1} $\simeq$ \msneu. In the gaugino
region, for every value of $M_2$
and $\mu$, an $m_0$ can be found  such that \MXC{1} $\simeq$ \msneu;
therefore the search for charginos cannot   be used to exclude regions in
the ($\mu$,$M_2$) plane if very small $m_0$  values are
allowed. The search for  selectron production
is used instead to put a limit on the sneutrino mass (and
thus on the chargino mass), the selectron
and the sneutrino masses being related by equations (1)-(3).
The selectron pair-production
cross-section is typically larger than the smuon pair-production
cross-section, due to the contribution of {\it t}-channel neutralino
exchange. However, at $|\mu| \leqsim$~200~\GeVcc\  the selectron production 
cross-section
tends to be small and the exclusion arises mainly from the search
for    neutralino pair-production.

Mixing between the left-handed and right-handed
sfermion states can be important
for the third family sfermions
and can lead to light \stone, \sbq$_{1}$ and \stq$_{1}$.  
Mass splitting terms 
at the Electroweak Scale proportional to $m_\tau$($A_{\tau}$$-$$\mu$\tanb),
$m_b$($A_{b}$$-$$\mu$\tanb), and $m_t$($A_{t}$$-$$\mu$/\tanb) were considered
for \stau, \sbq, and \stq\ respectively (see section~\ref{sec:constraining}). 
In the first instance  $A_{\tau}$=$A_{t}$=$A_{b}$=0 was assumed, then
the dependence of the results on  $A_{\tau}$ , $A_{b}$ and $A_{t}$   was  studied.
These terms  lead
to  \stone, $\sbq_{1}$ or $\stq_{1}$ being degenerate in mass
with $\XN{1}$,  or being the LSP for 
large values of $|\mu|$. 
The terms $A_{\tau}$,$A_{t}$, $A_{b}$ are arbitrary unless further
constraints on the model are imposed. 
To derive a conservative limit on the
LSP mass, which is valid for any stau mixing scenario, 
a model was used with no mixing
in the sbottom or stop sector, but only in the stau sector. This maximises the
region in the parameter space where the stau can be close in mass to the LSP.
Moreover the trilinear coupling $A_{\tau}$ was adjusted in every point of the
parameter space to get the stau close in mass to the LSP mass. For the stau to be
degenerate with the LSP in the interesting mass range, the corresponding
mixing  term has to be of the order of 12 000~\GeVcc. 

The results presented  here are
limited to the range  of the $\mu$ parameter where the lightest
neutralino is the LSP.

\subsection{Results}

\subsubsection { The LSP mass limit for large  \mbox{\boldmath${m_0}$}  \label{sec:hm0res}}

From charginos searches alone  a limit of 38.2~\GeVcc\ on the lightest 
neutralino mass is obtained, valid for
\tanb\ $\ge$ 1 and a heavy sneutrino ($\msnu\! >\! 300~\GeVcc$). 
The limit is reached for
\tanb~=~1, $\mu~=~-65.7$~\GeVcc, M$_2$~=~65.0~\GeVcc.  
This limit improves by $\sim$ 1 \GeVcc\  
due to the constraint from the search
for  neutralino production; thus
from chargino and neutralino searches the  LSP mass can be constrained
to be $\MXN{1}>39.2 $ \GeVcc, and the limit occurs at \tanb = 1.
Figure~\ref{fig:mumlsp} 
shows the  region in the  ($\mu$,$M_2$) plane
for \tanb=1 excluded by the chargino and neutralino
searches, relevant for the LSP mass limit at 
$m_0 =1000$ \GeVcc. The lowest value of \MXN{1} 
not excluded by chargino and neutralino searches   occurs
for   \tanb=1, $\mu=$~$-$76.5~\GeVcc\  and  $M_2$~=~67.0~\GeVcc.
For these parameters, $\XN{4} \XN{2}$ production  
and  chargino pair-production are important.
 
\begin{figure}[ht]
\begin{center}
\vskip 0.5 cm
\mbox{\epsfysize=14.0cm\epsfxsize=14cm\epsffile{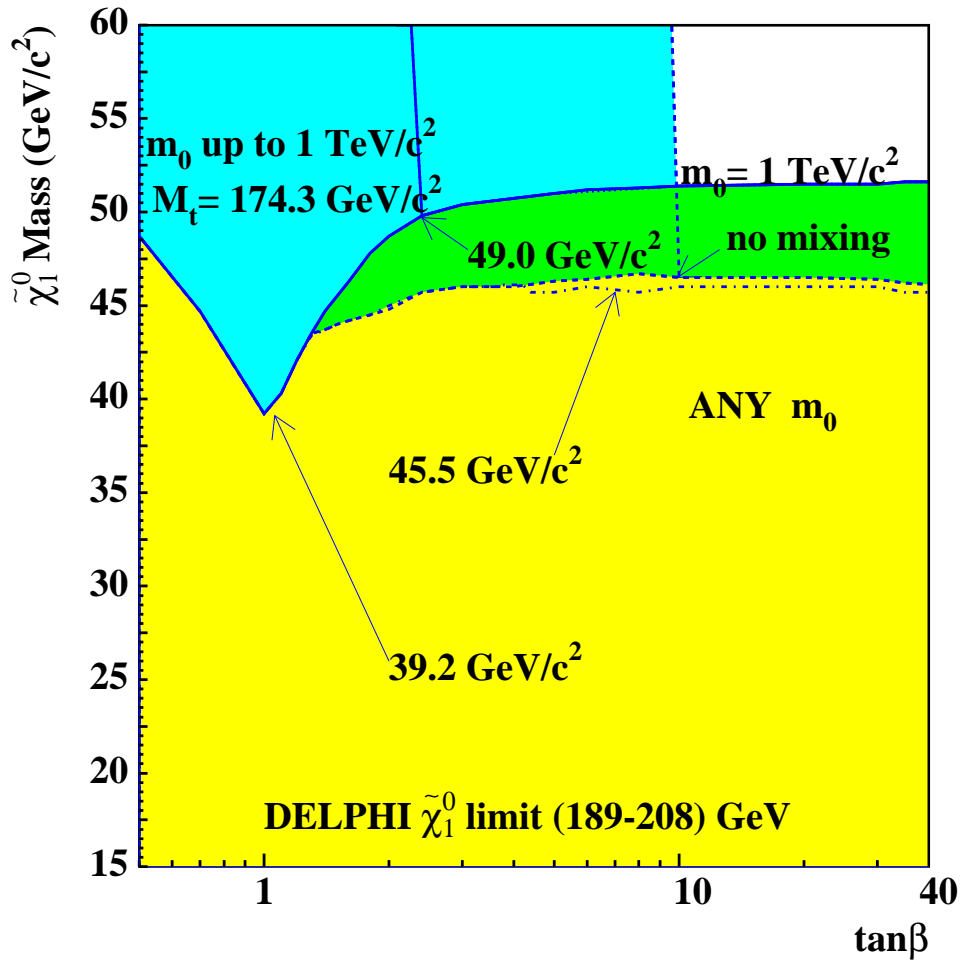}}
\caption[MSSM limits in ($\mu$,$M_2$) plane]{
The lower limit on the mass of the lightest
neutralino, \XN{1}, as a function of \tanb\ assuming a stable \XN{1}.
The solid curve shows the limit obtained for 
$m_0$~=1000~\GeVcc, the dashed curve shows the limit obtained
allowing for  any $m_0$ assuming that there is no mixing in the
third family
($A_{\tau}=\mu\tanb$, $A_{b}=\mu\tanb$, $A_{t}=\mu/\tanb$),
and the dash-dotted curve shows the limit obtained for
any $m_0$ allowing for mixing with 
$A_{\tau}$=$A_{b}$=$A_{t}$=0. 
The steep solid (dashed) line  shows the effect of the
searches for the Higgs boson
for the maximal $M_{\hn}$ scenario (no mixing scenario),  
$m_0 \le $ 1000~\GeVcc\ and
$M_t$= 174.3~\GeVcc, which amounts to excluding the region of 
$\tanb<2.36 (9.7)$. }
\label{fig:LSPLIM}
\end{center}
\end{figure}

Figure~\ref{fig:LSPLIM} gives the lower limit on $\MXN{1}$ as a function of 
\tanb.    
At $\tanb \geqsim~1.2$ the neutralino search no longer contributes,
the LSP limit is given by the chargino search, and its value reaches
about 
half the limit on the chargino mass at large \tanb, where the
isomass contours  of \XPM{1}\ and \XN{1}\
in the ($\mu$,$M_2$) plane 
become parallel. 
The rise of the LSP limit for small \tanb\
can be explained by the change of the shape of these contours
with \tanb, as illustrated in \cite{lsp,osakalsp}.
It should be noted that, because the chargino and neutralino
masses are invariant under the exchange $\tanb \leftrightarrow 1/\tanb$,
the  point $\tanb=1$ is the real minimum and the LSP limit for $\tanb<1$ can be
obtained by replacing $\tanb$  with $1/\tanb$ in figure~\ref{fig:LSPLIM}.

For  $M_{A}$ $\le$ 1000~\GeVcc,
$A_{t}$-$\mu/\tanb$=$\sqrt{6}$~\TeVcc\ (maximal  $M_{h^0}$ scenario  used in 
\protect{\cite{higgs}}), and $m_t =$ 174.3 \GeVcc,
the \tanb\ region  0.5$\le \tanb \le$ 2.36 is excluded by the Higgs searches. 
Thus  including
Higgs searches imposes a limit on  \MXN{1} (see figure~\ref{fig:LSPLIM}) of 
\MXN{1} $>$ 49.0 \GeVcc\
for \tanb $\ge$ 2.36 and   \MXN{1} $>$ 48.5 \GeVcc\ for \tanb $< $ 0.5.

Thus the lightest neutralino is constrained to have a mass:

\begin{center}
$\MXN{1}>49.0 $ \GeVcc
\end{center}

\noindent
for $m_0 =1000$ \GeVcc\  and  $M_2 \le 1000$ \GeVcc\ and
\tanb $\ge$ 1.0. The limit occurs at the edge of the \tanb\ exclusion given
by the searches for the Higgs boson.
However, if  $m_t = $ 179 \GeVcc, the \tanb\ area
excluded by searches for the  Higgs boson shrinks to 0.6$\le \tanb \le$ 2.0
and  these limits worsen to \MXN{1} $>$ 48.5 \GeVcc\ for \tanb $\ge$ 2.0 and 
\MXN{1} $>$ 47.0 \GeVcc\ for  \tanb $< $ 0.6. 
Thus the above limit worsens by 2 GeV if  $m_t = $ 179 \GeVcc\
and  the  \tanb $ < $ 1.0 region is included.

It should be noted that the \tanb\ region excluded by Higgs boson searches
is expected to shrink with the inclusion of complete one-loop corrections
\cite{frank}. The LSP mass limit degrades in this case. Moreover, if there
are sfermions lighter than $M_{h^0}/2$, the lightest Higgs boson will decay 
to them (see section \ref{subsub:mixing}). Such light sfermions are experimentally 
allowed only if they are degenerate in mass with
the LSP. In this case, 
the lightest Higgs boson decays ``invisibly''. 
The mass limit $M_{h^0} > 112.1$ \GeVcc\  set for an invisibly 
decaying Higgs boson \cite{higgsInv} can be used to exclude small
\tanb, and  
\MXN{1} $>$ 48.5 \GeVcc\ for \tanb $\ge$ 2.0.

\subsubsection{ The LSP  
mass limit  for any  \mbox{\boldmath${m_0}$} \label{sec:resanym0lsp}}

Figure~\ref{fig:LSPLIM} also 
gives the lower limit on $\MXN{1}$ as a function of 
\tanb\ for any $m_0$. The  
``any $m_0$'' limit resulting from chargino, neutralino and slepton
searches follows the
large $m_0$ limit up to \tanb=1.4; then it increases more  slowly due to the
opening of the possibility of the chargino-sneutrino degeneracy,  
reaching 46~\GeVcc\ at \tanb $\ge$ 2.36 (edge of the region excluded by Higgs boson searches);
finally, as discussed in more detail below,
it falls to its lowest
value, 45.5~\GeVcc,  at  \tanb $\ge$ 5, due to small  \mstone-\MXN{1},
if  mixing in the stau sector is of the form  $A_{\tau}-\mu\tanb$, and $A_{\tau}=0$.
This limit is set by searches
for $\XN{1}\XN{2}$
and $\XN{2}\XN{2}$ 
 production with $\XN{2} \to \stone \XN{1}$.

Thus

\begin{center}
$\MXN{1}>46.0$ \GeVcc
\end{center}

\noindent
independent of $m_0$, for $\tanb \ge 1$
if there is no mixing in the third 
family  
($A_{\tau}=\mu\tanb$, $A_{b}=\mu\tanb$, $A_{t}=\mu/\tanb$~), or if the mixing
parameters leading to $\mstone-\MXN{1} <$~6~\GeVcc\ are avoided. 
The Higgs boson search in the maximal  $M_{h^0}$  scenario
was used to exclude   0.5$\le \tanb \le$ 2.36. 

LSP mass limits obtained with various assumptions are summarised in
table~\ref{tab:limsum}.


\subsubsection{ 
The  dependence of the LSP limit on the mixing in the third family.
\label{subsub:mixing}}

Mixing in the third family affects in a complicated way the excluded regions obtained from Higgs,
chargino, neutralino and squark searches, and a consistent discussion of the
mixing is difficult.

For example, the maximal  $M_{h^0}$ scenario used to set \tanb\ 
limits from Higgs searches implies  
$A_{t}-\mu/\tanb=\sqrt{6}$~\TeVcc,
 thus a different $A_{t}$ for every  $\mu$. Such a scenario
implies as well that small $m_0$ values are forbidden for 
$M_2$ sufficiently small to be of interest from 
the point of view of the LSP mass limit
\footnote{\protect{To avoid ``tachyonic'' mass solutions for the squarks and
sleptons we must have  $m_{ll}+m_{rr}  > \sqrt{ (m_{ll}-m_{rr})^2 +4m_{lr}^2 }$
where $m_{lr}$ is the off-diagonal mixing term, and $m_{ll},m_{rr}$ are the diagonal 
mass terms. For example, for the stop we have:  $m_{lr}=m_{top}(A_t-\mu/\tanb)$ and
$m_{ll} \simeq m_0^2 + 9M_2^2 + m_{top}^2 + m_Z^2 \cos2\beta (0.5-2/3\sin^2\theta_W)$, 
$ m_{rr} \simeq m_0^2 + 8.3M_2^2 + m_{top}^2 + 2/3 m_Z^2 \cos2\beta\sin^2\theta_W $;
for an example value of $A_t-\mu/\tanb=\sqrt{6}$ \TeVcc, 
the condition above sets a lower limit
on a combination of $m_0^2$  and   $M_2^2$:
${m_0}^2 + 8.5{M_2}^2 > 0.39$ (\TeVc)$^2$.
Thus, if $M_2< 190$~\GeVcc\ we must have  $m_0> 300$~\GeVcc. 
}}. Thus for consistency  one should consider only the
large $m_0$ scenario for the LSP limit (thus higher limit), whenever  
\tanb\ limits from the 
maximal  $M_{h^0}$  scenario are used.

In the no-mixing scenario ($A_{\tau}=\mu\tanb$, $A_{b}=\mu\tanb$, $A_{t}=\mu/\tanb$~), the LSP limit
occurs in the chargino-sneutrino degeneracy region, at \tanb$>$ 1.5, where both
the chargino and sneutrino mass limits are given by the selectron exclusion (see section \ref{sec:resanym0rest}). 
If there is no mixing in the stop sector, the \tanb\ region excluded
by the Higgs searches grows to  \tanb $\le$ 9.4, compared to   0.5$\le \tanb \le$ 2.36
for the  maximal  $M_{h^0}$  scenario. However, there is no improvement of the LSP limit 
as compared
to  $\MXN{1}>46.0$ \GeVcc\ given above, as  the  \tanb\ dependence of the limit
``flattens out'' at large \tanb\ (see figure~\ref{fig:LSPLIM}).
 
As discussed in previous papers~\cite{lsp,osakalsp}, 
mixing in the stau sector can lead to a configuration where  \mstone$-$\MXN{1}
is small
enough to  make the 
\stone\ undetectable  and cause a blind spot both in   \stone\ and chargino
exclusion.
For  $A_{t}$=$A_{\tau}$=$A_{b}$=0, with  the present data,  
the LSP mass limit occurs  a) at large enough  $M_2$ that already for
\tanb $\ge$ 3  both \msbqone\ and \mstqone\
are pushed above \mstone\footnote{The ``mixing -independent'' (diagonal) 
terms of the mass matrices  grow faster with $M_2$ for squarks than for sleptons, and they  have different 
dependence on \tanb. For example, for  $A_{t}$=$A_{\tau}$=$A_{b}$=0,  $\mu$=0, and \tanb=1, both the
 \stq$_{1}$ and  \sbq$_{1}$ are  heavier than the  \stone; but they become  lighter  than the  \stone\ for  large $|\mu|$ values. 
The mass hierarchy between  \stone, \sbq$_{1}$, and \stq$_{1}$  depends on $M_2$, \tanb, $\mu$, and $m_0$.}, 
and  $\stone$ can  become  degenerate  in mass with $\XN{1}$, and b) at 
large enough  $m_0$ 
that  selectron and  sneutrino
pair-production are not 
allowed by  kinematics. 

For the above mixing assumptions, the  mass  of the lightest Higgs boson
corresponding to the LSP limit point varies between
109 and 120 \GeVcc, depending on the \tanb\ region. 
However, when the stau is light
the lightest Higgs boson decays predominantly to  $\stone \overline{\stone}$,
thus ``invisibly''. The mass limit $M_{h^0} > 112.1$ \GeVcc\  set for an invisibly 
decaying Higgs boson \cite{higgsInv} can be used to exclude some of 
these points, but it is enough
to change slightly the mixing in the stop sector  ($A_{t}$) to push the Higgs
mass above  112.1 \GeVcc.

For such a mixing configuration and for 
\tanb~$\geqsim$~3,  
the  LSP limit is therefore set
by the searches for stop and sbottom and those  
for $\XN{1} \XN{2}$ production with  $\XN{2} \to \stone \tau$, and 
it  falls to $\MXN{1}>$~45.5~\GeVcc\ 
(see dot-dashed line in figure~\ref{fig:LSPLIM}). This limit was verified to be
robust when varying independently $A_{t}$,$A_{\tau}$ and $A_{b}$ in the range $\pm$ 1000~\GeVcc.

\begin{figure}[ht]
\begin{center}
\vskip 0.5 cm

\mbox{\epsfysize=14.0cm\epsfxsize=14cm\epsffile{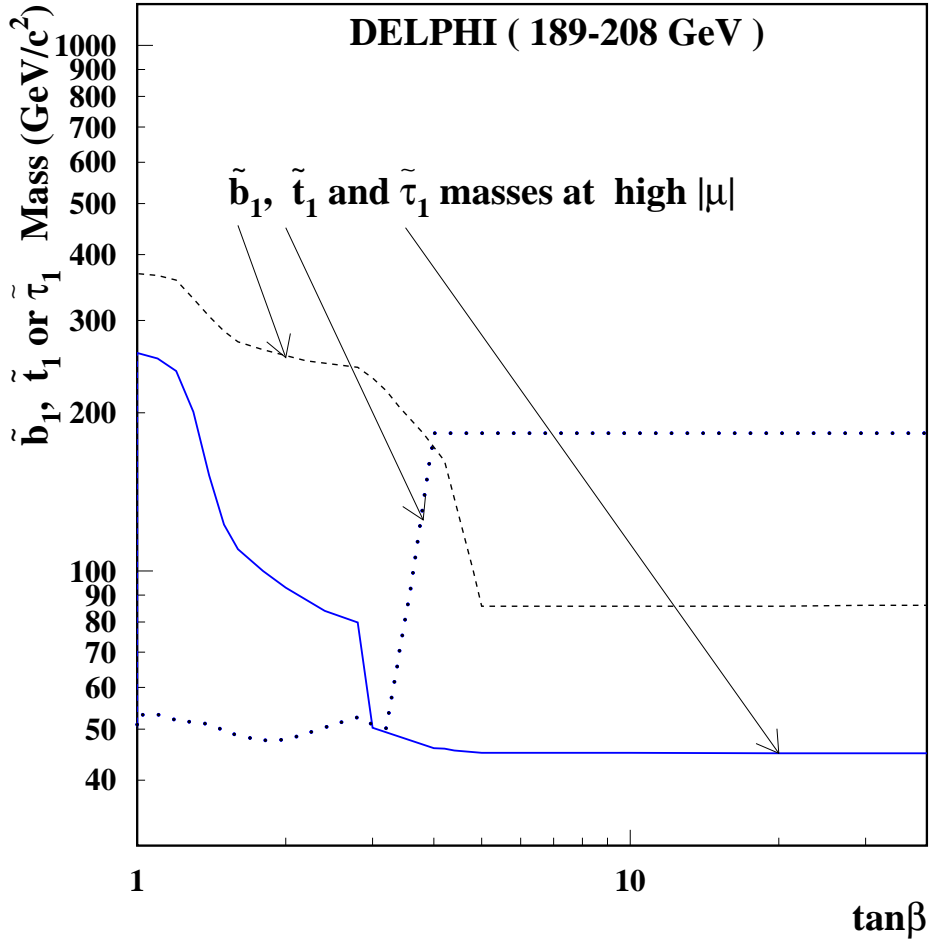}}

\caption[MSSM limits in ($\mu$,$M_2$) plane]{The 
masses  as a function of \tanb\
of the lightest stau (solid curve),
stop (dotted curve) and  sbottom (dashed curve),
at the largest allowed $|\mu|$ for the smallest
non-excluded $M_2$ value.
Mass splitting  in the stau (sbottom, stop) sector 
in the form  $A_{\tau}-\mu\tanb$ ( $A_{b}-\mu\tanb$,  $A_{t}-\mu/\tanb)$
was assumed, with $A_{\tau}=0$ ($A_{b}=A_{t}=$0).
}
\label{fig:higmu}
\end{center}
\end{figure}

\begin{figure}[htbp]
\begin{center}
\begin{tabular}{c}
\epsfxsize=16.0cm
\epsffile{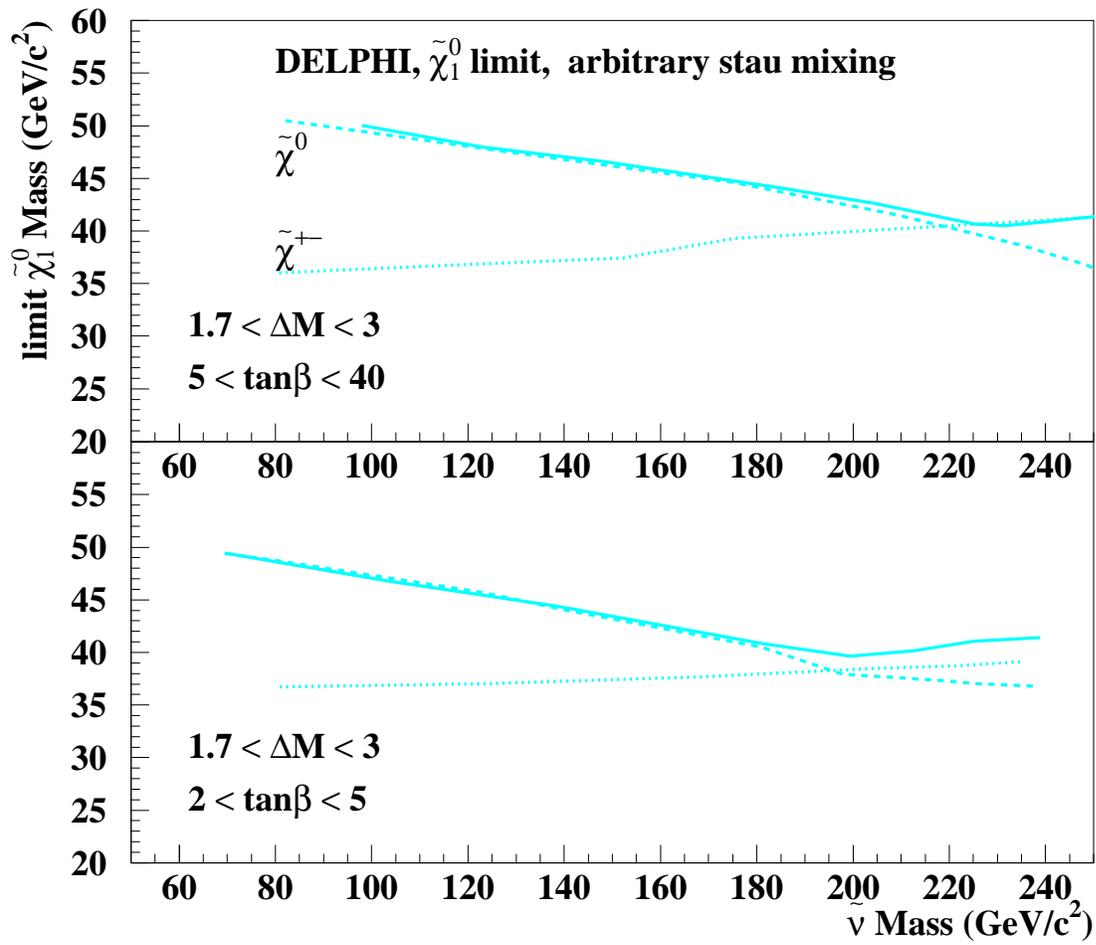}
\end{tabular}
\vspace*{0.5cm}
\caption[]{ The limit set on the LSP mass by chargino (dotted line) and neutralino
searches (dashed line) as a function of the sneutrino mass in the case when the lightest
stau is degenerate in mass with the lightest neutralino. The combined limit
is shown as solid line. These limits are valid for any
model of stau mixing. In any particular model, in which mixing in the stau
sector and in the sbottom and stop sector is related a more stringent limit
on the LSP mass can be set.
}
\label{fig:x1stauarb}
\end{center}
\end{figure}

As noted in previous papers
\cite{lsp,osakalsp}, 
 it is the appearance of the light sbottom
and stop which protects the stau from being degenerate in mass with the LSP
for very large $m_0$ values, where the  $\XN{1} \XN{2}$  and $\XN{2} \XN{2}$ 
production cross-sections
are small.
This is illustrated in figure~\ref{fig:higmu}, where the stop, sbottom,
and stau masses at the largest allowed $|\mu|$ value at the  lowest non-excluded
$M_2$ are plotted. At large \tanb, the largest $|\mu|$ value occurs when the sbottom
and stop masses have their lowest non-excluded (see section \ref{sec:sqresult})
values.
 In a pathological model where there is no 
mixing  in the sbottom or stop sector ($A_{b}=\mu\tanb$, $A_{t}=\mu/\tanb$)
but only in the stau sector, \stone\  can be made 
degenerate with \MXN{1}\ even  at large values of
$m_0$ and $|\mu|$ so that the  $\XN{1}\XN{2}$ and  $\XN{2}\XN{2}$  production 
cross-sections
at LEP are very small and the production of the Higgs boson and other
sfermions is not accessible kinematically. However, for large $m_0$ values the
$\XPM{1} \XPM{1}$ cross-section is sufficiently large to set a limit on the production of
``invisibly'' decaying charginos ($\XPM{1} \to \stone\ \nu$, \stone $ \to \XN{1} \tau$)
from the search with an ISR photon (see section \ref{sub:chadeg_res}).
The limit set on the neutralino mass in such a scenario is 39~\GeVcc\ for \tanb$>$2 and
is illustrated in figure \ref{fig:x1stauarb}.   

In mSUGRA (as defined in section  \ref{sec:framework}), $|\mu|^2$ is in the range
3.3~$m^2_{1/2}-0.5 m_Z ^2$~$ < \mu^2 < $~$m^2_{0}+3.8 m^2_{1/2}$ 
for \tanb$>$2 and
a light stau cannot be degenerate with neutralino
for large $m_0$. 
Neutralino production cross-section is thus large,
and neutralino searches   
set  a limit on  the LSP  mass  for small $\mstone -\MXN{1}$ which 
is close to the one obtained for  heavy sleptons (about 50 \GeVcc). 




\subsubsection{ 
The influence of radiative corrections to the gaugino mass relations on the LSP
limit.
\label{subsub:radiative}}
The relations between  chargino, neutralino and gluino masses and $|\mu|$
and $M_2$ are affected by radiative corrections of the order of 
2\%--20\% \cite{radcor}.
However,  only the relative relations between 
chargino, neutralino and gluino
masses  are important from the experimental point of view, 
and here the corrections
are much smaller. For example, the relation
\MXC{1}(\MXC{2})/\MXN{1} $\simeq$ 2 in the gaugino
region, which is 
exploited to set a limit on the LSP mass, 
receives corrections only of the order of  2\%.
The ratio $M_1/\alpha_1 : M_2 /\alpha_2$  can receive corrections
of similar order \cite{kribs}, thus affecting
\MXC{1}(\MXC{2})/\MXN{1} $\simeq$ 2 in the same way.

It is enough to lower the LSP
limits presented here  by 1~\GeVcc\  $(2\%)$ to
conservatively account for  the 2\%  change of the \MXC{1}(\MXC{2})/\MXN{1}
ratio. \\

\subsection{ 
\mbox{\boldmath$\tilde{\chi}^{\pm}_1$}
mass limits for any \mbox{\boldmath${m_0}$} }
\label{sec:resanym0rest}

Figure~\ref{fig:SNELIM} shows the chargino mass limit as
a function of \tanb\ for $M_2 <$ 200~\GeVcc. The lowest non-excluded
chargino mass is found at MSSM points very close to those
giving the LSP mass limit, and  the arguments  presented in 
section \ref{sec:resanym0lsp} 
also apply
to explain the dependence of the  chargino mass limit  on \tanb.
{For \tanb~$ \leqsim$~1.4} the limit occurs at large $m_0$ values. 
{For $1.4 \leqsim  \tanb \leqsim 3$} and $M_2 <$ 200~\GeVcc,
the limit occurs at  small $m_0$
in the chargino-sneutrino degeneracy region.
It falls
at {$\tanb \geqsim 4$}    
because of the small  $\DM\ =\mstau\ -\MXN{1}$. \\

\noindent
The lightest chargino  
is  constrained to have a  mass:

\begin{center}
$\MXC{1}>94 $ \GeVcc,

\end{center}

\noindent
independently of $m_0$, for $\tanb \le 40$,  $M_2 \le 1000$~\GeVcc, 
if either there is no mixing in the third 
family  
($A_{\tau}=\mu\tanb$, $A_{b}=\mu\tanb$, $A_{t}=\mu/\tanb$) or the mixing
parameters leading to $\mstone-\MXN{1} <$~6~\GeVcc\ are avoided. 

If  mixing in the stau sector is of the form  $A_{\tau}-\mu\tanb$, and $A_{\tau}=0$,
the limit falls to  90~\GeVcc,  at  \tanb $\ge$ 3, due to small  \mstone$-$\MXN{1}.
This limit is robust when 
varying $A_{t}$,$A_{\tau}$ and $A_{b}$ independently in the range $\pm$
1000~\GeVcc.
In the ``arbitrary'' mixing scenario described above, where there is no
mixing in the stop and sbottom sector, but only in the stau sector, this limit falls
to 80~\GeVcc\ and it is independent of the stau mixing.

The chargino mass limits for large $M_2$ values, where the chargino can be degenerate
in mass with the LSP 
are close to 75~\GeVcc\ (see section \ref{sub:chadeg_res}). 

Chargino mass limits obtained with various assumptions are summarised
in table \ref{tab:limsum}.

\subsection{  \mbox{\boldmath${\tilde\nu}$} and 
\mbox{\boldmath${\tilde{\mathrm e}_{\mathrm R}}$}
mass limits for any \mbox{\boldmath${m_0}$} \label{sub:m0rest}}
The sneutrino  and the \selr\ mass limits are:

\begin{center}
$\msnu>94$ \GeVcc\ and $\mselr>94$ \GeVcc.
\end{center}

\begin{figure}[hbt]
\begin{center}
\vskip 0.5 cm

\mbox{\epsfysize=14.0cm\epsfxsize=14cm\epsffile{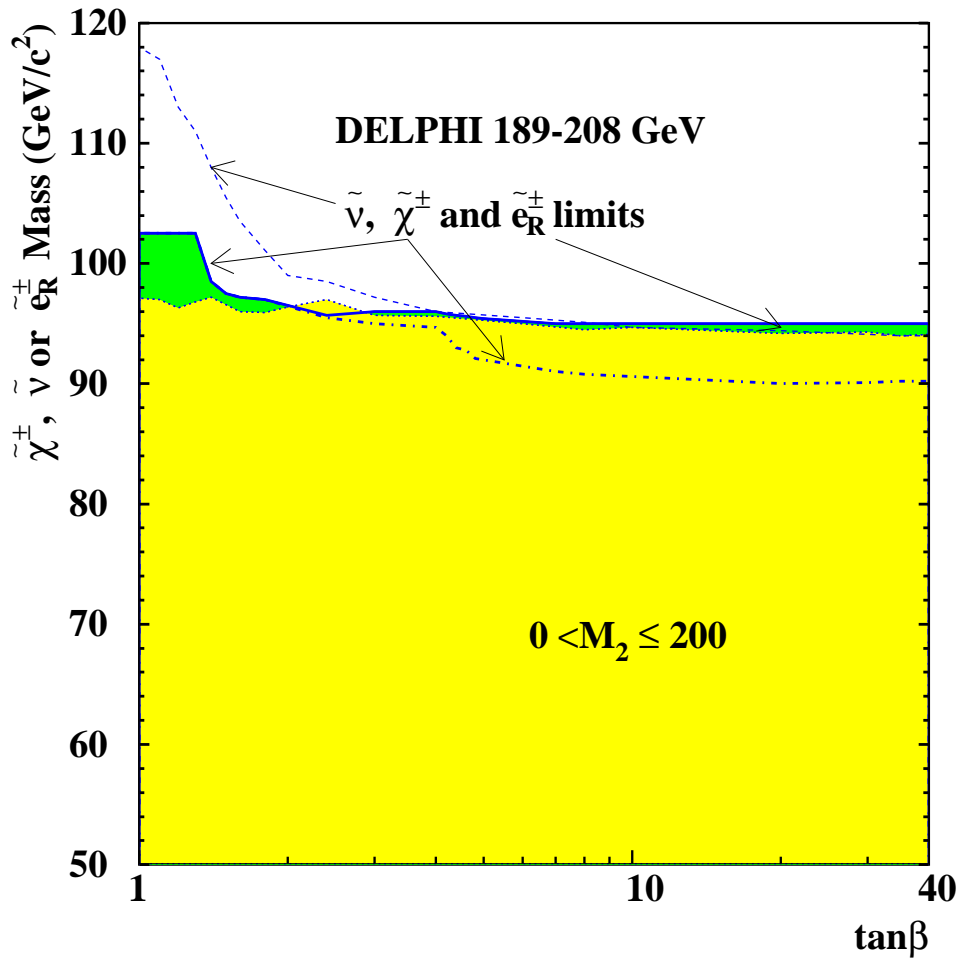}}

\caption[MSSM limits in ($\mu$,$M_2$) plane]{
The minimum
sneutrino mass  
(thin dashed curve) 
and the \selr\ mass (thin dotted curve and
light shading)
allowed by the slepton and neutralino
searches, as a function
of \tanb, together with the limits on the chargino mass (thick solid
curve and dark shading,
and thick dash-dotted curve). 
The chargino mass limit indicated by the solid curve and
the sneutrino and selectron mass limits  were
obtained assuming no mass splitting in the third sfermion family  
($A_{\tau}-\mu\tanb$=0 in particular). 
The selectron mass limit is valid for 
$\mselr- \MXN{1} > 10 $ \GeVcc.
The chargino mass limit indicated with the dash-dotted curve was obtained
allowing for mass splitting in the third sfermion family, 
with $A_{\tau}=A_{b}=A_{t}$=0.
}
\label{fig:SNELIM}
\end{center}
\end{figure}

\noindent
These limits, shown
in figure~\ref{fig:SNELIM},  were obtained assuming no
mass splitting in the third sfermion family ($A_{\tau}=\mu\tanb$),
 implying
\mselr=\mstaur=\mstone=\msmur, as this gives the lowest values. 

These limits result from 
the combination of slepton  and neutralino searches.
The selectron mass limit 
(see figure~\ref{fig:SNELIM}, dotted curve) is valid for
$-$1000~\GeVcc~$\leq~\mu~\leq~$~1000 \GeVcc\  
and $1 \le \tanb \le 40 $ provided that
\mselr~$-$~\MXN{1}~$>$~10~\GeVcc, and
it allows  a limit to be set on the sneutrino mass as shown
in figure~\ref{fig:SNELIM} (dashed curve). The sneutrino 
mass limit is expected to rise for $\tanb<1$,
the sneutrino being heavier than the  $\selr$ for small $\tanb$.
The  selectron mass limit for
$\tanb=1.5$ and $\mu=-200$~\GeVcc\ is presented in section \ref{sec:slep_res1}.

Slepton mass limits obtained with different assumptions are summarised
in table \ref{tab:limsum}.



\subsection{MSSM parameter space exclusion \label{sub:mssm}}

\begin{figure}[htbp]
\begin{center}
\begin{tabular}{c}
\epsfxsize=16.0cm
\epsffile{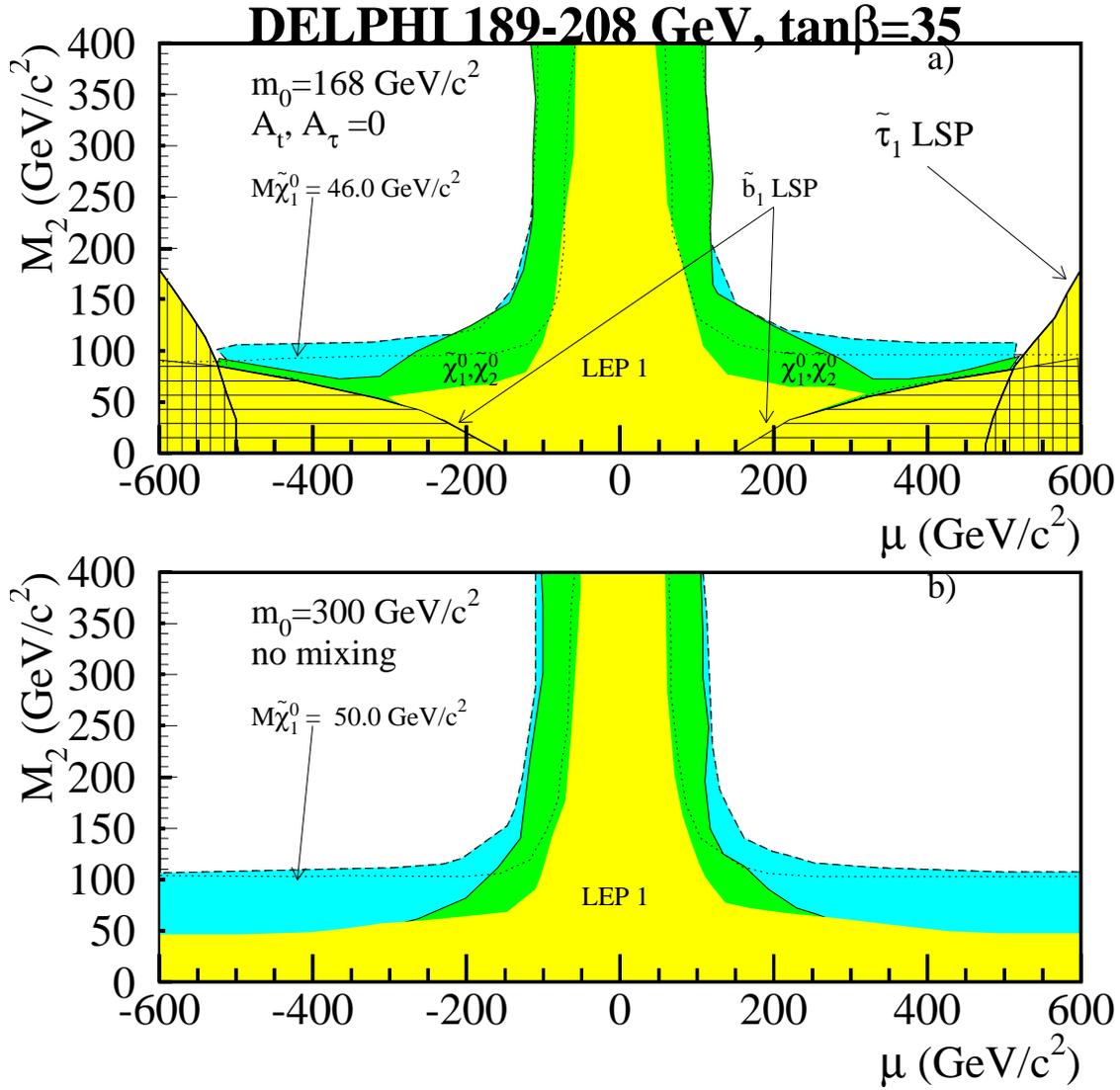}
\end{tabular}
\vspace*{0.5cm}
\caption[]{
Regions excluded at in the ($\mu$, M$_2$) 
plane for \tanb~=~ 35 and
different 
assumptions of mixing in the third family.
Light shaded regions are excluded by searches at LEP1 energies, darker shading
bounded by the solid line marks regions which are excluded by 
neutralino searches, while
intermediate shaded regions  bounded by the thin dashed line are excluded
by chargino searches.   
Mixing terms   
in the form $A_{\tau}-\mu\tanb$ ($A_{b}-\mu\tanb$,  $A_{t}-\mu/\tanb$)
were considered. For plot a) $A_{b}=A_{t}=A_{\tau}=0$ was assumed. 
The no-mixing scenario was used in b) ($A_{\tau}-\mu\tanb$=0, $A_{b}-\mu\tanb$=0,  
$A_{t}-\mu/\tanb$=0).
Plot a) is for the $m_0$ values giving the lowest non-excluded
LSP mass. In the vertically (horizontally) hatched areas the stau (the sbottom)
is the LSP.
Relevant isomass contours of the lightest neutralino
are also shown (\MXN{1}=46~\GeVcc\ in a) and \MXN{1}=50~\GeVcc\ in b)).}
\label{fig:neulim_m2m}
\end{center}
\end{figure}

The limits on Higgs, chargino, neutralino, slepton and squark production
used in the previous sections to set a mass limit on the LSP and other particles can be also
used to exclude regions in a parameter space of the CMSSM.
The excluded regions in the ($\mu,M_2$) plane for 
\tanb~=~35 and two assumptions
about $m_0$ and mixing
in the third family  are shown in figure~\ref{fig:neulim_m2m}.




\section{Summary}
\label{sec:summary}

Searches for charginos, neutralinos, sleptons and squarks in \ee\
collisions at centre-of-mass
energies up to 208 GeV were performed with the DELPHI detector at LEP. 
No evidence for a signal was found
in any of the channels and 95\% CL upper limits on the production 
cross-sections were derived.
Under assumptions that depend on the channel 
lower limits on the masses of SUSY particles were set.
In particular, in the framework of constrained  MSSM scenarios with gravity-induced
SUSY breaking, regions of the parameter space can be 
excluded, and these exclusions can be translated into limits on the 
masses of SUSY particles.
The combination of the results of the 
different search channels is crucial to ensure the best possible 
coverage of the parameter space.

A summary of the obtained lower limits on the masses of SUSY 
particles is shown in table~\ref{tab:limsum}.
The results presented extend and confirm previous exclusions set by 
DELPHI (see \cite{slep} to \cite{osakalsp}) 
and by the other LEP experiments~\cite{others}.

\newpage

\renewcommand\arraystretch{1.3}
\begin{table}[htb]
\begin{center}
\begin{tabular}{|c|c|c|}
\hline
  &   &   \\
Particle &  Validity conditions &  Mass limit \\
  &  &   (GeV/c$^2$)       \\
\hline
\hline

\selr&  tan$\beta$=1.5, $\mu$=-200, \DM$>$15  &  94   \\

\cline{2-2}

     & CMSSM, \DM$>$10 &  94     \\
\hline

\smur& BR(\smu$\rightarrow \mu \Xn$)=1   , \DM$>$5 & 88    \\

\cline{2-2}

   &   CMSSM, \DM$>$10 &     94       \\

\hline

\stau&  BR(\stau $\rightarrow \tau \Xn$)=1,  \DM$\ge m_\tau$ & 26
\\

\hline

\staur&  BR(\stau $\rightarrow \tau \Xn$)=1, \DM$>$15, no mixing  &  85     \\

\hline

\stau$_{min}$ & BR(\stau $\rightarrow \tau \Xn$)=1,  \DM$>$15, minimal cross-section  & 82   \\

\hline

\snu& CMSSM,  (\mselr-\MXN{1})$>$10  &  94   \\

\hline
\hline

\sbq   &   BR(\sbq $\rightarrow$ b \Xn)=1,\DM$>$7, no mixing    &  93  \\

\cline{2-2}

   &   BR(\sbq $\rightarrow$ b \Xn)=1, \DM$>$7, minimal cross-section &  76 \\

\hline

   &   BR(\stq $\rightarrow$ c \Xn)=1, \DM$>$10, no mixing     &  96     \\

\cline{2-2}

\stq&   BR(\stq $\rightarrow$ c \Xn)=1, \DM$>$2, no mixing    &  75 \\

\cline{2-2}

   &  BR(\stq $\rightarrow$ c \Xn)=1, \DM$>$10, minimal cross-section  &  92 \\

\cline{2-2}

  &    BR(\stq $\rightarrow$ c \Xn)=1, \DM$>$2, minimal cross-section    
&  71   \\

\hline
\hline

  & \msnu$>$1000, \DM$>$10, $M_1=\sim 0.5 M_2$,  & 102.7   \\

\cline{2-2}

\Xpm& M$_{{\tilde{f}}}>M_{\tilde{\chi^\pm}}$, \DM$>$3   & 97     \\

\cline{2-2}

   &  M$_{\tilde{f}}>M_{\tilde{\chi^\pm}}$, any \DM,  $M_1=\sim 0.5 M_2$ & 75 \\

\cline{2-2}

  &   \msnu$>$ 300, $| \mu | \ge $ M$_2$, no gaugino mass unification, any 
\DM\   &  70    \\

\cline{2-2}

  & CMSSM, \DM$>$3, any m$_0$, no mixing or \DM(\stau-\Xn)$>$6 &  94     \\

\cline{2-2}

  &  CMSSM, any m$_0$, any M$_2$, tan$\beta<$40, 
mixing A$_\tau$=A$_b$=A$_t$=0 &  90  
\\

\hline

   &  CMSSM, high m$_0$,  tan$\beta>$1, maximal mixing in $\tilde{t}$ sector   
& 49    \\
\cline{2-2}

\Xn&  CMSSM, any m$_0$,  tan$\beta<$40  no mixing or \DM(\stau-\Xn)$>$6 
    & 46    \\
\cline{2-2}

   & CMSSM, any m$_0$,  tan$\beta<$40,  mixing A$_\tau$=A$_b$=A$_t=0$  & 46 \\
\cline{2-2}

   & CMSSM, any m$_0$,  1$<$tan$\beta<$40,  mix. A$_\tau$=A$_b$=0,
A$_t=\sqrt{6}$ TeV/c$^2$ & 49

\\
\hline
\end{tabular}
\caption{\small Summary of mass limits for supersymmetric particles and their validity 
conditions.
In each line of the table \DM\ is the mass difference between the corresponding sparticle 
and the LSP. 
All masses and  \DM\ values are in \GeVcc. 
CMSSM  refers to a model with gauge and sfermion mass
unification, where $\mu$ however is a free parameter 
(see section 
\ref{sec:framework}). 
Neutralino mass limits should be lowered by 1~\GeVcc\ if
the radiative corrections of \protect{\cite{radcor}} are taken into account.}
\label{tab:limsum}
\end{center}
\end{table}
\renewcommand\arraystretch{1.0}

\subsection*{Acknowledgements}
\vskip 3 mm
 We are greatly indebted to our technical 
collaborators, to the members of the CERN-SL Division for the excellent 
performance of the LEP collider, and to the funding agencies for their

support in building and operating the DELPHI detector.\\
We acknowledge in particular the support of \\
Austrian Federal Ministry of Education, Science and Culture,
GZ 616.364/2-III/2a/98, \\
FNRS--FWO, Flanders Institute to encourage scientific and technological 
research in the industry (IWT), Federal Office for Scientific, Technical
and Cultural affairs (OSTC), Belgium,  \\
FINEP, CNPq, CAPES, FUJB and FAPERJ, Brazil, \\
Czech Ministry of Industry and Trade, GA CR 202/99/1362,\\
Commission of the European Communities (DG XII), \\
Direction des Sciences de la Mati$\grave{\mbox{\rm e}}$re, CEA, France, \\
Bundesministerium f$\ddot{\mbox{\rm u}}$r Bildung, Wissenschaft, Forschung 
und Technologie, Germany,\\
General Secretariat for Research and Technology, Greece, \\
National Science Foundation (NWO) and Foundation for Research on Matter (FOM),
The Netherlands, \\
Norwegian Research Council,  \\
State Committee for Scientific Research, Poland, SPUB-M/CERN/PO3/DZ296/2000,
SPUB-M/CERN/PO3/DZ297/2000 and 2P03B 104 19 and 2P03B 69 23(2002-2004)\\
JNICT--Junta Nacional de Investiga\c{c}\~{a}o Cient\'{\i}fica 
e Tecnol$\acute{\mbox{\rm o}}$gica, Portugal, \\
Vedecka grantova agentura MS SR, Slovakia, Nr. 95/5195/134, \\
Ministry of Science and Technology of the Republic of Slovenia, \\
CICYT, Spain, AEN99-0950 and AEN99-0761,  \\
The Swedish Natural Science Research Council,      \\
Particle Physics and Astronomy Research Council, UK, \\
Department of Energy, USA, DE-FG02-01ER41155, \\
EEC RTN contract HPRN-CT-00292-2002. \\


\clearpage

\clearpage

\clearpage

\clearpage

\newpage

\newpage

\newpage

\newpage

\newpage

\newpage

\newpage

\newpage

\newpage

\clearpage

\clearpage


\clearpage


\clearpage


\clearpage


\newpage

\newpage


\newpage

\newpage

\newpage

\newpage

\newpage
\newpage

\newpage

\newpage


\newpage

\newpage

\newpage

\newpage


\newpage

\newpage

\newpage

\newpage




\newpage

\newpage



\end{document}